%% file: main.tex
\documentclass[letterpaper,11pt]{article} %letter causes unused global option

\input{our-preamble.tex}
\input{ccl-defs.tex}
\usepackage{comment}
\usepackage{titling}
\usepackage{enumitem} % Provides the ability to easily modify the amount of white space in a list
\setlist{noitemsep,topsep=0pt,parsep=0pt} % Modify the amount of white space for all lists
\usepackage{subcaption}
\usepackage{tikz}
\numberwithin{equation}{section}
\usepackage[bookmarks=true,hypertexnames=false]{hyperref}
\hypersetup{colorlinks=true, citecolor=blue, linkcolor=red, urlcolor=blue}
\makeatletter

\newcommand{\Rmnum}[1]{\expandafter\@slowromancap\romannumeral #1@}
\makeatother

\newenvironment{remark}{\par\medskip{\bfseries \noindent Remark:}}{\par\medskip}{\par\medskip}
\newcommand{\EVAL}{\operatorname{EVAL}}
\newcommand{\EVALP}{\operatorname{EVALP}}
\newcommand{\simp}{\operatorname{simp}}
\def\MM{\mathbf{M}}
\def\ZZ{\mathbf{Z}}  
\def\fK{\mathfrak{K}}
\def\fL{\mathfrak{L}}

\def\fY{\mathfrak{Y}}
\def\fE{\mathfrak{E}}
\def\PiPi{\mathbf{\Pi}}
\renewcommand{\SS}{\mathbf{S}}
\newcommand{\ord}{\operatorname{ord}}
\renewcommand{\twt}{\operatorname{wt}}
\newcommand{\pur}{\underline}
\newcommand{\rank}{\operatorname{rank}}
\newcommand{\gm}{M}
\newcommand{\subgm}{M}
\def\multblrkone/{mult-brk-$1$}
\def\modblrkone/{mod-brk-$1$}
\def\blrkone/{brk$1$}
\newcommand{\firstdv}{u^*} %\egvone
\newcommand{\seconddv}{v^*} %\egvtwo
\newtheorem{argument}[theorem]{Argument}
\newtheorem{property}[theorem]{Property}
\newcommand{\llbracket}{[\![}
\newcommand{\rrbracket}{]\!]}
\begin{document}

%\title{{\bf A dichotomy for bounded degree graph homomorphisms with nonnegative weights}}
\title{\textbf{Dichotomy for Graph Homomorphisms with Complex Values on Bounded Degree Graphs}}

\vspace{0.3in}

\author{Jin-Yi Cai\thanks{Department of Computer Sciences, University of Wisconsin-Madison. Supported by NSF CCF-1714275.} \\ {\tt jyc@cs.wisc.edu} 
\and
Artem Govorov\thanks{Department of Computer Sciences, University of Wisconsin-Madison. Supported by NSF CCF-1714275. %\protect\\
 } \thanks{Artem Govorov is the author's preferred spelling
of his name,
rather than the official spelling Artsiom Hovarau.
 } \\ {\tt hovarau@cs.wisc.edu}
}
\date{}
\maketitle

\bibliographystyle{plainurl}

\begin{abstract}
The complexity of 
graph homomorphisms has been a subject of intense study~\cite{Dyer-Greenhill-2000, Dyer-Greenhill-corrig-2004,Bulatov-Grohe-2005,Thurley-2009, Grohe-Thurley-2011,Goldberg-et-al-2010,Cai-Chen-Lu-2013,Govorov-Cai-Dyer}.
The partition function $Z_\AA (\cdot)$ of graph homomorphism
is defined by a  symmetric matrix $\AA$ over $\mathbb{C}$.
We prove that the complexity dichotomy of~\cite{Cai-Chen-Lu-2013}
extends to bounded degree graphs. More precisely, we prove that either
 $G \mapsto Z_\AA (G)$ is computable in polynomial-time
for every $G$, or for some $\Delta >0$ it is  \#P-hard
over (simple)  graphs $G$ with maximum degree $\Delta(G) \le \Delta$. 
The  tractability criterion on $\AA$
for this dichotomy is explicit, and can be decided in polynomial-time in
the size of $\AA$.
We also show that the dichotomy is effective in that
 either a  P-time
algorithm for, or a reduction from \#SAT to, 
 $Z_\AA (\cdot)$  can be constructed from $\AA$, in the respective cases.
\end{abstract}
%\begin{abstract}
%We consider the complexity of counting weighted graph homomorphisms
%defined by a symmetric matrix $A$.
%% any symmetric matrix $A$.
%Each symmetric matrix $A$
%defines a graph homomorphism function $Z_A(\cdot)$,
%also known as the partition function.
%Dyer and Greenhill~\cite{Dyer-Greenhill-2000}
%established a complexity dichotomy of $Z_A(\cdot)$ for
%symmetric $\{0, 1 \}$-matrices $A$, and they further proved that its
%\#P-hardness part also holds for bounded degree graphs.
%Bulatov and Grohe~\cite{Bulatov-Grohe-2005} extended 
%the  Dyer-Greenhill dichotomy
%to nonnegative symmetric matrices $A$.
%However, their hardness proof requires graphs of arbitrarily large degree,
%and whether the
%bounded degree part of the  Dyer-Greenhill dichotomy can be 
%extended has been an open problem for 15 years.
%We resolve this open problem 
%and prove that for nonnegative symmetric $A$,
%either $Z_A(G)$ is in polynomial time for all graphs $G$,
%or it is \#P-hard for bounded degree (and simple) graphs $G$.
%We further extend the complexity dichotomy to include
%nonnegative vertex weights.
%Additionally, we prove that the \#P-hardness part
%of the dichotomy by Goldberg et al.~\cite{Goldberg-et-al-2010} for $Z_A(\cdot)$
%also holds for simple graphs,
%where $A$ is any real symmetric matrix.
%\end{abstract}

%\newpage
\thispagestyle{empty}
\clearpage
\setcounter{page}{1}

%%%-------
%\input{introduction}
%\input{background}
%\input{tractability}
%\input{technical-statements}
%\input{main-proof}
%\input{interpolation-via-stretchings}
%\input{bibliography}

\newpage
\input{introduction}

\input{organization}
\input{preliminaries}

\input{algebraic-preliminaries}

%\input{roots-of-unity}
%\input{exponential-polynomials}
%\input{technical-statements-complex}
%\input{vandermonde-argument}
%\input{some-preliminaries-ccl}
%\input{more-preliminaries}
%\input{purification-explained}
\input{very-high-level-description}
\input{non-mult-bl-rk-one-hardness}

\input{gadget-meta-thickening}

%\input{trivial-transformations}
%\input{pinning-avoidance}
\input{reduction-to-conneted-components}
\input{meta-argument-bounded-degree-graphs-outline}
\input{meta-argument-bounded-degree-graphs-outline-unipartite}
\input{regularization-step}
\input{gadget-cyclotomic-transformation-inside-meta}
\input{meta-argument-bounded-degree-graphs}
\input{tractability}

\input{meta-argument-bounded-degree-graphs-unipairite}
\input{tractability-unipartite}

\input{simple-gadget-meta-replacement}

\input{decidability}
\input{constructivity}
\bibliography{references}
\input{appendix1}
\input{appendix2}

%%%-------

%%%-------
%%\input{introduction}
%%\input{background}
%%\input{tractability}
%%\input{technical-statements}
%%\input{main-proof}
%%\input{interpolation-via-stretchings}
%%\input{bibliography}
%
%\newpage
%\input{introduction}
%\input{organization}
%\input{preliminaries}
%\input{algebraic-preliminaries}
%%\input{roots-of-unity}
%%\input{exponential-polynomials}
%%\input{technical-statements-complex}
%%\input{vandermonde-argument}
%%\input{some-preliminaries-ccl}
%%\input{more-preliminaries}
%%\input{purification-explained}
%\input{Very-High-Level-Description}
%%%%---
%%\input{non-mult-bl-rk-one-hardness}
%%\input{gadget-meta-thickening}
%%%\input{trivial-transformations}
%%%\input{pinning-avoidance}
%%\input{reduction-to-conneted-components}
%%\input{meta-argument-bounded-degree-graphs-outline}
%%\input{meta-argument-bounded-degree-graphs-outline-unipartite}
%%\input{regularization-step}
%%\input{gadget-cyclotomic-transformation-inside-meta}
%%\input{meta-argument-bounded-degree-graphs}
%%\input{tractability}
%%\input{meta-argument-bounded-degree-graphs-unipairite}
%%\input{tractability-unipartite}
%%\input{simple-gadget-meta-replacement}
%%\input{decidability}
%%\input{constructivity}
%%%%---
%\bibliography{references}
%%%%---
%%\input{appendix1}
%%\input{appendix2}
%%%%---
%
%\newpage
%\input{non-mult-bl-rk-one-hardness-short}
%%%-------

\end{document}

%% file: our-preamble.tex
\usepackage{amsmath, amsfonts, amssymb, amsthm, amscd, graphicx, 
enumerate, footmisc, mathtools, xcolor}
\providecommand{\U}[1]{\protect\rule{.1in}{.1in}}
\def\theenumi{\arabic{enumi}}

\def\theenumii{\alph{enumii}}
\def\p@enumii{\theenumi.}

\def\theenumiii{\arabic{enumiii}}
\def\p@enumiii{(\theenumi)(\theenumii)}

\def\p@enumiv{\p@enumiii.\theenumiii}
\newenvironment{lcases}
  {\left\lbrace\begin{aligned}}
  {\end{aligned}\right.}
\newcommand{\bbF}{\mathbb F}

\newcommand{\bbZ}{\mathbb Z}
\newcommand{\bbQ}{\mathbb Q}
\newcommand{\bbR}{\mathbb R}
\newcommand{\bbC}{\mathbb C}

\newcommand{\diag}{\operatorname{diag}}
\newcommand{\Aut}{\operatorname{Aut}}

\newcommand{\calPLG}{\mathcal{PLG}}

\newcommand{\calK}{\mathcal K}
\newcommand{\calP}{\mathcal P}

\newcommand{\calT}{\mathcal T}

\usepackage{fullpage}
\usepackage{mathrsfs}
\newtheorem{theorem}{Theorem}[section]
\newtheorem{claim}[theorem]{Claim}

\newtheorem{corollary}[theorem]{Corollary}

\newtheorem{definition}[theorem]{Definition}

\newtheorem{lemma}[theorem]{Lemma}

\numberwithin{equation}{section}

%% file: ccl-defs.tex
\def\II{\mathbf{I}}  
\newcommand{\nc}{\newcommand} \def\twt{\text{wt}} \def\fD{{\frak D}}
\nc{\Pf}{{\rm Pf}} \nc{\PfS}{{\rm PfS}} \def\uu{\mathbf{u}}
\nc{\GP}{Grassmann-Pl{\"u}cker } \nc{\adj}{{\rm adj}}
\def\AA{\mathbf{A}} \def\eval{\text{\sf EVAL}} \def\CC{\mathbf{C}}
\def\DD{\mathbf{D}}  \def\FF{\mathbf{F}}
\def\calP{\mathcal{P}} \def\calQ{\mathcal{Q}} \def\00{\mathbf{0}}
\def\HH{\mathbf{H}} \def\qq{\mathbf{q}} \def\calU{\mathcal{U}}
\def\calR{\mathcal{R}} \def\xx{\mathbf{x}} \def\lcm{\text{lcm}}
\def\calL{\mathcal{L}} \def\calD{\mathcal{D}} \def\yy{\mathbf{y}}
\def\kk{\mathbf{k}} \def\pp{\mathbf{p}} 
\def\PP{\mathbf{P}} \def\QQ{\mathbf{Q}} 
\def\GC{$\mathcal{GC}$} \def\11{\mathbf{1}} \def\oo{\omega}
\def\EE{\mathbf{E}} \def\RR{\mathbf{R}} \def\vv{\mathbf{v}}
  \def\GG{\mathbf{G}}
\def\dd{\mathbf{d}} \def\aa{\mathbf{a}} \def\bb{\mathbf{b}}
\def\calF{\mathcal{F}}  \def\bft{\mathbf{t}}
\def\zqt{\mathbb{Z}_{\fqq}}  \def\SS{\mathbf{S}}
  
\def\lcm{\text{{lcm}}}  
\def\ext{\textbf{ext}} \def\extt{\emph{\textbf{ext}}}
 \def\BB{\mathbf{B}} 
 \def\fqq{{\mathcal{Q}}} \def\lin#1{#1^{\text{lin}}}
\def\fq{\mathcal{Q}}  
 \def\fa{\boldsymbol{\frak a}}
\def\fb{\boldsymbol{\frak b}} 
\def\bcF{\boldsymbol{\mathcal{F}}} \def\cS{\mathcal{S}} \def\cT{\mathcal{T}}
\def\hq{\widehat{\pi}}  \def\COUNT{\text{\sf COUNT}}
 \def\dd{\mathbf{d}} 
 \def\mm{\mathbf{m}} \def\nn{\mathbf{n}}
\def\calT{\mathcal{T}} \def\calS{\mathcal{S}} \def\KK{\mathbf{K}}
\def\LL{\mathbf{L}} \def\calK{\mathcal{K}} \def\calL{\mathcal{L}}
\def\XX{\mathbf{X}} \def\YY{\mathbf{Y}} \def\JJ{\mathbf{J}}
\def\evalp{\text{\sf EVALP}}

\def\calW{\mathcal{W}} \def\WW{\mathbf{W}} 
\def\calZ{\mathcal{Z}}

%% file: introduction.tex
\section{Introduction}\label{sec:intro}

Given two graphs $G$ and $H$,
a graph homomorphism (GH) from $G$ to $H$ is a map $f$ from
the vertex set $V(G)$ to $V(H)$
such that, whenever $(u,v)$ is an edge in $G$,
$(f(u), f(v))$ is an edge in $H$~\cite{lovasz67,HellBook}.
In 1967, Lov\'{a}sz~\cite{lovasz67} proved that $H$ and $H'$ are isomorphic
iff for all $G$, the number of homomorphisms from $G$  to $H$ and
from  $G$ to $H'$ are the same.
More generally, one considers weighted graphs $H$ where
every edge of $H$ is given a weight. This is represented by
a symmetric matrix $\AA$ and the set of all
homomorphisms  from $G$  to $H$ can be aggregated
in a single sum-of-product expression called the
 partition function $Z_\AA(G)$~\cite{freedman-l-s}.  
The number of homomorphisms from $G$  to $H$ is the special
case where all edges of  $H$ have weight 1, and
$\AA$ is the 0-1 adjacency matrix of $H$.  
This  partition function $Z_\AA(G)$ provides
an elegant framework to express a wide-variety of \emph{graph
  properties}. These partition functions are also widely
studied in statistical physics representing spin systems~\cite{Baxter-book,Ising-1925,Potts-1952,Potts-Ward-1955,Jerrum-Sinclair-1993,Goldberg-Jerrum-2012,Goldberg-Guo-2017}.

We use the standard definition for graph homomorphism and its 
partition function $Z_\AA(G)$.
Our graphs $G$ and $H$ are undirected (unless otherwise specified).
$G$ is allowed to have multiple edges but no loops;
it is \emph{simple} if it has neither.
$H$ can have loops,
  multiple edges, and more generally, edge weights.
We allow edge weights to be arbitrary  complex numbers.~\footnote{To be
computable in the strict Turing model, they are algebraic complex numbers.}
Let $\AA = (A_{i,j})$ be an $m\times m$ symmetric matrix with entries $A_{i,j}
\in \mathbb{C}$,
we define \begin{equation}\label{Z-def}
Z_\AA(G)=\sum_{\xi:V\rightarrow [m]}\hspace{0.06cm} 
\prod_{(u,v)\in E} A_{\xi(u),\xi(v)}
\end{equation}
for every  undirected graph $G=(V,E)$.

%%%%%%%% maybe add a paragraph about what it can and cannot express.

The complexity of the partition function $Z_\AA(\cdot)$
has been shown to obey a dichotomy: Depending on $\AA$,
the computation
 $G \mapsto Z_\AA(G)$ is either in polynomial time or \#P-hard.
This has been proved for progressively more general
matrices $\AA$: In \cite{Dyer-Greenhill-2000,Dyer-Greenhill-corrig-2004}, Dyer and Greenhill 
first proved this complexity dichotomy for symmetric $\{0,1\}$-matrices $\AA$.
In this case, $Z_\AA(G)$ counts the number of graph homomorphisms without
weight.  Bulatov and Grohe~\cite{Bulatov-Grohe-2005,Thurley-2009, Grohe-Thurley-2011}
proved this  for $Z_\AA(\cdot)$
%a  complexity dichotomy for $Z_\AA(\cdot)$
where $\AA$ is any nonnegative symmetric  matrix.  
This was extended by Goldberg, Grohe, Jerrum and Thurley~\cite{Goldberg-et-al-2010}
to all real symmetric matrices.
Finally, Cai, Chen and Lu~\cite{Cai-Chen-Lu-2013} generalized this to
all complex  symmetric matrices.
Every subsequent complexity dichotomy subsumes the
previous one as a special case.
In each case, an explicit tractability criterion  on $\AA$ is given
such that if $\AA$ satisfies the criterion then  $Z_\AA(\cdot)$
is computable in P-time, otherwise it is \#P-hard.

In \cite{Dyer-Greenhill-2000} Dyer and Greenhill
established a stronger fact: 
if a $\{0, 1\}$-matrix $\AA$ fails the tractability condition then
 $Z_\AA(\cdot)$  is \#P-complete even when restricted to
bounded degree graphs.
We note that the complexity of GH for bounded degree graphs
is particularly interesting as much work has been done on the
approximate complexity of GH focusing on bounded degree graphs
 and approximate algorithms
are achieved for them~\cite{DyerFJ02,Weitz06,Sly10,SinclairST12,LiLY13,Barvinok-book,Barvinok-Soberon-2017,Peters-Regts-2018,HelmuthPR19}.
The corresponding complexity question for
 bounded degree graphs  remained open beyond the 0-1 case,
% until very recently
%for the other more general dichotomies,
until recently Govorov, Cai and Dyer~\cite{Govorov-Cai-Dyer}
were able to prove that the \#P-hardness part
of the  Bulatov-Grohe dichotomy for nonnegative weights
 holds  for  bounded degree graphs.
% to the  bounded degree case.
In this paper we finally extend the full complexity 
dichotomy for complex weights~\cite{Cai-Chen-Lu-2013} to the bounded degree case.

%{\color{red} (just formulated)}

Our first main theorem is the following.
\begin{theorem}\label{thm:main-dich-bd}
Let $\AA$ be a symmetric and algebraic complex matrix.
Then either $G \mapsto Z_\AA(G)$ can be computed in polynomial time on arbitrary graphs
$G$,
or for some $\Delta > 0$ depending on $\AA$,
it is \#P-hard on graphs $G$ of maximum degree  at most $\Delta$.
%of maximum degree $\Delta(G) \le \Delta$. 
\end{theorem}

The dichotomy criterion on $\AA$ is the same as in~\cite{Cai-Chen-Lu-2013}.
This complexity dichotomy has an explicit form, and given $\AA$,
it is decidable in polynomial time (in the size of $\AA$) whether
$\AA$ satisfies the criterion, and thus one can decide  which case the
partition function $Z_\AA(\cdot)$ belongs to.
However, there is a more demanding sense in which
the dichotomy of \cite{Cai-Chen-Lu-2013} is not constructive.
When $\AA$ satisfies the criterion, then an explicit
polynomial-time algorithm for  $G \mapsto Z_\AA(G)$ is given;
but when  $\AA$ does not satisfy  the criterion, it is only proved that
a polynomial time reduction from \#SAT to $Z_\AA(\cdot)$  \emph{exists} and not
given constructively.  In this paper we remedy this situation
and prove that the dichotomy in Theorem~\ref{thm:main-dich-bd}
can be made fully constructive.

By the standard definition of graph homomorphism,
the input graph $G$ is allowed to have multiple edges (but no loops).
Our polynomial-time algorithm for  $G \mapsto Z_\AA(G)$
in the tractable case of  Theorem~\ref{thm:main-dich-bd} works 
for graphs with multiple edges and loops. 
More importantly, we will prove that in the \#P-hard case,
we may restrict to \emph{simple} graphs $G$ 
(i.e., without  multiple edges and without loops) in addition to being 
of bounded degree.

We state this stronger form of Theorem~\ref{thm:main-dich-bd} next.
\begin{theorem}\label{thm:main-dich-bd-simp-effective}
The complexity dichotomy criterion in Theorem~\ref{thm:main-dich-bd}  is 
 polynomial-time decidable in the size of $\AA$.
If $\AA$ satisfies the criterion, then
 $G \mapsto Z_\AA(G)$ 
 is computable in polynomial-time by an explicit algorithm
for any $G$ (allowing multiple edges and  loops).
If $\AA$ does not satisfy  the criterion, then
$Z_\AA(G)$ is \#P-hard for bounded degree simple graphs $G$,
and a polynomial-time reduction from \#SAT to $Z_\AA(\cdot)$  
can be constructed from $\AA$.
\end{theorem}

Note that 
the P-time decidability of the dichotomy criterion is measured in the size of $\AA$.
%We should point out that,
When we say in the  \#P-hard case
a  polynomial-time reduction from \#SAT to $Z_\AA(\cdot)$  can be constructed,
this notion of polynomial-time is measured in terms of the size of
instances to \#SAT (and consequently the size of the graphs $G$  
produced  by the reduction as instances
to $Z_\AA(\cdot)$),
and $\AA$ is considered fixed---it defines the \#P-hard problem  $Z_\AA(\cdot)$.
In  the  \#P-hard case, Theorem~\ref{thm:main-dich-bd-simp-effective}
says that we give a constructive procedure which from $\AA$
produces a  P-time  reduction from \#SAT to $Z_\AA(\cdot)$.
However, we should point out that,
 in the  \#P-hard case, the construction procedure for the P-time \#P-hardness
reduction is  not in  polynomial time as measured in the size of $\AA$.

The proof in~\cite{Cai-Chen-Lu-2013} does not work for bounded degree graphs $G$.
The main structure of the proof in~\cite{Cai-Chen-Lu-2013} is a long sequence of
successively stringent conditions which a matrix $\AA$ must satisfy, or else
it is proved that  $Z_\AA(\cdot)$ is \#P-hard. This process
continues until the conditions on $\AA$ 
% that the  matrix $\AA$ must satisfy
imply that  $Z_\AA(\cdot)$ is computable in polynomial time. In each stage,
assuming  $\AA$ satisfies the condition of that stage,
 the matrix $\AA$ (or another matrix
$\AA'$ which has a better form, but  $Z_{\AA'}(\cdot)$ is
equivalent to  $Z_\AA(\cdot)$ in complexity)
% of $Z_{\AA'}(\cdot)$ 
%and $Z_\AA(\cdot)$ are equivalent) 
is passed on to the next stage.
%%%% -- What is A'? What do you mean by that.
%%% i don't want to get into technical issues of (C,D) etc.
The condition often gives some structural information
that allows for  a better representation of $\AA$,
which is not available otherwise.

However, close to the beginning of the 100-page proof 
 in~\cite{Cai-Chen-Lu-2013} (before Step 1.1, p.~949) there is a 
Lemma 7.1 which proves an equivalence of  $Z_\AA(\cdot)$ to another
problem called  $\COUNT(\AA)$. 
(This equivalence allows us to substitute $\AA$
with a  ``purified'' matrix $\BB$ which defines
an  ``equivalent''  problem  $Z_\BB(\cdot)$, but 
 $\BB$ has desirable structural properties without which
the proof in~\cite{Cai-Chen-Lu-2013} cannot continue.
 This is before all the substantive proof in~\cite{Cai-Chen-Lu-2013}
gets started.) Unfortunately, the proof of this Lemma 7.1 uses
graphs of unbounded degree, and we cannot find a way to modify
the proof to make it work for bounded degree graphs.
We also remark that the method in the recent proof~\cite{Govorov-Cai-Dyer}
 that extended the Bulatov-Grohe dichotomy for nonnegative weights
to bounded degree graphs is  also not sufficient here.
%  nilponent?
% thinkening kills "potency" ? (like 1 1\\ 1 -1 after square is trivial.) 
% 
However, a crucial construction, which is a refinement of a construction 
from~\cite{Govorov-Cai-Dyer},
is an important step in this paper.

In addition to this  crucial construction,
% from~\cite{Govorov-Cai-Dyer},
the main idea in this paper is \emph{algebraic} instead.
We will introduce a new  notion called \emph{multiplicative-block-rank-1},
and a related notion called \emph{modular-block-rank-1}.
These are weaker notions than the \emph{block-rank-1} condition
that was widely used in all previous dichotomies. We establish
a fundamental implication that if $\AA$ is not multiplicative-block-rank-1,
then  $Z_\AA(\cdot)$ is \#P-hard for bounded degree graphs.
%%% To Artem -- you say: of course, every non-modular-block-rank-1 is 
%%% non-multiplicative-block-rank-1, but CCL constructs the first one, 
%%% but it is OK to mention the second one because it's intro?
%%%
%%% I don't want to get into the detail of how these two notions are
%%% related, and CCL actually mostly establishes non-modular-block-rank-1
%%% which implies non-multiplicative-block-rank-1

We make the following important technical observation for
the long proof in~\cite{Cai-Chen-Lu-2013}. In every stage where
the matrix $\AA$ is proved to satisfy some additional conditions (or
else we get \#P-hardness), the proof actually constructively
establishes the following: by some explicit construction 
either we get some  specified conditions satisfied or we get a graph fragment which
defines a matrix that is \emph{non-multiplicative-block-rank-1}.
In this paper we show that in each case, this
property of non-multiplicative-block-rank-1 can be \emph{transferred}
from any subsequent stage to the previous stage.

Thus here is a very rough high-level outline of 
our proof: 

For the purification step we cannot simply substitute
$\AA$ by its purified form and move to the next stage.
Instead we will keep both $\AA$  and its purified form  $\pur{\AA}$,
and pass both to subsequent stages. It is only with respect to
%%% you say: -- we can use (the: reads bad) combinatrial gadget 
%constructions (from \cite{CCL}) -- or should we say it to be more specific?
%%% i think as is is fine.
the  purified form  $\pur{\AA}$
we can use combinatorial gadget constructions 
to conclude that the matrix has desirable properties. However, with a purely
algebraic argument we nevertheless ``transfer'' these
conclusions to the unpurified $\AA$. These  algebraic arguments
are in terms of properties of polynomials,
exponential polynomials, and properties of  finitely generated
subfields of $\mathbb{C}$. Ultimately most of the algebraic arguments
rely on a simple algebraic fact which we call the \emph{Vandermonde Argument}
(see Lemma~\ref{lem:simpleVandermonde}).
Then we go through the proof in~\cite{Cai-Chen-Lu-2013}
step-by-step. In each step, we show how to ``transfer'' 
the property of non-multiplicative-block-rank-1 of
a later  stage to the previous stage.  This task is
accomplished by three meta-arguments,
Arguments~\ref{meta1}, \ref{meta2}, and~\ref{meta3}, i.e., ({\sl Meta}$_1$), 
({\sl Meta}$_2$) and ({\sl Meta}$_3$). These formulate our transfer procedure.

The bulk of this paper will be dedicated to the proof of
Theorem~\ref{thm:main-dich-bd}.
After that we extend the \#P-hardness part to
simple graphs, and discuss the decidability and constructivity issues
in Theorem~\ref{thm:main-dich-bd-simp-effective}.

For the dichotomy of Goldberg et al.~\cite{Goldberg-et-al-2010}
for real symmetric matrices  $\AA$, it is proved in~\cite{Govorov-Cai-Dyer} 
that its \#P-hardness part can be made to hold for simple graphs.
This uses interpolation with stretchings.
For the dichotomy in~\cite{Cai-Chen-Lu-2013}
and  our Theorem~\ref{thm:main-dich-bd}
over complex symmetric matrices  $\AA$, 
this trick does not work.
While
real symmetric matrices can always be diagonalized,
for complex symmetric matrices this is not true, and
more importantly, the Jordan normal form
% of the complex symmetric matrix $\AA$
may contain nontrivial nilpotent blocks, i.e.,
blocks  of size greater than one and
% $k \times k$ where $k > 1$
corresponding to 
%the 
eigenvalue $0$.
 
%%%%%%%%%%%%%%%%%%%%%%%%%%%%%%%%%%%%%%%%%%%%%%%%%%%%%%%%%%%%%%%%%%%%
\begin{comment}

When trying the argument from~\cite{Govorov-Cai-Dyer} (an interpolation with stretchings)
we have to deal with the so-called exponential polynomials (see Section~\ref{sec:exp-poly})
but some bases $\lambda_i$ that correspond to the eigenvalues of $\AA$ can be zero
and $\deg p_i(x) \ge 1$ where $p_i(x)$ is the corresponding polynomial:
%this precisely corresponds to the case
%of a nonzero nilpotent block in the Jordan normal form of $\AA$.
this happens in the case that a nontrivial nilpotent block exists
in the Jordan normal form of $\AA$.
\end{comment}
%%%%%%%%%%%%%%%%%%%%%%%%%%%%%%%%%%%%%%%%%%%%%%%%%%%%%%%%%%%%%%%%%%%%

In this paper, we overcome this difficulty by not proving
a reduction from the case of bounded degree graphs
 to the case of bounded degree \emph{and} simple graphs.
Instead we use a transfer argument of the property of 
non-multiplicative-block-rank-$1$ constructions.
This is stated in Theorem~\ref{thm:gadget-simple-non-mult-bl-rk-1}.
In order to prove it, we will heavily make use of the results 
from~\cite{Cai-Govorov-GH-hom-type}.

%%%%%%%%%%%%%%%%%%%%%%%%%%%%%%%%%%%%%%%%%%%%%%%%%%%%%%%%%%%%%%%%%%%%
\begin{comment}

In fact our \#P-hardness for simple graphs in 
Theorem~\ref{thm:main-dich-bd-simp-edge-gadget}
results from applying this argument
on top of the \#P-hardness in Theorem~\ref{thm:main-dich-bd-edge-gadget}.

%so our proof is not reduction based (is gadget based).

Instead of trying to prove $\EVAL^{\Delta}(\AA) \le \EVAL_{\simp}^{\Delta'}(\AA)$,
where $\Delta > 0$ is given and $\Delta' > 0$ depends on $\Delta$,
we show that for any edge gadget $\Gamma$
such that $\gm_{\Gamma, \AA}$ is not multiplicative-block-rank-$1$
we can construct a simple edge gadget $\Gamma$ (i.e., no multiple edges and
edges between the distinguished vertices)
suh that $\gm_{\Gamma', \AA}$ is not multiplicative-block-rank-$1$.
This is stated Theorem~\ref{thm:gadget-simple-non-mult-bl-rk-1}.
In order to prove it, we will heavily make use of the results from~\cite{Cai-Govorov-GH-hom-type}.
In fact our \#P-hardness for simple graphs in Theorem~\ref{thm:main-dich-bd-simp-edge-gadget}
results from applying this argument
on top of the \#P-hardness in Theorem~\ref{thm:main-dich-bd-edge-gadget},
so our proof is not reduction based (is gadget based).

{\color{red} (added as you told me, raw material, we'll need to move it somewhere)}

\end{comment}
%%%%%%%%%%%%%%%%%%%%%%%%%%%%%%%%%%%%%%%%%%%%%%%%%%%%%%%%%%%%%%%%%%%%

The proof in~\cite{Cai-Chen-Lu-2013} starts
by reducing the matrix $\AA$ to its connected components,
and so we may assume that $\AA$ is connected.
This is achieved by means of the so-called
 first pinning lemma (Lemma 4.1 from~\cite{Cai-Chen-Lu-2013}).
(After that, 
$\AA$ undergoes the purification step defined in 
Section 7.1 from~\cite{Cai-Chen-Lu-2013}.)
However, even though the proof of this first pinning lemma 
does preserve degree boundedness,
the proof presented in~\cite{Cai-Chen-Lu-2013}
is not constructive and does not preserve simple graphs.
While both issues  can be tackled by using results from~\cite{Cai-Govorov-GH-hom-type},
this would introduce unnecessary complications.
%
%(We note that both problems can be tackled by using results from~\cite{Cai-Govorov-GH-hom-type},
%but this would introduce unnecessary complications at this particular point.
%
Instead we choose  a different route.
We will still use results from~\cite{Cai-Govorov-GH-hom-type} to extend
our dichotomy to simple graphs.
However in our paper, we replace the main theme of the proof in~\cite{Cai-Chen-Lu-2013}
from one that is  reduction based to one that relies on the three
meta-arguments
that transfer gadget constructions from one stage to another.
Thus we will not use any of the three pinning lemmas in~\cite{Cai-Chen-Lu-2013}.
In particular we
 use the transfer method to  move to any connected component of $\AA$
without actually performing a formal  complexity reduction.
Additionally, in all subsequent stages, we apply one of the 
meta-arguments, ({\sl Meta}$_1$), ({\sl Meta}$_2$) and ({\sl Meta}$_3$)
to obtain what is in effect a  complexity reduction without
the formal  complexity reduction.

%% file: organization.tex
\subsection*{Organization}

This paper is organized as follows.
After the introductory Section~\ref{sec:intro}, 
in Section~\ref{sec:preliminaries-main} we give some definitions of
$\EVAL$ problems, and also the basic \#P-hardness result Theorem~\ref{BulatovGrohe}
by Bulatov and Grohe. In Section~\ref{sec:algebraic-preliminaries}
%we introduce two new notions \multblrkone/ and \modblrkone/ which will
we introduce two new notions, multiplicative-block-rank-$1$ and modular-block-rank-$1$,
which play an important role in the entire proof. We also give some preliminary
results of an algebraic nature, in particular, the Vandermonde Argument.
In Section~\ref{highlevel-ours},
we give an overview of the proof structure. The proof is separated
into the bipartite and nonbipartite case, and in Section~\ref{highlevel-ours},
we concentrate on the bipartite case.

The real technical part of the proof starts with Section~\ref{sec:GCD-gadget},
where we prove Theorem~\ref{thm:non-mult-bl-rk-1-hardness-sym}
%that non-\multblrkone/ implies \#P-hardness for $\EVAL(\AA)$
that non-multiplicative-block-rank-$1$ implies \#P-hardness for $\EVAL(\AA)$
for bounded degree graphs. 
This uses a new gadget and the Vandermonde Argument.
In Section~\ref{sec:multblrkone-from-Abar-toA} 
we prove Theorem~\ref{thm:gadget-thickening-non-mult-bl-rk-1}, which is
again algebraic in nature.
In Section~\ref{sec:pinning-free} we describe the reduction
to connected components without using pinning.

Sections~\ref{outlinebipartite-ours}
and~\ref{outlinenonbip-ours} give a detailed proof outline
for the bipartite and the nonbipartite cases respectively.
This proof plan is carried out in the subsequent
Section~\ref{sec:Regularization} to Section~\ref{sec:Tractability-uni}.
We note in particular that from this point our proof is essentially a
meta-proof, i.e., we follow the proof in~\cite{Cai-Chen-Lu-2013}
closely, but use three meta-arguments, Arguments~\ref{meta1}, \ref{meta2}, and~\ref{meta3},
i.e., ({\sl Meta}$_1$), ({\sl Meta}$_2$) and ({\sl Meta}$_3$),
to carry out the details.  For the ease of readers we actually give
the full proof (without the meta-arguments) for Lemma~\ref{hahajaja-ours},
before introducing Argument~\ref{meta1} ({\sl Meta}$_1$).
This gives a concrete demonstration how such a proof is, without ({\sl Meta}$_1$),
and how ({\sl Meta}$_1$) transforms such a proof.
For all subsequent proofs from 
Section~\ref{sec:Regularization} to Section~\ref{sec:Tractability-uni}
we always apply ({\sl Meta}$_1$), ({\sl Meta}$_2$) and ({\sl Meta}$_3$).

In Section~\ref{sec:GH-dich-simp}
we prove that the dichotomy
in Theorem~\ref{thm:main-dich-bd} can be extended to
simple graphs in Theorem~\ref{thm:gadget-simple-non-mult-bl-rk-1}.
In Section~\ref{sec:dec-ours} (and in Appendix~\ref{appendixsubsec1:dec-ours}) we show the polynomial time
decidability of our dichotomies.
In Section~\ref{sec:constructivity}
we prove that our dichotomies are constructive in the sense of
Theorem~\ref{thm:constructivity-eval(A)}.
We also explain the changes that can be made to the proof
in~\cite{Cai-Chen-Lu-2013}
to make that proof constructive in Appendix~\ref{sec:constr-ccl}. %the Appendix.

An index of conditions and problem definitions is given in Figure~\ref{indexindex}.

\begin{figure*}[t] %originally without [t]
\begin{tabular}{ll ll ll} 
$(\calU_1)$\hspace{0.05cm}--\hspace{0.05cm}$(\calU_4)$ & p.\hspace{0.06cm}\pageref{CONDITIONU}\hspace{0.3cm} &
$(\calU_5)$ & p.\hspace{0.06cm}\pageref{CONDITIONU5}\hspace{0.3cm} &
$(\calR_1)$\hspace{0.05cm}--\hspace{0.05cm}$(\calR_3)$ & p.\hspace{0.06cm}\pageref{CONDITIONR} \\[0.7ex]
$(\calL_1)$\hspace{0.05cm}--\hspace{0.05cm}$(\calL_3)$ & p.\hspace{0.06cm}\pageref{CONDITIONL} &
$(\calD_1)$\hspace{0.05cm}--\hspace{0.05cm}$(\calD_4)$ & p.\hspace{0.06cm}\pageref{CONDITIOND} &
$(\calU'_1)$\hspace{0.05cm}--\hspace{0.05cm}$(\calU'_4)$ & p.\hspace{0.06cm}\pageref{CONDITIONUP} \\[0.7ex]
$(\calU'_5)$ & p.\hspace{0.06cm}\pageref{CONDITIONUP5} &
$(\calR_1')$\hspace{0.05cm}--\hspace{0.05cm}$(\calR_3')$ & p.\hspace{0.06cm}\pageref{CONDITIONRP} &
$(\calL'_1)$\hspace{0.05cm}--\hspace{0.05cm}$(\calL'_2)$ & p.\hspace{0.06cm}\pageref{CONDITIONLP} \\[0.7ex]
$(\calD'_1)$\hspace{0.05cm}--\hspace{0.05cm}$(\calD'_2)$ & p.\hspace{0.06cm}\pageref{CONDITIONDP} &
$(\calT_1)$\hspace{0.05cm}--\hspace{0.05cm}$(\calT_3)$ & p.\hspace{0.06cm}\pageref{CONDITIONT} &
$(\calS_1)$ & p.\hspace{0.06cm}\pageref{CONDITIONS1} \\[0.7ex]
$(\calS_2)$\hspace{0.05cm}--\hspace{0.05cm}$(\calS_3)$ & p.\hspace{0.06cm}\pageref{CONDITIONS23} &
({\sl Meta}$_1$) & p.\hspace{0.06cm}\pageref{meta1} &
({\sl Shape}$_1$)\hspace{0.05cm}--\hspace{0.05cm}({\sl Shape}$_5$) & p.\hspace{0.06cm}\pageref{SHAPECONDITION} \\[0.7ex]
({\sl Meta}$_2$) & p.\hspace{0.06cm}\pageref{meta2} &
({\sl Shape}$_6$) & p.\hspace{0.06cm}\pageref{SHAPECONDITION6} &
({\sl Meta}$_3$) & p.\hspace{0.06cm}\pageref{meta3} \\[0.7ex]
(\GC) & p.\hspace{0.06cm}\pageref{CONDITIONGC} &
$(\calF_1)$\hspace{0.05cm}--\hspace{0.05cm}$(\calF_4)$ & p.\hspace{0.06cm}\pageref{CONDITIONF14} &
$(\calS'_1)$\hspace{0.05cm}--\hspace{0.05cm}$(\calS_2')$ & p.\hspace{0.06cm}\pageref{CONDITIONSP} \\[0.7ex]
({\sl Shape}$_1'$)\hspace{0.05cm}--\hspace{0.05cm}({\sl Shape}$_6'$) & p.\hspace{0.06cm}\pageref{SHAPECONDITIONP} &
& &
& \\[1.6ex] \hline\hline
\end{tabular}\ \vspace{0.2cm}\centering
%%%---

%%%---table-2-sorted---
\begin{tabular}{ll ll}
$Z_\AA(G)$ and $\eval(\AA)$ & p.\hspace{0.06cm}\pageref{eqn:def-of-Z_A} & %\pageref{Z-def}
$Z_{\CC,\fD}(G)$ and $\eval(\CC,\fD)$ & p.\hspace{0.06cm}\pageref{ZCD} \\[0.7ex]
$Z_{\AA, \mathfrak D}(G)$ and $\EVAL(\AA, \fD)$ & p.\hspace{0.06cm}\pageref{ZAfrakD} &
$Z_{\CC, \fD}^\rightarrow(G)$ and $\EVAL^\rightarrow(\CC, \fD)$ & p.\hspace{0.06cm}\pageref{ZCDARROW} \\[0.7ex]
$Z_{\CC, \fD}^\leftarrow(G)$ and $\EVAL^\leftarrow(\CC, \fD)$ & p.\hspace{0.06cm}\pageref{ZCDARROW} &
$\EVAL^\leftrightarrow(\CC, \fD)$ & p.\hspace{0.06cm}\pageref{ZCDARROW} \\[0.7ex]
$M_{\Gamma, \AA}$ & p.\hspace{0.06cm}\pageref{M_{Gamma,A}} &
$M_{\Gamma, \CC, \fD}$ & p.\hspace{0.06cm}\pageref{M_{Gamma,C,fD}} \\[0.7ex]
$T_p(G)$ and $S_r(G)$ & p.\hspace{0.06cm}\pageref{thickening-stretching} &
$\AA^{\odot p}$ & p.\hspace{0.06cm}\pageref{A-to-odot-p} \\[0.7ex]
\multblrkone/ & p.\hspace{0.06cm}\pageref{def:mubr1} &
\modblrkone/ & p.\hspace{0.06cm}\pageref{def:mobr1} \\[0.7ex]
$\calP_{n, p, \ell}$ & p.\hspace{0.06cm}\pageref{Pnpl} &
$\calR_{d, n, p, \ell}$ & p.\hspace{0.06cm}\pageref{Rdnpl} \\[0.7ex]
$G_{n, p, \ell}$ & p.\hspace{0.06cm}\pageref{Gnpl} &
$Z_{\mathscr A, \mathscr D}$ and $\EVAL(\mathscr A, \mathscr D)$ & p.\hspace{0.06cm}\pageref{eqn:Z_{scrA,scrD}} \\[0.7ex] %\pageref{Z-scrA-scrD}
$Z_q(f)$ and $\eval(q)$ & p.\hspace{0.06cm}\pageref{ZQF} &
$\hom(G, H)$ & p.\hspace{0.06cm}\pageref{eqn:hom-def} \\[0.7ex] %\pageref{hom(G,H)}
$H_{\AA, \DD}$ & p.\hspace{0.06cm}\pageref{H_{A,D}} &
$H_\AA$ & p.\hspace{0.06cm}\pageref{H_A} \\[0.7ex]
$\hom_\phi(G, H)$ & p.\hspace{0.06cm}\pageref{eqn:partial-hom-dec} & %\pageref{hom_phi(G,H)}
$\hom_{(i, j)}(G, H)$ & p.\hspace{0.06cm}\pageref{hom_{(i,j)}(G,H)} \\[0.7ex]
$\calPLG[k]$ & p.\hspace{0.06cm}\pageref{PLG[k]} &
$\calPLG_{\simp}[k]$ & p.\hspace{0.06cm}\pageref{PLGsimp[k]}\vspace{0.1cm}
\end{tabular}
\caption{Index of Conditions and Problem Definitions}\vspace{-0.06cm}\label{indexindex}
\end{figure*}
%%%%%%%%%%%%

%% file: preliminaries.tex
\section{Preliminaries}\label{sec:preliminaries-main}

%We use $\mathbb{Q}$, $\mathbb{R}$, and $\mathbb{C}$
%  to denote the set of all rational, real, and
%  complex numbers, respectively.

We let $\mathbb{Q}$ denote the set of rational numbers,
  and let $\mathbb{R}$ and $\mathbb{C}$ denote the set
  of algebraic real and algebraic complex numbers, respectively.
  
\subsection{Some basic notation}
%\paragraph{Some basic notation}

For a positive integer $n$, we use $[n]$ to denote the set $\{1,
\ldots, n\}$ and $[0]=\emptyset$.
We use $[m:n]$, where $m\le n$, to denote 
$\{m,m+1,\ldots, n\}$.
We use $\11_n$ to denote the all-one vector of dimension $n$.
Sometimes we omit $n$ when the dimension is clear from the context. 
For a positive integer $N$, we let $\omega_N=e^{2\pi
i/N}$, a primitive $N$th root of unity.

Let $\xx,\yy$ be two vectors in $\mathbb{C}^n$. Then we use
  $\langle \xx,\yy\rangle$ to denote their inner product,
\[
\langle\xx,\yy\rangle=\sum_{i=1}^n x_i\cdot \overline{y_i},
\]
and $\xx\circ \yy\in \mathbb{C}^n$ to
  denote their Hadamard product,
  $(\xx\circ\yy)_i=x_i\cdot y_i$ for all $i\in [n]$. 

Let $\AA=(A_{i,j})$ be a  $k\times \ell$ matrix.
We use $\AA_{i,*}$ ($i\in [k]$) and $\AA_{*,j}$ ($j\in [\ell]$)
 to denote the $i${th} row vector and  $j${th}
  column vector of $\AA$, respectively.
  %and $\AA_{*,j}$, $j\in [\ell]$, to denote the $j${th}
  %column vector of $\AA$.
If $\AA=(A_{i,j})$ and  $\BB=(B_{s,t})$ are 
$k\times \ell$ and ${m\times n}$
matrices, respectively,   we 
let $\CC=\AA\otimes \BB$ denote their tensor product:
  $\CC$ is a $km\times \ell n$ matrix whose rows and columns are indexed by~$[k]\times [m]$
  and $[\ell]\times [n]$, respectively, such that
\[
C_{(i,s),(j,t)}=A_{i,j}\cdot B_{s,t},\ \ \ \ \text{for all
$i\in [k]$, $s\in [m]$, $j\in [\ell]$ and $t\in [n]$.}
\]

Given an $n\times n$ symmetric complex matrix $\AA$, 
  we use $H=(V,E)$ to denote the following undirected graph: 
  $V=[n]$ and $ij \in E$ iff $A_{i,j}\ne 0$.
We say $\AA$ is \emph{connected} if $H$ is connected,
  and we say $\AA$ has connected components
  $\AA_1,\ldots,\AA_s$ if the connected components of $H$
  are $V_1,\ldots,V_s$, and $\AA_i$ is the $|V_i|\times |V_i|$
  submatrix of $\AA$ restricted by $V_i\subseteq [n]$, for all $i\in [s]$.
Moreover, we say $\AA$ is \emph{bipartite}
  if $H$ is bipartite; otherwise, $\AA$ is \emph{nonbipartite}. 
Let $\Sigma$ and $\Pi$ be two permutations of $[n]$.
Then we use $\AA_{\Sigma,\Pi}$ to denote the $n\times n$ matrix
  whose $(i,j)${th} entry is $A_{\Sigma(i),\Pi(j)}$, $i,j\in [n]$. \vspace{0.01cm}

We say $\CC$ is the \emph{bipartization} of a matrix $\FF$ if 
\[
\CC=\left(\begin{matrix}\00& \FF\\ \FF^{\text{T}} & \00\end{matrix}\right). 
\]
%Note that $\CC$ is always symmetric. % matrix no matter whether
  %$\FF$ is or is not.
We usually use $D_i$ to 
  denote the $(i,i)$th entry of a diagonal matrix $\DD$. 

We say a problem is tractable if it can be solved in
  polynomial time.
Given two problems $\cal P$ and $\mathcal{Q}$,
  we say $\cal P$ is polynomial-time reducible to $\mathcal{Q}$,
  or $\mathcal{P}\le \mathcal{Q}$,
  if there is a polynomial-time algorithm that solves $\cal P$
  using an oracle for $\mathcal{Q}$.
These reductions are known as Cook reductions.
We also say $\calP$ is polynomial-time equivalent to $\calQ$, or
  $\calP\equiv \calQ$, if $\calP\le \calQ$ and $\calQ\le \calP$.

%\subsection[Model of Computation]{Model of Computation}
\paragraph*{Model of computation}

One technical issue is the \emph{model of computation} with algebraic numbers.
We adopt a standard model from \cite{Lenstra-1992} for computation in
  an algebraic number field as in~\cite{Cai-Chen-Lu-2013}.
This is precisely described in Section 2.2 in~\cite{Cai-Chen-Lu-2013},
see also~\cite{cai-chen-book}.

\subsection{Definitions of $\eval(\AA)$ and $\eval(\CC,\fD)$}\label{evalcd-ours}

Let $\AA = (A_{i,j}) \in \mathbb{C}^{m\times m}$ be a symme\-tric matrix.
  %with entries $(A_{i,j})$.
It defines a graph homomorphism problem $\eval(\AA)$ as follows:
Given
  an undirected graph $G=(V,E)$, compute
\begin{equation}\label{eqn:def-of-Z_A}
Z_\AA(G)=\sum_{\xi:V\rightarrow [m]} \text{wt}_\AA(\xi),\ \ \ \
\text{where\ \ \ \  $\text{wt}_\AA(\xi) =\prod_{(u,v)\in E}
A_{\xi(u),\xi(v)}$}.
\end{equation}
We call $\xi$ an \emph{assignment} to the vertices of $G$, and
  $\text{wt}_\AA(\xi)$ the \emph{weight} of $\xi$. 

This is the standard definition of the partition function
of graph homomorphism; here $G$ can have multiple edges but no loops,
while
incorporated into the weight in $\AA$ the underlying graph $H$ can
have multiple edges and loops.  Our tractability result will apply
to graphs $G$ with multiple edges and loops, while the \#P-hardness  
result will be valid even for simple graphs, i.e.,
%with neither multiple edges nor loops.
without multiple edges and without loops.
 This definition can be easily generalized to apply
to directed graphs $G$ when $\AA$ is not assumed to be symmetric,
namely we let the product for $\text{wt}_\AA(\xi)$
in (\ref{eqn:def-of-Z_A}) to range over directed egdes $(u,v)$ of $G$.

%{\color{red} This way, when $\AA$ is symmetric we have two definitions of $\EVAL(\AA)$:
%for undirected and directed input graphs but these problems are essentially the same
%since orientations of edges play no role. (verbiage added by me)}

We denote by  $\EVAL^{(\Delta)}(\AA)$ the problem 
 $\EVAL(\AA)$ when restricted to graphs $G$ with maximum degree $\Delta(G)
\le \Delta$.  Similar notations apply to the other $\EVAL$ problems introduced
below.

The problem  $\eval(\AA)$ can be generalized to include
vertex weights~\cite{Dyer-Greenhill-2000,Freedman-Lovasz-Schrijver-2007,Lovasz-2006}.
There are several versions that we will specify; the simplest is the following
where a single diagonal matrix $\DD$ specifies the weights
for every vertex.
\begin{definition}\label{def:EVAL(A,D)}
Let $\AA \in \bbC^{m \times m}$ be a symmetric matrix
and $\DD \in \bbC^{m \times m}$ a diagonal matrix.
The problem $\EVAL(\AA, \DD)$ is defined as follows:
Given an undirected graph $G = (V, E)$, compute
\[
%\label{eqn:Z_{A,D}}
Z_{\AA, \DD}(G) = \sum_{\xi: V \rightarrow [m]} \prod_{w \in V} D_{\xi(w)} \prod_{(u, v) \in E} A_{\xi(u), \xi(v)}.\label{ZAfrakD}
\]
\end{definition}
Note that $\EVAL(\AA)$ is the special case $\EVAL(\AA, \II_m)$.

It turns out  that for Theorem~\ref{thm:main-dich-bd}, the most important
 vertex weight dependent $\EVAL$ problem
% that depends on vertex weight
% dependent  $\EVAL$ problem
is the following problem $\eval(\CC,{\frak D})$.
\begin{definition}[Definition 2.1 from~\cite{Cai-Chen-Lu-2013}]\label{def:EVAL(C,frakD-modN)}
Let $\CC\in \mathbb{C}^{m\times m}$ be a symmetric matrix, and
\[
{\frak D}=\big(\DD^{[0]},\DD^{[1]}, \ldots,\DD^{[N-1]}\big)
\]
be a sequence of diagonal matrices in $\mathbb{C}^{m\times m}$
for some $N\ge 1$. %, where the $(i,i)${th} entry of $\DD^{[r]}$
  %is $D^{[r]}_{i}$.
We define the following problem $\eval(\CC,{\frak D})$:
Given an undirected graph $G=(V,E)$, compute
\begin{equation}
Z_{\CC,{\frak D}}(G)=\sum_{\xi:V\rightarrow [m]}\text{\rm
wt}_{\CC,\fD}(\xi),\label{ZCD}%\vspace{-0.22cm}\label{ZCD}
\end{equation}
where%\vspace{-0.05cm}
\begin{equation*}
\text{\rm wt}_{\CC,\fD}(\xi)=
\Bigg(\hspace{0.05cm} \prod_{v\in V}
  D^{[\text{\rm deg}(v)\bmod\hspace{-0.025cm}N]}_{\xi(v)}\Bigg)
\Bigg(\hspace{0.03cm}\prod_{(u,v)\in E}
  C_{\xi(u),\xi(v)}\Bigg)
\end{equation*}
and $\deg(v)$ denotes the degree of $v$ in $G$.
\end{definition}

One can also define a version where the dependence on
the degree is not subject to $\bmod N$. We use the same notation
$\EVAL(\AA, \mathfrak D)$  for this
problem, which is only used in Section~\ref{sec:GCD-gadget}.

\begin{definition}\label{def:EVAL(A,frakD)}
Let $\AA \in \bbC^{m \times m}$ be a symmetric matrix and
$\mathfrak D = \{ \DD^{\llbracket i \rrbracket} \}_{i = 0}^\infty$ a sequence of diagonal matrices in $\bbC^{m \times m}$.
The problem $\EVAL(\AA, \mathfrak D)$ is defined as follows:
Given an undirected graph $G = (V, E)$, compute
\[
%\label{eqn:Z_{A,frakD}}
Z_{\AA, \mathfrak D}(G) = \sum_{\xi: V \rightarrow [m]} \prod_{w \in V} D_{\xi(w)}^{\llbracket \deg(w) \rrbracket} \prod_{(u, v) \in E} A_{\xi(u), \xi(v)}.
\]
\end{definition}
Definition~\ref{def:EVAL(C,frakD-modN)} is
% can also be viewed as a
the special case of Definition~\ref{def:EVAL(A,frakD)}
where each $\DD^{\llbracket i \rrbracket} = \DD^{[i \bmod N]}$ for any $i \ge 0$.

%\vspace{.1in} --- commented out
Let $G$ be an undirected graph with 
  connected components $G_1,\ldots,G_s$. 

\begin{property}\label{connectedcomponents-ours} %[Property 2.2 from~\cite{Cai-Chen-Lu-2013}] %%%omitted
$Z_{\CC,\fD}(G)=Z_{\CC,\fD}(G_1)\times \cdots\times
Z_{\CC,\fD}(G_s)$. 
\end{property}

Property \ref{connectedcomponents-ours} implies that, 
%whether we need 
to design
  an algorithm for $\eval(\CC,\fD)$ or to reduce $\eval(\CC,\fD)$ to
  another problem,
% $\eval(\CC',\fD')$, 
it suffices 
  to consider connected input graphs. 
It also preserves bounded degree graphs
for any degree bound $\Delta \ge 0$, since $\Delta(G_i) \le \Delta(G)$ for each $i \in [s]$.
%{\color{red} (JYC: added for the proof of Theorem 5.1, you can shorten it)}
%%%
%if we want to reduce 
%  $\eval(\CC,\fD)$ to another problem $\eval(\CC',\fD')$,
%  we only need to consider input graphs that are connected.
%Also note that, since $\eval(\AA)$ is a special case of
As $\eval(\AA)$ is a special case of
  $\eval(\CC,\fD)$ in which every $\DD^{[i]}$
  is an identity matrix, Property \ref{connectedcomponents-ours} applies
to $\eval(\AA)$ as well.
%and the remarks above
%  apply to $\eval(\AA)$ as well. 

%\vspace{.1in} %commented out

\input{definition-of-arrowed-Z}

%\vspace{.1in} %commented out

In this paper, a crucial object is an \emph{edge gadget},
which will be used in different $\EVAL$ frameworks.
An edge gadget $\Gamma$ is simply an undirected graph $(V, E)$
with two 
%%% I think %distinct %different are both implied by "two ...ordered"
%distinct 
%different 
distinguished (ordered) vertices $\firstdv, \seconddv \in V$.

\begin{definition}\label{def:M(Gamma,A)}
Let $\AA \in \bbC^{m \times m}$ be a symmetric matrix
and let $\Gamma = (V, E)$ be an edge gadget
with distinguished vertices $\firstdv, \seconddv$ (in this order).
Define $\gm_{\Gamma, \AA} \in \bbC^{m \times m}$ to be the
edge weight matrix, or signature, of $\Gamma$
in the framework $\EVAL(\AA)$.
%More precisely, this means that every $e \in E$ is assigned $\AA$
%and for $i, j \in [m]$ we have
More precisely, we define $\gm_{\Gamma, \AA} \in \bbC^{m \times m}$ as follows:
for $i, j \in [m]$, let
\[
\gm_{\Gamma, \AA}(i, j)
= \sum_{\substack{\xi \colon V \to [m] \\ \xi(\firstdv) = i,~ \xi(\seconddv) = j}} \twt_{\Gamma, \AA}(\xi),
\quad \text{where} \quad \twt_{\Gamma, \AA}(\xi) = \prod_{(u, v) \in E} A_{\xi(u), \xi(v)}.\label{M_{Gamma,A}}
\]
\end{definition}

Note that while $\AA$ is symmetric, the matrix $\gm_{\Gamma, \AA}$ is not symmetric in general.
%
%%%%%%%%%%%%%%%%%%%%%%%%%%%%%%%%%%%%%%%%%%%%%%%%%%%%%%%%%%%%%%%%%%%%
\begin{comment}

{\color{red} (my old remark replaced by the above so to be removed)

In Definition~\ref{def:M(Gamma,A)}, $\gm_{\Gamma, \AA}$ does not have to be symmetric.
However, it can be used to simulate an edge weight matrix
for directed $\EVAL$ problem where the inputs are directed graphs (see the proof of Corollary~\ref{cor:non-mult-bl-rk-1-gadget-hardness-bd})
}

\end{comment}
%%%%%%%%%%%%%%%%%%%%%%%%%%%%%%%%%%%%%%%%%%%%%%%%%%%%%%%%%%%%%%%%%%%%
%%%%%% I will not define in preliminaries section the directed $\EVAL$ problem
% but only do so when it is needed, ie. locally as needed.
%
%More generally,
We give a more general definition. 
%{\color{red} (JYC: for the previous version it is not clear what more generally is because of the prev sentence, or just remove it)}
%
\begin{definition}\label{def:M(Gamma,C,D)}
Let $(\CC, \fD)$ be a pair from Definition~\ref{def:EVAL(C,frakD-modN)}
and let $\Gamma = (V, E)$ be an edge gadget
with distinguished vertices $\firstdv, \seconddv$ (in this order).
Define $\gm_{\Gamma, \CC, \fD} \in \bbC^{m \times m}$ to be the
edge weight matrix, or signature, of $\Gamma$
in the framework $\EVAL(\CC, \fD)$.
%More precisely, this means that every edge $e \in E$ is assigned $\CC$,
%every vertex $u \in V \setminus\{\firstdv, \seconddv\}$ is assigned $D^{[\text{\rm deg}(u)\bmod\hspace{-0.025cm}N]$
%and for $i, j \in [m]$ we have
More precisely, 
%we define $\gm_{\Gamma, \CC, \fD}$ as follows:
for $i, j \in [m]$, 
%let
\[
\gm_{\Gamma, \CC, \fD}(i, j)
= \sum_{\substack{\xi \colon V \to [m] \\ \xi(\firstdv) = i,~\xi(\seconddv) = j}} \twt_{\Gamma, \CC, \fD}(\xi)\label{M_{Gamma,C,fD}},
\]
where
\[
\twt_{\Gamma, \CC, \fD}(\xi)=
\Bigg(\hspace{0.05cm} \prod_{v\in V\setminus\{\firstdv, \seconddv\}}
  D^{[\text{\rm deg}(v)\bmod\hspace{-0.025cm}N]}_{\xi(v)}\Bigg)
\Bigg(\hspace{0.03cm}\prod_{(u,v)\in E}
  C_{\xi(u),\xi(v)}\Bigg).
\]
\end{definition}
Note that the vertex weights corresponding to $\firstdv$ and $\seconddv$
are excluded from the product in the definition of $\gm_{\Gamma, \CC, \fD}$.
Analogously, we can define the edge weight matrix of an edge gadget
%according to Definition~\ref{def:EVAL(A,frakD)}.
according to Definitions~\ref{def:EVAL(A,D)} and~\ref{def:EVAL(A,frakD)}.
%It will only be used locally in the main proof in Section~\ref{sec:GCD-gadget},
%and we will not need a special notation for it.
%{\color{red} (JYC: this is the shortest solution, or move this fragment to Section 5 somewhere)}

\begin{lemma}\label{lem:dif-components-rk1}
Let $(\CC, \fD)$ be as in Definition~\ref{def:EVAL(C,frakD-modN)}
and $\Gamma$ be an edge gadget with distinguished vertices $\firstdv, \seconddv$ (in this order).
If $\firstdv, \seconddv$ lie in different connected components of $\Gamma$,
then $\rank \gm_{\Gamma, \CC, \fD} \le 1$.
\end{lemma}
\begin{proof}
Let $\Gamma_1, \ldots, \Gamma_s$ ($s \ge 2$) be the connected components of $\Gamma$,
and suppose $\firstdv \in \Gamma_1$ and $ \seconddv \in \Gamma_2$.
For $w \in \{\firstdv, \seconddv \}$ from $\Gamma_\ell$
($\ell \in \{1,2\}$) and any $i \in [m]$, we write 
\[
Z_{\CC, \fD}(\Gamma_\ell, w, i)
= \sum_{\substack{\xi \colon V(\Gamma_\ell) \to [m] \\ \xi(w) = i}}
\Bigg( \prod_{v \in V(\Gamma_\ell) \setminus \{w\}} D^{[\text{\rm deg}(v)\bmod N]}_{\xi(v)} \Bigg)
 \Bigg( \prod_{(u, v) \in E(\Gamma_\ell)} C_{\xi(u),\xi(v)} \Bigg).
\]
Then
\[
\gm_{\Gamma, \CC, \fD}(i, j) = \left( \prod_{k = 3}^s Z_{\CC, \fD}(\Gamma_k) \right) Z_{\CC, \fD}(\Gamma_1, \firstdv, i) Z_{\CC, \fD}(\Gamma_2, \seconddv, j)
\]
for $i, j \in [m]$, and therefore $\rank \gm_{\Gamma, \CC, \fD} \le 1$.
\end{proof}
It is also obvious that if 
 $\firstdv, \seconddv$ belong to the same connected component
$\Gamma_1$, then
\[
\gm_{\Gamma, \CC, \fD} = \left( \prod_{k = 1}^s Z_{\CC, \fD}(\Gamma_k) \right) \gm_{\Gamma_1, \CC, \fD}.
\]
where $\prod_{k = 1}^s Z_{\CC, \fD}(\Gamma_k)$ is a scalar factor.
% multiplier which must be nonzero.

Since we will be only interested in edge gadgets with 
$\rank \gm_{\Gamma, \CC, \fD} > 1$, we may without loss of generality
assume $\Gamma$ is connected. 
 
In case $\CC$ is bipartite,
% the bipartization of  $\FF$.
% an $m \times n$ $\FF$.
we may assume any edge gadget $\Gamma$ in the framework of $\eval(\CC, \fD)$
to be (connected and) bipartite $\Gamma = (U \cup V, E)$,
because for nonbipartite $\Gamma$,  $\gm_{\Gamma, \CC, \fD} = 0$.
%
%%%%% assume $\CC$ is bipartite, each conn comp C_i is bipartite, 
%%%%  and \Gamma is connected and not bipartite,
%%%%  $\gm_{\Gamma, \CC, \fD} = sum over_i of \gm_{\Gamma, C_i, D} =0
% "Clearly, if $\Gamma$ is not bipartite, then $\gm_{\Gamma, \CC, \fD} = 0$"
%%%%%%%%%%%%%%%%%%%%%%%%%%%%%%%%%%%%%%%%%%%%%%%%%%%%%%%%%%%%%%%%%%%%%%%%%%%%%%%%%
If $\firstdv, \seconddv$ lie in the same bipartite part of $\Gamma$
(e.g., $\firstdv, \seconddv \in U$), then $\gm_{\Gamma, \CC, \fD}$ has the form
\[
\gm_{\Gamma, \CC, \fD} = \begin{pmatrix} \subgm^{[0]}_{\Gamma, \CC, \fD} & \00 \\ 
\00 & \subgm^{[1]}_{\Gamma, \CC, \fD} \end{pmatrix}.
\]
If  $\firstdv \in U$ and $\seconddv \in V$, then
\[
\gm_{\Gamma, \CC, \fD} = \begin{pmatrix} \00 & \subgm^{[0]}_{\Gamma, \CC, \fD} \\ 
\subgm^{[1]}_{\Gamma, \CC, \fD} & \00 \end{pmatrix}.
\]
Any proof in our paper for the first case can be easily
adapted to the second case. Therefore, for the bipartite case,
we will assume without loss of generality
 that $\firstdv$ and $\seconddv$ belong to the same bipartite part.

\begin{figure}[t]
\begin{subfigure}{0.5\textwidth}
\centering
\includegraphics{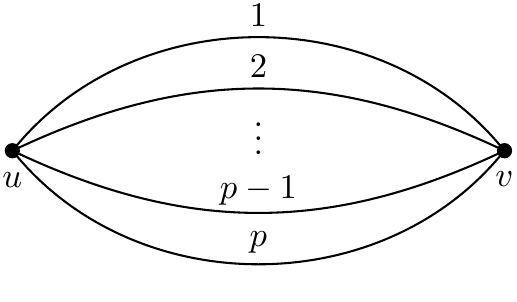} %[scale=1.2]
\end{subfigure}\hfill
\begin{subfigure}{0.5\textwidth}
\centering
\includegraphics{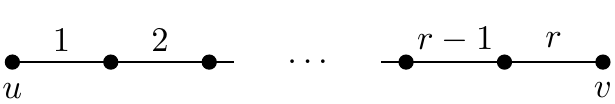} %[scale=1.2]
\end{subfigure}
\caption{\label{fig:thickening-stretching}The thickening $T_p e$ and the stretching $S_r e$ of an edge $e = (u, v)$.}
\end{figure}

\begin{figure}[t]
\begin{subfigure}{0.6\textwidth}
\centering
\includegraphics{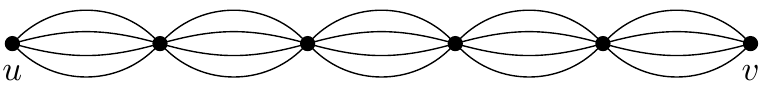} %[scale=1.2]
\end{subfigure}\hfill
\begin{subfigure}{0.4\textwidth}
\centering
\includegraphics{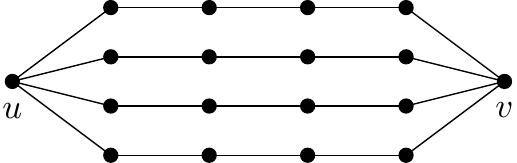} %[scale=1.2]
\end{subfigure}
\caption{\label{fig:thickenings-stretchings-composition}The graphs $T_4 S_5 e$ (on the left) and $S_5 T_4 e$ (on the right) where $e = (u, v)$.}
\end{figure}

Two simple
%but important
operations are known as \emph{thickening} and \emph{stretching}.
%The paper by Dyer and Greenhill~\cite{Dyer-Greenhill-2000}
%was crucially reliant on two important
%techniques called Stretchings and Thickenings.
%These operations are applied to the input graph $G$.
%
Let $p, r \ge 1$ be integers.
A $p$-\emph{thickening} of an edge
%$e = \{ u, v \}$
replaces it by $p$ parallel edges,
and
an $r$-\emph{stretching} replaces it by a path
of length $r$.
In both cases we retain the endpoints.
%The $p$-\emph{thickening} or $r$-\emph{stretching}
% of $G$ with respect to $F \subseteq E(G)$,
%denoted respectively by $T_p^{(F)}(G)$   and $S_r^{(F)}(G)$,
%are
%obtained by $p$-\emph{thickening} or  $r$-\emph{stretching}
% each edge from $F$, respectively.
%Other edges, if any, are unchanged in both cases.
%When $F = E(G)$, 
%we call them the $p$-\emph{thickening} and $r$-\emph{stretching} of $G$
We denote by $T_p(G)$, respectively $S_r(G)$, \label{thickening-stretching}
the graph obtained from $G$ by $p$-\emph{thickening}, respectively
 $r$-\emph{stretching}, 
every edge of $G$.
%and denote them by $T_p(G)$ and $S_r(G)$, respectively.
$T_p e$ and $S_r e$ are the special cases when the graph consists of
a single edge $e$.
See Figure \ref{fig:thickening-stretching} for an illustration.
Thickenings and stretchings  can be combined successively.
% in any order.
Examples are shown in Figure~\ref{fig:thickenings-stretchings-composition}.

%%%
%%%

%%%
%%%

For a matrix $\AA$, its Hadamard power  $\AA^{\odot p}$ \label{A-to-odot-p}
is the matrix
obtained by replacing each entry of $\AA$ with its $p$th power.
Clearly, $Z_\AA(T_p G) = Z_{\AA^{\odot p}}(G)$ and $Z_\AA(S_r G) = Z_{\AA^r}(G)$.
More generally, for the vertex-weighted case,
we have $Z_{\AA, \DD}(T_p G) = Z_{\AA^{\odot p}, \DD}(G)$ and $Z_{\AA, \DD}(S_r G) = Z_{\AA (\DD \AA)^{r - 1}, \DD}(G)$.
Here $(\DD \AA)^0 = \II_m$ if $\AA$ and $\DD$ are $m \times m$.

%%%%%%%%%%% above cut off Artem's verbiage galore

\subsection{Basic \#P-hardness}\label{subsec:basic-hardenss}

%%%%%%%%%%%%%%%%%%%%%%%%%%%%%%%%%%%%%%%%%%%%%%%%%%%%%%%%%%%%%%%%%%%%
\begin{comment}

{\color{red} We need to add (see also~\cite{Thurley-2009, Grohe-Thurley-2011}) somewhere, in GCD it was done in the introduction, but we can also do it here)}

\end{comment}
%%%%%%%%%%%%%%%%%%%%%%%%%%%%%%%%%%%%%%%%%%%%%%%%%%%%%%%%%%%%%%%%%%%%
We say
%a nonnegative
%a symmetric $m \times m$ matrix $\AA \in \bbC^{m \times m}$ is rectangular if
a symmetric matrix $\AA \in \bbC^{m \times m}$ is rectangular if
there are pairwise disjoint nonempty subsets of $[m]$:
$T_1, \ldots, T_r, P_1, \ldots, P_s, Q_1, \ldots, Q_s$,
for some $r, s\ge 0$, such that $A_{i, j} \not = 0$ iff
\[
(i, j) \in \bigcup_{k \in [r]} (T_k \times T_k) \cup \bigcup_{l \in [s]} [(P_l \times Q_l) \cup (Q_l \times P_l)].
\]
$T_k \times T_k, P_l \times Q_l$ and $Q_l \times P_l$ are called blocks of $\AA$.
Further,  we say  $\AA$ is \textit{block-rank-$1$}
% a nonnegative symmetric matrix $\AA$ {\color{red} (a symmetric $\AA \in \bbC^{m \times m}$)} is \textit{block-rank-$1$}
if $\AA$ is rectangular and every block of $\AA$ has rank one.
\begin{theorem}[Bulatov and Grohe \cite{Bulatov-Grohe-2005}]\label{BulatovGrohe}
Let $\AA$ be a symmetric matrix with nonnegative entries.
Then $\EVAL(\AA)$ is in polynomial time if $\AA$ is block-rank-$1$,
and is \#P-hard otherwise.
\end{theorem}

%% file: definition-of-arrowed-Z.tex
Next, suppose $\CC$ is the bipartization of an $m\times n$ 
matrix  $\FF$, so $\CC$ is $(m+n)\times (m+n)$.
Given a graph $G$ and a vertex $u$ in $G$,
  %(we sometimes use a triple to denote a bipartite graph, in which
  %$U\cup V$ is the vertex set and every edge $uv\in E$ has one vertex $u$
  %from $U$ and one vertex $v$ from $V$),
  %we define $Z_{\CC,\fD}^{\rightarrow}(G,u)$ and $Z_{\CC,\fD}^{\leftarrow}(G,u)$ as %follows.
  we use $\Xi_1$ to denote the set of $\xi\colon V\rightarrow [m+n]$ with
  $\xi(u)\in [m]$, and $\Xi_2$ to denote the set of
  $\xi$ with $\xi(u)\in [m+1:m+n]$. Then let
\[
Z_{\CC,\fD}^{\rightarrow}(G,u)=\sum_{\xi\in
\Xi_1}\text{wt}_{\CC,\fD}(\xi)\ \ \ \ \text{and} \ \ \ \
Z_{\CC,\fD}^{\leftarrow}(G,u)=\sum_{\xi\in
\Xi_2}\text{wt}_{\CC,\fD}(\xi).\label{ZCDARROW}
\]
The next property follows from the definitions.
  
\begin{property}\label{verytrivial-ours} %[Property 2.3 from~\cite{Cai-Chen-Lu-2013}}]  %%%omitted
$Z_{\CC,\fD}(G)
  =Z_{\CC,\fD}^\rightarrow(G,u)+Z_{\CC,\fD}^{\leftarrow}(G,u)$. 
\end{property}

We use these two new functions to express  the
partition function when the matrix is in a tensor product form
in the bipartite case.
The following is Lemma 2.4 of~\cite{Cai-Chen-Lu-2013}.

\begin{lemma}\label{tensorproduct-ours}
%[Lemma 2.4 from~\cite{Cai-Chen-Lu-2013}]\label{tensorproduct-ours}
For each $i\in \{0,1,2\}$, let $\FF^{[i]}$ be an $m_i\times n_i$ complex
matrix, where $m_0=m_1 m_2$ and $n_0=n_1 n_2$;
  let $\CC^{[i]}$ be the bipartization of $\FF^{[i]}$; and let
\[{\frak D}^{[i]}=\big(\DD^{[i,0]},\ldots,\DD^{[i,N-1]}\big)\] be
  a sequence of $(m_i+n_i)\times (m_i+n_i)$ diagonal matrices for
  some $N\ge 1$, where
\[
\DD^{[i,r]}=\left(\begin{matrix} \PP^{[i,r]} & \\
  & \QQ^{[i,r]}\end{matrix}\right)
\]
and $\PP^{[i,r]}$, $\QQ^{[i,r]}$ are $m_i\times m_i$, $n_i\times n_i$
   diagonal matrices, respectively.
%For each $i\in \{0,1,2\}$, $(\CC^{[i]},\fD^{[i]})$ satisfies
%  condition \emph{(Pinning)}.
Assume 
\[\FF^{[0]}=\FF^{[1]}\otimes \FF^{[2]},\ \ \ \PP^{[0,r]}=
  \PP^{[1,r]}\otimes \PP^{[2,r]},\ \ \ \text{and}\ \ \ \
  \QQ^{[0,r]}=\QQ^{[1,r]}\otimes \QQ^{[2,r]}\]
{for all $r\in [0:N-1]$.}
Then for any \emph{connected} graph $G$ and any vertex $u^*$ in $G$, 
\begin{align}
&Z_{\CC^{[0]},\fD^{[0]}}^\rightarrow
(G,u^*)=Z_{\CC^{[1]},\fD^{[1]}}^\rightarrow (G,u^*)
  \cdot Z_{\CC^{[2]},\fD^{[2]}}^\rightarrow (G,u^*) \text{\ \ \  and}\label{jujublabla}\\[0.2ex]
& \label{ttyyuuiioo} Z_{\CC^{[0]},\fD^{[0]}}^\leftarrow
(G,u^*)=Z_{\CC^{[1]},\fD^{[1]}}^\leftarrow (G,u^*)
  \cdot Z_{\CC^{[2]},\fD^{[2]}}^\leftarrow (G,u^*).%\\[-5.5ex] \nonumber %commented out
\end{align}
\end{lemma}

\begin{definition}\label{EVAL(C,D)-lr-one}
We define the problem $\EVAL^\rightarrow(\CC, \fD)$ (resp., $\EVAL^\leftarrow(\CC, \fD)$):
Given a pair $(G, u)$ where $G = (V, E)$ is an undirected graph and $u \in V$,
compute $Z^\rightarrow_{\CC, \fD}(G)$ (resp., $Z^\leftarrow_{\CC, \fD}(G)$).
%We denote the problem of computing $Z^\rightarrow_{\CC, \fD}$ and $Z^\leftarrow_{\CC, \fD}$
%by $\EVAL^\rightarrow_{\CC, \fD}$ and $\EVAL^\leftarrow_{\CC, \fD}$, respectively.
%\end{definition}
%
%\begin{definition}\label{EVAL(C,D)-lr-two}
%Next, we define the problem $\EVAL^\leftrightarrow_{\CC, \fD}$ as follows:
%We define the following problem $\EVAL^\leftrightarrow_{\CC, \fD}$:
%
Given a tuple $(\rightarrow, u, G)$ or $(\leftarrow, u, G)$,
the  problem $\EVAL^\leftrightarrow(\CC, \fD)$ is
to 
% where $G = (V, E)$ is an undirected graph and $u \in V$,
compute $Z^\rightarrow_{\CC, \fD}(G)$ or $Z^\leftarrow_{\CC, \fD}(G)$, respectively.
\end{definition}

Restricting the inputs of each of the problems
$\EVAL^\rightarrow(\CC, \fD)$, $\EVAL^\leftarrow(\CC, \fD)$ and $\EVAL^\leftrightarrow(\CC, \fD)$
to the pairs $(G, u)$ (for the first two), or the tuples $(\rightarrow, G, u)$ and $(\leftarrow, G, u)$ (for the latter), where $G = (V, E)$ is connected and $u \in V$,
we get a problem polynomial-time equivalent to the corresponding original problem,
so we only need to consider such pairs (tuples) as inputs,
and furthermore all these problems are polynomial-time equivalent.
This follows from the following: (1) a trivial extension of Property~\ref{connectedcomponents-ours}
for $Z_{\CC, \fD}^\rightarrow$ and $Z_{\CC, \fD}^\leftarrow$; (2) Property~\ref{verytrivial-ours};
(3) the fact that if $G = (V, E)$ is not bipartite and $u \in V$,
then $Z_{\CC, \fD}^\rightarrow(G, u) = Z_{\CC, \fD}^\leftarrow(G, u) = 0$;
and (4) the fact that if $G = (U \cup V, E)$ is connected and bipartite, $u \in U$ and $v \in V$, %,/with $u \in U$
then $Z_{\CC, \fD}^\rightarrow(G, u) = Z_{\CC, \fD}^\leftarrow(G, v)$ and $Z_{\CC, \fD}^\rightarrow(G, v) = Z_{\CC, \fD}^\leftarrow(G, u)$.
%It will be convenient for us to differentiate between these problems.

%{\color{red} (JYC: We explicitly use the fact that the reduction can be done in Steps 2.4, 2.5 and for the tractability for $\EVAL^\leftrightarrow_{\CC, \fD})$ for the bipartite case, otherwise it's incomplete; another alternative is to restrict the input in Definition 2.7 to connected graphs since we are only concerned with them, but I don't like it)}

We have the following trivial property.
\begin{property}\label{pr:EVAL(C,D)-to-EVAL(C,D)-lr-two}
$\EVAL(\CC, \fD) \le \EVAL^\leftrightarrow(\CC, \fD)$.
\end{property}

In~\cite{Cai-Chen-Lu-2013}, besides the first pinning lemma,
two more pinning lemmas are proved,
called the second and third pinning lemmas, which give the
reverse directions of Property~\ref{pr:EVAL(C,D)-to-EVAL(C,D)-lr-two}
(under certain technical conditions). However we are unable to prove
the second pinning lemma constructively. The proof in this paper
will be  made constructive. In order to avoid nonconstructive steps,
we will avoid using any of the three pinning lemmas, which
has the slight complication that we must use $\EVAL^\leftrightarrow(\CC, \fD)$
instead of  $\EVAL(\CC, \fD)$ in certain steps of the proof 
(see Theorem~\ref{bi-step-2-ours}).

%% file: algebraic-preliminaries.tex
\section{Algebraic preliminaries}\label{sec:algebraic-preliminaries}
We state some preliminaries of an algebraic nature.
%We need some facts from algebra.
\subsection{Finitely generated fields}

\begin{lemma}\label{lem:fin-many-roots-of-unity-full}
Any finitely generated field over $\bbQ$ 
contains finitely many roots of unity.
\end{lemma}
We will only need this lemma for finite dimensional extensions
over $\bbQ$ (since we work with algebraic numbers),
% for the consideration
%of strict Turing computability),
although the lemma
is true for not necessarily algebraic extensions. 
For an algebraic extension $K$, let $k = [K: \bbQ] < \infty$ 
be the degree of the extension. 
The  minimal polynomial over $\bbQ$
 of a (primitive)  root of unity of order $r$
is the $r$-th cyclotomic polynomial $\Phi_r(x)$ of degree $\varphi(r)$,
the Euler totient function,
each having at most $\varphi(r)$ roots in $K$.
Clearly
%It is easy to check that 
$\varphi(r) \ge \sqrt{r / 2}$, for $r \ge 1$.
Hence $r \le  2 (\varphi(r))^2 \le 2 k^2$.
Thus the number of roots of unity in $K$
is bounded by $\sum_{r = 1}^{2 k^2} \varphi(r)$.

The following lemma is a well-known fact.
%{\color{red} (short proof, why not)}

\begin{lemma}\label{lem:cyc-gr}
Let $K$ be a field. Assume $G \subset K$ is a finite multiplicative subgroup.
%{\color{red} (Assume $G$ is a finite multiplicative subgroup in $K$.)}
Then $G$ is cyclic.
\end{lemma}
%%%%%%%%%%%%%%%%%%%%%%%%%%%%%%%%%%%%%%%%%%%%%%%%%%%%%%%%%%%%%%%%%%%%
\begin{comment}

\begin{proof}
Let $|G| = n$. For any $a \in G$, clearly $\ord(a)$ divides $n$.
For any $d \mid n$, we let $C_d$ be the set of all elements $b$ in $G$ of order $d$. %of $G$ such that $b^d = 1$.
Note that any such element is a root of the polynomial $x^d - 1 \in K[x]$.
Take any $d \mid n$. If $b \in C_d$, then $\langle b \rangle = \{b^0 = 1, \ldots, b^{d - 1}\}$
is a cyclic subgroup of $G$ of order $d$ and any element from $\langle b \rangle$ is a root of $x^d - 1 \in K[x]$.
Since $K$ is a field, this polynomial has at most $d$ roots in $K$.
Hence $\langle b \rangle$ is precisely the set of these roots
and therefore $C_d \subseteq \langle d \rangle$.
Then is easy to see that $b^i$ where $0 \le i \le d - 1$
is of order $d$ iff $\gcd(i, d) = 1$.
In this case we conclude that $|C_d| = \varphi(d)$
where $\varphi(x)$ is Euler's totient function.
Therefore for any $d \mid n$,
either $C_d = \emptyset$ or $C_d = \varphi(d)$.
Since $G$ partitions into $\bigsqcup_{d \mid n} {C_d}$,
we have
\[
n = |G| = \sum_{d \mid n} |C_d| \le \sum_{d \mid n} \varphi(d) = n.
\]
(The last equality follows from Gauss's theorem (can omit it)).
Then it is clear that $|C_d| = \varphi(d)$ for each $d \mid n$.
In particular, $C_n \ne \emptyset$.
Taking any $a \in C_n$ we see that $a$, as an element of order $n$, generates the entire group $G$.
\end{proof}

\end{comment}
%%%%%%%%%%%%%%%%%%%%%%%%%%%%%%%%%%%%%%%%%%%%%%%%%%%%%%%%%%%%%%%%%%%%

The roots of unity in a field clearly form a multiplicative subgroup.
From Lemmas~\ref{lem:fin-many-roots-of-unity-full} and~\ref{lem:cyc-gr}
we infer the following corollary.
%{\color{red} (Without this corollary, we would have to always take the lcm or the orders of the roots of unity.
%Very unnatural, considering the corollary is true and is plain obvious as if we don't know it)}
\begin{corollary}\label{cor:fin-gen-field-roots-of-unity-full-fin-cyc-gr}
In any finitely generated field over $\bbQ$
%, i.e., of the form $\bbQ(\alpha_1, \ldots, \alpha_n)$,
the roots of unity form a finite cyclic group.
\end{corollary}

%%%%%%%%%%%%%%%%%%%%%%%%%%%%%%%%%%%%%%%%%%%%%%%%%%%%%%%%%%%%%%%%
\subsection{Multiplicative-block-rank-$1$ and modular-block-rank-$1$}\label{subsec:mult-brk1}
In this subsection we introduce a  new concept
called \emph{multiplicative-block-rank-1},
and its related notion \emph{modular-block-rank-1}.
They play an essential role in this paper. 

First, the definition of rectangularity of a matrix
can be easily extended to 
 not necessarily symmetric or even square matrices.
We say a matrix $\AA \in \bbC^{m \times n}$ is \textit{rectangular}
if its  rows and columns  can be permuted (separately)
so that it becomes a block-diagonal matrix where each block is
a matrix with no zero elements, with possibly one block being an all-$0$ matrix.
%%% I will not give a proof. it's kind of easy.
%%%
%%% for symmteric. first take connected components. it in fact the same permutation on
% rows and columns. etc. etc.
%
%Another way to state this is: for some $s \ge 0$,
%there exist $s$ pairwise disjoint and nonempty
%subsets $P_1, \ldots, P_s$ of $[m]$,
%and $s$ pairwise disjoint and nonempty
%subsets $Q_1, \ldots, Q_s$ of $[n]$,
%such that $A_{i, j} \ne 0$
%iff $i \in P_k$ and $j \in Q_k$ for some $k \in [s]$.
We say  a  matrix $\AA \in \bbC^{m \times n}$ is
 \textit{block-rank-$1$} 
%(\textit{\blrkone/})
%%% I decided it is best NOT to introduce another shortened name.
%%%% Just block-rank-$1$
if $\AA$ is rectangular and every (nonzero) block of $\AA$ has rank one.
(These notions coincide with the ones given in subsection~\ref{subsec:basic-hardenss}
 when  $\AA$ is symmetric.)
It is easy to see that  for every $p \ge 1$,
 $\AA$ is rectangular iff  $\AA^{\odot p}$  is rectangular,
 and if $\AA$ is block-rank-$1$,
then so is $\AA^{\odot p}$.
The converse of the latter statement is not true as shown by
the example $\begin{psmallmatrix} 1 & 1 \\ 1 & -1 \end{psmallmatrix}$. %used to be $\left[ \begin{smallmatrix} 1 & 1 \\ 1 & -1 \end{smallmatrix} \right]$.

It is easy to show that  $\AA \in \bbC^{m \times n}$ is 
rectangular iff the column locations of nonzero entries of every two rows either
coincide or are disjoint.
A symmetric statement holds when we exchange rows and columns.
It follows that  $\AA$ is not rectangular
iff for some $1 \le i < j \le m$ and $1 \le k < \ell < n$,
the $2 \times 2$ submatrix
\begin{equation}\label{eqn:2by2-submatrix}
\AA_{i, j; k, \ell} =
\begin{pmatrix}
A_{i, k} & A_{i, \ell} \\
A_{j, k} & A_{j, \ell}
\end{pmatrix}
\end{equation}
contains exactly one zero entry.
In this case we say that the tuple $(i, j, k, \ell)$
witnesses the non-rectangularity of $\AA$.
%{\color{red} (may remove it, only use it once)}
%the fact that $\AA$ is
% not rectangular,
%alternatively, the non-rectangularity of $\AA$.

\begin{definition}\label{def:mubr1}
We say $\AA \in \bbC^{m \times n}$ 
is \textit{multiplicative-block-rank-$1$} (\multblrkone/)
if there exists $k \ge 1$ such that $\AA^{\odot k}$ is block-rank-$1$.
\end{definition}
%%%
Clearly, every block-rank-$1$ matrix is \multblrkone/,
and  every \multblrkone/ matrix is rectangular.
%  block-rank-$1$ ==> take p=1. get \multblrkone/
% \multblrkone/ ==> for some k>=1,   $\AA^{\odot k}$ is  block-rank-$1$ 
% so $\AA^{\odot k}$ is rectangular. ==> A itself is rectangular.
%
If follows from the definition that if $\AA \in \bbC^{m \times n}$ is not 
\multblrkone/,
%multiplicative-block-rank-$1$,
then for every $p \ge 1$, $\AA^{\odot p}$ is not \multblrkone/.
%multiplicative-block-rank-$1$.
It is the notion of \emph{non-\multblrkone/} that will be central to
our proof in this paper.

\begin{definition}\label{def:mobr1}
We say $\AA \in \bbC^{m \times n}$ is modular-block-rank-$1$ (\modblrkone/)
if the matrix  $(|A_{i, j}|)_{i, j = 1}^{m, n}$
obtained from $\AA$ by taking the complex norm entrywise
 is block-rank-$1$.
\end{definition}

Clearly, a \multblrkone/ matrix is \modblrkone/;
we will often use the contrapositive:
every non-\modblrkone/ matrix is non-\multblrkone/.
We also note that for a nonnegative real matrix $\AA \in \bbR^{m \times n}$,
$\AA$ is block-rank-$1$ iff \multblrkone/ iff $\AA$ is \modblrkone/.
%\blrkone/ iff \multblrkone/ iff $\AA$ is \modblrkone/.

Next, it is easy to see that the following hold:
\begin{enumerate}
\item $\AA \in \bbC^{m \times n}$
is not block-rank-$1$
%\blrkone/
iff either $\AA$ is not rectangular or
$\AA$ is rectangular but for some $1 \le i < j \le m$
and $1 \le k < \ell \le n$,
the $2 \times 2$ submatrix $\AA_{i, j; k, \ell}$
in (\ref{eqn:2by2-submatrix})
%\[
%\AA_{i, j; k, \ell} =
%\begin{pmatrix}
%A_{i, k} & A_{i, \ell} \\
%A_{j, k} & A_{j, \ell}
%\end{pmatrix}
%\]
has no zero entries and is nondegenerate.
This is equivalent to saying that
$\AA \in \bbC^{m \times n}$ is not block-rank-$1$
%\blrkone/
iff for some  $1 \le i < j \le m$
and $1 \le k < \ell \le n$,
the $2 \times 2$ submatrix $\AA_{i, j; k, \ell}$
%\[
%\AA_{i, j; k, \ell} =
%\begin{pmatrix}
%A_{i, k} & A_{i, \ell} \\
%A_{j, k} & A_{j, \ell}
%\end{pmatrix}
%\]
has at most one zero entry and is nondegenerate.

%%%In this case we say that  $(i, j, k, \ell)$
%%%witnesses the fact that $\AA$ is not \blrkone/.
%%%%or the non-\blrkone/-ness of $\AA$.

\item $\AA \in \bbC^{m \times n}$
is not \multblrkone/
iff either $\AA$ is not rectangular or
$\AA$ is rectangular but for some $1 \le i < j \le m$
and $1 \le k < \ell \le n$,
the $2 \times 2$ 
 submatrix $\AA_{i, j; k, \ell}$
in (\ref{eqn:2by2-submatrix})
%\[
%\AA_{i, j; k, \ell} =
%\begin{pmatrix}
%A_{i, k} & A_{i, \ell} \\
%A_{j, k} & A_{j, \ell}
%\end{pmatrix}
%\]
has no zero entries and its $p$th Hadamard power is nondegenerate
 for all $p \ge 1$.
This is equivalent to saying that
$\AA \in \bbC^{m \times n}$ is not \multblrkone/
iff for some  $1 \le i < j \le m$
and $1 \le k < \ell \le n$,
the $2 \times 2$ submatrix $\AA_{i, j; k, \ell}$
%\[
%\AA_{i, j; k, \ell} =
%\begin{pmatrix}
%A_{i, k} & A_{i, \ell} \\
%A_{j, k} & A_{j, \ell}
%\end{pmatrix}
%\]
has at most one zero entry and
$(\AA_{i, j; k, \ell})^{\odot p}$
is nondegenerate for all $p \ge 1$.

%%In this case we say that  $(i, j, k, \ell)$
%%witnesses the fact that $\AA$ is not \multblrkone/.
%%%or the non-\multblrkone/-ness of $\AA$.

\item $\AA \in \bbC^{m \times n}$
is not \modblrkone/
iff either $\AA$ is not rectangular or
$\AA$ is rectangular but for some $1 \le i < j \le m$
and $1 \le k < \ell \le n$,
the $2 \times 2$ submatrix
%\[
%\AA_{i, j; k, \ell} =
%\begin{pmatrix}
%A_{i, k} & A_{i, \ell} \\
%A_{j, k} & A_{j, \ell}
%\end{pmatrix}
%\]
%has no zero entries and 
\[
|\AA|_{i, j; k, \ell} =
\begin{pmatrix}
|A_{i, k}| & |A_{i, \ell}| \\
|A_{j, k}| & |A_{j, \ell}|
\end{pmatrix}
\]
%{\color{red} of $(|A_{i, j}|)_{i, j = 1}^m$ (may add, btw is the notation $|\AA|_{i, j; k, \ell} $ fine?)}
has no zero entries and
is nondegenerate.
This is equivalent to saying that
$\AA \in \bbC^{m \times n}$ is not \modblrkone/
iff for some  $1 \le i < j \le m$
and $1 \le k < \ell \le n$,
the $2 \times 2$ submatrix $|\AA|_{i, j; k, \ell}$
%\[
%\AA_{i, j; k, \ell} =
%\begin{pmatrix}
%A_{i, k} & A_{i, \ell} \\
%A_{j, k} & A_{j, \ell}
%\end{pmatrix}
%\]
%has at most one zero entry and the matrix
%\[
%\begin{pmatrix}
%|A_{i_1, j_1}| & |A_{i_1, j_2}| \\
%|A_{i_2, j_2}| & |A_{i_2, j_2}|
%\end{pmatrix}
%\]
has at most one zero entry and
is nondegenerate.

%%In this case we say that  $(i, j, k, \ell)$
%%witnesses the fact that $\AA$ is not \modblrkone/.
%%%or the non-\modblrkone/-ness of $\AA$.
\end{enumerate}

%In each of the respective cases, we say that  $(i, j, k, \ell)$
%witnesses the fact that $\AA$ is not \blrkone/,
%or  not \multblrkone/,
%or
%not \modblrkone/.

In each case, we say that  $(i, j, k, \ell)$
witnesses the respective property.
%{\color{red} (JYC: may not need it)}

The statement regarding non-\multblrkone/ requires a slight
justification. Suppose $\AA$ is non-\multblrkone/ but rectangular.
Then for all $p \ge 1$,
$\AA^{\odot p}$ is not block-rank-$1$.
%\blrkone/.
Thus there exists  $(i, j, k, \ell)$ (depending on $p$)
such that $\AA_{i, j; k, \ell}$ has no zero entries and $(\AA_{i, j; k, \ell})^{\odot p}$
is nondegenerate.
If for every $(i, j, k, \ell)$ for which $\AA_{i, j; k, \ell}$ has no zero entries,
there exists some $p \ge 1$ (depending on $(i, j, k, \ell)$)
such that $(\AA_{i, j; k, \ell})^{\odot p}$ is
degenerate, then $(\AA_{i, j; k, \ell})^{\odot pq}$
is also degenerate for all $q \ge 1$.
Thus if we take a common multiplier of these $p$'s we reach a contradiction.

Any $\AA = (A_{i, j})_{i, j = 1}^m \in \bbC^{m \times m}$ has finitely many
entries, so they are contained in a finitely generated field
$K = \bbQ(\{A_{i, j}\}_{i, j = 1}^m)$ over $\bbQ$.
By Corollary~\ref{cor:fin-gen-field-roots-of-unity-full-fin-cyc-gr},
the roots of unity in $K$ form a finite cyclic group.
Let  $R$ be (or any positive multiple of) the  order
 of this finite cyclic group, and let $\BB \in K^{m \times m}$
with entries from $K$.
Then we have the following.

\begin{lemma}\label{lem:mubr1-br1-equivalence}
$\BB$ is \multblrkone/ iff $\BB^{\odot R}$ is block-rank-1.
% \blrkone/.
\end{lemma}
\begin{proof}
We prove the forward implication; the converse is trivial.
Assume $\BB$ is \multblrkone/.
Then $\BB$ is rectangular, and hence so is $\BB^{\odot R}$.
By definition, there exists  $N \ge 1$ such that $\BB^{\odot N}$ is block-rank-$1$.
% \blrkone/.
We show that $\BB^{\odot R}$ is block-rank-$1$.
%\blrkone/.
%Clearly, $\BB^{\odot R}$ is rectangular.
%More generally, but $\BB$, $\BB^N$ and $\BB^R$ have the same support.
It suffices to show that for any  $i < j$ and $k < \ell$,
if
\[
\BB_{i, j; k, \ell} =
\begin{pmatrix}
B_{i, k} & B_{i, \ell} \\
B_{j, k} & B_{j, \ell}
\end{pmatrix}
\]
has no zero entries,
then $(\BB_{i, j; k, \ell})^{\odot R}$ has rank $1$.
Since $(\BB_{i, j; k, \ell})^{\odot N}$ has rank $1$,
we have $B_{i, k}^N B_{j, \ell}^N = B_{i, \ell}^N B_{j, k}^N$.
%So $((B_{i, k} B_{j, \ell}) / (B_{i, \ell} B_{j, k}))^N = 1$.
%Therefore
So $(B_{i, k} B_{j, \ell}) / (B_{i, \ell} B_{j, k})$ is a root of unity
in $K$; thus it belongs to a cyclic group of order  $R$ (or of order dividing $R$).
It follows that
% $((B_{i, k} B_{j, \ell}) / (B_{i, \ell} B_{j, k}))^R = 1$.
$B_{i, k}^R B_{j, \ell}^R = B_{i, \ell}^R B_{j, k}^R$.
\end{proof}
%\begin{lemma}\label{lem:mubr1-br1-equivalence}
%$\AA$ is \multblrkone/ iff $\AA^{\odot R}$ is block-rank-1.
%% \blrkone/.
%\end{lemma}
%\begin{proof}
%We prove the forward implication; the converse is trivial.
%Assume $\AA$ is \multblrkone/.
%Then $\AA$ is rectangular, and hence so is $\AA^{\odot R}$.
%By definition, there exists  $N \ge 1$ such that $\AA^{\odot N}$ is block-rank-$1$.
%% \blrkone/.
%We show that $\AA^{\odot R}$ is block-rank-$1$.
%%\blrkone/.
%%Clearly, $\AA^{\odot R}$ is rectangular.
%%More generally, but $\AA$, $\AA^N$ and $\AA^R$ have the same support.
%It suffices to show that for any  $i < j$ and $k < \ell$,
%if
%\[
%\AA_{i, j; k, \ell} =
%\begin{pmatrix}
%A_{i, k} & A_{i, \ell} \\
%A_{j, k} & A_{j, \ell}
%\end{pmatrix}
%\]
%has no zero entries,
%then $(\AA_{i, j; k, \ell})^{\odot R}$ has rank $1$.
%Since $(\AA_{i, j; k, \ell})^{\odot N}$ has rank $1$,
%we have $A_{i, k}^N A_{j, \ell}^N = A_{i, \ell}^N A_{j, k}^N$.
%%So $((A_{i, k} A_{j, \ell}) / (A_{i, \ell} A_{j, k}))^N = 1$.
%%Therefore
%So $(A_{i, k} A_{j, \ell}) / (A_{i, \ell} A_{j, k})$ is a root of unity
%in $K$, {\color{red};} thus {\color{red} it} belongs to a cyclic group of order  $R$ (or of order dividing $R$).
%It follows that
%% $((A_{i, k} A_{j, \ell}) / (A_{i, \ell} A_{j, k}))^R = 1$.
%$A_{i, k}^R A_{j, \ell}^R = A_{i, \ell}^R A_{j, k}^R$.
%\end{proof}
%\begin{remark}
%It is easy to see that the statement of Lemma~\ref{lem:mubr1-br1-equivalence}
%holds with $R$ replaced by $R r$ where $r \ge 1$.
%{\color{red} (for constructivity, needed if $R$ can't be computed)}
%\end{remark}

The following statements can be easily checked.
\begin{property}\label{prop:equiv-direct-sum-mbrk1}
Let $\AA_1, \ldots, \AA_n$ be complex square matrices (where $n \ge 1$).
Then $\diag(\AA_1, \ldots, \AA_n)$ is \multblrkone/
iff each $\AA_i$ is \multblrkone/.
The same statement holds with \modblrkone/.
\end{property}
%\begin{lemma}\label{lem:equiv-direct-sum-mbrk1}
%Let $\AA, \BB$ be complex matrices.
%Then
%\[
%\diag(\AA, \BB) = \begin{pmatrix} \AA & 0 \\ 0 & \BB \end{pmatrix}
%\]
%is \multblrkone/ (\modblrkone/)
%iff both $\AA, \BB$ are \multblrkone/ (\modblrkone/).
%\end{lemma}

\begin{property}\label{prop:equiv-direct-sum-mbrk1-perm}
Let $\AA, \BB$ be complex matrices.
Then $\begin{psmallmatrix} \00 & \AA \\ \BB & \00 \end{psmallmatrix}$
%\[
%\begin{pmatrix} \00 & \AA \\ \BB & \00 \end{pmatrix}
%\]
is \multblrkone/
iff both $\AA, \BB$ are \multblrkone/.
The same statement holds with \modblrkone/.
\end{property}

\begin{property}\label{prop:equiv-tensor-product-mbrk1}
Let $\AA, \BB$ be complex matrices.
Then $\AA \otimes \BB$ is \multblrkone/ 
iff either $\AA$ or $\BB$ is a zero matrix,
or both $\AA$ and $\BB$ are \multblrkone/.
The same statement holds with \modblrkone/.
\end{property}

\begin{lemma}\label{lem:edge-gadget-component-nmrk1}
Let $(\CC, \fD)$ be as in Definition~\ref{def:EVAL(C,frakD-modN)}
and $\Gamma$ be an edge gadget such that $\gm_{\Gamma, \CC, \fD}$ is not \multblrkone/. %(not \modblrkone/).
% (not \modblrkone/).
Then the distinguished vertices $\firstdv, \seconddv$ of $\Gamma$ are in 
the same connected component $\Gamma_0$ of $\Gamma$.
%Moreover, if $\Gamma_0$ is this connected component viewed as a gadget,
% with the same pair of distinguished vertices in the same order),
%then 
Furthermore $\gm_{\Gamma_0, \CC, \fD}$ is not \multblrkone/. %(not \modblrkone/).
% (not \modblrkone/).
%%% since \modblrkone/ ==> \multblrkone/. i think no need to have this
%% case spelled out any more.  (ie transition from 
% $\gm_{\Gamma, \CC, \fD}$ is not \moltblrkone/ to
% $\gm_{\Gamma_0, \CC, \fD}$ is not \multblrkone/. is enough
%{\color{red} (added not \modblrkone/ because it's used in the nonbipartite case)}
%{\color{red} (The same holds with \modblrkone/. (or say instead))}
The same holds with not \modblrkone/.
\end{lemma}
\begin{proof}
If $\firstdv, \seconddv$
are in different connected components of $\Gamma$,
then $\rank \gm_{\Gamma, \CC, \fD} \le 1$ by Lemma~\ref{lem:dif-components-rk1}
and therefore $\gm_{\Gamma, \CC, \fD}$ is \multblrkone/ (\modblrkone/),
%(not \modblrkone/)
which is a contradiction.
For the second part, it suffices to notice that
the other connected components of $\Gamma$, if there is any, %{\color{red} (if there is any)}
contribute a scalar multiplier to $\gm_{\Gamma, \CC, \fD}$.
This scalar multiplier is nonzero because 
otherwise $\gm_{\Gamma, \CC_, \fD} = 0$, clearly a contradiction.
%More precisely, this is because if $\Gamma_1, \ldots, \Gamma_s$ ($s \ge 0$)
%are the other connected components of $\Gamma$, then
%\[
%\gm_{\Gamma, \CC, \fD} = \left( \prod_{k = 1}^s Z_{\CC, \fD}(\Gamma_k) \right) \gm_{\Gamma_0, \CC, \fD}
%\]
%so $\prod_{k = 1}^s Z_{\CC, \fD}(\Gamma_k)$ is this scalar multiplier which must be nonzero.
%%Then we see that
%Because $\gm_{\Gamma_0, \CC, \fD}$
%differs from $\gm_{\Gamma, \CC, \fD}$ by
%%this nonzero multiplier
%%and so
%we also conclude that $\gm_{\Gamma_0, \CC, \fD}$ is not \multblrkone/ (not \modblrkone/)
%which proves the second part.
\end{proof}
\begin{remark}
%{\color{red} (Not sure if I talking about $\eval(\AA)$ captures gadgetry, which is why it's formulated like this)}
%
%
%Again, note that 
Since $\eval(\AA)$ is a special case of $\eval(\CC, \fD)$
with  every $\DD^{[i]}$ being the identity matrix,
Lemma~\ref{lem:edge-gadget-component-nmrk1}
applies to $\EVAL(\AA)$.
%(which includes $\gm_{\Gamma, \AA}$ being the same as $\gm_{\Gamma, \CC, \fD}$ for every edge gadget $\Gamma$)
%in which every $\DD^{[i]}$ is the identity matrix,
%a similar statement to Lemma~\ref{lem:edge-gadget-component-nmrk1}
%%holds for the signature matrices of the form $\gm_{\Gamma, \AA}$ (of edge gadgets the framework $\EVAL(\AA)$).
\end{remark}

\subsection{Purification}\label{subsec:purification-ours}

As described in Introduction (Section~\ref{sec:intro}), %Section~\ref{sec:intro}
an important first step in the proof of the dichotomy in~\cite{Cai-Chen-Lu-2013}
is a process called \emph{purification}. In this paper we cannot 
directly substitute $\AA$ by its purification $\pur \AA$ because that
reduction in~\cite{Cai-Chen-Lu-2013}
does not preserve the complexity in the bounded degree case.
However, purification is still important for the proof in this paper.

%Let  $\mathscr{A}= \{A_{i,j}:i,j\in [m] \}$ denote the finite set of algebraic numbers
%  from the entries of $\AA \in \mathbb{C}^{m\times m}$.
%
\begin{definition}[Definition 7.2 from~\cite{Cai-Chen-Lu-2013}]\label{def-generating-ours}
Let $\mathscr{A}=\{a_1,\ldots,a_n\}$ be a set of $n$ nonzero algebraic numbers
  for some $n \ge 1$.
We say $\{g_1,\ldots,g_d\}$ for some $d \ge 0$
 is a \emph{generating set} of $\mathscr{A}$ if
\begin{enumerate}
\item every $g_i$ is a nonzero algebraic number in $\mathbb{Q}(\mathscr{A})$; and
\item for every $a\in \mathscr{A}$, there exists a \emph{unique}
  tuple $(k_1,\ldots,k_d)\in
  \mathbb{Z}^d$ such that
\[
\frac{a}{g_1^{k_1}\cdots g_d^{k_d}}\ \text{is a root of unity.}
\]
\end{enumerate}
\end{definition}
Clearly $d=0$ iff the set $\mathscr{A}$ consists of roots of unity only.
It follows from the definition that
$g_1^{k_1}\cdots g_d^{k_d}$ of
  any nonzero $(k_1,\ldots,k_d)\in \mathbb{Z}^d$ cannot be a root of unity.
%We prove the following lemma. {\color{red} (replace by) 
The following lemma is Lemma 7.3 from~\cite{Cai-Chen-Lu-2013}.
%}
%
\begin{lemma}%[{\color{red}Lemma 7.3 from~\cite{Cai-Chen-Lu-2013}}]
\label{generating-set-exists-ours}
Every set $\mathscr{A}$ %=\{a_j\}_{j\in [n]}$
  of nonzero algebraic numbers has a generating set.
\end{lemma}

We apply Lemma~\ref{generating-set-exists-ours}
to  $\mathscr{A}= \{A_{i,j}:i,j\in [m], A_{i,j}\ne 0 \}$,
 the set of nonzero 
  entries of $\AA \in \mathbb{C}^{m\times m}$.
The purification matrix $\pur \AA$ of $\AA$ is constructed by essentially
 replacing every $g_i$ by the $i$th smallest prime $p_i$.
More precisely, we assume a particular generating set
 $(g_1, \ldots, g_d)$ has been chosen for $\mathscr{A}$.
% (although there are no unique
%canonical choice).   all this will be discussed later when discuss decidability etc.
Let $p_1< \ldots <p_d$ denote the $d$ smallest primes.
For every $i,j\in [m]$,
  let ${\pur A}_{i,j}=0$ if $A_{i,j}=0$.
Suppose  $A_{i,j} \not = 0$.
  Let $(k_1,\ldots,k_d)$ be the  unique tuple of integers
such that
\[
\zeta_{i,j}=\frac{A_{i,j}}{g_1^{k_1}\cdots g_d^{k_d}}\]
{is a root of unity}.
Then we define ${\pur A}_{i,j} =p_1^{k_1}\cdots p_d^{k_d}\cdot \zeta_{i,j}$.
%$B_{i,j}$ is well-defined by the uniqueness of
%$(k_1,\ldots,k_d)\in\mathbb{Z}^d$;
By taking the prime
factorization of $|{\pur A}_{i,j}|$ we can recover $(k_1,\ldots,k_d)$
uniquely, and recover $A_{i,j}$ by
\[A_{i,j} =
 g_1^{k_1}\cdots g_d^{k_d}\cdot  \frac{\pur A_{i,j}}{p_1^{k_1}\cdots p_d^{k_d}}.\]
This matrix ${\pur \AA}$ will be called the \emph{purification} of $\AA$ 
obtained by going from $(g_1, \ldots, g_d)$ to $(p_1, \ldots, p_d)$. %{\color{red} (emphasize starting from the purification?)}
We can also adapt this purification process to be applied to
any finite set of algebraic numbers.
%% this take care of  doing it  for a vector, 
%(as in your old def 10.4 (a_1...) --> (b_1...)
% or doing it for a factororized form of a matrix as in the regulartized form.

Clearly, $\AA$ is connected iff $\pur \AA$ is connected;
$\AA$ is rectangular iff $\pur \AA$ is rectangular.
% make the next sentence a part of the lm
%and $\AA$ is \multblrkone/ iff $\pur \AA$ is \multblrkone/.
Furthermore, we have the following lemma.

\begin{lemma}
 $\AA$ is \multblrkone/ iff  $\pur \AA$ is \multblrkone/
iff  $\pur \AA$ is \modblrkone/.
\end{lemma}
\begin{proof}
We may assume $\AA$ is rectangular for otherwise all sides are false.
%%% otherwise both sides are false.
Suppose  $\AA^{\odot N}$ is block-rank-$1$ for some $N \ge 1$.
Then for any  $2 \times 2$ submatrix
$\begin{psmallmatrix} a & b \\ c & d \end{psmallmatrix}$ of $\AA$ without zero entries,
we have $a^Nd^N = b^Nc^N$.
We can write it in terms of the generators $g_1, \ldots, g_d$.
Replacing $a$ by $a' \zeta $ where $a'$ is
 a product of integer powers of $g_1, \ldots, g_d$, and similarly for
$b,c$ and $d$, we get
 %$(a,b,c,d)$ by $(a'\zeta_1, b'\zeta_2, c'\zeta_3, d'\zeta_4)$
%where each $a', b', c', d'$ is a product of integer powers of $g_1, \lgots, g_d$,
$(a'^Nd'^N)/(b'^Nc'^N)$ is a root of unity, and thus so is
$(a'd')/(b'c')$. By the property of generators, $(a'd')/(b'c') =1$.
%\sout{By purification of $\AA$ to $\pur \AA$  going from $(g_1, \ldots, g_d)$
%to primes $(p_1, \ldots, p_d)$}
Since $\pur \AA$ is the purification of $\AA$ obtained by going from $(g_1, \ldots, g_d)$
to the primes $(p_1, \ldots, p_d)$, we get $\pur \AA$ is  \multblrkone/ and \modblrkone/.
%\sout{In the other direction, assume $\pur \AA$ is  \multblrkone/, thus also  \modblrkone/.}
For the other directions, it is easy to see that for $\pur \AA$,
the statements that $\pur \AA$ is \modblrkone/ and \multblrkone/ are equivalent,
so we can assume they both hold.
%\sout{Then $(a'd')/(b'c') =1$, and $(ad)/(bc)$ is
%a  root of unity of some order. Taking $N$ to be
%a common multiple of these orders, we have $\AA^{\odot N}$ is block-rank-$1$.}
Following the previous substitution procedure,
for any  $2 \times 2$ submatrix
$\begin{psmallmatrix} a & b \\ c & d \end{psmallmatrix}$ of $\AA$ without zero entries,
we have $(a'd')/(b'c') =1$, and $(ad)/(bc)$ is
a  root of unity of some order. Taking $N$ to be
a common multiple of these orders, we have $\AA^{\odot N}$ is block-rank-$1$.
\end{proof}

The next lemma shows that, at least for unbounded degree
graphs, the purification replacement does not affect the complexity.
We emphasize that the stated equivalence in the next lemma
\emph{is not} claimed for bounded degree graphs. %\emph{does not} claim 
%{\color{red} (JYC: one point, maybe we should say that the proof hold for not necessarily connected matrices)}
%
\begin{lemma}[Lemma 7.4 from~\cite{Cai-Chen-Lu-2013}]\label{lem:eval-A-B-equivalence}
%%% what do you mean by "extended"?
%[Lemma 7.4 from~\cite{Cai-Chen-Lu-2013} extended]\label{lem:eval-A-B-equivalence}
Let $\AA\in\mathbb{C}^{m\times m}$ be a symmetric matrix with algebraic entries.
  %and let $\BB$ be the $m\times m$ matrix constructed above.
Then $\eval(\pur \AA)\equiv \eval(\AA).$
\end{lemma}

%The following corollary will be used to prove the 
%stronger Theorem~\ref{thm:non-mult-bl-rk-1-hardness-sym}.

%%%%%%%%%%%%%%%%%%%%%%%%%%%%%%%%%%%%%%%%%%%%%%%%%%%%%%%%%%%%%%%%%%%%
\begin{comment}

Next, will give a reorganized proof of an important statement which will be used
in the proof of the stronger Theorem~\ref{thm:non-mult-bl-rk-1-hardness-sym}.
We note that this statement follows from the proof of the main dichotomy in~\cite{Cai-Chen-Lu-2013}.
However, in order to get this statement by following the proof~\cite{Cai-Chen-Lu-2013},
we first need to apply a nonconstructive step such as the first pinning lemma (Lemma 4.1 from~\cite{Cai-Chen-Lu-2013})
to reduce to connected components (Section 4) and we would like to avoid it.

\end{comment}
%%%%%%%%%%%%%%%%%%%%%%%%%%%%%%%%%%%%%%%%%%%%%%%%%%%%%%%%%%%%%%%%%%%%

The following corollary will be used to prove the
stronger Theorem~\ref{thm:non-mult-bl-rk-1-hardness-sym}.
Note that the stated \#P-hardness of $\EVAL(\AA)$ in
in this corollary is for graphs %Corollary~\ref{cor:non-mubr1-hard-unbounded}}
without degree bound.

%%% is A here meant to be symmetric???
%%% if not claiming for unsymmetric, then 
%%% Theorem~\ref{thm:non-mult-bl-rk-1-hardness-sym}. is not stronger

\begin{corollary}\label{cor:non-mubr1-hard-unbounded}
If a symmetric matrix $\AA \in \bbC^{m \times m}$ is not \multblrkone/,
then $\EVAL(\AA)$ is \#P-hard.
% {\color{red} (unbounded degree)}
\end{corollary}
\begin{proof}
By Lemma~\ref{lem:eval-A-B-equivalence}, $\EVAL(\AA) \equiv \EVAL(\pur \AA)$
(for graphs with unbounded degree). % {\color{red} for unbounded degree graphs}
Let $N$ be  the least common multiple of the orders of $\zeta_{i j}$
for which $\AA_{i,j} \ne 0$.
% (equiv. $\BB_{i, j} \ne 0$).
Since $\AA$ is not \multblrkone/, $\pur \AA$ is also not \multblrkone/.
Then ${\pur \AA}^{\odot N}$ is not \multblrkone/,
and since ${\pur \AA}^{\odot N}$  in nonnegative,
 it is not block-rank-$1$.
% \blrkone/.
%Besides, $\BB^{\odot N}$ is a nonnegative matrix
By the Bulatov-Grohe dichotomy, Theorem~\ref{BulatovGrohe},
%{basicsharp-ours},
$\EVAL({\pur \AA}^{\odot N})$ is \#P-hard.
We note that $Z_{{\pur \AA}^{\odot N}}(G) = Z_{{\pur \AA}}(T_N(G))$ 
for any graph $G = (V, E)$ and its $N$-thickening $T_N(G)$,
which proves that $\EVAL({\pur \AA}^{\odot N}) \le \EVAL({\pur \AA})$.
It follows that $\EVAL({\pur \AA})$ and so $\EVAL(\AA)$ is \#P-hard.
\end{proof}

\subsection{Vandermonde Argument}\label{subsec:Vandermonde}
We start with an exceedingly simple lemma, which  ultimately 
underlies a lot of our algebraic reasonings in this paper.
We will call this lemma and its corollaries
the \emph{Vandermonde Argument}.

\begin{lemma}\label{lem:simpleVandermonde}
%Fix any field $\bbF$.
%Since the next lemmas are are all for C. i changed  $\bbF$ to C.
Let $n \ge 0$, and $a_i, x_i \in \bbC$ for $1 \le i \le n$.
If
\begin{equation}\label{eq:simple-Vandermonde-conditoon}
\sum_{i = 1}^n a_i x_i^j = 0, ~~~~\text{for all}~~ 0 \le j < n,
\end{equation}
then for any function $f \colon \bbC \to \bbC$, we have
$\sum_{i = 1}^n a_i f(x_i) = 0$.
If (\ref{eq:simple-Vandermonde-conditoon}) is true
for $1 \le j \le n$, then the same conclusion holds
for any  function $f$ satisfying $f(0) =0$.
\end{lemma}

\begin{proof}
The statement is vacuously true if $n=0$, since an empty
sum is 0.
Assume $n \ge 1$.
We partition $[n]$ into a disjoint union $\bigcup_{\ell=1}^{p} I_\ell$
such that $i, i'$ belong to
the same $I_\ell$ iff $x_i = x_{i'}$. Then (\ref{eq:simple-Vandermonde-conditoon})
is a Vandermonde system of rank $p$ with a solution
$(\sum_{i \in I_\ell} a_i)_{\ell \in [p]}$.
Thus $\sum_{i \in I_\ell} a_i =0$ for all $1 \le \ell \le p$.
It follows that $\sum_{i = 1}^n a_i f(x_i) = 0$
for any function $f \colon \bbC \to \bbC$.
If  (\ref{eq:simple-Vandermonde-conditoon})
is true for  $1 \le j \le n$, then the same proof works except
when some $x_i=0$. In that case, we can separate out
the term $\sum_{i \in I_{\ell_0}} a_i$ for the unique $I_{\ell_0}$ that contains
such $i$, and we get a Vandermonde system of rank $p-1$ on the other terms
$(\sum_{i \in I_\ell} a_i)_{\ell \in [p], \ell \ne \ell_0}$, which must be all zero.
\end{proof}

The next lemma is a multivariate version of Lemma~\ref{lem:simpleVandermonde}. %{\color{red} (JYC: not entirely true, we need to mention conjugations)}

\begin{lemma}\label{lem:vand-conj-simp}
Let $m \ge 0$,   $g(x_1, \ldots, x_m) = \sum_{(i_1, \ldots, i_m) \in I} a_{i_1, \ldots, i_m} \prod_{j = 1}^m x_j^{i_j} \in \bbC[x_1, \ldots, x_m]$
and $\lambda_1, \ldots, \lambda_m \in \bbC$.
If $g(\lambda_1^k, \ldots, \lambda_m^k) = 0$ for $1 \le k \le |I|$,
then $g(\overline{\lambda_1}, \ldots, \overline{\lambda_m}) = 0$.
\end{lemma}
\begin{proof}
If $m = 0$, then $g$ is a constant polynomial.
So if $g \ne 0$ then $|I| = 1$ and the condition at $k=1$ is non-vacuous, %If
which leads to $g = 0$.
Let $m \ge 1$.
%%%
We have
\[
\sum_{(i_1, \ldots, i_m) \in I} a_{i_1, \ldots, i_m} (\prod_{j = 1}^m \lambda_j^{i_j})^k = 0
\]
for $1 \le k \le |I|$.
Let $f \colon \bbC \to \bbC$ be  the conjugation function
$f(z) = \overline z$ for $z \in \bbC$.
Note that $f(0) = 0$.
Then by 
Lemma~\ref{lem:simpleVandermonde},
we have
\begin{align*}
g(\overline{\lambda_1}, \ldots, \overline{\lambda_m}) &= \sum_{(i_1, \ldots, i_m) \in I} a_{i_1, \ldots, i_m} \prod_{j = 1}^m (\overline{\lambda_j})^{i_j} = \sum_{(i_1, \ldots, i_m) \in I} a_{i_1, \ldots, i_m} \overline{\left(\prod_{j = 1}^m \lambda_j^{i_j}\right)} \\
&= \sum_{(i_1, \ldots, i_m) \in I} a_{i_1, \ldots, i_m} f(\prod_{j = 1}^m \lambda_j^{i_j}) = 0.
\end{align*}
\end{proof}

\begin{corollary}\label{lem:vand-conj}
Let $m \ge 0$, $g_i(x_1, \ldots, x_m) \in \bbC[x_1, \ldots, x_m]$ 
where $1 \le i \le n$,
let $g(x_1, \ldots, x_m) = \prod_{i = 1}^n g_i(x_1, \ldots, x_m)$.
Let $\lambda_1, \ldots, \lambda_m \in \bbC$.
Assume $g_i(\overline{\lambda_1}, \ldots, \overline{\lambda_m}) \ne 0$ for 
all $1 \le i \le n$.
Then for some $1 \le k \le |I|$, here $|I|$ is the number of terms 
in $g$,
we have $g_i(\lambda_1^k, \ldots, \lambda_m^k) \ne 0$ for all $1 \le i \le n$. %{\color{red} where or here? Just to check with you, I like here more actually}
\end{corollary}
\begin{proof}
%Suppose the contrary.
Suppose otherwise. Then for any $1 \le k \le |I|$, there exists $1 \le i \le n$
such that $g_i(\lambda_1^k, \ldots, \lambda_m^k) = 0$.
%Let $g(x_1, \ldots, x_m) = \prod_{i = 1}^n g_i(x_1, \ldots, x_m)$.
% = \sum_{(i_1, \ldots, i_m) \in I} a_{i_1, \ldots, i_m} \prod_{j = 1}^m x_j^{i_j}$.
Then  for any $1 \le k \le |I|$, $g(\lambda_1^k, \ldots, \lambda_m^k) = 0$.
Applying Lemma~\ref{lem:vand-conj-simp},
we have $g(\overline{\lambda_1}, \ldots, \overline{\lambda_m}) = 0$
contradicting $g_i(\overline{\lambda_1}, \ldots, \overline{\lambda_m}) \ne 0$ for
all $1 \le i \le n$.
\end{proof}

\begin{definition}\label{def:mult-ind}
Let $m \ge 0$. We call $\{b_i \mid i \in [m]\}
\subset \bbC \setminus \{0\}$
\textit{multiplicatively independent} if $\prod_{j = 1}^m b_j^{i_j} = 1$
where $i_j \in \bbZ$ ($j \in [m]$) implies $i_j = 0$ for all $j \in [m]$.
\end{definition}
\begin{remark}
For $m=0$, the empty set $\{b_i \mid i \in [m]\}$
is multiplicatively independent, as the statement $i_j = 0$ for all $j \in [m]$
is vacuously true because $[m] = \emptyset$.
Also we note that the generating set for any set  of nonzero algebraic numbers
is multiplicatively independent.
\end{remark}

\begin{lemma}\label{lem:mult-ind-plus-root-uniqueness}
Let $m \ge 0$, let $ \{b_i \mid i \in [m]\}
\subset \bbC \setminus \{0\}$ be multiplicatively independent
and let $\varepsilon_N \in \bbC$ be a root of unity of order $N$ where $N \ge 1$.
Assume $\varepsilon_N^{i_0} \prod_{j = 1}^m b_j^{i_j} = 
\varepsilon_N^{i'_0} \prod_{j = 1}^m b_j^{i'_j}$
where $0 \le i_0, i'_0 < N$ and 
$i_j, i_j' \in \bbZ$ for $1 \le j \le m$.
Then $i_j = i'_j$ for $0 \le j \le m$.
\end{lemma}
\begin{proof}
Raising both sides of the given equality to the $N$th power,
we get $\prod_{j = 1}^m b_j^{N i_j} = \prod_{j = 1}^m b_j^{N i'_j}$
so $\prod_{j = 1}^m b_j^{N (i_j - i'_j)} = 1$.
Since $b_1, \ldots, b_m$ are multiplicatively independent,
we get $N (i_j - i'_j) = 0$, so $i_j = i'_j$ for $1 \le j \le m$.
Then from the original equality
we conclude that $\varepsilon_N^{i_{0}} = \varepsilon_N^{i'_{0}}$.
But $\varepsilon_N$ is a root of unity of order $N$ and $0 \le i_{0}, i'_{0} < N$
so $i_0 = i'_0$.
\end{proof}
%%%

\begin{lemma}\label{lem:vand-mult-ind-simp}
Let $m \ge 0$ and $g(x_0, x_1, \ldots, x_{m}) = \sum_{(i_0, \ldots, i_{m}) 
\in I} a_{i_0, \ldots, i_{m}} \prod_{j = 0}^{m} x_j^{i_j} \in \bbC[x_0^{\pm 1}, \ldots, x_{m}^{\pm 1}]$.
Next, let $b_1, \ldots, b_m \in \bbC \setminus \{ 0 \}$ be multiplicatively independent
and $\varepsilon_N \in \bbC$ be a root of unity of order $N$ where $N \ge 1$.
If $g(\varepsilon_N^k, b_1^k, \ldots, b_m^k) = 0$ for $1 \le k \le |I|$,
then for any $c_1, \ldots, c_m \in \bbC$ and $t \in \bbZ$, $g(\varepsilon_N^t,
c_1, \ldots, c_m) = 0$.
\end{lemma}
\begin{proof}
We have
\[
\sum_{(i_0, \ldots, i_{m}) \in I} a_{i_0, \ldots, i_{m}} 
\left( \varepsilon_N^{i_0} \prod_{j = 1}^m b_j^{i_j}\right)^k = 0
\]
for $1 \le k \le |I|$.
Let $f \colon \bbC \to \bbC$ be defined as
 $f( \varepsilon_N^{i_0} \prod_{j = 1}^m b_j^{i_j}) 
= \varepsilon_N^{ t i_0} \prod_{j = 1}^m c_j^{i_j}$,
%{\color{red} ($f(x) = (\prod_{j = 1}^m c_j^{i_j}) \varepsilon_N^{t i_{m + 1}}$ if $x = (\prod_{j = 1}^m b_j^{i_j}) \varepsilon_N^{i_{m + 1}}$)}
where $0 \le i_0 < N$ and $i_j \in \bbZ$ for $1 \le j \le m$,
and $f(x) = 0$ if $x$ is not of this form.
% {\color{red} (and define $f$ to be $0$ everywhere else)}.
Since $b_1, \ldots, b_m$ are multiplicatively independent,
Lemma~\ref{lem:mult-ind-plus-root-uniqueness} implies that
 $f$ is well-defined.
Note that $f(0) = 0$.
Then by 
%the remark after the statement of 
Lemma~\ref{lem:simpleVandermonde}
% {\color{red} (Then by Lemma~\ref{lem:simpleVandermonde})},
we have
%\begin{align*}
%g(c_1, \ldots, c_m, \varepsilon_N) &= \sum_{(i_1, \ldots, i_m) \in I} a_{i_1, \ldots, i_m} (\prod_{j = 1}^m c_j^{i_j}) \varepsilon_N^{i_{m + 1}} = \sum_{(i_1, \ldots, i_m) \in I)} a_{i_1, \ldots, i_m} f((\prod_{j = 1}^m b_j^{i_j}) \varepsilon_N^{i_{m + 1}}) \\
%&= \sum_{(i_1, \ldots, i_m) \in I} a_{i_1, \ldots, i_m} f((\prod_{j = 1}^m b_j^{i_j}) \varepsilon_N^{i_{m + 1}}) = 0.
%\end{align*}
\[
g(\varepsilon_N^t, c_1, \ldots, c_m) 
= \sum_{(i_0, \ldots, i_{m}) \in I} a_{i_0, \ldots, i_{m}} 
 \varepsilon_N^{t i_{0}} \prod_{j = 1}^m c_j^{i_j} 
= \sum_{(i_0, \ldots, i_{m}) \in I} a_{i_0, \ldots, i_{m}} 
f\left( \varepsilon_N^{i_0} \prod_{j = 1}^m b_j^{i_j} \right) = 0.
\]
\end{proof}

\begin{corollary}\label{cor:vand-pur-ind-simp}
Let $m \ge 0$, $b_1, \ldots, b_m \in \bbC$,
let  $\{g_1, \ldots, g_d\}$, where $d \ge 0$, be a generating set for
the set of nonzero entries of $(b_1, \ldots, b_m)$,
and let  $(c_1, \ldots, c_m)$ be the purification of $(b_1, \ldots, b_m)$
obtained by going from $(g_1, \ldots, g_d)$ to the $d$ smallest primes
 $(p_1, \ldots, p_d)$. %, here $p_1 < \ldots < p_d$. %, where
 %$(b_1, \ldots, b_m), (c_1, \ldots, c_m) \in \bbC^m$ where $m \ge 1$,
Let
\[f(x_1, \ldots, x_m) = \sum_{(i_1, \ldots, i_m) \in I} a_{i_1, \ldots, i_m} \prod_{j = 1}^m x_j^{i_j} \in \bbC[x_1, \ldots, x_m].\]
If $f(c_1, \ldots, c_m) \ne 0$, then $f(b_1^\ell, \ldots, b_m^\ell) \ne 0$ for some $1 \le \ell \le |I|$.
\end{corollary}
%%%%%%%%%%%%%%%%%%%%%%%%%%%%%%%%%%%%%%%
\begin{proof}
Clearly, $b_i = 0$ iff $c_i = 0$ for $i \in [m]$.
Let $J = \{i \in [m] \mid b_i = c_i = 0\}$.
If $J = [m]$ then the corollary is obviously true.
We may assume $[m] \setminus J \ne \emptyset$.
For each $i \in [m]$, we have the following:
if $i \in J$, then $b_i = c_i = 0$,
and if $i \notin J$, then
\[
b_i =  \zeta_i g_1^{k_{i, 1}} \cdots g_d^{k_{i, d}}, \quad c_i = \zeta_i p_1^{k_{i, 1}} \cdots p_d^{k_{i, d}},
\]
where $(k_{i, 1}, \ldots, k_{i, d}) \in \bbZ^d$ and $\zeta_i$ is a root of unity,
both of which must be then uniquely determined.
Let $N$ be the least common multiple of the orders
of all $\zeta_i$ where $i \in [m] \setminus J$,
and let $\varepsilon_N \in \bbC$ be a root of unity of order $N$.
Then for each $i \in [m] \setminus J$, we have some $0 \le k_{i,0}  <N$ such that
\[
b_i = \varepsilon_N^{k_{i,0}} g_1^{k_{i, 1}} \cdots g_d^{k_{i, d}}, 
\quad c_i = \varepsilon_N^{k_{i,0}}  p_1^{k_{i, 1}} \cdots p_d^{k_{i, d}}.
\]
Now for each $i \in [m]$, we define $h_i(y_0, y_1, \ldots, y_{d}) \in \bbC[y_0, y_1^{\pm 1}, \ldots, y_{d}^{\pm 1}]$
as follows:
if $i \in J$, we let $h_i(y_0, y_1, \ldots, y_{d}) = 0$,
and if $i \notin J$,
we let $h_i(y_0, y_1, \ldots, y_{d}) = y_0^{k_{i,0}}
 y_1^{k_{i, 1}} \ldots y_{d}^{k_{i, d}}$.
Next, let
\[
g(y_0, y_1, \ldots, y_{d}) = 
f(h_1(y_0, y_1, \ldots, y_{d}), \ldots, h_m(y_0, y_1, \ldots, y_{d})).
\]
%%%
Expanding the above expression, combining terms with equal exponents %{\color{red} of / with}
and removing all zero terms, we get
\begin{equation}\label{eq:poly-subs}
\begin{split}
g(y_0, y_1, \ldots, y_{d}) &= \sum_{(i_1, \ldots, i_m) \in I} a_{i_1, \ldots, i_m} 
\prod_{j = 1}^m (h_j(y_0, y_1, \ldots, y_{d}))^{i_j} \\
&= \sum_{(j_0, j_1, \ldots, j_d) \in I'} a'_{j_0, j_1, \ldots, j_d} 
y_0^{j_0} y_1^{j_1} \cdots y_{d}^{j_d}
\end{split}
\end{equation}
for some $I'$.
Notice that each $h_i$ is either $0$ or a ratio of two monomials,
so we have $|I'| \le |I|$.

Now we observe that
 $g(\varepsilon_N^\ell, g_1^\ell, \ldots, g_d^\ell) = f(b_1^\ell, \ldots, b_m^\ell)$ 
for all $\ell \ge 1$,
and $g(\varepsilon_N, p_1, \ldots, p_d) = f(c_1, \ldots, c_m) \ne 0$.
Since $\{g_1, \ldots, g_d\}$ is a generating set for the set of nonzero entries 
of $(b_1, \ldots, b_m)$,
$g_1, \ldots, g_d$ are multiplicatively independent.
Then by Lemma~\ref{lem:vand-mult-ind-simp},
$g(\varepsilon_N^\ell, g_1^\ell, \ldots, g_d^\ell) \ne 0$ for some $1 \le \ell \le
|I'| \le |I|$,
which completes the proof.
\end{proof}

\subsection{Exponential polynomials}\label{subsec:exp-poly}
\input{input-to-alg-pre-exponential-polynomials}

\subsection{Two technical lemmas}\label{subsec:two-technical-lemmas}
\input{input-twotechnical-statements}

%% file: input-to-alg-pre-exponential-polynomials.tex
%\section{Exponential polynomials}\label{sec:exp-poly}

We call a symbolic expression
$P(x) = \sum_{i = 1}^k \lambda_i^x p_i(x)$
 an \emph{exponential polynomial} (over $\bbC$),
where $k \ge 0$, $\lambda_i \in \bbC$ and $p_i(x) = \sum_{0 \le j < s_i} a_{i j} x^j \in \bbC[x]$ with $s_i \ge 1$,
for $1 \le i \le k$.
Here $\deg p_i(x) < s_i$ and we assume the 0 polynomial has degree $-\infty$.
% where $0 \in \bbC[x]$.
Note that we can define $P(n) = \sum_{i = 1}^k \lambda_i^n p_i(n)$ 
for integers $n \ge 0$, and if all $\lambda_i \ne 0$ this definition extends to $\bbZ$.
Clearly, all exponential polynomials over $\bbC$
form a vector space over $\bbC$.
%{\color{red} (even a commutative algebra, no need to mention)}
%
%{\color{red} (might need to replace $t + 1 \le n \le t + \sum_{i = 1}^k (s_i + 1)$
%by $n \in [t + 1 : t + \sum_{i = 1}^k (s_i + 1)]$)}
%
%{\color{red} (JYC: might be better to use $s + 1 = \sum_{i = 1}^k (s_i + 1)$, more logical than $s = \sum_{i = 1}^k (s_i + 1)$)}
%
%{\color{red} (not sure if I should say all $a_{i j} = 0$ or all $p_i(x) = 0$.}

%%%%%%%%% now p_i has s_i (symbolic) terms

\begin{lemma}\label{lem:exp-poly-nonzero-value}
%Let $\Char \bbC = 0$.
Let $t \in \bbZ$.
For any exponential polynomial $P(x)$ with all $\lambda_i \in \bbC \setminus \{ 0 \}$
and pairwise distinct,
if $P(n) = 0$ for $t < n \le t + \sum_{i = 1}^k s_i$,
then for all $1 \le i \le k$, $p_i(x) = 0$ is the zero polynomial,
i.e.,  for all  $1 \le i \le k$ and $0 \le j < s_i$ we have $a_{i j} = 0$.
\end{lemma}
\begin{proof}
%When $k = 0$, the statement is trivial.
%Assume $k \ge 1$.
We induct on $S = \sum_{i = 1}^k s_i$.
%If $\sum_{i = 1}^k s_i = 0$,
If $S =  0$,
as all $s_i \ge 1$, this is only possible when $k=0$.
So $P(x) = 0$ and the statement is trivial.
Let $S  \ge 1$, so in particular $k \ge 1$.
%$\sum_{i = 1}^k s_i \ge 1$, so in particular $k \ge 1$.
Without loss of generality,
we may assume that $s_i \ge 1$, and 
%$a_{i s_i} \ne 0$
$\deg p_i = s_i -1 \ge 0$ for all $1 \le i \le k$,
%%% can assume s_i>=1 . ow that p_i is 0, shouldn't be there . but 
% if all p_i not there, then we are back at S=0 case.
as we can discard zero terms in $p_i(x)$, or zero polynomials $p_i$;
%  the leading  corresponding terms $a_{i s_i} x^i$
%from $Q(x)$ for which $a_{i s_i} = 0$,
%thereby decreasing $\sum_{i = 1}^k (s_i + 1)$
this decreases $S = \sum_{i = 1}^k s_i$ and we are done by induction.
%(and even $k$ when some $a_{i s_i} = 0$ where $s_i = 0$)
%so we can apply the induction hypothesis.
%After this reduction, our goal is to show that $\sum_{i = 1}^k (s_i + 1) = 0$,
%(eqivalently $k = 0$) so that the case $\sum_{i = 1}^k (s_i + 1) \ge 1$ is impossible
%resulting in a contradiction.
%Assume still $\sum_{i = 1}^k (s_i + 1) \ge 1$ so $k \ge 1$.

Let $Q(x) = P(x + 1) - \lambda_1 P(x) = \sum_{i = 1}^k \lambda_i^x q_i(x)$
where $q_i(x) = \lambda_i p_i(x + 1) - \lambda_1 p_i(x)$ for $1 \le i \le k$.
Note that either  $\deg q_1 = \deg p_1 -1$ or $q_1 =0$.
%Then we can write $q_1(x) = \sum_{j = 0}^{s_1 - 1} b_{1 j} x^j$
%(if $s_1 = 1$, this sum for q_1 is empty so $q_1(x) = 0$)
%and $q_i(x) = \sum_{j = 0}^{s_i} b_{i j} x^j$ for $2 \le i \le k$.
Also $\deg q_i = \deg p_i$ for $2 \le i \le k$.

By the given condition, we have $Q(n) = P(n + 1) - \lambda_1 P(n) = 0$
for $t < n  \le (t + \sum_{i = 1}^k s_i) -1$.
%Here if $s_1 = 0$, then $q_1(x) = 0$
%so we further write $Q(x) = \sum_{i = 2}^k \lambda_i^x q_i(x)$.
%%% why is some k-1 on the upper limit of the sum creeped in???
%and use the fact that $s_1 + \sum_{i = 2}^{k - 1} (s_i + 1) = \sum_{i = 2}^k (s_i + 1)$.
%If $s_1 \ge 1$, then $s_1 + \sum_{i = 2}^{k - 1} (s_i + 1) = (s_1 - 1 + 1) +\sum_{i = 2}^{k - 1} (s_i + 1)$.
By the induction hypothesis,
$q_i(x) = 0$
% (all $b_{i j} = 0$) 
for $1 \le i \le k$.
%%%
%However if $k > 1$ then
%the leading coefficient  of $q_2$ is
%$(\lambda_2 - \lambda_1) a_{2, s_2 -1} \ne 0$, so $q_2(x) \ne 0$,
%a contradiction.
%%%
However if $k > 1$ then $\deg q_2 = \deg p_2 = s_2 - 1 \ge 0 > -\infty$
(as the leading coefficient  of $q_2$ is
$(\lambda_2 - \lambda_1) a_{2, s_2 -1} \ne 0$), so $q_2(x) \ne 0$,
a contradiction. %{\color{red} (JYC: a little alternative, not forcing it)}
%%%
Therefore $k = 1$.
And $q_1 =0$ and $\lambda_1 \ne 0$ imply that $s_1 =1$, and $p_1(x)$ is a constant.
But from $P(t+1) = 0$  we get
 $p_1(t+1) = 0$. This shows that it is the zero constant,
contradicting $\deg p_1 = s_1 -1 = 0 > -\infty$.
\end{proof}

\begin{corollary}\label{lem:exp-poly-nondeg-matrix}
%Let $\Char \bbC = 0$.
Let $s_i \ge 1$,
%%% To Artem: i changed this. from >=0 to >=1
$\lambda_i \in \bbC \setminus \{ 0 \}$ be pairwise distinct, 
 where $1 \le i \le k$. Let 
$t \in \bbZ$, and $S = \sum_{i = 1}^k s_i$.
Define
 $\MM = (M_{n, (i, j)})$ to be an
% $\sum_{i = 1}^k s_i \times \sum_{i = 1}^k s_i$ 
$S \times S$ matrix
as follows: the rows are indexed by $n$ with $t < n \le t + S$,
%\sum_{i = 1}^k s_i$,
the columns are indexed by the pairs $(i, j)$ where $1 \le i \le k$ 
and $0 \le j < s_i$
and the $(n, (i, j))$ entry of $\MM$ is $M_{n, (i, j)} = \lambda_i^n n^j$.
Then $\det \MM \ne 0$.
%$\MM$ is nondegenerate ($\det \MM \ne 0$).
\end{corollary}
\begin{proof}
Assume $\sum_{1 \le i \le k} \sum_{0 \le j < s_i} a_{i j} \MM_{*, (i, j)} = 0$
for some $a_{i j} \in \bbC$.
% ($a_{i j} \in \bbC$ for $1 \le i \le k$ and $0 \le j \le s_i$).
In other words, 
\[\sum_{i=1}^k \sum_{0 \le j < s_i} a_{i j} \lambda_i^n n^j = 0,\]
for $t < n \le t + S$.
By Lemma~\ref{lem:exp-poly-nonzero-value}, all $a_{i j} = 0$.
Hence the columns of $\MM$ are linearly independent
and therefore $\det \MM \ne 0$.
%\MM$ is nondegenerate.
\end{proof}

%% file: input-twotechnical-statements.tex
%\section{Two technical lemmas}
%\subsection{Two technical lemmas}

The following lemmas are (essentially) from  \cite{Dyer-Greenhill-2000}
(see Lemma 3.6 and Theorem 3.1).

\begin{lemma}\label{lem:ADconjA-pairwise-lin-ind}
Let $\AA$ and $\DD$ be $m \times m$ matrices,
where $\AA$ is complex symmetric
with all columns nonzero and pairwise linearly independent,
and $\DD$ is positive diagonal.
Then all columns of $\AA \DD \overline \AA$ are nonzero and pairwise linearly independent.
\end{lemma}
\begin{proof}
The case $m = 1$ is trivial. Assume $m \ge 2$. 
Let $\DD = \diag(\alpha_i)_{i = 1}^m$, and  $\PiPi =  \diag(\sqrt{\alpha_i})_{i = 1}^m$.
Then $\PiPi^2 = \DD$.
We have  $\AA \DD \overline \AA = \QQ^T \overline \QQ$, where $\QQ = \PiPi \AA$.
Then $\QQ$ has pairwise linearly independent columns.
Let $\qq_i$ denote the $i$th column of $\QQ$.
By the Cauchy-Schwartz inequality,
%\[
%|\qq_i^T \overline{\qq_j}| < \left( (\qq_i^T \overline{\qq_i}) (\qq_j^T \overline{\qq_j}) \right)^{1 / 2},
%\]
%{\color{red}
\[
| \langle \qq_i, \qq_j \rangle | < \Vert \qq_i \Vert \cdot \Vert \qq_j \Vert,
\]
%}
whenever $i \ne j$, since $\qq_i$  and $\qq_j$  are
linearly independent,  where the 2-norm $\Vert \qq \Vert = 
\sqrt{\langle \qq, \qq \rangle}$.
Then for any $1 \le i < j \le m$, the $i$th and $j$th columns of 
$\AA \DD \overline \AA$
contain a submatrix
\[
\begin{pmatrix} %used to be bmatrix
\langle \qq_i, \qq_i  \rangle & \langle \qq_i, \qq_j   \rangle\\
\langle \qq_j, \qq_i  \rangle &  \langle \qq_j, \qq_j   \rangle
%\qq_i^T \overline{\qq_i} & \qq_i^T \overline{\qq_j} \\
%\qq_j^T \overline{\qq_i} & \qq_j^T \overline{\qq_j}
\end{pmatrix},
\]
so they are linearly independent.
\end{proof}

%The following is also adapted from~\cite{Dyer-Greenhill-2000} (Theorem 3.1). 
\begin{lemma}\label{lem:ADconjA-nondeg-thick}
Let $\AA$ and $\DD$ be $m \times m$ matrices,
where $\AA$ is complex symmetric
with all columns nonzero and pairwise linearly independent, 
and $\DD$ is positive diagonal.
Then for all sufficiently large positive integers 
$p$, the matrix $\BB = (\AA \DD \overline \AA)^{\odot p}$ is nondegenerate.
%corresponding to the edge gadget in Figure~\ref{fig:ADA-to-p-gadget-advanced} is nondegenerate.
\end{lemma}
\begin{proof}
If $m = 1$, then any $p \ge 1$ works. Let $m \ge 2$.
Following the proof of Lemma~\ref{lem:ADconjA-pairwise-lin-ind},
we have
$|\langle \qq_i, \qq_j \rangle| < \Vert \qq_i \Vert \cdot \Vert \qq_j \Vert$,
 for all $1 \le i < j \le m$. Let
\[
\gamma = \max_{1 \le i < j \le m} \frac{|\langle \qq_i, \qq_j \rangle|}{\Vert \qq_i \Vert \cdot \Vert \qq_j \Vert}  < 1.
\]
Let $\AA' = \AA \DD \overline \AA = \QQ^T \overline \QQ$ so $A'_{i j} 
= \langle \qq_i, \qq_j \rangle$.
Then $|A'_{i j}| \le \gamma \sqrt{A'_{i i} A'_{j j}}$ for all $i \ne j$.
%{\color{red} (Note that all $A'_{i i} = \qq_i^T \overline{\qq_i} \in \bbR_{\ge 0}$.)}
%Consider the determinant of $\AA'$.
Each term of $\det(\AA')$ has the form
\[
\pm \prod_{i = 1}^m A'_{i \sigma(i)},
\]
where $\sigma$ is a permutation of $[m]$.
Denote $t(\sigma) = |\{ i \mid \sigma(i) \ne i \}|$. Then
\[
\left|\prod_{i = 1}^m A'_{i \sigma(i)}\right| \le \gamma^{t(\sigma)} \prod_{i = 1}^m \sqrt{A'_{i i}} \prod_{i = 1}^m \sqrt{A'_{\sigma(i) \sigma(i)}} = \gamma^{t(\sigma)} \prod_{i = 1}^m A'_{i i}.
\]
%Consider the  $p$-thickening of $\AA'$ for $p \ge 1$.
Each term of $\det\left((\AA')^{\odot p}\right)$ has the form 
$\pm \prod_{i = 1}^m A'^{~p}_{i \sigma(i)}$
for some permutation $\sigma$ of $[m]$.
Now
\[
| \{ \sigma \mid t(\sigma) = j \} | \le \binom{m}{j} j! \le m^j,
\]
for $0 \le j \le m$.
By separating out the identity permutation
and all other terms, for $p \ge \lfloor \ln(2 m) / \ln(1 / \gamma) \rfloor + 1$,
we have $2 m \gamma^p < 1$, and  
\begin{align*}
\left|\det\left((\AA')^{\odot p}\right)\right| &\ge \left( \prod_{i = 1}^m A'_{i i} \right)^p - \left( \prod_{i = 1}^m A'_{i i} \right)^p \sum_{j = 1}^m m^j \gamma^{p j} \\
&\ge \left( \prod_{i = 1}^m A'_{i i} \right)^p \left( 1 - \frac{m \gamma^p}{1 - m \gamma^p} \right) = \left( \prod_{i = 1}^m A'_{i i} \right)^p \left( \frac{1 - 2 m \gamma^p}{1 - m \gamma^p} \right) > 0.
\end{align*}
\end{proof}

%\begin{figure}
%\centering
%%\input{pics/ADA-to-p-gadget-advanced}
%\includegraphics[scale=1.2]{pics/ADA-to-p-gadget-advanced.pdf}
%\caption{\label{fig:AD(A-to-l)-to-p-gadget-advanced}The edge gadget with the edge weight matrix $(A D A^{\odot \ell})^{\odot p}$
%where the thickened line corresponds to $\ell$ parallel edges.
%{\color{red} The rows and columns of the matrix correspond respectively to $u$ and $v$.}
%%\caption{\label{fig:ADA-to-p-gadget-advanced}The edge gadget $S_2 T_p e,\, e = (u, v)$ with the edge weight matrix $(A D A)^{\odot p}$.
%%{\color{red} The edge gadget where each edge} 
%} %corresponding to the matrix $\tau_p (A D A) = (A D A)^{\odot p}$. from Theorem~\ref{thm:ADA-nondeg-thick}
%\end{figure}
%%%

%% file: very-high-level-description.tex
\section{A high-level description of the proof of Theorem~\ref{thm:main-dich-bd}}\label{highlevel-ours}
In the  proof of the dichotomy theorem in~\cite{Cai-Chen-Lu-2013}
a first preliminary step is to 
reduce the problem to connected graphs and matrices.
This is stated as Lemma~4.6 (p.~940) in~\cite{Cai-Chen-Lu-2013}
and its proof is accomplished by applying the  so-called first
pinning lemma (Lemma~4.1 (p.~937)). 
The proof of this first pinning lemma uses interpolation.
Although  the proof of this lemma in~\cite{Cai-Chen-Lu-2013}
 can be done for bounded degree
graphs,  it 
involves a noncontructive step that we want to avoid in this paper.
Therefore we will make the transition to connected graphs
by another technique that is based on transforming gadgets.

An important theorem in this paper is Theorem~\ref{thm:non-mult-bl-rk-1-hardness-sym}
which shows that if a complex symmetric matrix $\AA$ is not \multblrkone/,
then $\EVAL(\AA)$ remains \#P-hard even restricted to
bounded degree graphs. 
Indeed, in Corollary~\ref{cor:non-mult-bl-rk-1-gadget-hardness-bd}
it is shown that if we have an edge gadget  $\Gamma$
with a signature matrix $\gm_{\Gamma, \AA}$ that is
not \multblrkone/,
then for some $\Delta > 0$, the problem $\EVAL^{(\Delta)}(\AA)$ is \#P-hard.
Using this gadget-based approach, we make the 
transition to connected components.
% as shown in Lemma~\ref{connected}.
%%% this statement of Lemma~\ref{connected} is not for bounded degree.
%% there are no clear such statement for bounded degree, because everything
% was formualted officllay for gadgets.
We will prove that if  $\AA$  has connected components $\{\AA_i\}_{i \in [s]}$, %{\color{red} (see Section~\ref{sec:pinning-free})}
then
\begin{enumerate}
\item Either  $\eval(\AA_i)$ is polynomial-time computable for every $i$,
and this implies that $\eval(\AA)$ is also polynomial-time computable;
\item Or for some $i \in [s]$,
we have an edge gadget $\Gamma$ such that $\gm_{\Gamma, \AA_i}$ is not \multblrkone/,
from which we can get an  edge gadget $\Gamma'$ such that $\gm_{\Gamma', \AA}$
is not \multblrkone/, and therefore by 
Corollary~\ref{cor:non-mult-bl-rk-1-gadget-hardness-bd},
 %Theorem~\ref{thm:non-mult-bl-rk-1-hardness-sym}
%%% the cor is more direct here
we get \#P-hardness for $\EVAL^{(\Delta)}(\AA)$, for some $\Delta > 0$.
\end{enumerate} 

After this preliminary step, we restrict to 
  \emph{connected} and {symmetric} $\AA$.
Our tractable cases are the same as in~\cite{Cai-Chen-Lu-2013}
and so our description will focus on how to prove
 \#P-hardness for bounded degree graphs. 
As in~\cite{Cai-Chen-Lu-2013},
the difficulty starts with {gadget constructions}.
With a graph gadget, one can take any input undirected graph $G$ and produce
  a modified graph ${G}^*$ by replacing each edge of $G$ with the gadget.
Moreover, one can define a suitable modified matrix $\AA^*$
  from the given matrix  $\AA$ and the gadget
  such that $Z_{ {\AA}^*} (G) = Z_{\AA} ( {G}^*)$, 
  {for all undirected graphs $G$.} 
This gives a reduction from
  $\eval({\AA^*})$ to $\eval(\AA)$.
If the gadget has bounded degree $k$, it also
gives a reduction from
  $\eval^{(\Delta)}({\AA^*})$ to $\eval^{(k\Delta)}(\AA)$ for any $\Delta \ge 0$.
If the gadget were to produce nonnegative symmetric
matrices ${\AA}^*$, then one could apply 
%the Bulatov--Grohe dichotomy,
Theorem~\ref{BulatovGrohe}
and its extension to bounded degree graphs~\cite{Govorov-Cai-Dyer}
to ${\AA}^*$.

However, for complex matrices $\AA$,
any graph gadget
  will only produce a matrix $ {\AA}^*$ whose entries  are
polynomials of the entries of $\AA$, as they are  obtained  
by arithmetic operations $+$ and $\times$.
There are no nonconstant polynomials on $\bbC$ that always produce
nonnegative output.
Pointedly, \emph{conjugation} is not an
arithmetic operation.  However, it is clear that for roots of unity,
%%%JYC:
one {\it can} produce conjugation by multiplication. 

Thus, as in~\cite{Cai-Chen-Lu-2013} we wish to replace our
matrix $\AA$ by its purification matrix.
It is here the proof in~\cite{Cai-Chen-Lu-2013} fundamentally does not
go through for  bounded degree graphs.
An essential observation of this paper is that, in each of the steps in
the proof of~\cite{Cai-Chen-Lu-2013} we in fact can prove the
following: Either the matrix $\AA$ satisfies some additional
conditions, or we can produce an 
 edge gadget $\Gamma$ such that $\gm_{\Gamma, \AA}$ is not \multblrkone/.
We state three meta-arguments,
Arguments~\ref{meta1}, \ref{meta2}, and~\ref{meta3}, i.e., ({\sl Meta}$_1$), ({\sl Meta}$_2$) and ({\sl Meta}$_3$), to formalize this ability
to transfer such gadgets from one step of the proof 
to a previous step. Thus in the last step (when $\AA$ does not satisfy
all the tractability conditions) we have such a gadget whose signature matrix is
not \multblrkone/,  then in a finite number of
steps we can get such a gadget in the initial setting,
and then invoke Corollary~\ref{cor:non-mult-bl-rk-1-gadget-hardness-bd}.
%  Theorem~\ref{thm:non-mult-bl-rk-1-hardness-sym}.

To carry out this plan, we must separate out
the cases where $\AA$ is bipartite or nonbipartite.
For a (nonzero) symmetric, connected and nonbipartite  $\AA$,
it is  \multblrkone/ iff it has the form
$\AA = (A'_{i,j} \zeta_{i,j})$ where
$\AA' = (A'_{i,j})$ is symmetric, has no zero entries and
 has rank 1, and $\zeta_{i,j}$ are roots of unity.  
For a (nonzero) symmetric, connected and bipartite  $\AA$,
it is  \multblrkone/ iff it is the bipartization of a rectangular matrix  $\BB$
of the form $(B'_{i,j}\zeta_{i,j})$, where $\BB' = (B'_{i,j})$ has no zero entries and 
 has rank 1, and $\zeta_{i,j}$ are roots of unity.   
For convenience,  in this section
we will only describe the bipartite case  in the discussion below;
with some minor changes, similar statements hold
for  nonbipartite matrices $\AA$.

In the bipartite case, if $\AA$ is  \multblrkone/,
we have the rank one matrix $(B'_{i,j})$ which has the form 
$(\mu_i \nu_j)$
%_{\tiny \substack{1 \le i \le k\\
% k < j \le m}}$
for some nonzero $\mu_i, \nu_j$.
% where $1 \le i \le k$ and $k+1 \le  j \le m$.
%%%% seems no need to talk about this  $\mu_i, \nu_j$ belong to ...
%%% in this very high level discription
 %If we apply the purification to 
%$\AA = \left( \begin{smallmatrix} {\bf 0} & \BB \\ \BB^{T} & {\bf 0}
 %\end{smallmatrix} \right)$ using a generating set $\{g_1, \ldots, g_d\}$,
  %we may assume the entries $\mu_i, \nu_j$
%all  belong  to the multiplicative group generated by
%$\{g_1, \ldots, g_d\}$.
%
Thus $\BB$ has the form
\[
\BB\hspace{0.015cm} = \left(\begin{matrix}
\mu_1\hspace{-0.05cm} \\
& \hspace{-0.05cm}\mu_2\hspace{-0.05cm} \\
& & \hspace{-0.05cm}\ddots\hspace{-0.05cm}\\
& & & \hspace{-0.05cm}\mu_k
\end{matrix}\right)
\left( \begin{matrix}
\zeta_{1,1} & \zeta_{1,2} & \ldots & \zeta_{1,m-k} \\
\zeta_{2,1} & \zeta_{2,2} & \ldots & \zeta_{2,m-k} \\
\vdots      & \vdots      & \ddots & \vdots \\
\zeta_{k,1} & \zeta_{k,2} & \ldots & \zeta_{k,m-k}
\end{matrix}\right)
\left(\begin{matrix}
\nu_{1}\hspace{-0.05cm} \\
& \hspace{-0.05cm}\nu_{2}\hspace{-0.05cm} \\
& & \hspace{-0.05cm}\ddots\hspace{-0.05cm}\\
& & & \hspace{-0.05cm}\nu_{m-k}
\end{matrix}\right),\vspace{0.16cm}\]
for some $1\le k<m$, in which $\mu_i, \nu_j$ are nonzero,
  and every $\zeta_{i,j}$ is a root of unity.
The claim is that, for every (nonzero)
symmetric, connected, and bipartite matrix  $\AA\in \mathbb{C}^{m\times m}$,
either we can already prove the $\#$P-hardness of
  $\eval^{(\Delta)}(\AA)$ for some $\Delta >0$, % {\color{red} we can show that there exists an edge gadget $\Gamma$ such that $\gm_{\Gamma, \AA}$ is not \multblrkone/}
  or we may assume
 $\AA$ is the bipartization of $\BB$ of the above form.
In the latter case we pass \emph{both} $\AA$ and its purification
$\pur \AA$  to the next step.

Continuing now with  $\AA$ and  $\pur \AA$,
the next step is to {further regularize}
its entries.  In particular we need to combine those rows and
columns of the matrix  where they
%the rows (and columns)
are essentially the same, apart from a multiple of a root of unity.
This process is called
\emph{cyclotomic reduction}. % and \emph{Inverse Cyclotomic} Reduction.
To carry out this process, we need to use the more general form
$\eval(\CC,\fD)$ defined earlier in Section \ref{evalcd-ours}.
%Let $\CC\in \mathbb{C}^{m\times m}$ be a
%  symmetric  matrix,
%  and ${\frak D}=\{\DD^{[0]},\DD^{[1]}, \ldots,\DD^{[N-1]}\}$ be
%  a sequence of diagonal matrices in $\mathbb{C}^{m\times m}$
%  for some $N\ge 1$. We denote the $(i,i)^{th}$ entry of $\DD^{[t]}$
% by
%  $D^{[t]}_{i}$.
%We define the following problem $\eval(\CC,{\frak D})$:
%Given an
%  undirected graph $G=(V,E)$,
%\begin{equation*}
%Z_{\CC,{\frak D}}(G)=\sum_{\xi:V\rightarrow [m]}\left(\hspace{0.03cm}\prod_{(u,v)\in E}
%  %A_{\xi(u),\xi(v)}\right)
%C_{\xi(u),\xi(v)}\right)
%%%% Xi, pl check ?
%\left( \prod_{i=0}^{N-1} \left(\hspace{0.05cm} \prod_{v\in V,\hspace{0.1cm} \text{\rm deg}(v)
%  \equiv i \hspace{-0.2cm}\mod\hspace{-0.03cm}N} D^{[i]}_{\xi(v)}\right)\right).
%\end{equation*}
Introduced in~\cite{Cai-Chen-Lu-2013},
the following type of matrices are called
\emph{discrete unitary} matrices. 

\begin{definition}[discrete unitary matrix, Definition 3.1 from~\cite{Cai-Chen-Lu-2013}]
Let $\FF\in \mathbb{C}^{m\times m}$ be a
  \emph{(}not necessarily symmetric\emph{)} matrix with entries $(F_{i,j})$.
We call $\FF$ an \emph{$M$-discrete unitary matrix}, for some positive
  integer $M$, if it satisfies the following conditions:
\begin{enumerate}
\item Every entry $F_{i,j}$ of $\FF$ is a root of unity,
  and $F_{1,i}=F_{i,1}=1$ for all $i\in [m]$.
\item $M$ is the least common multiple \emph{(}$\lcm$\emph{)} 
  of orders of all the entries $F_{i,j}$ of $\FF$. 
\item 
 For all $i\ne j\in [m]$, we have
$\langle\FF_{i,*},\FF_{j,*}\rangle=0$ {and} 
$\langle \FF_{*,i},\FF_{*,j}\rangle=0$.
\end{enumerate}
\end{definition}

Some of the simplest examples of discrete unitary matrices are
as follows:
\[
%{\bf H} = 
\left(\begin{matrix} 1 & 1 \\
             1 & -1 \end{matrix}\right),
~~
%{\bf H_4} = 
\left(\begin{matrix} 1 & 1 & 1 & 1\\            1 & 1 & -1 & -1 \\
            1 & -1 & 1 & -1 \\
            1 & -1 & -1 & 1 \end{matrix}\right),
~~
%\FF_{3}  = 
\left(\begin{matrix} 1 & 1 & 1 \\
                             1 & \omega & \omega^2 \\
                 1 & \omega^2 & \omega
                  \end{matrix}\right),
~~
%\FF_{5}  = 
\left(\begin{matrix} 1 & 1 & 1 & 1 & 1\\
                                 1 & \zeta&\zeta^{-1} & \zeta^2 & \zeta^{-2}\\
                                 1 & \zeta^2&\zeta^{-2} & \zeta^{-1} & \zeta\\
                 1 & \zeta^{-1}& \zeta & \zeta^{-2}&\zeta^2 \\
                                 1 & \zeta^{-2}&\zeta^2 & \zeta&\zeta^{-1}
                  \end{matrix}\right),
\]
where $\omega = e^{2 \pi i /3}$ and $\zeta = e^{2 \pi i /5}$.
Tensor products of discrete unitary matrices
  are also discrete unitary matrices.
%These matrices play a major role in our proof. 

Coming back to the proof outline,
we show that either there exists an edge gadget $\Gamma$
such that $\gm_{\Gamma, \AA}$ is not \multblrkone/
(which implies that $\eval^{(\Delta)}(\AA)$ is \#P-hard for some $\Delta > 0$) %%% no sout
or $\eval^{(\Delta)}(\AA)$ is equivalent to some
$\eval^{(\Delta)}(\CC,{\frak D})$, and the pair $(\CC,{\frak D})$
satisfies some stringent conditions. In fact one can show that
either  $\eval^{(\Delta)}(\AA)$ is \#P-hard  for some $\Delta > 0$, or
the pair  $(\CC,{\frak D})$ has a tensor product form,
and the problem $\eval(\CC,{\frak D})$ (and also for bounded degree graphs)
can be expressed as a product of an \emph{outer problem} 
$\eval(\CC',{\frak K})$ and an \emph{inner problem} $\eval(\CC'', {\frak L})$,
where $\eval(\CC',{\frak K})$ is tractable, and thus we will focus on
the inner problem $\eval(\CC'', {\frak L})$. We rename
$(\CC,{\frak D})$ as the pair $(\CC'', {\frak L})$.
We show that 
 $\CC$ is the bipartization of a discrete unitary matrix $\FF$. 
In addition, there are further stringent requirements for ${\frak D}$.
%The detailed statements can be found in Theorem~\ref{bi-step-2-ours}
% and \ref{step30-ours}, summarized in properties $(\calU_1)$ to $(\calU_5)$.
Roughly speaking, the first matrix $\DD^{[0]}$ in $\fD$ must be the identity matrix;
%each $\DD^{[r]}$, for $1 \le r \le N-1$, must also  take certain
%normal forms and are inter-connected by roots of
%unity in a precise way;
and for any matrix $\DD^{[r]}$ in~$\fD$, each entry of $\DD^{[r]}$
  is either zero or a root of unity.
%Most strikingly, it is condition $(\calU_5)$.
%It states that
We call these conditions, with some abuse of terminology, 
  the \emph{discrete unitary
%requirements. The proof of these requirements is involved
%%%JYC: changed a word
requirements}. The proof for these requirements in~\cite{Cai-Chen-Lu-2013}
  is demanding 
  and among the most difficult in that paper;
but here we will use the   meta-arguments,
Arguments~\ref{meta2} and~\ref{meta3}, i.e., ({\sl Meta}$_2$) and ({\sl Meta}$_3$), 
to observe that essentially the same proof can be cast
in terms of transforming non-\multblrkone/ gadgets
from one setting to another.

%The proof of the properties on $\DD^{[r]}$ is involved.
%One particularly striking property we must prove along the way
%is the following lemma
%********
%Put statement of Lemma 8.8 here.:
%For each $\DD^{[r]}$, it is rank 1.
%********
%Later we will give a part  of the proof of this Lemma.
Next, assume that we have a problem $\eval(\CC,\fD)$ satisfying
the discrete unitary requirements with $\CC$ being
  the bipartization of $\FF$. Recall that $\omega_q = e^{2 \pi i /q}$. 
\begin{definition}[Definition 3.2 from~\cite{Cai-Chen-Lu-2013}]
Let $q>1$ be a prime power.
The following $q\times q$ matrix $\bcF_q$ is called the \emph{$q$-Fourier matrix}:
The $(x,y)$th entry of $\bcF_q$ is $\omega_q^{xy}$, $x,y\in [0:q-1]$.
%is $\omega_q^{xy}$,
%%%JYC: added a def'n of \omega_q
%  and $k\ge 1$ be an integer such that $\gcd\hspace{0.04cm}(k,q)=1$.
%We call the following $q\times q$
%  matrix $\bcF_{q,k}$ a $(q,k)$-\emph{Fourier matrix}:
% \emph{(}In particular,
%  when $k=1$, we use $\bcF_{q}$ to denote $\bcF_{q,1}$ for short\emph{)}:
%  The $(x,y)^{th}$ entry, where $x,y\in [0:q-1]$, is\vspace{-0.07cm}
%\[
%\omega_{q}^{kxy}=e^{2\pi i\big({kxy}/{q}\big)}.\vspace{0.03cm}
%\]
%In particular, when $k=1$, we use $\bcF_{q}$ to denote $\bcF_{q,1}$ for short.
\end{definition}

We show that, either  there exists an  edge gadget $\Gamma$ 
such that $\gm_{\Gamma, \CC, \fD}$
is not \modblrkone/ (which implies that it is not \multblrkone/ either),
or after a permutation of rows and columns,
$\FF$ becomes the \emph{tensor product} of a collection of suitable Fourier matrices:
\[
\bcF_{q_1}\otimes \bcF_{q_2}\otimes \cdots\otimes \bcF_{q_d},\ \ \ \ \text{where
  $d\ge 1$ and every $q_i$ is a prime power.}
\]
Basically, we show that even with the stringent conditions imposed
  on the pair $(\CC,\fD)$ by the discrete unitary requirements,
  we still get \#P-hardness  for $\eval^{(\Delta)}(\CC,\fD)$,
for some $\Delta > 0$, unless $\FF$ is the %{\color{red} we still get an edge gadget $\Gamma$ such that $M_{\Gamma, \CC, \fD}$ is not \modblrkone/}
  tensor product of Fourier matrices.
On the other hand, the tensor product decomposition into Fourier
  matrices finally  gives us a canonical way of writing the 
entries of $\FF$ in a closed form.
More exactly, we index the rows and columns of $\FF$ using
  $\xx=(x_1,\ldots,x_d) \in \mathbb{Z}_{q_1}\times \cdots
  \times \mathbb{Z}_{q_d}$ so that
\[
F_{\xx,\yy}=\prod_{i\in [d]} \omega_{q_i}^{x_iy_i},\ \ \ \ \text{for any 
  $\xx,\yy\in \mathbb{Z}_{q_1}\times \cdots\times\mathbb{Z}_{q_d}$.}
\]
%In the bipartite case this expression takes a
% relatively
%unsurprising form. However for non-bipartite case we must also
%prove a similar statement, and in that case there are some
%complications.  The expression must involve a non-degenerate quadratic
%form.
Assume $q_1,\ldots,q_d$ are powers of $s\le d$ distinct primes $p_1,\ldots,p_s$.
We can also lump together all prime powers of
the same prime $p_i$, and view the set of indices as
$\mathbb{Z}_{q_1}\times \cdots \times \mathbb{Z}_{q_d}=
  G_1\times \cdots \times G_s,$
where $G_i$ is the finite Abelian group which is the direct product of all 
  groups $\mathbb{Z}_{q_j}$ in the list
  with $q_j$ being a power of $p_i$. 

This canonical tensor product decomposition of $\FF$ gives
  a natural way to index the rows and columns of $\CC$ and the diagonal
  matrices in $\fD$.
% using $\xx$.
More exactly,
for $\xx \in  \mathbb{Z}_{q_1}\times \cdots\times \mathbb{Z}_{q_d}$,
 we index the first half of the rows and columns of $\CC$
  and every $\DD^{[r]}$ in $\fD$ using $(0,\xx)$, and index
  the second half of the rows and columns using $(1,\xx)$. 
  %$\xx\in \mathbb{Z}_{q_1}\times \cdots\times \mathbb{Z}_{q_d}$. 

With this canonical expression of $\FF$ and $\CC$, we further 
  inquire into
the structure of $\fD$.
%Here one more substantial difficulty awaits us.
%It turns out that there
%The canonical tensor product decomposition gives us a way to
%index the rows and columns of $\CC$ and every diagonal matrix in
%$\fD$ using elements from a suitable Abelian group.  More precisely,
%  the first half and the second half of the entries
%of each $\DD^{[r]}$ in $\fD$ are indexed by $(0,\xx)$ and $(1, \xx)$ respectively,
%where $\xx$ is from the group.
There are two more properties that we must demand of
  those diagonal matrices in $\fD$.
If $\fD$ does not satisfy these additional properties, then
  $\eval^{(\Delta)}({\CC,\fD})$ is \#P-hard for some $\Delta>0$. %{\color{red} there exists an edge gadget $\Gamma$ such that $M_{\Gamma, \CC, \fD}$ is not \modblrkone/}

First, for each $r$,
we define $\Lambda_r$
 and $\Delta_r$
to be the support of $\DD^{[r]}$,
where $\Lambda_r$  refers to the first half of
the entries and  $\Delta_r$  refers to the second half of
the entries (here we 
%follow the convention of using
use $D_i$ to denote
  the $(i,i)$th entry of a diagonal matrix $\DD$): 
\begin{eqnarray*}
\Lambda_r=\big\{\xx: D^{[r]}_{(0,\xx)}\
\ne 0\big\}\ \ \ \ \text{and}\ \ \ \
\Delta_r=\big\{\xx: D^{[r]}_{(1,\xx)}\ne 0\big\}.
\end{eqnarray*}
We let $\cS$ denote the set of subscripts $r$ such that $\Lambda_r\ne \emptyset$
  and $\cT$ denote the set of $r$ such that $\Delta_r\ne \emptyset$.
We can prove that for each
$r\in \cS$,
the support set  $\Lambda_r$
must be a direct product of cosets,
$\Lambda_r=\prod_{i=1}^s
  \hspace{0.03cm}\Lambda_{r,i}$,  where $\Lambda_{r,i}$ are cosets
 in the Abelian
groups $G_i$, $i=1,\ldots,s$, corresponding
  to the constituent prime powers of the group;
and for each $r\in \cT$, $\Delta_r=\prod_{i=1}^s\hspace{0.03cm}
  \Delta_{r,i}$ is a direct product of cosets in the same
 Abelian groups. Otherwise,
$\eval^{(\Delta)}({\CC,\fD})$ is \#P-hard for some $\Delta>0$;
more precisely, there is an edge gadget $\Gamma$
such that $\gm_{\Gamma, \CC, \fD}$
is not \modblrkone/. %\multblrkone/.

%%%%%%%%%% I know you Artem would like me to describe it in terms of
%% getting a non mult-brk-1 gadget etc. but in the outline, it is hard and verbose
% to state it that way.

Second, we show that for each $r \in \cS$ and $r \in \cT$, respectively,
 $\DD^{[r]}$ on its support $\Lambda_{r}$ for the first half of its
entries
and on $\Delta_{r}$ for the second half of its entries, respectively,
  possesses a {\it quadratic} structure; otherwise $\eval^{(\Delta)}({\CC,\fD})$ is \#P-hard for some $\Delta>0$.
  %otherwise there is an edge gadget $\Gamma$ such that $M_{\Gamma, \CC, \fD}$ is not \modblrkone/
The  quadratic structure
is expressed 
as \emph{a set of exponential difference equations} over
bases which are appropriate roots of unity of orders equal to
various prime powers. These proof steps in~\cite{Cai-Chen-Lu-2013}
are the most demanding in that paper; however here we 
 apply  the   meta-arguments to the same proof 
and observe that they let us transform  non-\multblrkone/ gadgets
from one setting to another.

After all these necessary conditions, we finally show that,
if $\CC$ and $\fD$ satisfy all these requirements,
there is a polynomial-time
algorithm for $\eval({\CC,\fD})$ (to be precise, for $\eval^\leftrightarrow(\CC, \fD)$) and thus,
  $\eval(\AA)$ is also in polynomial time.
The tractability part of the proof is almost identical to that of~\cite{Cai-Chen-Lu-2013}.

%% file: non-mult-bl-rk-one-hardness.tex
\section{Non-\multblrkone/ implies bounded degree hardness}\label{sec:GCD-gadget}
%\section{Non-multiplicative-block-rank-1 implies bounded degree hardness}\label{sec:GCD-gadget}
%\section{Non-multiplicative-block-rank-1 implies bounded degree hardness}\label{sec:GCD-gadget} 
%\section{Hardness on bounded degree graphs for non-multiplicatively-block-rank-$1$ matrices} %Block-rank-$1$ multiplicativity
%\section{Hardness for $Z_\AA(\cdot)$ on bounded degrees graphs for non-\multblrkone/ matrices $\AA$}

This section is dedicated to the proof 
of Theorem~\ref{thm:non-mult-bl-rk-1-hardness-sym}.

\begin{theorem}\label{thm:non-mult-bl-rk-1-hardness-sym}
%Let $\AA$ be a complex symmetric $m \times m$ matrix.
Let $\AA \in \bbC^{m \times m}$ be a symmetric matrix.
If $\AA$ is not \multblrkone/,
then for some $\Delta > 0$, the problem $\EVAL^{(\Delta)}(\AA)$ is \#P-hard.
\end{theorem}

%We dedicate this section to proving
%Theorem~\ref{thm:non-mult-bl-rk-1-hardness-sym}.
% will be proved shortly in 
%this section.
%Section~\ref{sec:nmubr1-bd-hardness-proof}.

It will be convenient to state Theorem~\ref{thm:non-mult-bl-rk-1-hardness-sym}
for directed graphs as well. 
%input to $\EVAL$.
For a directed graph $G = (V, E)$
we let $\deg(u)$ denote the sum of its in-degree and out-degree for $u \in V$,
%$\deg(u)  = \deg^+(u) + \deg^-(u)$ for $u \in V$,
%%% we never defined \deg^+(u) \deg^-(u) notation. so not used. 
and $\Delta(G) = \max_{u \in V} \deg(u)$. %$\Delta(G) = \max_{u \in V} (\deg(u))$

\begin{corollary}\label{cor:non-mult-bl-rk-1-hardness}
%Let $\AA$ be a complex $m \times m$ matrix.
Let $\AA \in \bbC^{m \times m}$ (not necessary symmetric).
If $\AA$ is not \multblrkone/,
then for some $\Delta > 0$, the problem $\EVAL^{(\Delta)}(\AA)$ is \#P-hard.
%(Here the inputs are directed graphs.)
\end{corollary}
\begin{proof}
Recall that for not necessarily symmetric $\AA$, $\EVAL(\AA)$ is defined for directed graphs $G$.
Let $\AA' = \begin{pmatrix} 0 & \AA \\ \AA^T & 0 \end{pmatrix}$
be the bipartization of the matrix $\AA$.
%Consider an undirected graph $G$ with $\Delta(G) \le \Delta$.
Now let $G$ be an undirected connected graph.
%%% JYC I think i need connected
%%% only need to consider connected, had been justified before.
If $G$ is not bipartite,
then $Z_{\AA'}(G) = 0$.
Assume $G = (U \cup V, E)$ is
bipartite with $U \cup V$ being
a bipartization of (the vertices of) $G$.
Let  $\overrightarrow G$ and $\overleftarrow G$ be the directed graphs
obtained by orienting all edges of $G$ from $U$ to $V$, 
and from $V$ to $U$, respectively.
Then $Z_{\AA'}(G) = Z_\AA(\overrightarrow G) + Z_\AA(\overleftarrow G)$.
Note that $\Delta(\overrightarrow G) = \Delta(\overleftarrow G) = \Delta(G) $.
Therefore $\EVAL^{(\Delta)}(\AA') \le \EVAL^{(\Delta)}(\AA)$ for any $\Delta \ge 0$.
By Property~\ref{prop:equiv-direct-sum-mbrk1-perm},
$\AA'$ is not \multblrkone/.
Then by Theorem~\ref{thm:non-mult-bl-rk-1-hardness-sym},
for some $\Delta > 0$, $\EVAL^{(\Delta)}(\AA')$ is \#P-hard
and so is $\EVAL^{(\Delta)}(\AA)$.
\end{proof}

\begin{corollary}\label{cor:non-mult-bl-rk-1-gadget-hardness-bd}
%Let $\AA$ be a complex (symmetric) $m \times m$ matrix
Let $\AA \in \bbC^{m \times m}$ be a symmetric matrix
and  let $\Gamma$ be an edge gadget.
If $\gm_{\Gamma, \AA}$ is not \multblrkone/
(which is true if $\gm_{\Gamma, \AA}$ is not \modblrkone/),
then for some $\Delta > 0$, the problem $\EVAL^{(\Delta)}(\AA)$ is \#P-hard.
\end{corollary}
\begin{proof}
By Corollary~\ref{cor:non-mult-bl-rk-1-hardness},
even if $\gm_{\Gamma, \AA}$ is not symmetric,
the problem $\EVAL^{(\Delta)}(\gm_{\Gamma, \AA})$ is \#P-hard,
 %(taking directed input graphs)
for some $\Delta > 0$.
Here the inputs to $\EVAL^{(\Delta)}(\gm_{\Gamma, \AA})$ are directed graphs.
%as $\gm_{\Gamma, \AA}$ may not be symmetric.
Let $\firstdv, \seconddv$ be the first and second distinguished vertices of $\Gamma$, 
respectively.
%correspondingly.
%(corresponding to the rows and the columns of $\gm_{\Gamma, \AA}$, respectively).
Given a directed graph $G$ of $\Delta(G) \le \Delta$,
we construct an undirected graph $G'$ 
by replacing every directed edge $e = (u, v) \in E(G)$
by a gadget $\Gamma$, attaching $\firstdv, \seconddv$ to $u, v$, respectively.
Clearly, $G'$ can be constructed in polynomial time from $G$
and $Z_\AA(G') = Z_{\gm_{\Gamma, \AA}}(G)$.
Also note that $\Delta(G') \le \Delta(\Gamma) \Delta(G) \le \Delta(\Gamma) \Delta$.
%(In particular, if $\Delta(G) = 0$, then $G$ consists of isolated vertices so that $G' = G$ and so $\Delta(G') = 0$.)
%{\color{red} (alternatively we can use)
%\[
%\Delta(G') \le \max(\Delta' \cdot \max(d_{\Gamma, u^*}, d_{\Gamma, v^*}), \max_{u \in V(\Gamma) \setminus \{u^*, v^*\}} d_{\Gamma, u}).
%\]
%}
Let $\Delta' = \Delta(\Gamma) \Delta$.
%Clearly, $\Delta' > 0$ as $\Delta(\Gamma) > 0$ (which is implied by Lemma~\ref{lem:edge-gadget-component-nmrk1}).
It follows that $\EVAL^{(\Delta)}(\gm_{\Gamma, \AA}) \le \EVAL^{(\Delta')}(\AA)$
and therefore $\EVAL^{(\Delta')}(\AA)$ is \#P-hard.
%If $\Delta$ becomes $0$, just replace $\Delta$ by $1$,
%the conclusion does not change.
%Now we can use $\Delta'$ instead of $\Delta$
%in the statement of the corollary and we are done.
\end{proof}

%\subsection{Proof of Theorem~\ref{thm:non-mult-bl-rk-1-hardness-sym}}\label{sec:nmubr1-bd-hardness-proof}

\subsection{Pairwise independent rows and columns}

We now turn to the proof of Theorem~\ref{thm:non-mult-bl-rk-1-hardness-sym}.
This proof adapts a gadget design from~\cite{Govorov-Cai-Dyer}
which extends the Bulatov-Grohe dichotomy,  Theorem~\ref{BulatovGrohe},
to bounded degree and simple graphs.

Let $\AA \in \bbC^{m \times m}$ be a symmetric matrix
but not \multblrkone/.
%although the non-block-rank-$1$-ness will only be used at the end of the proof.
%For now, we do not put this restriction on $A$.
%
The first step is to eliminate pairwise linearly dependent rows and columns
of $\AA$.
This step will naturally create nontrivial vertex weights
even though we initially start with the vertex unweighted case $\DD=\II_m$
 (see Definition~\ref{def:EVAL(A,D)}).

%{\color{red} (not sure where to put this text)
%
%Thus to prove \#P-hardness, it suffices to consider input graphs $G$
%that are connected and have $|V(G)| > 1$.
%}

If $\AA$ has a zero row or column $i$, then
for any connected input graph $G$ other than a single isolated vertex,
no map $\xi: V(G) \rightarrow [m]$ having
 a nonzero contribution to $Z_\AA(G)$
can map any vertex of $G$ to $i$.  
So, by crossing out all zero rows and columns (they have the same index
set since $\AA$ is symmetric) 
we can express the problem $\EVAL^{(\Delta)}(\AA)$ for $\Delta \ge 0$ 
on a smaller domain, so 
we may assume that $\AA$ has no zero rows or columns.
%for otherwise we can cross out these rows and columns from $A$,
%both indexed by the same subset since $A$ is symmetric,
%%%%%%%%%%%%%%%%%%%%%%%%%%%%%%%%%%%%%%%%%%%%%%%%%%%%%%%%%%%%%%%%%%%%
\begin{comment}

{\color{red}
We then delete the same set of rows and columns from $D$,
thereby expressing the problem
%$\EVAL(A, D)$ on a smaller domain.
$\EVAL_{\simp}^{(\Delta)}(A, D)$ for $\Delta \ge 0$ on a smaller domain. (old)
}
\end{comment}
%%%%%%%%%%%%%%%%%%%%%%%%%%%%%%%%%%%%%%%%%%%%%%%%%%%%%%%%%%%%%%%%%%%%
Also permuting the rows and columns of $\AA$ simultaneously by the same permutation
does not change the value of $Z_\AA(\cdot)$,
and so it does not change the complexity of
$\EVAL^{(\Delta)}(\AA)$ for $\Delta \ge 0$ either.
%$\EVAL(A, D)$ either.
%$\EVAL_{\simp}^{(\Delta)}(A, D)$ for $\Delta \ge 0$ either.
%(It also holds for $\EVAL_{\simp}^{(\Delta)}(A, D)$ for $\Delta \ge 0$,
%where the subscript $\simp$ can be included or excluded.)
Having no zero rows and columns implies that pairwise linear dependence
is an equivalence relation, and so
we may assume that the pairwise linearly dependent rows and
columns of $\AA$ are contiguously arranged.
%The same is true for the rows as $A$ is symmetric.
Then, after renaming the indices, the entries of $\AA$ are of the following form:
$A_{(i, j), (i', j')} = \mu_{i j} \mu_{i' j'} A'_{i, i'}$,
where $\AA'$ is a complex symmetric $s \times s$ matrix
with all columns nonzero and pairwise linearly independent,
 $1 \le i, i' \le s$, $1 \le j \le m_i$, $1 \le j' \le m_{i'}$,
 $\sum_{i = 1}^s m_i = m$,
and all $\mu_{i j} \ne 0$.
%Applying the same row and column permutation
%to the matrix $D$ and renaming the indices, we can assume
%We also rename the indices of the matrix $D$ so that
%the diagonal entries of $D$
%are of the following form:
%$D_{(i, j), (i, j)} = \alpha_{i j} > 0$
%for $1 \le i \le s$ and $1 \le j \le m_i$.
%So it is also possible to assume that $\mu_{i 1} = 1$ for $1 \le i \le s$
%but we will not use this assumption.
As $m \ge 1$ we get $s \ge 1$.

Then the partition function $Z_\AA(\cdot)$ can be written in 
a compressed form
%\begin{align*}
\[
Z_\AA(G)  =
%&= \sum_{\xi: V(G) \rightarrow [m]} \prod_{u \in V(G)} D_{\xi(u)} \prod_{\{v, w\} \in E(G)} A_{\xi(v), \xi(w)} \\
%&= 
\sum_{\zeta: V(G) \rightarrow [s]} \left( \prod_{w \in V(G)} \sum_{j = 1}^{m_{\zeta(w)}} %\alpha_{\zeta(w) j} 
\mu_{\zeta(w) j}^{\deg(w)} \right) \prod_{(u, v) \in E(G)} A'_{\zeta(u), \zeta(v)} = 
Z_{\AA', \mathfrak D}(G),
\]
%\end{align*}
%%% we start with no vertex wright now. so no alphas'
where $\mathfrak D = \{ \DD^{\llbracket k \rrbracket}\}_{k = 0}^\infty$
consists of diagonal matrices, and
 $D^{\llbracket k \rrbracket}_i = \sum_{j = 1}^{m_i} \mu_{i j}^k$ for $k \ge 0$ and $1 \le i \le s$.
Note the dependence on the vertex degree $\deg(w)$ for $w \in V(G)$.
Since the underlying graph $G$ remains unchanged,
this way we obtain the equivalence
$\EVAL^{(\Delta)}(\AA) \equiv \EVAL^{(\Delta)}(\AA', \mathfrak D)$ for any $\Delta \ge 0$.
%{\color{red} Here the subscript $\simp$ can be included or excluded,
%and the same is true for the superscript $(\Delta)$,
%%in the same way as the superscript $(\Delta)$,
%the statement remains true in all cases. (old)}
Here the superscript $(\Delta)$ can be included or excluded,
%in the same way as the superscript $(\Delta)$,
the statement remains true in both cases.
We also point out that
 the entries of the matrices $\DD^{\llbracket k \rrbracket} \in \mathfrak D$
are computable in polynomial time
in the input size of $\AA$ as well as in $k$.

\subsection{Operation $W_\ell$}

\begin{figure}
\centering
\includegraphics{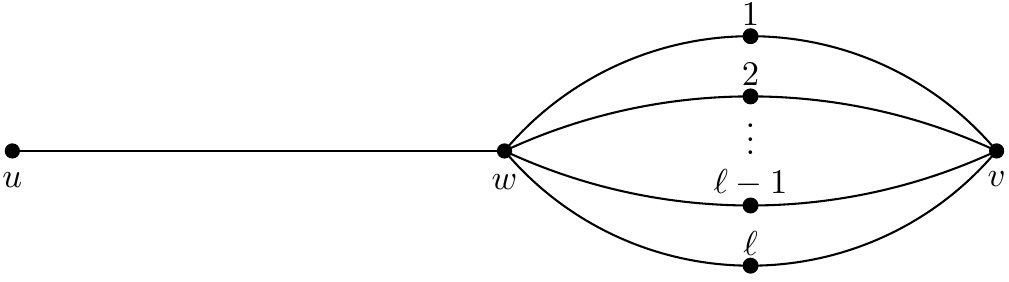} % [scale=1.2]
\caption{\label{fig:W-operation}The operation (edge gadget) $W_\ell(e)$ where $e = (u, v)$ is a directed edge from $u$ to $v$} 
\end{figure}

%This operation $\mathcal P_{n, p, \ell}$ and $\mathcal R_{d, n, p, \ell}$ in Section~\ref{sec:Gadget-Rdnpl}.
For each $\ell \ge 1$, we define an operation $W_\ell$
on directed edges.
Let $e = (u, v)$ be a directed edge (from $u$ to $v$) as follows.
First, we apply a $2$-stretching $S_2$ on $e = (u, v)$ viewed as an undirected edge.
We get a  path  $(u, w, v)$  of length $2$, after this operation.
We then $\ell$-thicken the edge $(w, v)$ in $(u, w, v)$. This is our
$W_\ell$ operation on  $e$.
This operation produces the graph $W_\ell(e)$ 
in Figure~\ref{fig:W-operation}.

\subsection{Gadgets $\mathcal P_{n, p, \ell}$ and $\mathcal R_{d, n, p, \ell}$}\label{sec:Gadget-Rdnpl}

%%%%%%%%%%%%%%%%%%%%%%%%%%%%%%%%%%%%%%%%%%%%%%%%%%%%%%%%%%%%%%%%%%%%
\begin{comment}

{\color{red} (JYC: not sure it's a good idea to define thickening and strecthing on directed edges, I don't here, but make these constructions feel intuitive;
after thinking for a while: actually it could be a good step to simplify notation, up to you)}

%{\color{red} We first introduce the  \emph{edge gadget}
% $\mathcal P_{n, p, \ell}$, for all $p, n, \ell \ge 1$.
%It is  obtained by replacing each edge
%of a  path of length $n$ by  the gadget in
%Figure~\ref{fig:ADA-to-p-gadget-advanced}
%from Lemma~\ref{lem:ADA-nondeg-thick}.
%More succinctly  $\mathcal P_{n, p, \ell}$ is  $S_2 T_p S_n e$, where $e$ is an edge. (old)}

\end{comment}
%%%%%%%%%%%%%%%%%%%%%%%%%%%%%%%%%%%%%%%%%%%%%%%%%%%%%%%%%%%%%%%%%%%%

\begin{figure}
\centering
\includegraphics[scale=1.2]{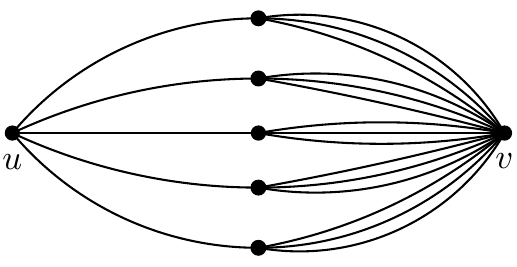}
\caption{\label{fig:component-gadget-1}An edge gadget $W_3 T_5(e)$.
The gadget $\mathcal P_{n, p, \ell}$ is an  $n$-chain of the edge gadget
$W_\ell T_p(e)$.}
\end{figure}

We first introduce the  \emph{edge gadget}
 $\mathcal P_{n, p, \ell}$, for all $n, p, \ell \ge 1$. \label{Pnpl}
It is obtained as follows.
Given a  directed edge  $e = (u, v)$,
let  $S_n e = (u_0 = u, u_1, \ldots, u_n =v)$ be a  path of length $n$
 from $u$ to $v$.
In $S_n e$ we orient every edge $(u_i, u_{i + 1})$ from $u_i$ to $u_{i + 1}$ (for $0 \le i \le n - 1$).
Then we consider $T_p S_n e$, which is obtained by applying $T_p$ on our $S_n e$,
while preserving the orientation of each edge.
After that, on every directed edge of $T_p S_n e$ we apply $W_\ell$
which results in an undirected graph containing 
the original vertices $u$ and $v$.
We make $u$ and $v$ the first and second distingushed vertices, respectively,
and we denote the resulting edge gadget by $\mathcal P_{n, p, \ell}$.
Succinctly, we can write $\mathcal P_{n, p, \ell} =
W_\ell T_p S_n e$, where $e = (u, v)$ is a directed edge,
$T_p$ and $S_n$ respect edge directions, and 
$\mathcal P_{n, p, \ell}$ has $u, v$
as its first and second distinguished vertices, respectively.
Note that  $\mathcal P_{n, p, \ell}$ contains only undirected edges,
while the roles of $u$ and $v$ are not symmetric;
the specification of the  direction of  edges is only for the purpose of 
describing the construction.
(As an example, $\mathcal P_{n, 5, 3}$ is an $n$-chain of the edge gadget $W_3 T_5(e)$
depicted in Figure~\ref{fig:component-gadget-1}.) 
%(As an example, $\mathcal P_{n, 5, 3}$ is an $n$-chain of the edge gadget 
%depicted in Figure~\ref{fig:component-gadget-1}.) 
%

%{\color{red} Note that $\mathcal P_{n, p, \ell}$ contains only undirected edges. (added)}
%Alternatively, we can think of $\mathcal P_{n, p, \ell}$
%as a concatenation of the edge gadgets $T_p S_n e$.
%\begin{figure}
%\centering
%\includegraphics{pictures/component-gadget-(p,l)=(5,3)-arc-based.pdf} %[scale=1.2]
%\caption{\label{fig:component-gadget}The edge gadget with the edge weight matrix $(A D A^{\odot 3})^{\odot 5}$
%where the thickened line corresponds to $\ell$ parallel edges.
%{\color{red} The rows and columns of the matrix correspond respectively to $u$ and $v$.}} 
%\end{figure}

%{\color{red}  (old)
%
%To define the gadget $\mathcal R_{d, n, p, \ell}$, for all $d, p, n, \ell \ge 1$, we start
%with a cycle on $d$ vertices $F_1, \ldots, F_d$ (call it a $d$-cycle),
%replace every edge of the  $d$-cycle by a copy of  $\mathcal P_{n, p, \ell}$, 
%and append a dangling
%edge at each vertex $F_i$ of the $d$-cycle.
%To be specific, a $2$-cycle has
% two vertices with $2$ parallel edges between them,
%and a $1$-cycle
%is a loop on one vertex. The gadget $\mathcal R_{d, n, p, \ell}$
%always has $d$ dangling edges.}

\begin{figure}[t]
\centering
\includegraphics[scale=1.2]{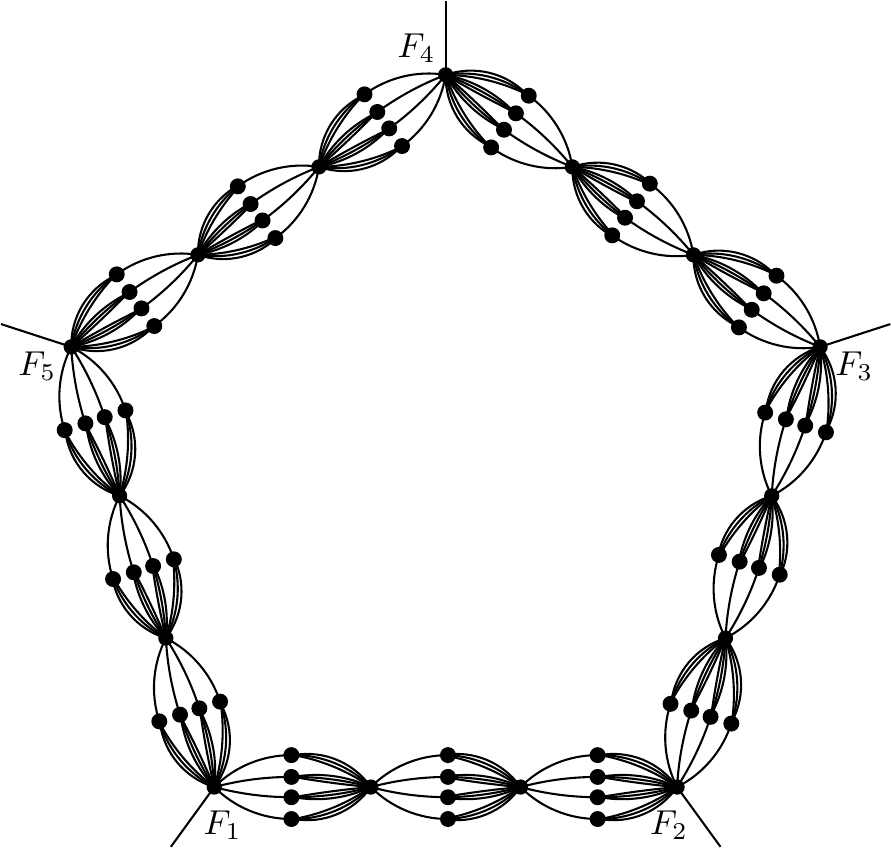}
\caption{\label{fig:d-gon-gagdet-simplified}The gadget $\mathcal R_{5, 3, 4, 3}$.}
%{\color{red} The dangling edges are to be $(\ell + 1)$-thickened.}
\end{figure}

To define the gadget $\mathcal R_{d, n, p, \ell}$, for all $d, n, p, \ell \ge 1$, we start \label{Rdnpl}
with a (directed) cycle on $d$ vertices $F_1, \ldots, F_d$ (call it a $d$-cycle),
in which we orient $F_i$ to $F_{i + 1}$ for each $i \in [d]$ (here $F_{d + 1} = F_1$).
Then 
replace every edge $(F_i, F_{i + 1})$ of the  $d$-cycle by a copy of  $\mathcal P_{n, p, \ell}$,
whereby we identify the first and second distinguished vertices
of $\mathcal P_{n, p, \ell}$ with $F_i$ and $F_{i + 1}$, respectively. Finally
we append a dangling
edge at each vertex $F_i$ of the $d$-cycle.
For the specific cases of $d=1$ and  $d=2$, 
 a $2$-cycle has
 two vertices with $2$ parallel directed edges of opposite orientations between them,
and a $1$-cycle
is a directed loop on one vertex. The gadget $\mathcal R_{d, n, p, \ell}$
 has $d$ dangling edges in total.
Note that all
 $\mathcal R_{d, n, p, \ell}$ are undirected loopless graphs,
 for $d, n, p, \ell \ge 1$.
%{\color{red} Also note that $\mathcal P_{n, p, \ell}$ contains only undirected edges. (added)}
%and so is the resulting graph
%from $G$ after the substitution of gadgets.
%
An example of a gadget $\mathcal R_{d, n, p, \ell}$
is shown in Figure~\ref{fig:d-gon-gagdet-simplified}.
%although we do not require the vertices $F_1, \ldots, F_d$
%to be arranged in a regular $d$-gon by any means.
For the special cases $d = 1$ and $d = 2$, %$d = 1$ and $2$
examples of gadgets $\mathcal R_{d, n, p, \ell}$
can be seen in Figure~\ref{fig:d=1,2-gadgets}.

%%%

\begin{figure}[t]
%\setbox1=\hbox{\includegraphics[scale=0.9]{pictures/d-gon-gadget-extended-complex-ew-modified-(d,n,p,l)=(1,5,5,3),dang=black.pdf}}
%\setbox2=\hbox{\includegraphics[scale=0.9]{pictures/d-gon-gadget-extended-complex-ew-modified-(d,n,p,l)=(2,4,3,4),dang=black.pdf}} % The smaller image
\setbox1=\hbox{\includegraphics[scale=0.9]{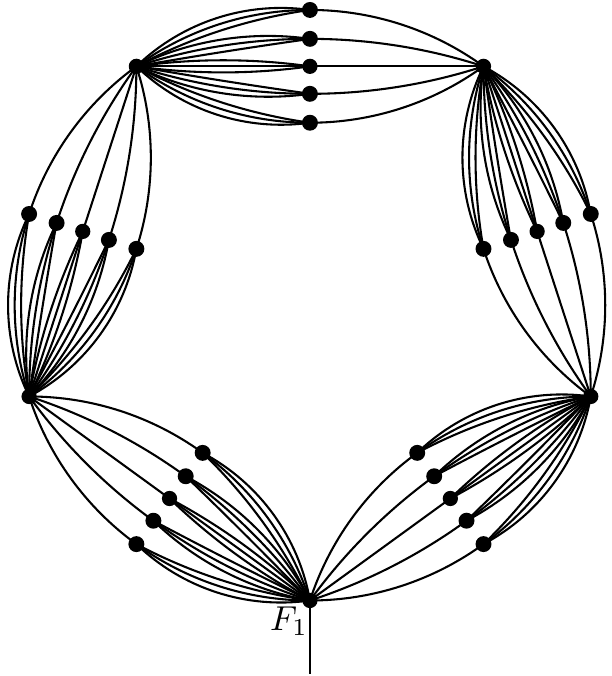}}
\setbox2=\hbox{\includegraphics[scale=0.9]{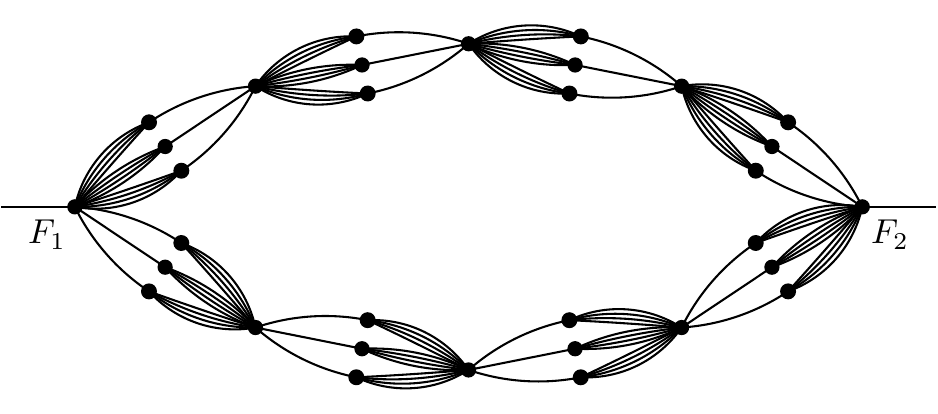}} % The smaller image
\begin{subfigure}[t]{0.4\textwidth} %{0.4\textwidth} %{0.5\textwidth}
\centering
\includegraphics[scale=0.9]{fig8.pdf} %[scale=0.95] %[scale=0.73] %[scale=0.8]
\subcaption{$\mathcal R_{1, 5, 5, 3}$\label{subfig-1:d=1}}
\end{subfigure}\hfill
\begin{subfigure}[t]{0.6\textwidth} %{0.6\textwidth} %{0.5\textwidth} 
\centering
\raisebox{0.5\ht1}{\raisebox{-0.5\ht2}{
\includegraphics[scale=0.9]{fig9.pdf}
}}
\subcaption{$\mathcal R_{2, 4, 3, 4}$\label{subfig-2:d=2}}
\end{subfigure}
\caption{\label{fig:d=1,2-gadgets}Examples of gadgets $\mathcal R_{d, n, p, \ell}$ for $d = 1, 2$}
\end{figure}

%%%
%\begin{figure}[t]
%\begin{subfigure}{0.4\textwidth} %{0.5\textwidth}
%\centering
%\includegraphics[scale=1]{pictures/d-gon-gadget-extended-complex-ew-modified-(d,n,p,l)=(1,5,5,3),dang=red.pdf} %[scale=0.73]
%\subcaption{$\mathcal R_{1, 5, 5, 3}$\label{subfig-1:d=1}}
%\end{subfigure}\hfill
%\begin{subfigure}{0.6\textwidth} %{0.5\textwidth}
%\centering
%\includegraphics[scale=1]{pictures/d-gon-gadget-extended-complex-ew-modified-(d,n,p,l)=(2,4,3,4),dang=red.pdf} %[scale=0.73]
%\subcaption{$\mathcal R_{2, 4, 3, 4}$\label{subfig-2:d=2}}
%\end{subfigure}
%\caption{\label{fig:d=1,2-gadgets}Examples of gadgets $\mathcal R_{d, n, p, \ell}$ for $d = 1, 2$.
%{\color{red} The dangling edges are to be $(\ell + 1)$-thickened.}}
%\end{figure}
%%%

We note that
vertices in $\mathcal P_{n, p, \ell}$ have degrees at most $p \ell + p$,
and vertices in $\mathcal R_{d, n, p, \ell}$  have degrees at most
% $p + 1 + p \ell + \ell$,
% this is $(p+1)(\ell +1)$,  but counts \ell +1 external danging edges at each F_i
$p \ell + p + 1$,
taking into account the dangling edges. 
These are independent of $d$ and  $n$.
Clearly
$|V({\mathcal R_{d, n, p, \ell}})| = d n (p+1)$ and 
$|E({\mathcal R_{d, n, p, \ell}})| = d ((\ell + 1)np+1)$, including the  
$d$ dangling edges.

If we attach $\ell +1$ dangling edges at each $F_i$, as we eventually will do,
then the degree bound for  $\mathcal R_{d, n, p, \ell}$ is $(p+1)(\ell +1)$, 
which is still independent of $d$ and  $n$.
\subsection{Construction of $G_{n, p, \ell}$ using $\mathcal R_{d, n, p, \ell}$}\label{subsec:Construction-G_{npell}}

Assume for now that $G$ does not contain isolated vertices.
We will  replace every vertex $u \in V(G)$ of degree  $d = d_u = \deg(u) \ge 1$ 
by a copy of  $\mathcal R_{d, n, p, \ell}$  and then \emph{$(\ell + 1)$-thicken}
the edges corresponding to $E(G)$, for all $n, p, \ell \ge 1$.
This defines the (undirected) graph $G_{n, p, \ell}$. \label{Gnpl}
The replacement operation 
%of every $u \in V(G)$ by
%$\mathcal R_{d, n, p}$ 
can
be described in three steps: In step one, each $u \in V(G)$  is replaced
by a (directed) $d$-cycle on vertices $F_1^u, \ldots, F_d^u$,
each having a dangling edge attached.
Here we orient $F_i^u$ to $F_{i + 1}^u$ for each $i \in [d]$ (and
we set $F_{d + 1}^u = F_1^u$). 
%Then in step two, 
The  $d$ dangling edges will be identified one-to-one with
the $d$ incident edges at $u$ in $G$.
If $u$ and $v$ are adjacent vertices in $G$, then
the edge $(u, v)$ in $G$ will be replaced by merging a
pair of dangling edges, one from the $d_u$-cycle at $u$
and one from the $d_v$-cycle at $v$.
This edge remains undirected.
Thus in step one we obtain a graph $\widehat G$, which basically replaces every
vertex $u \in V(G)$  by a (directed) cycle of $\deg(u)$ vertices.
% so it has arity $d$,
Then in step two, for every (directed) cycle in $\widehat G$  that
corresponds to some $u \in V(G)$ we replace each (directed) edge
% $F_i F_{i + 1}$
on the cycle by a copy of the edge gadget
%\emph{edge gadget}
 $\mathcal P_{n, p, \ell}$,
 whereby we respectively identify the first and second distinguished vertices
of $\mathcal P_{n, p, \ell}$ with the tail and head of this (directed) edge.
% for $1 \le i \le d= d_u$ (here $F_{d+1}$ is $F_1$).
Finally, in step three, we $(\ell + 1)$-thicken all the edges obtained
by merging pairs of dangling edges (these edges of $\widehat G$
 can be identified with 
edges from $E(G)$).

The construction  above defines an undirected graph $G_{n, p, \ell}$;
later we will also  refer to  $\widehat G$.
Since all gadgets $\mathcal R_{d, n, p, \ell}$ are loopless graphs,
so are $G_{n, p, \ell}$ for all $n, p, \ell \ge 1$.
%%% I think it is better not to consider $G$ having loops.
% this just causes confusion.
% this is in the #P hardness part of the proof. for that we don't
% need to consider G having loops. much less multi loops.
%
%(even if $G$ had multiloops, 
%if we view a loop as adding degree $2$ to the incident vertex).
As a technical remark, if $G$ contains vertices of degree $1$, 
then the intermediate graph $\widehat G$ has (directed) loops but all graphs $G_{n, p, \ell}$
($n, p, \ell \ge 1$) do not.
Also note that all vertices in $G_{n, p, \ell}$ have degree
at most
% $\max(2, 2 p, 2 p + 1) = 
$(p + 1)(\ell + 1)$,
% by $p \ge 1$,
which is independent of $n$ and $G$.

Next, it is not hard to see that
%\begin{align*}
%\begingroup
%\allowdisplaybreaks
\begin{gather*}
|V(G_{n, p, \ell})| = \sum_{u \in V(G)} d_u n (p + 1)  
%=
%\Bigg(
%n (p + 1) \sum_{u \in V(G)} d_u 
%\Bigg) (p + 1) n 
= 2 n (p + 1) |E(G)|, \\
|E(G_{n, p, \ell})| = (\ell + 1) |E(G)| + \sum_{u \in V(G)}
d_u (\ell + 1) n p   
= (\ell + 1) (2 n p + 1) |E(G)|.
%correcting an error, only one d_u
%|E(G_{n, p})| = |E(G)| + \sum_{u \in V(G)} d_u \cdot 2 d_u n p = |E(G)| + 2 \sum_{u \in V(G)} d_u^2 n p \le |E(G)| +
%8 np |E(G)|^2.
%% I use \sum d_u^2 <= (\sum d_u)^2 = (2 |E|)^2
%% 2 |V(G)| |E(G)|^2 n p.
\end{gather*}
%\endgroup
%\end{align*}
%In the latter, we used the fact that in a multigraph
%any vertex can be adjacent to at most all of its edges
%so $d_u \le |E(G)|$ for $u \in V(G)$.
%Then it immediately follows that the size of 
Hence  the size of 
the graphs $G_{n, p, \ell}$ is polynomially bounded in the size of $G$, $n$, $p$ and $\ell$.
%In fact, as noted before all graphs $G_{n, p}$ are simple
%so only the bound on the number of vertices $|V(G_{n, p})|$ is enough
%to bound the size of $G_{n, p}$.
%We still included both for completeness
%and because in our framework we generally deal with multigraphs
%we do not put an emphasis on this fact in this argument.

%%% reasoning template
Soon we will choose a fixed $p$,
%Since we chose a fixed $p$,
and then a fixed $\ell$, all depending only on $\AA$. Then 
we will choose $n$ to be bounded by a polynomial in the size of $G$;
whenever something is computable in polynomial time in $n$,
it is also computable in polynomial time in the size of $G$
(we will simply say in polynomial time).

\subsection{Picking $p$ and then picking $\ell$}

Consider $Z_{\AA', \mathfrak D}(G_{n, p, \ell})$.
We first find suitable $p \ge 1$, and then $\ell \ge 1$,
so that our construction allows us to compute
a value of interest by  interpolation.
More precisely, we show that we can find $p, \ell \ge 1$
such that
the matrix $\BB = (\AA' \DD^{\llbracket \ell +1 \rrbracket}
 (\AA')^{\odot \ell})^{\odot p}$ is nondegenerate,
and all diagonal entries in $\DD^{\llbracket p(\ell + 1) \rrbracket}$ 
and $\DD^{\llbracket (p + 1)(\ell +1) \rrbracket}$
are nonzero.
We note that $\BB$ is the  signature matrix of the edge gadget
$W_\ell T_p (e)$ (which, by definition, excludes the vertex weights of 
the two distinguished vertices of $W_\ell T_p (e)$ ),
and $\mathcal P_{n, p, \ell}$ is just a  chain of $n$ copies of this
edge gadget.
In 
$\BB = (\AA' \DD^{\llbracket \ell +1 \rrbracket}
 (\AA')^{\odot \ell})^{\odot p}$,
the superscript $\llbracket \ell +1  \rrbracket$
is from the operator $W_\ell$ which creates those degree
$\ell+ 1$ vertices, the superscript $\odot \ell$ is also from $W_\ell$,
 and the  superscript $\odot p$ is
 from the thickening operator $T_p$.
%(This is illustrated in Figure~\ref{fig:W-operation}).
%An example  of $W_\ell T_p (e)$ is shown
%with signature matrix in Figure~\ref{fig:component-gadget}.
An example  of $W_\ell T_p (e)$ is shown, with $\ell = 3$ and $p=5$,
in Figure~\ref{fig:component-gadget-1}, with the
edge weight matrix $(\AA' \DD^{\llbracket 4 \rrbracket} (\AA')^{\odot 3})^{\odot 5}$
in the framework $\EVAL(\AA', \mathfrak D)$.

%\begin{figure}
%\centering
%\includegraphics[scale=1.2]{pictures/component-gadget-(p,l)=(5,3)-arc-based.pdf}
%\caption{\label{fig:component-gadget} The gadget $W_3 T_5(e)$
%has edge weight matrix $(\AA' \DD^{\llbracket 4 \rrbracket} (\AA')^{\odot 3})^{\odot 5}$
%in the framework $\EVAL(\AA', \mathfrak D)$.}
%\end{figure}

\subsubsection{Picking $p$}
%\paragraph{Picking $p$.}
Let $\HH \in \bbC^{s \times s}$
be a diagonal matrix with
the $i$th diagonal entry
 $H_i = \sum_{j = 1}^{m_i} \mu_{i j} \overline{\mu_{i j}} > 0$ for $1 \le i \le s$.
So $\HH$ is positive diagonal.
%(In other words $\HH = \DD^{\llbracket 1 \rrbracket} \circ \overline{\DD^{\llbracket 1 \rrbracket}}$
%is the Hadamard product of $\DD^{\llbracket 1 \rrbracket}$ and $\overline{\DD^{\llbracket 1 \rrbracket}}$.)
By Lemma~\ref{lem:ADconjA-nondeg-thick},
we can fix some $p \ge 1$ such that $(\AA' \HH \overline{\AA'})^{\odot p}$ is nondegenerate.
%, where the superscript $\llbracket 2 \rrbracket$
%is from the stretching operator $S_2$ which creates those degree
%$2$ vertices, and the  superscript $\odot p$ is
% from the thickening operator $T_p$, followed by $S_2$, which creates those
%parallel paths of length $2$.
Also, clearly we have
 $\sum_{j = 1}^{m_i} \mu_{i j}^p (\overline{\mu_{i j}})^p > 0$
and $\sum_{j = 1}^{m_i} \mu_{i j}^{p + 1} (\overline{\mu_{i j}})^{p + 1} > 0$ for $1 \le i \le s$.
Thus, we have
\begin{equation}\label{eq:cong-system}
\begin{lcases}
&\det\left(\left(\AA' \HH \overline{\AA'}\right)^{\odot p}\right) \ne 0, \\
&\sum_{j = 1}^{m_i} \mu_{i j}^p (\overline{\mu_{i j}})^p \ne 0, \quad 1 \le i \le s, \\
&\sum_{j = 1}^{m_i} \mu_{i j}^{p + 1} (\overline{\mu_{i j}})^{p + 1} \ne 0, \quad 1 \le i \le s.
\end{lcases}
\end{equation}
Now consider the following infinite sequence of
 systems of conditions indexed by $\ell \ge 1$: 
\begin{equation}\label{eq:par-by-l-system}
\begin{lcases}
&\det\left( (\AA' \DD^{\llbracket \ell +1 \rrbracket} (\AA')^{\odot  \ell})^{\odot p} \right) \ne 0, \\
&\sum_{j = 1}^{m_i} \mu_{i j}^{p + p \ell} \ne 0, \quad 1 \le i \le s, \\
&\sum_{j = 1}^{m_i} \mu_{i j}^{(p + 1) + (p+1) \ell } \ne 0, \quad 1 \le i \le s.
\end{lcases}
\end{equation}

\subsubsection{Picking $\ell$}
%\paragraph{Picking $\ell$}
Our next goal is to find $\ell \ge 1$
such that each condition in the system (\ref{eq:par-by-l-system})
indexed by $\ell$ is satisfied.
This will involve a Vandermonde Argument 
from subsecton~\ref{subsec:Vandermonde}.

\begin{itemize}
\item
Let $\XX = (X_{i, j})_{i, j = 1}^s$
be an $s \times s$ matrix
whose entries are indeterminates $X_{i, j}$
for $1 \le i, j \le s$;
\item
Let $\yy = (y_{i j})_{i = 1, j = 1}^{s, m_i}$ be a tuple 
of indeterminates $y_{i j}$ 
%for $1 \le i \le s$ and $1 \le j \le m_i$
(containing $m = \sum_{i = 1}^s m_i$ elements);
\item
For each $1 \le i \le s$, denote by $\yy_{i, *} = (y_{i j})_{j = 1}^{m_i}$
the subtuple of $\yy$
whose entries are indeterminates $y_{i j}$ for $1 \le j \le m_i$;
\item
Let
 $\ZZ = \ZZ(\yy)$ be a diagonal $s \times s$ matrix
whose entries are $Z_i(\yy) = \sum_{j = 1}^{m_i} \mu_{i j} y_{i j}$
 for $1 \le i \le s$;
% {\color{red} (or use $Z((\yy))_i$)}
\item
Let $\boldsymbol{\mu} = (\mu_{i j})_{i = 1, j = 1}^{s, m_i}$  (containing $m$ elements),
and for each $1 \le i \le s$, let $\boldsymbol{\mu}_{i, *} = (\mu_{i j})_{j = 1}^{m_i}$;
\item
Finally, let 
\begin{align*}
f_1(\XX, \yy) &= \det\left((\AA' \ZZ(\yy) \XX)^{\odot p}\right),\\
f_{2, i}(\yy_{i, *}) &= \sum_{j = 1}^{m_i} \mu_{i j}^p y_{i j}^p, 
~~~~~~\mbox{$1 \le i \le s$, and } \\
%$ %where $1 \le i \le s$ 
%and 
f_{3, i}(\yy_{i, *})  &=  \sum_{j = 1}^{m_i} \mu_{i j}^{p + 1} y_{i j}^{p + 1}
~~\mbox{$1 \le i \le s$}.
\end{align*}
\end{itemize}
We treat the expressions $f_1(\XX, \yy)$, $f_{2, i}(\yy_{i, *})$ and $f_{3, i}(\yy_{i, *})$
where $1 \le i \le s$ as polynomials in $X_{i, j}$, where $1 \le i, j \le s$,
and $y_{i j}$, where $1 \le i \le s$ and $1 \le j \le m_i$,
even if some of these variables are missing in some polynomials.

Note that $\ZZ(\overline{\boldsymbol{\mu}}) = \HH$, so
by (\ref{eq:cong-system})
we have
\[%\begin{equation*}
\begin{lcases}
&f_1\left(\overline{\AA'}, \overline{\boldsymbol{\mu}}\right) = \det\left((\AA' \HH \overline{\AA'})^{\odot p}\right) \ne 0, \\
&f_{2, i}\left(\overline{\boldsymbol{\mu}_{i, *}}\right) = \sum_{j = 1}^{m_i} \mu_{i j}^p (\overline{\mu_{i j}})^p \ne 0, \quad 1 \le i \le s, \\
&f_{3, i}\left(\overline{\boldsymbol{\mu}_{i, *}}\right) = \sum_{j = 1}^{m_i} \mu_{i j}^{p + 1} (\overline{\mu_{i j}})^{p + 1} \ne 0, \quad 1 \le i \le s,
\end{lcases}
\]%\end{equation*}
and, since   $\ZZ(\boldsymbol{\mu}^{\odot \ell}) = \DD^{\llbracket \ell + 1 \rrbracket}$,
\[%\begin{equation*}
\begin{lcases}
&f_1\left((\AA')^{\odot \ell}, \boldsymbol{\mu}^{\odot \ell}\right) = \det\left((\AA' \DD^{\llbracket 1 + \ell \rrbracket} (\AA')^{\odot  \ell})^{\odot p}\right), \, \ell \ge 1, \\
&f_{2, i}\left(\boldsymbol{\mu}_{i, *}^{\odot \ell}\right) = \sum_{j = 1}^{m_i} \mu_{i j}^{p + p \ell}, \quad 1 \le i \le s, \, \ell \ge 1, \\
&f_{3, i}\left(\boldsymbol{\mu}_{i, *}^{\odot \ell}\right) = \sum_{j = 1}^{m_i} \mu_{i j}^{p + 1 + (p +1) \ell}, \quad 1 \le i \le s, \, \ell \ge 1.
\end{lcases}
\]%\end{equation*}

Now by Corollary~\ref{lem:vand-conj},
we get that for some $\ell \ge 1$,
each condition in the system (\ref{eq:par-by-l-system}) indexed by $\ell$
is satisfied. So we fix such an $\ell \ge 1$.
%(More precisely, we can take $1 \le \ell \le d$ where $d$ is the sum of the \emph{total degrees}
%$\deg(f_1) + \sum_{i = 1}^s \deg(f_{2, i}) + \sum_{i = 1}^s \deg(f_{3, i})$.)
%(More precisely, we can take $1 \le \ell \le d$ where $d$ is the number of terms
%in the expansion of $f_1(\XX, \yy) f_2(\XX, \yy) f_3(\XX, \yy)$.)
%Now let $\BB = (\AA' \DD^{\llbracket 1 + \ell \rrbracket} (\AA')^{[\ell]})^{\odot p}$.
%Then $\BB$ is nondegenerate.
From (\ref{eq:par-by-l-system}) we get that $\BB$ is nondegenerate and
all diagonal entries in $\DD^{\llbracket p + p \ell \rrbracket}$ and $\DD^{\llbracket p + 1 + (p+1) \ell \rrbracket}$
are nonzero, so $\DD^{\llbracket p + p \ell \rrbracket}$ 
and $\DD^{\llbracket p + 1 + (p +1) \ell \rrbracket}$
are nondegenerate as well.

%{\color{red} (end of insertion)}
%
%{\color{red} 

\subsection{Interpolation using $\LL^{(n)}$}\label{subsec:Interpolation-using-LL}

We now analyze the partition function value
$Z_{\AA', \mathfrak D}(G_{n, p, \ell})$.
The edge gadget $\mathcal P_{n, p, \ell}$ has the edge weight matrix
\begin{align}
%\label{Pnpl-edgeweightmatrix}
\LL^{(n)} &=
\underbrace{\BB \DD^{\llbracket p + p \ell \rrbracket} \BB \ldots \BB \DD^{\llbracket p + p \ell \rrbracket} \BB}_{\DD^{\llbracket p + p \ell \rrbracket} \text{ appears } n - 1 ~\ge~ 0 \text{ times}} = \BB (\DD^{\llbracket p + p \ell \rrbracket} \BB)^{n - 1} \label{Pnpl-edgeweightmatrix1} \\
&= (\DD^{\llbracket p + p \ell \rrbracket})^{-1 / 2} ((\DD^{\llbracket p + p \ell \rrbracket})^{1 / 2} \BB (\DD^{\llbracket p + p \ell \rrbracket})^{1 / 2})^n (\DD^{\llbracket p + p \ell \rrbracket})^{-1 / 2},
\label{Pnpl-edgeweightmatrix2}
\end{align}
%goto
where in the notation $\LL^{(n)}$
we suppress the indices $p, \ell$
%since they will be fixed soon.
since they are fixed in what follows.
The  $n-1$ occurrences of
%the superscript
 $\DD^{\llbracket p + p \ell \rrbracket}$ in (\ref{Pnpl-edgeweightmatrix1}) are due to
those $n-1$ vertices of degree $p + p \ell$.
Here $(\DD^{\llbracket p + p \ell \rrbracket})^{1 / 2}$ is a diagonal matrix
with arbitrarily chosen square roots
of the corresponding entries
of $\DD^{\llbracket p + p \ell \rrbracket}$ on the main diagonal,
and $(\DD^{\llbracket p + p \ell \rrbracket})^{-1 / 2}$ is its inverse.
In $G_{n, p, \ell}$, all the vertices $F_i^u$ on various cycles
have degree $(p + 1)(\ell +1)$ each,
including both the incident edges internal to the gadget $\mathcal R_{d, n, p, \ell}$
and the $(\ell + 1)$-thickened merged dangling edges.
%%%
%including both incident edges on the cycle
%and the $(\ell +1)$-thickened merged dangling edges {\color{red} (explain)}. 
%%%
These are the end vertices of the edge gadgets $\mathcal P_{n, p, \ell}$,
but the contributions by their vertex weights are
not included in $\LL^{(n)}$.
In $Z_{\AA', \mathfrak D}(G_{n, p, \ell})$ they
must be accounted for separately.

Let $\widetilde \BB = (\DD^{\llbracket p + p \ell \rrbracket})^{1 / 2} \BB (\DD^{\llbracket p + p \ell \rrbracket})^{1 / 2}$.
Write the Jordan normal form of $\widetilde \BB$
as $\widetilde \BB = \SS^{-1} \JJ \SS$,
 where $\SS$ is a nondegenate complex matrix
and $\JJ = \diag(\JJ_{\lambda_i, s_i})_{i = 1}^q$ is 
the  Jordan  normal form matrix of $\widetilde \BB$.
Here $q \ge 1$ is the number of Jordan blocks, 
each $s_i \ge 1$, $\sum_{i = 1}^q s_i = s$ and 
each Jordan block
 $\JJ_{\lambda_i, s_i}$ is an $s_i \times s_i$ 
(upper triangular) matrix, and the $\lambda_i$'s are
the eigenvalues of $\widetilde \BB$. 
The eigenvalues $\lambda_i$'s may not be all distinct,
but are all nonzero since $\widetilde \BB$ is nondegenerate.
\[
\JJ_{\lambda_i, s_i}=
\begin{bmatrix}
\lambda_i & 1 & 0 & \ldots & 0 & 0 \\
0 & \lambda_i & 1 & \ldots & 0 & 0 \\
0 & 0 & \lambda_i & \ldots & 0 & 0 \\
\vdots & \vdots & \vdots & \ddots & \vdots  & \vdots \\
0 & 0 & 0 & \ldots & \lambda_i & 1 \\
0 & 0 & 0 & \ldots & 0 & \lambda_i
\end{bmatrix}.
\]
%%%
Then ${\widetilde \BB}^n = \SS^{-1} \JJ^n \SS$,
 so the edge weight matrix for 
 $\mathcal P_{n, p, \ell}$ becomes
\[
\LL^{(n)} 
%= B (D^{\llbracket p + p \ell \rrbracket} B)^{n - 1}
= (\DD^{\llbracket p + p \ell \rrbracket})^{-1 / 2} {\widetilde \BB}^n (\DD^{\llbracket p + p \ell \rrbracket})^{-1 / 2}
= (\DD^{\llbracket p + p \ell \rrbracket})^{-1 / 2} \SS^{-1} \JJ^n \SS (\DD^{\llbracket p + p \ell \rrbracket})^{-1 / 2}.
\]
Note that $\LL^{(n)}$ as a matrix is formally defined for any $n \ge 0$,
and $\LL^{(0)} = (\DD^{\llbracket p + p \ell \rrbracket})^{-1}$.
This setting $n=0$ does not correspond to any actual gadget,
which might be called $\mathcal P_{0, p, \ell}$, 
but we will ``realize'' this virtual
gadget by interpolation in what follows.

Clearly, $\widetilde \BB$ is nondegenerate as $\BB$ and $(\DD^{\llbracket p + p \ell \rrbracket})^{1/2}$ both are, and so is $\JJ$.
All $\lambda_i \ne 0$.
If we write out the closed form expression for the entries of
$\JJ_{\lambda_i, s_i}^n$ we get $\binom{n}{j} \lambda_i^{n-j}$ for $0 \le j < s_i$.
Hence
we can write the $(i,j)${th} entry of
$\LL^{(n)}$ as
 $L_{i j}^{(n)} = \sum_{l = 1}^q \lambda_l^n p_{i j l}(n)$
for every $n \ge 0$ and some polynomials $p_{i j l}(x) \in \bbC[x]$
 with $\deg p_{i j l}(x) \le s_l - 1$,
whose coefficients depend on $\SS$,
 $\DD^{\llbracket p + p \ell \rrbracket}$, $\JJ$, but not on $n$,
for all $1 \le i, j \le s$, and $1 \le  l \le q$.

Note that for all $n, p, \ell \ge 1$,
the gadget $\mathcal R_{d_v, n, p, \ell}$ for $v \in V(G)$ employs  exactly
$d_v$ copies of $\mathcal P_{n, p, \ell}$. 
Let
$t = \sum_{v \in V(G)} d_v = 2 |E(G)|$; this  is precisely the number of
edge gadgets $\calP_{n, p, \ell}$ in $G_{n, p, \ell}$.
In the evaluation of the partition function $Z_{\AA', \mathfrak D}(G_{n, p, \ell})$,
we stratify the vertex assignments in $G_{n, p, \ell}$ as follows.
Denote by $\kappa = (k_{i j})_{1 \le i, j \le s}$
a tuple of nonnegative integers, where the indexing is over all
$s^2$ (ordered) pairs $(i, j)$. There are
a total of $\binom{t + s^2 - 1}{s^2 - 1}$
such tuples that satisfy $\sum_{1 \le i, j \le s}  k_{i j} = t$.
For a fixed $s$, this is a polynomial in $t$, and thus
a polynomial in the size of $G$. 
Denote by $\mathcal K$ the set of all such 
tuples $\kappa$.
We will stratify all vertex assignments in $G_{n, p, \ell}$ 
by $\kappa\in  \mathcal K$, namely all assignments
such that there are exactly $k_{i j}$ many constituent edge gadgets $\mathcal P_{n, p, \ell}$ with the two ordered end points
 assigned $i$ and $j$, respectively.

For each $\kappa \in \mathcal K$, the edge gadgets $\mathcal P_{n, p, \ell}$
in total contribute $\prod_{1 \le i, j \le s} (L_{i j}^{(n)})^{k_{i j}}$
to the partition function $Z_{\AA', \mathfrak D}(G_{n, p, \ell})$.
If we factor this product out for each $\kappa \in \mathcal K$, we can express
$Z_{\AA', \mathfrak D}(G_{n, p, \ell})$ as a linear
combination of these products over all $\kappa \in \mathcal K$,
with  polynomially many coefficient values  $c_\kappa$
that are independent of all edge gadgets $\mathcal P_{n, p, \ell}$.
Another way to define these coefficients $c_\kappa$ is to think in terms of
$\widehat G$: For any $\kappa = (k_{i j})_{1 \le i \le j \le s}
 \in \mathcal K$,
we say a vertex assignment on $\widehat G$ is consistent with $\kappa$
if it assigns exactly  $k_{i j}$ many directed cycle edges  of  $\widehat G$
(i.e., those that belong to the directed cycles that replaced vertices in $G$)
as ordered
%%% I don't have to specify for 1-cycle, ie loop in G, the i=j
% this is because i already said it is a vertex assignment in \widehat G
pairs of vertices to the value $(i, j)$.
%(For any (directed) loop in $G'$, as a (directed) cycle of length $1$ that came from 
%a degree $1$ vertex of $G$, it can only be assigned $(i,i)$ for some
% $1 \le i \le s$.)
%{\color{red} (recall that these cycle edges are directed. (added))}
Let  $\LL'$ be any (directed) edge signature to be
assigned on each of these (directed)
cycle edges in $\widehat G$,
and keep  the edge signature $\AA'$ on each edge of $\widehat G$ 
obtained by $(\ell + 1)$-thickening of all the
merged dangling edges between any two such cycles,
% {\color{red} (these edges are undirected (added))}
and 
%the suitable vertex
%weights specified by $\mathfrak D$, namely 
each vertex receives its vertex weight according to  
$\DD^{\llbracket p + 1 + (p +1) \ell \rrbracket}$.
%  specified by $\mathfrak D$.
Then $c_\kappa$
is the sum, over all assignments
consistent with $\kappa$, of the products of  all 
edge weights and vertex weights \emph{other than}
the contributions by $\LL'$, in the evaluation of
the partition function.
% on $\widehat G$.
In other words, for each $\kappa \in \mathcal K$,
\[
c_\kappa = \sum_{\substack{\zeta \colon V(\widehat G) \to [s] \\ \zeta \text{ is consistent with } \kappa}} \prod_{w \in V(\widehat G)} D_{\zeta(w)}^{\llbracket p + 1 +  (p +1)  \ell \rrbracket} \prod_{(u, v) \in \widetilde E} (A'_{\zeta(u), \zeta(v)})^{\ell + 1},
\]
where $\widetilde E \subseteq E({\widehat G})$ are the non-cycle edges of $\widehat G$ that are in $1$-$1$ correspondence with $E(G)$.
In particular, $|\widetilde E| = |E(G)|$. 
Importantly, the values $c_\kappa$ are independent of $n$.
Thus for some polynomially many values $c_\kappa$, where $\kappa
\in \mathcal K$, we have for all $n \ge 1$,
\begin{equation}\label{stratification-isolating-L-exp-poly}
%\[
Z_{\AA', \mathfrak D}(G_{n, p, \ell}) = \sum_{\kappa \in \mathcal K} c_\kappa \prod_{1 \le i, j \le s} (L_{i j}^{(n)})^{k_{i j}}
= \sum_{\kappa \in \mathcal K} c_\kappa \prod_{1 \le i, j \le s} (\sum_{\ell = 1}^q \lambda_\ell^n p_{i j \ell}(n))^{k_{i j}}. 
%\]
\end{equation}
%Note that $(L^{(n)})_{i j} = (L^{(n)})_{j i}$ so everything is well-defined.
Expanding out the last sum and rearranging the terms, for some %values $b_{i_1, \ldots, i_q}$ and
polynomials $f_{i_1, \ldots, i_q}(x) \in \bbC[x]$ independent of $n$
with $\deg f_{i_1, \ldots, i_q}(x) < s t$, we get
%%% I will just use < st. since each was < s_i -1 beofre multiplied by t
% comng from sum of k_{i j}
\[
Z_{\AA', \mathfrak D}(G_{n, p, \ell})
= \sum_{\substack{i_1 + \ldots + i_q = t \\ i_1, \ldots, i_q \ge 0}} ( \prod_{j = 1}^q \lambda_j^{i_j} )^n f_{i_1, \ldots, i_q}(n) %b_{i_1, \ldots, i_q} 
\]
for all $n \ge 1$.
%{\color{red} (The inequality $\deg f_{i_1, \ldots, i_q}(x) \le (s - q) t$ is true because each $\deg p_{i j \ell}(x) = s_\ell - 1 \le \sum_{k = 1}^q (s_k - 1) = s - q$ as each $s_k \ge 1$, and because $\sum k_{i j} = t$ for each $\kappa \in \mathcal K$.) (added, can upper bound $(s - q) t \le s t$ and use the $s t$ onwards instead of $(s - q) t$)}

Let there be exactly $r$ 
pairwise distinct values among $\prod_{j = 1}^q \lambda_j^{i_j}$,
denoted by $\chi_i$, where $1 \le i \le r$.
%where $i_1 + \ldots + i_q = t$ and $i_1, \ldots, i_q \ge 0$
%and let $\chi_i$, where $1 \le i \le r$, be these values without repetitions.
Note that all $\chi_i \ne 0$ since all $\lambda_j \ne 0$.
It is also clear that $1 \le r \le \binom{t + q - 1}{q - 1}$.
Rearranging the terms in the previous sum,
for some polynomials $g_i(x) = \sum_{j = 0}^{s t -1} b_{i j} x^j \in \bbC[x]$
 where $1 \le i \le r$,
we get for all $n \ge 1$,
\begin{equation}\label{eqn:exp-poly-expression-for-Z-A'}
Z_{\AA', \mathfrak D}(G_{n, p, \ell})
= \sum_{i = 1}^r \chi_i^n g_i(n) = \sum_{i = 1}^r \sum_{j = 0}^{s t -1} b_{i j} \chi_i^n n^j %b_{i_1, \ldots, i_q}.
\end{equation}

If we are given $Z_{\AA', \mathfrak D}(G_{n, p, \ell})$
for polynomially many $n \ge 1$,
(\ref{eqn:exp-poly-expression-for-Z-A'})
represents a linear system with the unknowns $b_{i j}$.
%with the rows indexed by $n$. % Vandermonde
%an expression
The number of unknowns is clearly $\le r s t$,
% $r ((s - q) t + 1)$
which is polynomial in the size of the input graph $G$ since
$r \le \binom{t + q - 1}{q - 1}$, $t = 2 |E(G)|$,  and $s$ is a constant.
The values $\chi_i^n n^j$ where $1 \le i \le r$ and $0 \le j < s  t$
can be clearly computed in polynomial time.

We show how to compute the value
\[\sum_{i = 1}^r b_{i 0}\]
from the values $Z_{\AA', \mathfrak D}(G_{n, p, \ell}),\, n \ge 1$ in polynomial time.
Since all $\chi_i \ne 0$ and are pairwise distinct,
by Corollary~\ref{lem:exp-poly-nondeg-matrix},
the square submatrix of the linear 
system (\ref{eqn:exp-poly-expression-for-Z-A'}) indexed by rows $n= 1, 
\ldots, rst$
is nondegenerate.
%by Lemma~\ref{lem:exp-poly-lin-ind} the polynomials $g_i(x)$ where $1 \le i \le r$
%are linearly independent (over $\bbC$).
It follows that rows $1, \ldots, rst$ of this linear system
are linearly independent. 
Therefore 
%by looking only at rows $1, \ldots, r (s + 1)$
we can solve this system in polynomial time
and find all the values $b_{i j}$,
and after that compute $\sum_{i = 1}^r b_{i 0}$.

From (\ref{eqn:exp-poly-expression-for-Z-A'}),
this value is formally
$Z_{\AA', \mathfrak D}(G_{n, p, \ell})$ at $n=0$.

\subsection{The problem $\EVAL(\mathscr A, \mathscr D)$}
We need to define a general $\EVAL$ problem,
where the vertices and edges can individually take specific weights.
Let $\mathscr A$ be a set of (edge weight) $m \times m$ matrices and
$\mathscr D$ a set of diagonal (vertex weight) $m \times m$ matrices.
A GH-grid $\Omega = (G, \rho)$ consists of
a graph $G = (V, E)$ with possibly both directed and undirected edges,
and 
%{\color{red} (directed and undirected, possibly multiple)}
 loops,
and $\rho$ assigns to each edge $e \in E$ or loop an
$\AA^{(e)} \in \mathscr A$
and  to each vertex $v \in V$
a $\DD^{(v)} \in \mathscr D$.
(A loop is considered an edge of the form $(v,v)$.)
%{\color{red} (A (directed or undirected) loop is just a (respectively directed or undirected) edge of the form $(v,v)$.) (new)}
 If $e \in E$
is a directed edge then the tail and head correspond to rows and columns of
$\AA^{(e)}$, respectively;
if  $e \in E$
is an undirected edge then $\AA^{(e)}$ must be
symmetric.
%and assigns its endpoints to its input variables:
%the rows and columns correspond to the start and end columns;
%$\rho$ also assigns to each vertex $v \in V$
%a (vertex weight) matrix $D^{(v)} \in \mathscr D$.

%%% did we ever use $\EVAL(\mathscr A, \mathscr D)$ for directed graphs?

\begin{definition}\label{def:EVAL(scrA,scrD)}
The problem $\EVAL(\mathscr A, \mathscr D)$ is defined as follows:
Given a GH-grid $\Omega = \Omega(G)$, compute
\[
\label{eqn:Z_{scrA,scrD}}
Z_{\mathscr A, \mathscr D}(\Omega) =
\sum_{\xi \colon V \to [m]} \prod_{w \in V} D_{\xi(w)}^{(w)} \prod_{e = (u, v) \in E} A_{\xi(u), \xi(v)}^{(e)}
\]
\end{definition}
If $\mathscr A = \{ \AA \}$ or $\mathscr D = \{ \DD \}$, then
we simply write $Z_{\AA, \mathscr D}(\cdot)$ or $Z_{\mathscr A, \DD}(\cdot)$, respectively.
We remark that
the problem $\EVAL(\mathscr A, \mathscr D)$ generalizes both
problems $\EVAL(\AA)$ and $\EVAL(\AA, \DD)$, by taking $\mathscr A$ to
be a single symmetric matrix, and
by taking $\mathscr D$ to be a single diagonal matrix.
But $\EVAL(\AA, \mathfrak D)$ is not naturally
expressible as  $\EVAL(\mathscr A, \mathscr D)$ because the
latter does not force the
vertex-weight matrix on a vertex according to its degree.

Now we will consider a problem in the framework of $Z_{\mathscr A, \mathscr D}$
according to Definition~\ref{def:EVAL(scrA,scrD)}.
Let $G_{0, p, \ell}$ be the 
%(containing both directed and undirected edges) 
GH-grid,
with the underlying graph $\widehat G$,
%{\color{red}
and every edge of the directed cycle in $\widehat G$  
is assigned
the edge weight matrix $(\DD^{\llbracket p + p \ell \rrbracket})^{-1}$
(which is $\LL^{(0)}$ even though we do not have an actual gadget for this),
and we keep the vertex-weight matrices $\DD^{\llbracket p + 1 + (p+1) \ell \rrbracket}$ at all vertices.
% $F_i$.
%%% I think on $\widehat G$ all vertices are of this type.
Note that even though these cycle edges are directed,
the matrix $(\DD^{\llbracket  p + p \ell \rrbracket})^{-1}$ is symmetric.
%}
%obtained by replacing the constituent edge gadgets $\mathcal P_{n, p, \ell}$
%in any fixed $G_{n, p, \ell}$ (does not matter which, say $n=1$) each by an edge with the
%edge weight matrix $(D^{\llbracket p + p \ell \rrbracket})^{-1}$ assigned to it,
%and keeping the vertex-weight matrices $D^{\llbracket 2 p + 1 \rrbracket}$ at all vertices $F_i$.
The other edges, i.e.,
those undirected edges that
came from the $(\ell+1)$-thickenings of the original edges of $G$, 
will each be assigned  
the edge weight matrix $\AA'$.
So, $\mathscr A = \{(\DD^{\llbracket p + p \ell \rrbracket})^{-1}, \AA'\}$,
and $\mathscr D = \{\DD^{\llbracket p + 1 + (p+1) \ell  \rrbracket}\}$
for the specification in the problem 
 $\EVAL(\mathscr A, \mathscr D)$.
We note that $\widehat G$ may have (directed) loops, and
Definition~\ref{def:EVAL(scrA,scrD)} specifically allows this.
Then (recall that $0^0 = 1$)
\[
Z_{\{ (\DD^{\llbracket p + p \ell \rrbracket})^{-1}, \AA' \}, \DD^{\llbracket p + 1 +
 (p+1) \ell \rrbracket}}(G_{0, p, \ell})
= \sum_{i = 1}^r \chi_i^0 g_i(0)
= \sum_{i = 1}^r \sum_{j = 0}^{s t -1} b_{i j} \chi_i^0 0^j
= \sum_{i = 1}^r b_{i 0}
\]
and we have just computed this value in polynomial time
in the size of $G$ from the values $Z_{\AA', \mathfrak D}(G_{n, p, \ell})$, for $n \ge 1$.
In other words, we have achieved it by querying
the oracle $\EVAL(\AA', \mathfrak D)$ on the 
instances $G_{n, p, \ell}$ (all of bounded degree), for $n \ge 1$,
in polynomial time.

Equivalently, we have shown that we can simulate
a virtual ``gadget'' $\mathcal R_{d, 0, p, \ell}$
replacing every occurrence of $\mathcal R_{d, n, p, \ell}$
in  $G_{n, p, \ell}$ in polynomial time.
% simultaneously in polynomial time,
The virtual gadget $\mathcal R_{d, 0, p, \ell}$  has
%(together with the $d$-dangling edges each of which 
%was $(\ell + 1)$-thickened) has
the edge signature $(\DD^{\llbracket p + p \ell \rrbracket})^{-1}$ in place of
$(\DD^{\llbracket p + p \ell \rrbracket})^{-1 / 2} {\widetilde \BB}^n
(\DD^{\llbracket p + p \ell \rrbracket})^{-1 / 2}$ in each $\mathcal P_{n, p, \ell}$, since
\[
(\DD^{\llbracket p + p \ell \rrbracket})^{-1 / 2} {\widetilde \BB}^0
(\DD^{\llbracket p + p \ell \rrbracket})^{-1 / 2}
%(D^{\llbracket p + p \ell \rrbracket})^{-1 / 2} ((D^{\llbracket p + p \ell \rrbracket})^{1 / 2} B (D^{\llbracket p + p \ell \rrbracket})^{1 / 2})^0 (D^{\llbracket p + p \ell \rrbracket})^{-1 / 2} 
= (\DD^{\llbracket p + p \ell \rrbracket})^{-1 / 2} \II_s (\DD^{\llbracket p + p \ell \rrbracket})^{-1 / 2} = (\DD^{\llbracket p + p \ell \rrbracket})^{-1}.
\]
%in any fixed $\mathcal R_{d, n, p, \ell}, \, n \ge 1$.
%Informally, an edge with such an edge weight matrix is precisely
%the ``edge gadget'' $\mathcal P_{0, p}$.
Additionally, each vertex $F_i^u$ retains the vertex-weight contribution with the matrix $\DD^{\llbracket p + 1 + (p +1) \ell \rrbracket}$ 
in $\mathcal R_{d, 0, p, \ell}$
(recall that to get $G_{n, p, \ell}$,
 each merged dangling edge was $(\ell + 1)$-thickened, resulting
in each vertex $F_i^u$ having degree $(p +1)(\ell+1)$).
%We view it as having ``virtual'' degree $p + 1 + (p +1) \ell$.
This precisely results in the GH-grid $G_{0, p, \ell}$.

However, even though  $G_{0, p, \ell}$
still retains the cycles,
since $(\DD^{\llbracket p + p \ell \rrbracket})^{-1}$ is a diagonal matrix, each
vertex $F_i^u$
in a cycle is forced to receive the same vertex 
assignment value in the domain set $[s]$;
all other vertex assignments contribute
 zero in the evaluation of $Z_{\{ (\DD^{\llbracket p + p \ell \rrbracket})^{-1}, \AA' \}, \DD^{\llbracket p + 1 + (p+1) \ell \rrbracket}}(G_{0, p, \ell})$.
This can be easily seen by traversing the vertices  $F_1^u, \ldots, F_d^u$
in a cycle that corresponds to a vertex $u$ in $G$, where $d = \deg(u)$.
Hence we can view each cycle employing the
virtual gadget $\mathcal R_{d, 0, p, \ell}$ as a single vertex
that contributes only a diagonal matrix of nonzero vertex weights
$\PP^{\llbracket d \rrbracket} = (\DD^{\llbracket p + 1 + (p+1) \ell \rrbracket} (\DD^{\llbracket p + p \ell \rrbracket})^{-1})^d$, where $d$ is  the vertex degree in $G$.
Contracting each cycle
%the gadgets $\mathcal R_{d, 0, p}$
%where $d = \deg(u), u \in V(G)$ 
to a single vertex,
we arrive at the $(\ell + 1)$-thickening $T_{\ell + 1}(G)$
of the original graph $G$.
For each edge $e$ of $G$ we can further collapse its $(\ell + 1)$-thickening
back by assigning to $e$ the edge weight matrix $(\AA')^{\odot(\ell + 1)}$.
We still have to keep the vertex weight matrices at each vertex 
of $T_{\ell + 1}(G)$: 
if a vertex in $T_{\ell + 1}(G)$ has degree $d (\ell + 1)$,
then the corresponding vertex in $G$ of  degree $d$
 must keep  the vertex weight matrix
$(\DD^{\llbracket p + 1 + (p+1) \ell \rrbracket} 
(\DD^{\llbracket p + p \ell \rrbracket})^{-1})^d$.
After this step we arrive at the original graph $G$,
and the value of the corresponding partition function on $G$ is $Z_{\{ (\DD^{\llbracket p + p \ell \rrbracket})^{-1}, \AA' \}, \DD^{\llbracket p + 1 + (p+1) \ell \rrbracket}}(G_{0, p, \ell})$.
More formally, we have the following.
Let $\mathfrak P = \{ \PP^{\llbracket i \rrbracket} \}_{i = 0}^\infty$,
where we let $\PP^{\llbracket 0 \rrbracket} = \II_s$, and for $i>0$,
we have  $P^{\llbracket i \rrbracket}_j = w_j^i$
where $w_j = \sum_{k = 1}^{m_j} \mu_{j k}^{p + 1 + (p+1) \ell} / \sum_{k = 1}^{m_j} \mu_{j k}^{p + p \ell} \ne 0$
for $1 \le j \le q$ (each $w_j$ is well-defined and is nonzero by (\ref{eq:par-by-l-system})).
%We can also define $P^{\llbracket 0 \rrbracket} = I_s$.
This shows that we now can interpolate
the value $Z_{(\AA')^{\odot (\ell + 1)}, \mathfrak P}(G) = Z_{\{ (\DD^{\llbracket p + p \ell \rrbracket})^{-1}, \AA' \}, \DD^{\llbracket p + 1 + p \ell + \ell \rrbracket}}(G_{0, p, \ell})$
using the values $Z_{\AA', \mathfrak D}(G_{n, p, \ell})$
in polynomial time in the size of $G$.

In the above, the
graph $G$ is arbitrary, except it has no isolated vertices.
The case when $G$ has isolated vertices can be handled easily as follows.

Given an arbitrary graph $G$, assume it has $h \ge 0$ isolated vertices.
Let $G^*$ denote the graph obtained from $G$ by their removal.
Then $G^*$ is of size not larger than $G$ and $h \le |V(G)|$.
Obviously, $Z_{(\AA')^{\odot (\ell + 1)}, \mathfrak P}(G) = (\sum_{i = 1}^s P_i^{\llbracket 0 \rrbracket})^h Z_{(\AA')^{\odot (\ell + 1)}, \mathfrak P}(G^*) = s^h Z_{(\AA')^{\odot (\ell + 1)}, \mathfrak P}(G^*)$.
Here the integer $s \ge 1$ is a constant, so the factor $s^h$ can be easily computed.
% in polynomial time.
Thus, knowing the value $Z_{(\AA')^{\odot (\ell + 1)}, \mathfrak P}(G^*)$
we can compute the value $Z_{(\AA')^{\odot (\ell + 1)}, \mathfrak P}(G)$ in polynomial time.
Further, since we only use the graphs $G_{n, p, \ell}$,
for $n \ge 1$, during the interpolation,
each being of degree at most $(p + 1)(\ell +1)$,
combining it with the possible isolated vertex removal step,
we conclude $\EVAL((\AA')^{\odot (\ell + 1)}, \mathfrak P) \le \EVAL^{(p + 1)(\ell +1)}(\AA', \mathfrak D)$.

Next, it is easy to see that for an arbitrary graph $G$
\begin{align*}
Z_{(\AA')^{\odot(\ell + 1)}, \mathfrak P}(G) &= \sum_{\zeta: V(G) \rightarrow [s]} \prod_{z \in V(G)} P^{\llbracket \deg(z) \rrbracket}_{\zeta(z)} \prod_{(u, v) \in E(G)} (\AA')^{\odot(\ell + 1)}_{\zeta(u), \zeta(v)} \\
&= \sum_{\zeta: V(G) \rightarrow [s]} \prod_{z \in V(G)} w_{\zeta(z)}^{\deg(z)} \prod_{(u, v) \in E(G)} (\AA')^{\odot(\ell + 1)}_{\zeta(u), \zeta(v)} \\
&= \sum_{\zeta: V(G) \rightarrow [s]} \prod_{(u, v) \in E(G)} w_{\zeta(u)} w_{\zeta(v)} (\AA')^{\odot(\ell + 1)}_{\zeta(u), \zeta(v)} \\
&= \sum_{\zeta: V(G) \rightarrow [s]} \prod_{(u, v) \in E(G)} C_{\zeta(u), \zeta(v)} = Z_\CC(G).
\end{align*}
Here $\CC$ is an $s \times s$ matrix with
the entries $C_{i, j} = (A'_{i, j})^{\ell + 1} w_i w_j$ where $1 \le i, j \le s$.
Clearly, $\CC$ is a symmetric matrix.
In the above chain of equalities,
we were able to \emph{redistribute the weights $w_i$
and $w_j$ into the edge weights} $(A'_{i, j})^{\ell + 1}$
which resulted in the edge weights $C_{i, j}$,
so that precisely each edge $(u,v)$ in $G$ gets two factors
$w_{\zeta(u)}$ and $w_{\zeta(v)}$ since the vertex weights at  $u$ and $v$
were $w_{\zeta(u)}^{\deg(u)}$ and $w_{\zeta(v)}^{\deg(v)}$, respectively.
%(This is a crucial step in our proof.)
This step is the final objective in our proof 
of Theorem~\ref{thm:non-mult-bl-rk-1-hardness-sym};
all preceding gadget constructions and interpolation steps are 
in preparation for being able to carry out this step.
Because the underlying graph $G$ is arbitrary, it follows that
$\EVAL((\AA')^{\odot (\ell + 1)}, \mathfrak P) 
\equiv \EVAL(\CC)$.
%Putting $\Delta = 2 p + 1$ everywhere
%and combining it with the previous $\EVAL$-reductions or equivalences,
%we obtain $\EVAL(C) \le_{\mathrm T}^{\mathrm P} \EVAL_{\simp}^{(2 p + 1)}(A, D)$.
%{\color{red} (plz proofread)
Combining this with the previous $\EVAL$-reductions and equivalences,
%($\Delta = 2 p + 1$), 
we obtain
\[
\EVAL(\CC) \equiv \EVAL((\AA')^{\odot (\ell + 1)}, \mathfrak P) \le \EVAL^{((p + 1)(\ell +1))}(\AA', \mathfrak D) \equiv \EVAL^{((p + 1)(\ell +1))}(\AA),
\]
so that $\EVAL(\CC) \le \EVAL^{(\Delta)}(\AA)$,
by taking $\Delta = (p + 1)(\ell +1)$.

Remembering that our goal is to prove
the \#P-hardness for the matrix $\AA$
that is not \multblrkone/,
we finally use this assumption.
We first note that when we condensed $\AA$ to $\AA'$, all $\mu_{i j} \ne 0$.
Thus up to nonzero row and column
multipliers, any $2$ by $2$ submatrix witnessing non-\multblrkone/ for $\AA$ is also
a $2$ by $2$ submatrix of $\AA'$
 witnessing non-\multblrkone/ for $\AA'$. %{\color{red} (JYC: this is the only time the term witness is used)}
Hence  $\AA'$ is also not \multblrkone/.
Therefore so is $(\AA')^{\odot(\ell + 1)}$.
Finally, because all $w_i \ne 0$, $\CC$ is also not \multblrkone/.
Hence $\EVAL(\CC)$ is \#P-hard by Corollary~\ref{cor:non-mubr1-hard-unbounded}.
We conclude that $\EVAL^{(\Delta)}(\AA)$ is also \#P-hard,
for  $\Delta = (p + 1)(\ell +1)$.
This completes the proof of Theorem~\ref{thm:non-mult-bl-rk-1-hardness-sym}.

%% file: gadget-meta-thickening.tex
\section{Non-multipicative-block-rank-$1$ from $\pur \AA$ to $\AA$}\label{sec:multblrkone-from-Abar-toA} 
%\section{Non-\multblrkone/ from $\pur \AA$ to $\AA$}\label{sec:multblrkone-from-Abar-toA}
%\section{From non-\modblrkone/ to non-\multblrkone/}
%\section{Obtaining non-multiplicatively-block-rank-$1$ gadgets}

%{\color{red} (should move Corollary~\ref{cor:vand-pur-ind-simp} here, good place for it)}

\begin{theorem}\label{thm:gadget-thickening-non-mult-bl-rk-1}
%Let $\AA \in \bbC^{m \times m}$ be a symmetric matrix,
%let $\pur \AA \in \bbC^{m \times m}$ be the purification of $\AA$
%by going from the generating set 
% $(g_1, \ldots, g_d)$ to primes $(p_1, \ldots, p_d)$, and 
% let $\Gamma$ be an edge gadget.
%If $\gm_{\Gamma, \pur \AA}$ is not \multblrkone/
%% multipicatively-block-rank-$1$
%(which is true if $\gm_{\Gamma, {\pur \AA}}$ is not \modblrkone/),
%%{\color{red}(which is true if $\gm_{\Gamma, \BB}$ is not \modblrkone/ (maybe it's better to commingle to save space instead of making a corollary))},
%then for some $p \ge 1$,
%the matrix $\gm_{T_p(\Gamma), \AA} = \gm_{\Gamma, \AA^{\odot p}}$
%is not \multblrkone/. {\color{red} (the generating or a generating, already fixed?)}
%%%
Let $\AA \in \bbC^{m \times m}$ be a symmetric matrix,
let $\{g_1, \ldots, g_d\}$, where $d \ge 0$, be a generating set of nonzero entries of $\AA$,
%let $p_1 < \ldots < p_d$ be $d$ smallest primes,
let $\pur \AA \in \bbC^{m \times m}$ be the purification of $\AA$ obtained
by going from $(g_1, \ldots, g_d)$ to the $d$ smallest primes $(p_1, \ldots, p_d)$, and 
 let $\Gamma$ be an edge gadget.
If $\gm_{\Gamma, \pur \AA}$ is not \multblrkone/
% multipicatively-block-rank-$1$
(which is true if $\gm_{\Gamma, {\pur \AA}}$ is not \modblrkone/),
%{\color{red}(which is true if $\gm_{\Gamma, \BB}$ is not \modblrkone/ (maybe it's better to commingle to save space instead of making a corollary))},
then for some $p \ge 1$,
the matrix $\gm_{T_p(\Gamma), \AA} = \gm_{\Gamma, \AA^{\odot p}}$
is not \multblrkone/.
\end{theorem}
\begin{proof}
Let $\CC = \gm_{\Gamma, \pur \AA}$,
and let $\BB_n = \gm_{\Gamma, \AA^{\odot n}}$ for $n \ge 1$.
%
%Since $\CC$ is not \multblrkone/,
%there exist $1 \le i_1 < i_2 \le m$ and $1 \le j_1 < j_2 \le m$ such that
%$(i_1, i_2, j_1, j_2)$ witness this fact.
%This means that the $2 \times 2$ submatrix
Since $\CC$ is not \multblrkone/,
there exist $1 \le i_1 < i_2 \le m$ and $1 \le j_1 < j_2 \le m$ such that
the $2 \times 2$ submatrix
\[
\CC_{i_1, i_2; j_1, j_2} =
\begin{pmatrix}
C_{i_1, j_1} & C_{i_1, j_2} \\
C_{i_2, j_1} & C_{i_2, j_2}
\end{pmatrix}
\]
contains at least three nonzero entries
and for every $n \ge 1$, $\CC_{i_1, i_2; j_1, j_2}^{\odot n}$
is nondegenerate, i.e.,
\begin{equation}\label{eqn:2by2C-nondegereate}
C_{i_1, j_1}^n C_{i_2, j_2}^n - C_{i_1, j_2}^n C_{i_2, j_1}^n \ne 0.
\end{equation}
By Corollary~\ref{cor:fin-gen-field-roots-of-unity-full-fin-cyc-gr},
the multiplicative
group of roots of unity in the field  $\mathbb F = \bbQ(\{A_{i, j}\}_{i, j = 1}^m)$ %$\bbQ(\{A_{i, j}\}_{i, j = 1}^m), ${\mathbb F}, K
is a  finite cyclic group.
Let $R$ be (or any positive multiple of) the order of this group.
%Let $R$ be the order and $\varepsilon_R$
%be a generator of this group.
Next, let $I = \{i_1, i_2\} \times \{j_1, j_2\}$ and
for each $(i, j) \in I$, consider the polynomial
\[
% probably
%    x_{i_1, j_2}, x_{i_2, j_1}, x_{i_1, j_2}, x_{i_2, j_1}
p_{i, j}(x_{i_1, j_1}, x_{i_1, j_2}, x_{i_2, j_1}, x_{i_2, j_2}) = ( \prod_{\substack{(i', j') \in I \\ (i', j') \ne (i, j)}} x_{i', j'} ) (x_{i_1, j_1}^R x_{i_2, j_2}^R -
 x_{i_1, j_2}^R x_{i_2, j_1}^R).
\]
Since there are at least three nonzero entries in 
$\CC_{i_1, i_2; j_1, j_2}$ and by (\ref{eqn:2by2C-nondegereate}), 
for some $(a, b) \in I$,
%The previous reasoning implies that for some $(a, b) \in I$,
\begin{equation}\label{eq:pur-nonzero}
p_{a, b}(C_{i_1, j_1}, C_{i_1, j_2}, C_{i_2, j_1}, C_{i_2, j_2}) \ne 0.
\end{equation}
Let $\XX = (X_{k, \ell})_{k, \ell = 1}^m$ be a symmetric matrix of indeterminates
in which $X_{k, \ell}$ and $X_{\ell, k}$ are identified (i.e., $X_{k, \ell} = X_{\ell, k}$)
for $k, \ell \in [m]$.
Consider the matrix $\gm_{\Gamma, \XX}$.
While we only defined $\gm_{\Gamma, \XX}$ where the entries of $\XX$ 
are complex numbers,
the definition extends to arbitrary commutative rings.
For the matrix  $\gm_{\Gamma, \XX}$,
every edge in $\Gamma$ is  assigned the matrix $\XX$,
and therefore  the entries of $\gm_{\Gamma, \XX}$
are complex polynomials in  $\XX$.
In other words, $\gm_{\Gamma, \XX} =
(f_{i, j}(\XX))_{i, j = 1}^m$
for some $f_{i, j}(\XX) \in \bbC[\XX]$, 
% (f_{i, j}((X_{k, \ell})_{k, \ell = 1}^m))_{i, j = 1}^m$
%for some $f_{i, j}((X_{k, \ell})_{k, \ell = 1}^m) \in \bbC[(X_{k, \ell})_{k, \ell = 1}^m]$
where $i, j \in [m]$. (Here we view $\XX = (X_{k, \ell})_{k, \ell = 1}^m$ as s list of entries.)
More precisely, if $\firstdv, \seconddv$ are the distinguished vertices of $\Gamma$ (in this order),
then for each $i, j \in [m]$, we can write
\[
f_{i, j}((X_{k, \ell})_{k, \ell = 1}^m) = \sum_{\substack{\xi: V(\Gamma) \rightarrow [m] \\ \xi(\firstdv) = i, \xi(\seconddv) = j}} \prod_{(u, v) \in E(\Gamma)} X_{\xi(u), \xi(v)}.
\]
Clearly, $\gm_{\Gamma, \AA^{\odot n}} = 
(f_{i, j}(\AA^{\odot n}))_{i, j = 1}^m$
%(f_{i, j}(((A_{k, \ell})^{n})_{k, \ell = 1}^m))_{i, j = 1}^m$
so the entries of $\gm_{\Gamma, \AA^{\odot n}}$
belong to $\mathbb F$, for $n \ge 1$. %$K$ $\mathbf F$ $\bbQ(\{A_{i, j}\}_{i, j = 1}^m)$
% $\bbQ(\{A_{i, j}\}_{i, j = 1}^m)$ for $n \ge 1$.

%Let $\BB_n = \gm_{\Gamma, \AA^{\odot n}}$ for $n \ge 1$.
Since $\BB_n = \gm_{\Gamma, \AA^{\odot n}}$,
%we have $B_{n; i, j} = f_{i, j}((A_{k, \ell}^n)_{k, \ell = 1}^m)$
we have $B_{n; i, j} = f_{i, j}(\AA^{\odot n})$
for $i, j \in [m]$ and $n \ge 1$.
Because $\CC = \gm_{\Gamma, \pur \AA}$
%we also have $C_{i, j} = f_{i j}((\pur A_{k, \ell})_{k, \ell = 1}^m)$
we also have $C_{i, j} = f_{i, j}(\pur \AA)$
for $i, j \in [m]$.
%{\color{red} (JYC: stupid question, maybe it's better to use $1 \le i, j \le m$ to be consistent with Section~\ref{sec:GH-dich-simp}, and $1 \le i_1 < i_2 \le m$ at the beginning, and even the prev section.)}

Let $q_{a, b}((X_{i, j})_{i, j = 1}^m)$ be a complex polynomial defined as
%\[
%q_{a, b}((X_{i, j})_{i, j = 1}^m) = p_{a, b}(f_{i_1, j_1}((X_{i, j})_{i, j = 1}^m), f_{i_1, j_2}((X_{i, j})_{i, j = 1}^m), f_{i_2, j_1}((X_{i, j})_{i, j = 1}^m), f_{i_2, j_2}((X_{i, j})_{i, j = 1}^m)).
%\]
\[
q_{a, b}(\XX) = p_{a, b}(f_{i_1, j_1}(\XX) , f_{i_1, j_2}(\XX) ,  f_{i_2, j_1}(\XX) , 
f_{i_2, j_2}(\XX)).
\]
Then (\ref{eq:pur-nonzero}) rewrites as
\[
q_{a, b}(\pur \AA) \ne 0.
\]
Since $\pur \AA$ is the purification of $\AA$ obtained
by going from $(g_1, \ldots, g_d)$ to $(p_1, \ldots, p_d)$,
by Corollary~\ref{cor:vand-pur-ind-simp}, we have
%%%%%%% NO: this q_{a, b} does not take only 4 input entries. it takes the whole
% matrix
%\[
%q_{a, b}(A_{i_1, j_1}^p, A_{i_1, j_2}^p, A_{i_2, j_1}^p, A_{i_2, j_2}^p) \ne 0
%\]
\[q_{a, b}(\AA^{\odot p}) \ne 0
\]
for some $p \ge 1$ (bounded by the number of terms in the expansion of $q_{a, b}(\XX)$).
This is the same as
\[
p_{a, b}(B_{p; i_1, j_1}, B_{p; i_1, j_2}, B_{p; i_2, j_1}, B_{p; i_2, j_2}) = ( \prod_{\substack{(i, j) \in I \\ (i, j) \ne (a, b)}} B_{p; i, j} ) 
(B_{p; i_1, j_1}^R B_{p; i_2, j_2}^R - B_{p; i_1, j_2}^R B_{p; i_2, j_1}^R) \ne 0.
%%% corrected 3-3-2020
\]
It follows that the matrix
\[
\BB_{p; i_1, i_2; j_1, j_2} =
\begin{pmatrix}
B_{p; i_1, j_1} & B_{p; i_1, j_2} \\
B_{p; i_2, j_1} & B_{p; i_2, j_2}
\end{pmatrix}
\]
has at most one zero entry (which can only be $B_{p; a, b}$),
and
\begin{equation}\label{eq:non-deg-hadamard-power}
B_{p; i_1, j_1}^R B_{p; i_2, j_2}^R - B_{p; i_1, j_2}^R B_{p; i_2, j_1}^R \ne 0.
\end{equation}
If $\BB_{p; i_1, i_2; j_1, j_2}$ has precisely one zero entry, i.e., if $\BB_{p; a, b} = 0$,
then clearly $\BB_p$ is not rectangular so neither is $\gm_{\Gamma, \AA^{\odot p}}^{\odot R} = \BB_p^{\odot R}$
implying that the latter is not block-rank-$1$.
Assume $\BB_{p; i_1, i_2; j_1, j_2}$ has no zero entries.
In this case, (\ref{eq:non-deg-hadamard-power})
means that $\BB_{p; i_1, i_2; j_1, j_2}^{\odot R}$ is nondegenerate
and we conclude that $\gm_{\Gamma, \AA^{\odot p}}^{\odot R} = \BB_p^{\odot R}$ is not block-rank-$1$.

Finally, by Lemma~\ref{lem:mubr1-br1-equivalence},
 $\gm_{\Gamma, \AA^{\odot p}}^{\odot R}$ is not block-rank-$1$
implies that
$\gm_{\Gamma, \AA^{\odot p}}$ is not \multblrkone/.
\end{proof}

%%%%%%%%%%%%%%%%%%%%%%%%%%%%%%%%%%%%%%%%%%%%%%%%%%%%%%%%%%%%%%%%%%%%
\begin{comment}

\begin{remark}
The proof of Theorem~\ref{thm:gadget-thickening-non-mult-bl-rk-1},
can be easily modified so as to show that $\gm_{\Gamma, \AA^{\odot p}}^{\odot (R r)}$
is not block-rank-$1$ for any fixed $r \ge 1$, which also implies $\gm_{\Gamma, \AA^{\odot p}}$ is not \multblrkone/.
See the remark after Lemma~\ref{lem:mubr1-br1-equivalence}.
This is achieved by using $R r$ in place of $R$ after the latter is introduced.
In particular, $p \ge 1$ remains bounded by the number of terms
in the expansion of $q_{a, b}((X_{i, j})_{i, j = 1}^m)$
defined after this substitution.
This will be used for the constructivity of our dichotomy in Section~\ref{sec:constructivity}.
{\color{red} (might need later for constructivity if $R$ can't be computed)}
\end{remark}

\end{comment}
%%%%%%%%%%%%%%%%%%%%%%%%%%%%%%%%%%%%%%%%%%%%%%%%%%%%%%%%%%%%%%%%%%%%

%% file: reduction-to-conneted-components.tex
\section{Reduction to connected components without pinning}\label{sec:pinning-free}

In this section we show that the complexity of $\eval(\AA)$
for bounded
degree graphs
can, just as in~\cite{Cai-Chen-Lu-2013}, be 
reduced to connected $\AA$.
We will do so by a gadget based approach without invoking
any pinning lemma in order to avoid some nonconstructive steps,
which will be important later when we make claims about the effectiveness
of the dichotomy theorem proved in this paper.

For any symmetric matrix $\AA \in \bbC^{m \times m}$,
it is obvious that $Z_\AA(\cdot)$ is unchanged
by a simultaneous row and column permutation on $\AA$ by the same permutation,
which amounts to renaming the elements in $[m]$.
Also, for any edge gadget $\Gamma$,
  the property of $\gm_{\Gamma, \AA}$ being \multblrkone/ (\modblrkone/) is unchanged.
%1st because the sum of product is unchanged, except names for i, j
% then
% %(because it involves a similarity transformation of $\gm_{\Gamma, \AA}$
%by the corresponding permutation matrix).
By a similar reasoning, we can freely multiply the matrix $\AA$ by a
nonzero scalar.
A
%More generally, a 
similar remark holds for a pair 
$(\CC, \fD)$ from Definition~\ref{def:EVAL(C,frakD-modN)}.
%and more generally
%any of the previously defined $\eval$ framework.
We often do these steps implicitly.
% and omit it as it will not affect
%the results of this paper.

%\subsection{Reduction to connected matrices}

%We will show that it suffices to consider connected matrices.
%Unlike Section 4 in~\cite{Cai-Chen-Lu-2013},
%Our reasoning will allow us to avoid using
%the first pinning lemma (Lemma 4.1) from~\cite{Cai-Chen-Lu-2013}.
%Our reasoning will be based on the assumption
%that will follows from the statements formulated
%in Sections~\ref{outlinebipartite-ours} and~\ref{sec:outlinenonbipartite}.
%{\color{red} We note that the proof of the first pinning lemma (in fact, the proofs of the three pinning lemmas)
%from~\cite{Cai-Chen-Lu-2013} preserves degree-boundedness (up to an addition of constant parameter)
%and by the results of~\cite{Cai-Govorov-GH-hom-type} the first pinning lemma can be adapted to also preserve simple graphs,
%but our arguments below will allow us to bypass this step very easily. (not sure if we need to say it, we can say it somewhere in intro, I'd remove it)}

The following lemma allows us to focus on the connected components of $\AA$;
it is a gadget based version of Lemma 4.6 from~\cite{Cai-Chen-Lu-2013}
(without using the first pinning lemma (Lemma 4.1) of~\cite{Cai-Chen-Lu-2013}).

\begin{lemma}\label{connected-ours}
%[{\color{red}modification of Lemma 4.6 from~\cite{Cai-Chen-Lu-2013}}]\label{connected-ours}
Let $\AA\in \mathbb{C}^{m\times m}$ be a symmetric
  matrix with components $\{\AA_i\}_{i \in [s]}$. 
\begin{enumerate}
\item\label{item:part1} If $\Gamma$ is an edge gadget such that $\gm_{\Gamma, \AA_i}$ is not \multblrkone/ for some $i \in [s]$,
then there is 
%we can construct 
an edge gadget $\Gamma_0$
% in polynomial time 
such that $\gm_{\Gamma_0, \AA}$ is not \multblrkone/.
\item\label{item:part2} If $\eval(\AA_i)$ is polynomial-time computable for every $i$,
then so is $\eval(\AA)$.
\end{enumerate}
\end{lemma}
\begin{proof}
\begin{enumerate}
%We first prove~\ref{item:part1}.
\item By Lemma~\ref{lem:edge-gadget-component-nmrk1}, 
we may replace $\Gamma$ by the connected component $\Gamma_0$
containing  the two distinguished vertices of $\Gamma$,
and $\gm_{\Gamma_0, \AA_i}$ is not \multblrkone/.
Since $\Gamma_0$ is connected, it is easy to see that $\gm_{\Gamma_0, \AA} = \diag(\gm_{\Gamma_0, \AA_i})_{i \in [s]}$.
Then  by Property~\ref{prop:equiv-direct-sum-mbrk1},
$\gm_{\Gamma_0, \AA}$ is not \multblrkone/.
% and we are done.

%Now we prove~\ref{item:part2}.
\item By Property~\ref{connectedcomponents-ours},
it suffices to restrict the input to connected graphs.
For any connected $G$ we have $Z_{\AA}(G) = \sum_{i = 1}^s Z_{\AA_i}(G)$
which shows that $\EVAL(\AA)$ is polynomial-time computable.
\end{enumerate}
\end{proof}

The theorems stated in Sections~\ref{outlinebipartite-ours} 
and~\ref{outlinenonbip-ours}
will show that for connected $\AA$, either there is a gadget $\Gamma$
such that $\gm_{\Gamma, \AA}$ is not \multblrkone/ (which leads to
 $\EVAL^{(\Delta)}(\AA)$  is \#P-hard for
some $\Delta > 0$), % {\color{red} (do we really need it)}
or $\EVAL(\AA)$ is tractable (without degree restriction).
Lemma~\ref{connected-ours} allows us to reach the same conclusion
for general $\AA$ without assuming it is connected.

%Lemma~\ref{connected-ours} allows us to assume that $\AA$ is connected,
%since from the sequences of Theorems from Sections~\ref{outlinebipartite-ours} and~\ref{sec:outlinenonbipartite},
%it will follow that assuming so, either we can always construct a gadget $\Gamma$
%such that $\gm_{\Gamma, \AA}$ is not \multblrkone/ or show that $\EVAL(\AA)$ is tractable.

%% file: meta-argument-bounded-degree-graphs-outline.tex
\section{Proof outline of the case: $\AA$ is bipartite}\label{outlinebipartite-ours}

We now give a proof outline  of Theorem~\ref{thm:main-dich-bd}
for the case when $\AA \in \mathbb{C}^{m\times m}$ is connected and bipartite.
For $m=1$, the only bipartite graph on one vertex is an isolated vertex
and $\eval(\AA)$
is trivially computable:  $Z_\AA(G) =0$ if $G$ contains any edge,
% including loop. but we officailly does'nt allow G to have loops
and $1$ otherwise.
For $m=2$, any connected bipartite  graph consists of a single edge,
and again  $\eval(\AA)$
clearly is  tractable. %{\color{red} (It refers to the RHS graph, a bit confusing, and with the previous sentence)}
%%% addded
More precisely, $Z_\AA(G)$ is $0$ unless $G$ is bipartite;
for connected and bipartite $G = (V, E)$, there are at most two assignments
$\xi\colon V \rightarrow \{0,1\}$ which could yield nonzero values;
finally, if $G$ has connected components $G_i$,
then  $Z_\AA(G)$
  is the product of $Z_\AA(G_i)$'s.
%%%
So we assume $m >2$.

In each of the steps below, we consider an $\eval$ problem
passed down by the previous step (Step~1 starts with $\eval(\AA)$ itself)
and show that
\begin{enumerate}[wide]
%\item either we can construct an edge gadget
%whose signature is not \multblrkone/
%and so the $\eval$-problem is \#P-hard for bounded degree graphs, or
\item either there is an edge gadget
whose signature is not \multblrkone/, or
\item the matrix that defines the problem
safisfies certain structural properties, or
\item there is another $\eval$ problem inheriting all the structural
conditions such that
if for the latter there is an edge gadget
whose signature is not \multblrkone/,
% multiplicatitively-block-rank-$1$,
then for the former there is also an
%(possibly different) 
edge gadget
whose signature is not \multblrkone/.
%multiplicatitively-block-rank-$1$.
\end{enumerate}
%{\color{red} (Recall that every non-\modblrkone/ matrix
%is automatically non-\multblrkone/.
%Later in some steps whenever we produce an edge gadget whose signature is
%not \modblrkone/, we will use implicitly this contrapositive.) (added)}

%One can view these three steps as three filters that remove \#P-hard
%  $\eval^{\Delta}(\AA)$ (for some $\Delta > 0$ depending on $\AA$) using different arguments.
%(More concretely, the removed problems are those for which there exists an edge gadget $\Gamma$
%  such that $\gm_{\Gamma, \AA}$ is not \multblrkone/.)
%One can view these three steps as three filters that remove matrices $\AA$
%for which there exists an edge gadget $\Gamma$
%  such that $\gm_{\Gamma, \AA}$ is not \multblrkone/.
%%%% I don't think this "filters" makes much sense now.
%%%% we never filter out if it has  a  edge gadget not \modblrkone/,
%%% we just pass "on" to the next stage, that also has this property,
% to be shown in the end.

Finally, in the last step, we show that if all the structural conditions 
are satisfied, then the problem  $\EVAL(\AA)$ is polynomial-time solvable.

\subsection{Step 1: Regularized form of matrix $\AA$ and its purification}

We start with $\eval(\AA)$, where $\AA\in \mathbb{C}^{m\times m}$
  is a fixed symmetric, connected, and bipartite matrix with \emph{algebraic} entries.
%It is easy to see that if $m=1$, then $\eval(\AA)$ is tractable.
%So in the discussion below, we always assume $m>1$.
%% now m > 2
%In this step, we show that $\eval(\AA)$ is either {$\#$P-hard} or {polynomial-time
%  equivalent} to $\eval(\AA')$, in which $\AA'$ is also an
%  $m\times m$ matrix but has a very nice structure.
In this step, we show that either $\AA$ is not \multblrkone/
(so $\EVAL^\Delta(\AA)$ {$\#$P-hard} for some $\Delta > 0$)
or $\AA$ has a \emph{regularized} form.

\begin{definition}[modification of Definition 5.1 from~\cite{Cai-Chen-Lu-2013}]\label{def:regularized}
Let $\AA\in \mathbb{C}^{m\times m}$ be a symmetric, connected, and bipartite matrix.
We say it is a \emph{regularized bipartite matrix} if
%  there exist $m$ positive real numbers $c_1,c_2,\ldots,c_m$
%%% not say m numbers since they may not be distinct.
there exist nonzero complex numbers $\mu_1, \ldots, \mu_m$
that generate a torsion-free multiplicative group (i.e., that contains no roots of unity other that $1$)
  and an integer $1\le k<m$ such that
%  \textbf{\emph{(number of bits to describe $c_i$? Matrix form of $\AA$)}}
\begin{enumerate}[wide]
\item $A_{i,j}=0$ for all $i,j\in [k]$;
  $A_{i,j}=0$ for all $i,j\in [k+1:m]$; and 
\item $A_{i,j}/(\mu_i\mu_j)=A_{j,i}/(\mu_i\mu_j)$ is a root of unity for all
  $i\in [k]$, $j\in [k+1:m]$.
\end{enumerate}
We say
$\AA$  is a \emph{purified bipartite matrix} if
$\mu_1, \ldots,\mu_m$
are
positive rational numbers.
\end{definition}
In other words, $\AA$ is regularized bipartite 
if there exists a $k \times (m-k)$ matrix
$\BB$ of the form
\begin{equation}\label{eqn:regularized-form-bipartite}
\BB = \left(\begin{matrix}
\mu_1 \\
& \mu_2 \\
& & \ddots\\
& & & \mu_k
\end{matrix}\right)
\left( \begin{matrix}
\zeta_{1,1} & \zeta_{1,2} & \ldots & \zeta_{1,m-k} \\
\zeta_{2,1} & \zeta_{2,2} & \ldots & \zeta_{2,m-k} \\
\vdots      & \vdots      & \ddots & \vdots \\
\zeta_{k,1} & \zeta_{k,2} & \ldots & \zeta_{k,m-k}
\end{matrix}\right)
\left(\begin{matrix}
\mu_{k+1} \\
& \mu_{k+2} \\
& & \ddots\\
& & & \mu_{m}
\end{matrix}\right),
\end{equation}
where every $\mu_i$ is a nonzero complex number
such that $\mu_1, \ldots, \mu_m$ generate a torsion-free multiplicative group
  and every $\zeta_{i,j}$ is a root of unity, %\sout{and} {\color{red}, (reads better, this such that is too long so it's as if it captures the part after and too)} 
and $\AA$ is the bipartization of $\BB$.
If further $\mu_1, \ldots,\mu_m$
are
positive rational numbers, then $\AA$  is purified bipartite.

\begin{theorem}[modification of Theorem 5.2 from~\cite{Cai-Chen-Lu-2013}]\label{bi-step-1-ours}
Let $\AA\in \mathbb{C}^{m\times m}$
  be a symmetric, connected and bipartite matrix with algebraic entries. %, %for some $m>1$,
Then $\AA$ is \multblrkone/ iff $\AA$ is a regularized bipartite matrix.
In that case, if  $\{g_1, \ldots, g_d\}$ is a generating set %{\color{red} of the set of}
of nonzero entries of $\AA$, then we can choose $\mu_1, \ldots, \mu_m$ 
to  belong to the multiplicative
subgroup generated by $\{g_1, \ldots, g_d\}$.

As a consequence, either $\AA$ is not \multblrkone/
(a fortiori, $\EVAL^{(\Delta)}(\AA)$ is \#P-hard for some $\Delta > 0$)
or $\AA$ is a reqularized bipartite matrix.
\end{theorem}

%As a consequence, either $\AA$ is not \multblrkone/
%(a fortiori, $\EVAL^{(\Delta)}(\AA)$ is \#P-hard for some $\Delta > 0$)
%or $\AA$ is a reqularized bipartite matrix.
%In the statement of Theorem~\ref{bi-step-1-ours},
%In Theorem~\ref{bi-step-1-ours}, 
Note that if
 $\AA$ is not a regularized bipartite matrix,
then an edge $e = (\firstdv, \seconddv)$ 
forms an edge gadget with the distinguished vertices $\firstdv, \seconddv$
such that its signature $M_{e, \AA} = \AA$ is not \multblrkone/
by Theorem~\ref{bi-step-1-ours}. 

\subsection{Step 2: Reduction to discrete unitary matrix}

Now let $\AA\in \mathbb{C}^{m\times m}$ denote a
  regularized bipartite matrix.
We show that either there is  an edge gadget $\Gamma$
% we can construct an edge gadget $\Gamma$
such that $\gm_{\Gamma, \AA}$ is not \multblrkone/
(so $\EVAL^{(\Delta)}(\AA)$ is \#P-hard for some $\Delta > 0$)
or  there is  a pair $(\CC, \fD)$,
%or we can construct a pair $(\CC, \fD)$,
where the matrix $\CC$ is the bipartization of a discrete unitary matrix
(see Section~\ref{highlevel-ours} for the definition),
such that $(\CC, \fD)$ satisfies the following property:
(A) If there is an edge gadget $\Gamma$
such that $\gm_{\Gamma, \CC, \fD}$
is not \multblrkone/,
%we can construct an edge gadget $\Gamma'$ (in polynomial time)
then there is an  edge gadget $\Gamma'$ 
such that $\gm_{\Gamma', \AA}$
is not \multblrkone/;
 (B) If  $\eval^\leftrightarrow(\CC,\fD)$ is tractable then so is $\eval(\AA)$. %{\color{red} (better than $\EVAL(\AA) \le \EVAL^\leftrightarrow(\CC, \fD)$ because the former doesn't follow from the latter)}

%{\color{red} Also note that the tensor product of two discrete unitary matrices
%is also discrete unitary. (JYC: just want to confirm with you that it's not needed; there was a comment "seems out of place to say  tensor product  here", not sure if it's mine or yours, there was no JYC initials)}

%%% seems out of place to say  tensor product  here
%
%Also note that the tensor product of two discrete unitary matrices
%is also discrete unitary.

\begin{theorem}[modification of Theorem 5.3 from~\cite{Cai-Chen-Lu-2013}]\label{bi-step-2-ours}
Given a regularized bipartite matrix $\AA\in \mathbb{C}^{m\times m}$, 
%either 1. $\EVAL(\AA)$ is tractable
either
(I) there exists an edge gadget $\Gamma$
such that $\gm_{\Gamma, \AA}$ is not \multblrkone/
(a fortiori,  $\eval^{(\Delta)}(\AA)$ is \#P-hard for some $\Delta > 0$)
or (II) there exists a triple
  $((M,N),\CC,\fD)$ such that 
\begin{enumerate}[wide]
\item
%2.1. 
for any edge gadget $\Gamma$
such that $\gm_{\Gamma, \CC, \fD}$
is not \multblrkone/, %(not \modblrkone/)
we can construct an edge gadget $\Gamma'$
% (in polynomial time)
%there is an edge gadget $\Gamma'$ 
so that $\gm_{\Gamma', \AA}$
is not \multblrkone/;
\item
%2.2.
 $\EVAL(\AA) \le \EVAL^\leftrightarrow(\CC, \fD)$;
% %{\color{red} (constructively (can be removed))}
% $\EVAL(\AA) \le \EVAL(\CC, \fD)$;
and
\item
%and 3.3.
 $((M,N),\CC,\fD)$ satisfies the following 
  conditions:\label{CONDITIONU}
\begin{enumerate}[wide]
\item[$(\calU_1)$]
$\CC\in \mathbb{C}^{2n\times 2n}$ for some $n\ge 1$, and
  \[\fD=\big(\DD^{[0]}, \DD^{[1]},\ldots,\DD^{[N-1]}\big)\] is a
  sequence of $N$ $2n\times 2n$ diagonal matrices over $\mathbb{C}$
  for some \emph{even} $N>1$.
  
\item[$(\calU_2)$]
$\CC$ is the bipartization of an $M$-\emph{discrete unitary
  matrix} $\FF\in \mathbb{C}^{n\times n}$,
  where\\ $M\ge 1$ and $M\hspace{0.05cm}|\hspace{0.05cm} N$. {(}Note that 
  $\CC$  and $\FF$
uniquely determine each other.)
\item[$(\calU_3)$]
$\DD^{[0]}$ is the $2n\times 2n$ identity matrix,  
  and for every $r\in [N-1]$ we have
\begin{align*}
\exists\hspace{0.04cm} i\in [n],\hspace{0.06cm} D_{i}^{[r]}
  \ne 0\ &\Longrightarrow\ \exists\hspace{0.04cm} i'\in [n],
  \hspace{0.06cm}D_{i'}^{[r]}=1,\ \ \ \ \ \ \ \ \text{and}\\[0.2ex]
\exists\hspace{0.04cm} i\in [n+1:2n],\hspace{0.06cm}
  D_{i}^{[r]}\ne 0\ &\Longrightarrow\ 
  \exists\hspace{0.04cm} i'\in [n+1:2n],\hspace{0.06cm}D_{i'}^{[r]}=1. 
\end{align*}

\item[$(\calU_4)$] For all $r \in [N-1]$ and all $i\in [2n]$,
  $D^{[r]}_i\in \mathbb{Q}(\oo_N)$ and $|D_i^{[r]}|\in \{0,1\}$.
\end{enumerate}
\end{enumerate}
\end{theorem}
%{\color{red} (my addition)}
%We note that in Theorem~\ref{bi-step-2-ours},
%we slightly abuse the notation for convenience
%as was done in~\cite{Cai-Chen-Lu-2013},
%and the pair $(\CC, \fD)$ in the statement of Theorem~\ref{bi-step-2-ours}
%is not the pair also denoted by $(\CC, \fD)$ and used in the derivation of this theorem.

In fact there are two levels of $(\CC, \fD)$ involved in proving 
Theorem~\ref{bi-step-2-ours}. 
Assuming  $\gm_{\Gamma, \AA}$ is \multblrkone/ for every edge gadget $\Gamma$,
 the problem $\EVAL(\AA)$  is first shown to be equivalent to
some $\EVAL(\CC, \fD)$, which must further be factorizable
as a tensor product 
of an \emph{outer problem}
$\eval(\CC',{\frak K})$ and an \emph{inner problem}
 $\eval(\CC'', {\frak L})$, where $\eval(\CC',{\frak K})$ is tractable. %{\color{red} (JYC: very imprecise, but whatever, it's still an outline)}
It is the inner $(\CC'', {\frak L})$  we rename as $\EVAL(\CC, \fD)$
in the conclusion of Theorem~\ref{bi-step-2-ours}.

In addition to $(\CC, \fD)$,
we will also need to introduce a purified pair $(\pur \CC, \fD)$,
where $\pur \CC$ is a purification of $\CC$, as an auxiliary tool
that will be used to relate to the purification matrix $\pur \AA$.

\subsection{Step 3: Canonical form of $\CC$, $\FF$ and $\fD$}

After the first two steps, the original problem $\eval(\AA)$ is
  shown to be either tractable,
  %or we can construct an edge gadget $\Gamma$
or there is  an edge gadget $\Gamma$
  such that $\gm_{\Gamma, \AA}$ is not \multblrkone/
  (a fortiori, $\EVAL^{(\Delta)}(\AA)$ is \#P-hard for some $\Delta > 0$), or
  %we can construct a pair $(\CC, \fD)$ such that
there is a pair $(\CC, \fD)$ such that
   for any an edge gadget $\Gamma$
  such that $\gm_{\Gamma, \CC, \fD}$
  is not \multblrkone/, %(not \modblrkone/)
  %we can construct (there exists) an edge gadget $\Gamma'$ (in polynomial time)
there is  an edge gadget $\Gamma'$ 
  so that $\gm_{\Gamma', \AA}$
  is not \multblrkone/;
%  and $\EVAL(\AA) \le \EVAL^\leftrightarrow(\CC, \fD)$. %(constructively)
and $\EVAL(\AA) \le \EVAL^\leftrightarrow(\CC, \fD)$.
There are also positive integers $M$ and $N$ such
  that $((M,N),\CC,\fD)$ satisfies conditions $(\calU_1)$--$(\calU_4)$.

For convenience, we use $2m$ to denote the number of
%%% i don't call it use 2m again! because earlier m was the whole A dimension 
% more as 2m here.
  rows of $\CC$ and $\DD^{[r]}$, though it should be noted
  that this new $m$ is indeed the $n$ in Theorem \ref{bi-step-2-ours},
  which is different from the $m$ used in the first two steps.
We also denote the upper-right $m\times m$ block of $\CC$ by $\FF$.
%This m >1 is stated later.... so now i remove it...
%We may still assume the new $m > 1$ for otherwise $\EVAL(\CC, \fD)$ is
%tractable.
%%% I added this: again, for connected G, and bipartite.
% there are only two maps.. so tractable.

In this step, we adopt the following convention: Given an
  $n\times n$ matrix, we use $[0:n-1]$, instead of $[n]$,
  to index its rows and columns.
For example, we index the rows of $\FF$ using $[0:m-1]$
  and index the rows of $\CC$ using $[0:2m-1]$.

%{\color{blue}
%We start with the special case when $M=1$.
%As $\FF$ is $M$-discrete unitary, we must have $m=1$.
%It is easy to check that $\eval^\leftrightarrow(\CC,\fD)$
%  is tractable: $\CC$ is a $2\times 2$ matrix %with $C_{1,1}=C_{2,2}=0$  and $C_{1,2}=C_{2,1}=1$;
%\[
%\left( \begin{matrix} 0 & 1 \\ 1 & 0 \end{matrix} \right); 
%\]
%$Z_{\CC,{\frak D}}(G)$ is $0$ unless $G$ is bipartite;
%for connected and bipartite $G = (V, E)$, there are at most two assignments
%$\xi\colon V \rightarrow \{0,1\}$ which could yield nonzero values;
%finally, if $G$ has connected components $G_i$,
%then  $Z_{\CC,\fD}(G)$
%  is the product of $Z_{\CC,\fD}(G_i)$'s.%\vspace{0.005cm}
%(old incorrect)}

We start with the special case when $M=1$.
As $\FF$ is $M$-discrete unitary, we must have $m=1$.
It is easy to check that $\eval^\leftrightarrow(\CC,\fD)$
  is tractable: $\CC$ is a $2\times 2$ matrix %with $C_{1,1}=C_{2,2}=0$  and $C_{1,2}=C_{2,1}=1$;
\[
\left( \begin{matrix} 0 & 1 \\ 1 & 0 \end{matrix} \right); 
\]
$Z_{\CC,{\frak D}}^\rightarrow(G)$ and $Z_{\CC,{\frak D}}^\leftarrow(G)$ are $0$ unless $G$ is bipartite;
for connected and bipartite $G$, there are at most two assignments
$\xi\colon V \rightarrow \{0,1\}$ which could yield nonzero values: at most one for $Z_{\CC,{\frak D}}^\rightarrow(G)$ and at most one for $Z_{\CC,{\frak D}}^\leftarrow(G)$;
finally, if $G$ has connected components $G_i$,
then the proof is similar.
%then  $Z_{\CC,\fD}(G)$
%  is the product of $Z_{\CC,\fD}(G_i)$'s. (not sure how to say the last sentence, or should I just generalize Property 2.4)

%%%%%%%%%% where do we state the easy statement that
%%%% for un-connected A,
%%% the Z_A(G) is the product over all connected components of G.

%%% JYC
For the general case when the parameter $M>1$ we further investigate the structure of
  $\FF$ as well as the diagonal matrices in
 $\fD$, and derive three necessary conditions on them
 so that any violation of these conditions will %of which
 let us construct an edge gadget $\Gamma$
 such that $\gm_{\Gamma, \CC_, \fD}$ is not \modblrkone/. %(and so not \multblrkone/ either)
%  $\eval(\CC,\fD)$
%  to not be {$\#$P-hard}.
In the tractability part, we prove that these conditions
  are actually sufficient for $\EVAL^\leftrightarrow(\CC, \fD)$ to be polynomial-time computable.

%%%%%%%%%%%%%%%%%%%%%%%%%%%%%%%%%%%%%%%%%%%%%%%%%%%%%%%%%%%%%%%%%%%%
\begin{comment}
{\color{blue} (old text)
\subsubsection*{Step 3.1: Entries of $\DD^{[r]}$ are Either $0$ or Powers of $\oo_N$}

%Suppose $((M,N),\CC,\fD)$ satisfies conditions $(\calU_1)$--$(\calU_4)$ and $M>1$.
%In the first step, we show that either $\eval(\CC,\fD)$ is \#P-hard
%  or every entry of $\DD^{[r]}$ (in $\fD$), $r\in [N-1]$, is either $0$
%  or a power of $\oo_N$.

In the first step, we prove the following theorem:

\begin{theorem}\label{step30}
Suppose $((M,N), \CC,\fD)$ satisfies $(\calU_1)$--$(\calU_4)$ with $M>1$.
Then~either $\eval(\CC,\fD)$~is \#P-hard or $((M,N),\CC,\fD)$
  satisfies the following condition $(\calU_5)$: 
\begin{enumerate}
\item[$(\calU_5)$] For all $r\in [N-1]$ and $i\in [0:2n-1]$,
   $D^{[r]}_i$ is either $0$ or a power of $\oo_N$.\label{CONDITIONU5}
\end{enumerate}
\end{theorem}
}

{\color{red} (new text)}
\end{comment}
%%%%%%%%%%%%%%%%%%%%%%%%%%%%%%%%%%%%%%%%%%%%%%%%%%%%%%%%%%%%%%%%%%%%
\subsubsection{Step 3.1: Entries of $\DD^{[r]}$ are either $0$ or powers of $\oo_N$}

%Suppose $((M,N),\CC,\fD)$ satisfies conditions $(\calU_1)$--$(\calU_4)$ and $M>1$.
%In the first step, we show that either $\eval(\CC,\fD)$ is \#P-hard
%  or every entry of $\DD^{[r]}$ (in $\fD$), $r\in [N-1]$, is either $0$
%  or a power of $\oo_N$.

In the first step within Step 3, we prove the following theorem:

\begin{theorem}[modification of Theorem 5.4 from~\cite{Cai-Chen-Lu-2013}]\label{step30-ours}
Suppose $((M,N), \CC,\fD)$ satisfies $(\calU_1)$--$(\calU_4)$ with $M>1$.
Then~either there is an edge gadget $\Gamma$
%we can construct (there exists) an edge gadget $\Gamma$
such that $\gm_{\Gamma, \CC, \fD}$ is not \modblrkone/ %(and so not \multblrkone/ either)
or $((M,N),\CC,\fD)$
  satisfies the following condition $(\calU_5)$: 
\begin{enumerate}[wide]
\item[$(\calU_5)$] For all $r\in [N-1]$ and $i\in [0:2n-1]$,
   $D^{[r]}_i$ is either $0$ or a power of $\oo_N$.\label{CONDITIONU5}
\end{enumerate}
\end{theorem}

\subsubsection{Step 3.2: Fourier decomposition}\label{subsec:Fourier-decomposition}

Second, we show that either there exists
% we can construct (there exists)
an edge gadget $\Gamma$ such that $\gm_{\Gamma, \CC, \fD}$
is not \modblrkone/, %(and so not \multblrkone/ either)
  or we can permute
  the rows and columns of $\FF$, so that the new $\FF$ is the tensor
  product of a collection of \emph{Fourier matrices} defined below:

\begin{definition}[Definition 5.5 from~\cite{Cai-Chen-Lu-2013}]\label{FourierMatrix-ours}
Let $q>1$ be a prime power,
  and $k\ge 1$ be an integer such that $\gcd\hspace{0.04cm}(k,q)=1$.
We call the following $q\times q$
  matrix $\bcF_{q,k}$ a $(q,k)$-\emph{Fourier matrix}:
% \emph{(}In particular,
%  when $k=1$, we use $\bcF_{q}$ to denote $\bcF_{q,1}$ for short\emph{)}:
  The $(x,y)${th} entry of $\bcF_{q,k}$, where $x,y\in [0:q-1]$, is
\[
\omega_{q}^{kxy}=e^{2\pi i\big({kxy}/{q}\big)}.
\]
In particular, when $k=1$, we use $\bcF_{q}$ to denote $\bcF_{q,1}$ for short.
\end{definition}

\begin{theorem}[modification of Theorem 5.6 from~\cite{Cai-Chen-Lu-2013}]\label{bi-step-3-ours}
Assume $((M,N),\CC,\fD)$ satisfies conditions $(\calU_1)$--$(\calU_5)$ and
  $M>1$.
Then either 
%we can construct 
there exists an edge gadget $\Gamma$
such that $\gm_{\Gamma, \CC, \fD}$ is not \modblrkone/ or there exist %(and so not \multblrkone/ either)
  permutations $\Sigma$ and $\Pi$ of $[0:m-1]$ and 
  a sequence $q_1,q_2,\ldots,q_d$ of $d$ prime powers, for some $d\ge 1$,
such that 
\begin{equation}\label{normalform-ours}
\FF_{\Sigma,\Pi}=
\bigotimes_{i\in [d]}\hspace{0.06cm} \bcF_{q_i}.
\end{equation}\end{theorem}

Suppose there do exist permutations $\Sigma, \Pi$ and prime powers
  $ q_1,\ldots,q_d$  such that
  $\FF_{\Sigma,\Pi}$ satisfies (\ref{normalform-ours}). Then we
  let $\CC_{\Sigma,\Pi}$ denote the bipartization of $\FF_{\Sigma,\Pi}$
  and let $\fD_{\Sigma,\Pi}$ denote a sequence of $N$ $2m\times 2m$
  diagonal matrices in~which the $r${th} matrix is
\[
\left(\begin{matrix}
D^{[r]}_{\Sigma(0)}\\
%& D^t_{\Sigma(2)}\\
& \ddots\\
& & D^{[r]}_{\Sigma(m-1)}\\
& & & D^{[r]}_{\Pi(0)+m}\\
%& & & & D^t_{\Pi(2)+m,\Pi(2)+m}\\
& & & & \ddots\\
& & & & & D^{[r]}_{\Pi(m-1)+m}
\end{matrix}\right),\ \ \ \text{$r\in [0:N-1]$.}
\]
Since 
%%%% don't we officially have to say it preserves the gadget nonblckrk1 stuff?
%permuting the rows and
%columns of $\CC$ and $\DD^{[r]}$
%by the same permutation pair 
%  does not affect the complexity of $\eval(\CC,\fD)$,
  %$\eval(\CC_{\Sigma,\Pi},\fD_{\Sigma,\Pi})\equiv\eval(\CC,\fD)$,
$\eval(\CC_{\Sigma,\Pi},\fD_{\Sigma,\Pi})$ and $\eval(\CC,\fD)$
are really the same problem, %{\color{red} (enough? what about $\leftrightarrow$, I guess this sentences is more like saying that the frameworks are the same)}
%From now on, 
we will let $\FF,\CC$ and $\fD$ denote $\FF_{\Sigma,\Pi},
  \CC_{\Sigma,\Pi}$ and $\fD_{\Sigma,\Pi}$, respectively, with
%By (\ref{normalform}), the new $\FF$ satisfies
\begin{equation}\label{normalnew-ours}
\FF=\bigotimes_{i\in [d]}\hspace{0.06cm} \bcF_{q_i}.
\end{equation}

Before moving forward, we rearrange the prime powers $q_1,q_2,\ldots,q_d$
  and divide them into groups according to different primes.
We need the following notation.
%\begin{definition}
%Let $t\ge 1$ be a positive integer, and $q>1$ be a prime power.
%We call $f$ a \emph{$2t$-variable non-degene\-rate
%  quadratic polynomial over $\mathbb{Z}_q$} if
%  $$f(x_1,\ldots,x_t,y_1,\ldots,y_t)=\sum_{i,j\in [t]}a_{i,j}x_iy_j,$$
%  where $x_i,y_j$ are variables in $\mathbb{Z}_q$ and $a_{i,j}\in \mathbb{Z}_q$
%  for all $i,j\in [t]$,
%  and
%\begin{itemize}
%\item For all nonzero $\aa=(a_1,a_2,\ldots,a_t)\in \mathbb{Z}_q^t$, $f(\aa,\yy)$ is a
%  nonzero linear form of $\yy$; and
%\item For all nonzero $\bb=(b_1,b_2,\ldots,b_t)\in \mathbb{Z}_q^t$, $f(\xx,\bb)$ is a
%  nonzero linear form of $\xx$.
%\end{itemize}
%\end{definition}
Let $\pp=(p_1,\ldots,p_s)$ be a strictly increasing
  sequence of primes % such that $p_1<p_2<
%  \hspace{-0.05cm}\ldots\hspace{-0.05cm}<p_s$
  and $\bft=(t_1,\ldots,t_s)$ be a sequence of positive integers.
Let $\fqq=\{\qq_{i}:
  i\in [s]\}$ be a set of $s$ sequences in which
  each $\qq_i$ is a nonincreasing 
  sequence $(q_{i,1},\ldots,q_{i,t_i})$ of powers of $p_i$. % such
%  that $q_{i,1}\ge \ldots\ge q_{i,t_i}$.
We let $q_i$ denote $q_{i,1}$ for all $i\in [s]$, let
\[\mathbb{Z}_{\qq_i}= \prod_{j\in [t_i]}\mathbb{Z}_{q_{i,j}}
= \mathbb{Z}_{q_{i,1}} \times \cdots \times \mathbb{Z}_{q_{i,t_i}},
\]
for all $i\in [s]$, and let 
%%\text{Matrix form of $\mathbb{Z}_{\fqq}$}\ \
\[
\mathbb{Z}_{\fqq}  = \prod_{i\in [s],j\in [t_i]}
  \hspace{-0.34cm}\mathbb{Z}_{q_{i,j}} = \prod_{i\in [s]}\mathbb{Z}_{\qq_i}
=\mathbb{Z}_{q_{1,1}} \times \cdots \times \mathbb{Z}_{q_{1,t_1}}
 \times\cdots
%\mathbb{Z}_{q_{2,1}} \times \cdots %\times \mathbb{Z}_{q_{2,t_2}}
 \times \mathbb{Z}_{q_{s,1}} \times \cdots \times \mathbb{Z}_{q_{s,t_s}}
\]
be the Cartesian products of the respective finite Abelian groups.
Both $\mathbb{Z}_\fqq$ and $\mathbb{Z}_{\qq_i}$
  are finite Abelian groups, under componentwise operations.
% under these operations.
This implies that both $\mathbb{Z}_\fqq$ and $\mathbb{Z}_{\qq_i}$ are
  $\mathbb{Z}$-modules and thus $k\xx$ is well defined for
 all $k\in \mathbb{Z}$
  and $\xx$ in $\mathbb{Z}_\fqq$ or $\mathbb{Z}_{\qq_i}$.
As $\mathbb{Z}$-modules,
we can also refer to their members as ``vectors''.
%\ \ \
%\text{and}\ \ \
%\mathbb{Z}_{\qq_i}\equiv \prod_{j\in [t_i]}\mathbb{Z}_{q_{i,j}},\ \ \text{
%for all $i\in [s]$.}
%$$
When we use $\xx$ to denote a vector in
  $\mathbb{Z}_{\fqq}$, we denote its $(i,j)${th} entry
  by $x_{i,j}\in \mathbb{Z}_{q_{i,j}}$. We
  use $\xx_i$ to denote $(x_{i,j}: j\in [t_i])
  \in \mathbb{Z}_{\qq_i}$, so $\xx=(\xx_1,\ldots,\xx_s)$.
Given $\xx,\yy\in \mathbb{Z}_{\fqq}$,
  we let $\xx\pm \yy$ denote the vector in $\mathbb{Z}_{\fqq}$
  whose $(i,j)${th} entry is
 $x_{i,j}\pm y_{i,j}{\pmod{q_{i,j}}}.$
Similarly, for each $i\in [s]$, we can define $\xx\pm \yy$
  for vectors $\xx,\yy\in \mathbb{Z}_{\qq_i}$. 
%One can check that both $\mathbb{Z}_\fqq$ and $\mathbb{Z}_{\qq_i}$
%  are finite Abelian groups under these operations.
%This implies that both $\mathbb{Z}_\fqq$ and $\mathbb{Z}_{\qq_i}$ are
%  $\mathbb{Z}$-modules and thus, $k\xx$ is
%well defined for all $k\in \mathbb{Z}$
%  and $\xx$ in $\mathbb{Z}_\fqq$ or $\mathbb{Z}_{\qq_i}$.

From (\ref{normalnew-ours}), there exist $\pp,\bft,\fq$ such that
  $((M,N),\CC,\fD,(\pp,\bft,\fq))$ satisfies the 
  following three conditions
  $(\calR_1)$--$(\calR_3)$, which we will refer to combined 
  as $(\calR)$.\label{CONDITIONR}
\begin{enumerate}[wide]
\item[$(\calR_1)$]
 $\pp=(p_1,\ldots,p_s)$ is a strictly increasing sequence of primes;
  % such that $p_1<\cdots<p_s$;
  $\bft=(t_1,\ldots,t_s)$ is a sequence of positive integers;
  $\fq=\{\qq_{i}:i\in [s]\}$ is a collection of $s$ sequences,
  in which each $\qq_i=(q_{i,1},\ldots,q_{i,t_i})$ 
  is a nonincreasing sequence of
  powers of $p_i$. %such that $q_{i,1}\ge \cdots\ge q_{i,t_i}$. 

%%% To Xi: do you like this: $\wp =(\qq_{i},i\in [s])$

\item[$(\calR_2)$] $\CC$ is the bipartization of $\FF\in
  \mathbb{C}^{m\times m}$ and $((M,N),\CC,\fD)$ satisfies
  $(\calU_{1})$--$(\calU_5)$.

\item[$(\calR_3)$] There is a bijection $\rho\colon
  [0:m-1]\rightarrow \mathbb{Z}_{\fq}$ (so $m=\prod_{i,j} q_{i,j}$) such that
\begin{equation}\label{jjj1-ours}
F_{a,b}=\prod_{i\in [s],j\in [t_i]}
\omega_{q_{i,j}}^{x_{i,j}\hspace{0.03cm}y_{i,j}},
  \quad\text{for all\ $a,b\in [0:m-1]$},
\end{equation}
where $(x_{i,j}:i\in [s],\hspace{0.03cm}j\in [t_i])=\xx=\rho(a)$ and
  $(y_{i,j}:i\in [s],j\in [t_i])=\yy=\rho(b)$.
Note that
  (\ref{jjj1-ours}) also gives us an expression of $M$ using $\fq$.
It is the product of the largest prime powers $q_{i} = q_{i,1}$
 for each distinct prime $p_i$:
$M= q_1q_2\cdots q_s.$ 
\end{enumerate}

For convenience, we from now on use $\xx\in \mathbb{Z}_{\fq}$ to
  index the rows and columns of $\FF$:
\begin{equation}\label{new-R3-with-F-in-bold-face-index-ours}
F_{\xx,\yy}= F_{\rho^{-1}(\xx),\rho^{-1}(\yy)}
=\prod_{i\in [s],j\in [t_i]}
\omega_{q_{i,j}}^{x_{i,j}\hspace{0.03cm}y_{i,j}},
  \ \ \ \text{for all\ $\xx,\yy\in \zqt$},
\end{equation}
whenever we have a tuple
  $((M,N),\CC,\fD,(\pp,\bft,\fq))$ that is known to satisfy condition ($\calR$).
We assume that $\FF$ is indexed by $(\xx,\yy)\in \mathbb{Z}_\calQ^2$
rather than $(a,b)\in [0:m-1]^2$, and $(\calR_3)$
refers to (\ref{new-R3-with-F-in-bold-face-index-ours}).
Correspondingly,
  we use $\{0,1\} \times \mathbb{Z}_{\fqq}$ to index the entries of the matrices $\CC$ and $\DD^{[r]}$:
$(0,\xx)$ refers to the $(\rho^{-1}(\xx))${th} row or column,
  and $(1,\xx)$ refers to the $(m+\rho^{-1}(\xx))${th} row or column.

\subsubsection{Step 3.3: Affine support for $\fD$}\label{subsec:Step-3.3}

Now we have a $4$-tuple $((M,N),\CC,\fD,$ $(\pp,\bft,\fq))$
  that satisfies ($\calR$).
In this step, we prove for every $r\in [N-1]$ (recall that $\DD^{[0]}$ is
  already known to be the identity matrix),
  the nonzero entries of the $r${th} matrix $\DD^{[r]}$
  in $\fD$ must have a very nice \emph{coset} structure;
  %otherwise we can construct (there exists) an edge gadget $\Gamma$
otherwise there exists an edge gadget $\Gamma$
  such that $\gm_{\Gamma, \CC_, \fD}$ is not \modblrkone/. %(and so not \multblrkone/ either)

For every $r\in [N-1]$, we define $\Lambda_r \subseteq \zqt$
 and $\Delta_r\subseteq \zqt$ as 
\[
\Lambda_r=\big\{\xx\in \zqt:
  D^{[r]}_{(0,\xx)}\ne 0\big\}\ \ \ \text{and}\ \ \
\Delta_r=\big\{\xx\in \zqt:
  D^{[r]}_{(1,\xx)}\ne 0\big\}. 
\]
We use $\cS$ to denote the set of $r\in [N-1]$ such that $\Lambda_r\ne \emptyset$
  and $\cT$ to denote the set of $r\in [N-1]$ such that $\Delta_r\ne \emptyset$.
We recall the following standard definition of a coset of a group,
specialized to our situation. 

\begin{definition}[Definition 5.7 from~\cite{Cai-Chen-Lu-2013}]
Let $\Phi$ be a nonempty subset of $\mathbb{Z}_{\fqq}$ \emph{(}or $\mathbb{Z}_{
  \qq_i}$ for some $i\in [s]$\emph{)}.
We say $\Phi$ is a \emph{coset} in $\mathbb{Z}_{\fqq}$ \emph{(}or $\mathbb{Z}_{
  \qq_i}$\emph{)} if there is a
  vector $\xx_0\in \Phi$ such that
$
 \{\xx-\xx_0\hspace{0.05cm} |\hspace{0.05cm} \xx\in \Phi \}
$
is a subgroup of $\mathbb{Z}_{\fqq}$ \emph{(}or $\mathbb{Z}_{\qq_i}$\emph{)}.
Given a coset $\Phi$ \emph{(}in $\mathbb{Z}_\calQ$ or $\mathbb{Z}_{\qq_i}$\emph{)},
  we use $\Phi^{\text{\rm lin}}$ to denote its corresponding
  subgroup $\{\xx-\xx'\hspace{0.05cm}|\hspace{0.05cm} \xx,\xx'\in \Phi\}$. 
%Being a subgroup, clearly
% $\Phi^{\text{\rm lin}}=\{\xx-\xx'\hspace{0.05cm}|\hspace{0.05cm} \xx,\xx'\in \Phi\}
%= \{\xx-\xx_0\hspace{0.05cm} |\hspace{0.05cm} \xx\in \Phi \}$,
%  for any $\xx_0\in \Phi$. 
\end{definition}

\begin{theorem}[modification of Theorem 5.8 from~\cite{Cai-Chen-Lu-2013}]\label{bi-step-4-ours}
Let $((M,N),\CC,\fD,(\pp,\bft,\fq))$ be a $4$-tuple that satisfies
  $(\calR)$. Then either 
%we can construct an gadget $\Gamma$ such that $\gm_{\Gamma, \CC, \fD}$ is not \modblrkone/; or
there is an edge gadget $\Gamma$ such that $\gm_{\Gamma, \CC, \fD}$ is not \modblrkone/ or
  $\Lambda_r,\Delta_r\subseteq \zqt$ satisfy the
  following condition $(\calL)$:
\begin{enumerate}[wide]
\item[$(\calL_1)$] For every $r\in \cS$, $\Lambda_r=\prod_{i=1}^s
  \Lambda_{r,i}$, where $\Lambda_{r,i}$
  is a coset in $\mathbb{Z}_{\qq_i}$, $i\in [s]$.
\item[$(\calL_2)$] For every $r\in \cT$, $\Delta_r=\prod_{i=1}^s
  \Delta_{r,i}$, where $\Delta_{r,i}$
  is a coset in $\mathbb{Z}_{\qq_i}$, $i\in [s]$.%\vspace{0.016cm}
\end{enumerate}
\end{theorem}

Suppose for any gadget $\Gamma$,
$\gm_{\Gamma, \CC, \fD}$ is \modblrkone/.
%%% waitijg for Artem clarification
% (this implies that itand \multblrkone/).
Then by Theorem \ref{bi-step-4-ours},
  $((M,N),\CC,\fD, (\pp,\bft,$ $\fq))$ satisfies
  not only ($\calR$) but also ($\calL$).
Actually, by ($\calU_3$), $\fD$ also satisfies the following:\label{CONDITIONL}
%For every $r\in \cS$, we arbitrarily pick a vector from
%  $\Lambda_r$ and call it $\boldsymbol{\frak a}^{[r]}$.
%We also pick a vector $\boldsymbol{\frak b}^{[r]}$
%  from $\Delta_r$, for every $r\in \cT$.
%It is safe to assume that our $4$-tuple $((M,N),\CC,\fD,(\pp,\bft,\fq ))$
%  satisfies the following additional condition:
\begin{enumerate}[wide]
\item[$(\calL_3)$] 
There exist an $\fa^{[r]}\in \Lambda_r$ for each $r\in \cS$,
  a $\fb^{[r]}\in \Delta_r$ for each $r\in \cT$ such that
 %For every $r\in \cS$, $\exists\hspace{0.06cm}
\[D^{[r]}_{(0,\fa^{[r]})}= 
  D^{[r]}_{(1,\fb^{[r]})}=1.\]
\end{enumerate}
%Indeed, if some $\DD^{[r]}$ does not satisfy ($\calL_3$), then
%by a global factor to $Z_{\CC,{\frak D}}(G)$ we can normalize every
%such $\DD^{[r]}$
%  to a matrix that satisfies $(\calL_3)$. This does not affect
%  the complexity of problem $\eval(\CC,\fD)$.
%We can simultaneously normalize every $\DD^{[r]}$
%for every $r \in \cS$ and every
%$r \in \cT$.  These simultaneous normalizations do not
%interfere with each other.
From now on, when we say condition $(\calL)$, we mean all three conditions
  $(\calL_1)$--$(\calL_3)$.

\subsubsection{Step 3.4: Quadratic structure}

In this final step within
Step 3, we prove that for every $r\in [N-1]$, the nonzero entries of
  $\DD^{[r]}$ must have a \emph{quadratic}
  structure; otherwise there is an edge gadget $\Gamma$
%otherwise we can construct an edge gadget $\Gamma$
   such that $\gm_{\Gamma, \CC_, \fD}$ is not \modblrkone/. %(and so not \multblrkone/ either)
We start with some notation.

%We need the following notation: Given a coset $\Phi$ in $\zqt$ (or $\mathbb{Z}_{\qq_i}$),
%  we let $\Phi^{\text{lin}}$ denote
%  $$\lin{\Phi}\equiv\{\xx-\xx'\mid \xx,\xx'\in \Phi\},$$
%which is a subgroup in $\zqt$ (or $\mathbb{Z}_{\qq_i}$).
Given $\xx$ in $\mathbb{Z}_{\qq_i}$ for some $i\in [s]$,
    we use $\ext_r(\xx)$ (extension of $\xx$ for short), 
    where $r\in \cS$, to denote the following unique vector: 
\[\left(\fa^{[r]}_1,\ldots,\fa^{[r]}_{i-1},\xx,\fa^{[r]}_{i+1},\ldots,
  \fa^{[r]}_s\right)\in \zqt.\] %such that
%in the\vspace{-0.03cm} expression $\xx' = (\xx_1', \ldots, \xx_s')
%\in \mathbb{Z}_{\fqq} = \prod_{i \in [s]} \mathbb{Z}_{\qq_i}$,
%its $i${th} component
%    $\xx_i'=\xx$, the vector given in  $\mathbb{Z}_{\qq_i}$,\vspace{-0.09cm}
%  and $$\xx_j'=\fa^{[r]}_j,\ \ \text{for all $j\ne i$}.$$
%Recall that $\fa^{[r]}$ is a vector we picked from $\Lambda_r$ in condition %$(\calL_3)$.
Similarly we let $\ext_r'(\xx)$, where $r\in \cT$,
  denote the following unique vector: 
\[\left(\fb^{[r]}_1,\ldots,\fb^{[r]}_{i-1},\xx,
  \fb^{[r]}_{i+1},\ldots,\fb^{[r]}_s\right) \in \zqt.\] %such that
%  $\xx'_i=\xx$ and $$\xx'_j=\fb^{[r]}_j,\ \ \text{for all $j\ne i$.}$$
Let $\aa$ be a vector in $\mathbb{Z}_{\qq_i}$ for some $i\in [s]$.
Then we use $\widetilde{\aa}$ to denote the vector $\bb\in \zqt$
  such that $\bb_i=\aa$ and $\bb_j=\00$ for all other $j\ne i$.
Also recall that $q_k=q_{k,1}$. 

\begin{theorem}[modification of Theorem 5.9 from~\cite{Cai-Chen-Lu-2013}]\label{bi-step-5-ours}
Let $((M,N),\CC,\fD,(\pp,\bft,\fq))$ be a tuple that satisfies
  both $(\calR)$ and $(\calL)$.
Then either there is an edge gadget $\Gamma$
% we can construct an edge gadget $\Gamma$
   such that $\gm_{\Gamma, \CC_, \fD}$ is not \modblrkone/, or %(and so not \multblrkone/ either)
  $\fD$ satisfies the following condition $(\calD)$:
\begin{enumerate}[wide]
\item[$(\calD_1)$] For all $r\in \cS$ and
  $\xx\in \Lambda_r$, we have\label{CONDITIOND}
\begin{equation}\label{D1eqn-bi-step-5-ours}
D^{[r]}_{(0,\xx)}=D^{[r]}_{(0,\extt_r({\xx_1}))} D^{[r]}_{(0,\extt_r({\xx_2}))} \cdots
   D^{[r]}_{(0,\extt_r({\xx_{s}}))}. 
\end{equation}
%Let $\xx_k$ be a vector in $\Lambda_{r,k}$, where $k\in [s]$,
%  then $D^{[r]}_{(0,\extt_r(\xx_k))}$ is a power of $\oo_{q_{k,1}}$.

\item[$(\calD_2)$] For all $r\in \cT$ and $\xx\in \Delta_r$, we have
\begin{equation}\label{D2eqn-ours}
D^{[r]}_{(1,\xx)}=D^{[r]}_{(1,\extt_r'({\xx_1}))} D^{[r]}_{(1,\extt_r'({\xx_2}))} \cdots
  D^{[r]}_{(1,\extt_r'({\xx_{s}}))}.
\end{equation}
\item[$(\calD_3)$] For all $r\in \cS$, $k\in [s]$, and $\aa\in
  {\Lambda_{r,k}^{\text{\emph{lin}}}}$, %\subseteq\mathbb{Z}_{\qq_k}$,
  there are $\bb\in \mathbb{Z}_{\qq_k}$
  and $\alpha\in \mathbb{Z}_{N}$ such that
\begin{equation}\label{targetprove1-ours}
\oo_{N}^\alpha\cdot F_{\xx,\widetilde{\bb}}=D^{[r]}_{(0,\xx+\widetilde{\aa})}\cdot
  \overline{D^{[r]}_{(0,\xx)}}\ \ \ \text{for all\ $\xx\in\Lambda_r$.}
\end{equation}
\item[$(\calD_4)$] For all $r\in \cT$, $k\in [s]$, and $\aa\in
  {\Delta_{r,k}^{\text{\emph{lin}}}}$, %\subseteq \mathbb{Z}_{\qq_k}$,
  there are $\bb\in \mathbb{Z}_{\qq_k}$ and $\alpha\in
  \mathbb{Z}_{N}$ such that
\begin{equation}\label{targetprove2-ours}
\oo_{N}^\alpha\cdot F_{\widetilde{\bb},\xx}=D^{[r]}_{(1,\xx+\widetilde{\aa})}\cdot
  \overline{D^{[r]}_{(1,\xx)}}\ \ \ \text{for all\ $\xx\in\Delta_r$.}
\end{equation}
\end{enumerate}
\end{theorem}
Note that in $(\calD_3)$ and $(\calD_4)$,
the expressions on the left-hand side do not depend
on all other components of $\xx$ except the $k${th} component
$\xx_k$, since all other components of $\widetilde{\bb}$
are $\00$.
The statements in conditions $(\calD_3)$--$(\calD_4)$
are a technically precise way to express the idea that
there is a quadratic structure on the support of each diagonal matrix $\DD^{[r]}$.
We express it in terms of an exponential difference equation.

\subsection{Tractability}
Now we can state a theorem of tractability.

\begin{theorem}[modification of Theorem 5.10 from~\cite{Cai-Chen-Lu-2013}]\label{tractable-1-ours}
Suppose that $((M,N),\CC,\fD,(\pp,\bft,\fq))$ satisfies
 $(\calR),(\calL)$, and $(\calD)$. Then the problem
  $\eval^\leftrightarrow(\CC,\fD)$ can be solved in polynomial time.
\end{theorem}

%% file: meta-argument-bounded-degree-graphs-outline-unipartite.tex
%%%%%%%%%%%%%%%%%%%%%%%%%%%%%%%%%%%%%%%%%%%%%%%%%%%%%%%%%%%%%%%%%%%%
\begin{comment}
{\color{blue} (old text)

\section*{Proof Outline of the Case: $\AA$ is not Bipartite}\label{outlinenonbip}

Both the definitions and theorems of the case when the fixed matrix $\AA$ is not bipartite are
  similar to, but also have significant differences from, those of the bipartite case.
%We will list these theorems.
}

{\color{red} (new text)}
\end{comment}
%%%%%%%%%%%%%%%%%%%%%%%%%%%%%%%%%%%%%%%%%%%%%%%%%%%%%%%%%%%%%%%%%%%%
\section{Proof outline of the case: $\AA$ is not bipartite}\label{outlinenonbip-ours}

The definitions and theorems of the
 case for nonbipartite $\AA$ is similar to the bipartite case.
However, there are some nontrivial differences.

\subsection{Step 1: Regularized form of matrix $\AA$ and its purification}\label{sec:outlinenonbipartite}

We start with $\AA\in \mathbb{C}^{m\times m}$,
  a symmetric, connected, and nonbipartite matrix with algebraic entries.
The problem $\eval(\AA)$ is clearly tractable if $m=1$;
in the discussion below, we assume $m>1$.

\begin{definition}[modification of Definition 6.1 from~\cite{Cai-Chen-Lu-2013}]\label{def:regularized-nonbipartite}
Let $\AA\in \mathbb{C}^{m\times m}$ be a symmetric, connected, and nonbipartite matrix.
We say $\AA$
%, after multiplying by a nonzero scalar, 
  is a \emph{regularized nonbipartite matrix} if
 % there exist $m$ positive real numbers $c_1,\ldots,c_m$
there exist nonzero complex numbers $\mu_1,\ldots,\mu_m$
%that generate a torsion-free multiplicative group
% (i.e., that contains no roots of unity other that $1$) and
  such that $A_{i,j}/(\mu_i\mu_j)$ is a root of unity for all $i,j\in [m]$.
We may pre-multiply $\AA$ by a nonzero scalar, and require that
$\mu_1,\ldots,\mu_m$ generate a torsion-free multiplicative group.
%\textbf{\emph{(number of bits to describe $c_i$?)}}
We say $\AA$ is a \emph{purified nonbipartite matrix} if
$\mu_1, \ldots,\mu_m$
are
positive rational numbers.
\end{definition}
Thus, 
%In other words, 
% i don't say In other words, 
% because I don't want to repeat this pre-multiply scalar...
% and torsion-free multiplicative group.
 $\AA$ is regularized nonbipartite
if %{\color{red} up to multiplication by a nonzero scalar}
$\AA$ has the form
\begin{equation}\label{eqn:non-bipartite-regularized}
\AA = \left(\begin{matrix}
\mu_1 \\
& \mu_2 \\
& & \ddots\\
& & & \mu_m
\end{matrix}\right)
\left( \begin{matrix}
\zeta_{1,1} & \zeta_{1,2} & \ldots & \zeta_{1,m} \\
\zeta_{2,1} & \zeta_{2,2} & \ldots & \zeta_{2,m} \\
\vdots      & \vdots      & \ddots & \vdots \\
\zeta_{m,1} & \zeta_{m,2} & \ldots & \zeta_{m,m}
\end{matrix}\right)
\left(\begin{matrix}
\mu_1 \\
& \mu_2 \\
& & \ddots\\
& & & \mu_m
\end{matrix}\right),\vspace{0.05cm}
\end{equation}
where every $\mu_i$ is a nonzero complex number such that $\mu_1, \ldots, \mu_n$ generate a torsion-free multiplicative group,
%such that $\mu_1, \ldots, \mu_m$ generate a torsion-free subgroup in $\bbC^*$ 
and
$\zeta_{i,j} = \zeta_{j,i}$ are all roots
of unity.
If further $\mu_1, \ldots,\mu_m$
are
positive rational numbers, then $\AA$  is purified nonbipartite.

When we go from $\AA$ to its purified form $\pur \AA$,
we may pre-multiply $\AA$ by a nonzero constant 
(in the multiplicative group $\frak G$ generated by $\{g_1, \ldots, g_d\}$)
and then assume
that the same generators $\{g_1, \ldots, g_d\}$ are chosen
for going from $\AA$ to $\pur \AA$,
as for going from $(\mu_1, \ldots, \mu_m)$ to its purification $({\pur \mu_1},  \ldots, {\pur \mu_m})$ %{\color{red} (first time occuring, not defined yet)},
and that  $\mu_1, \ldots, \mu_m$ belong to $\frak G$. Thus,
$\pur \AA$ has the form
\begin{equation}\label{eqn:non-bipartite-purified}
{\pur \AA} = \left(\begin{matrix}
{\pur \mu_1} \\
& {\pur \mu_2} \\
& & \ddots\\
& & & {\pur \mu_m}
\end{matrix}\right)
\left( \begin{matrix}
\zeta_{1,1} & \zeta_{1,2} & \ldots & \zeta_{1,m} \\
\zeta_{2,1} & \zeta_{2,2} & \ldots & \zeta_{2,m} \\
\vdots      & \vdots      & \ddots & \vdots \\
\zeta_{m,1} & \zeta_{m,2} & \ldots & \zeta_{m,m}
\end{matrix}\right)
\left(\begin{matrix}
{\pur \mu_1} \\
& {\pur \mu_2} \\
& & \ddots\\
& & & {\pur \mu_m}
\end{matrix}\right).\vspace{0.05cm}
\end{equation}

We prove the following theorem. 

\begin{theorem}[modification of Theorem 6.2 from~\cite{Cai-Chen-Lu-2013}]\label{t-step-1-ours}
Let $\AA\in \mathbb{C}^{m\times m}$ be a symmetric, connected and
  nonbipartite~matrix, where $m>1$. 
Then $\AA$ is \multblrkone/ iff $\AA$ is a regularized nonbipartite matrix. %{\color{red} (up to multiplication by a nonzero scalar) (JYC: you already said premultiply before, but does it make this theorem fall into the scope of this assumption?)}.
%(up to multiplication by a nonzero scalar in the direct implication).
In that case, if $\{g_1, \ldots, g_d\}$ is a generating set of nonzero entries of $\AA$, %of the set
then we can choose $\mu_1, \ldots, m_m$ to belong to the multiplicative group generated by $\{g_1, \ldots, g_d\}$.
As a consequence, either $\AA$ is not \multblrkone/
(a fortiori, $\EVAL^{(\Delta)}(\AA)$ is \#P-hard for some $\Delta > 0$)
or $\AA$ is a reqularized nonbipartite matrix.
%up to multiplication by a nonzero scalar.
%Moreover, if $\AA$ is a regularized nonbipartite matrix
%and $\{g_1, \ldots, g_d\}$ is a generating set of the set of nonzero entries of $\AA$,
%then we can choose $\mu_1, \ldots, \mu_m$ in Definition~\ref{def:regularized}
%so that $\mu_1, \ldots, \mu_m$ belong to the multiplicative
%subgroup generated by $\{g_1, \ldots, g_d\}$. {\color{red} (well the entries are all nonzero in this case, but this is how we define generator sets of interest)}
%%Moreover, in case $\AA$ is a regularized nonbipartite matrix,
%%we can choose $\mu_1, \ldots, \mu_m$ in Definition~\ref{def:regularized}
%%so that $\mu_1, \ldots, \mu_m$ belong to the multiplicative
%%subgroup generated by a set $\{g_1, \ldots, g_d\}$
%%that is a generator of the set of nonzero entries of $\AA$.
%%are generated by the same generating set.
%%{\color{red} (or so that $\mu_1, \ldots, \mu_m$ in Definition~\ref{def:regularized}
%%are in the multiplicative subgroup generated by
%%the nonzero entries of $\AA$. (no need to define a generating set at this point.))}
%%{\color{red} (up to a simultaneous row and column permutation by the same permutation)}.
\end{theorem}
%In the statement of Theorem~\ref{t-step-1-ours},
%an edge $e = (\firstdv, \seconddv)$ with the distinguished endpoints $\firstdv, \seconddv$
%plays a role of a gadget whose signature matrix $M_{e, \AA}$
%is not \multblrkone/.
%In the statement of Theorem~\ref{t-step-1-ours},
%in case $\AA$ is not a regularized bipartite matrix,
%for an edge $e = (\firstdv, \seconddv)$ with the distinguished endpoints $\firstdv, \seconddv$,
%we have that $M_{e, \AA} = \AA$ is not \multblrkone/
%which is then by using Lemma~\ref{connected-ours} used to conclude
%Theorem~\ref{thm:main-dich-bd-edge-gadget}
%(for the original $\AA$: the current $\AA$ is some connected $\AA_i$ where $i \in [s]$ in that context).

If
 $\AA$ is not a regularized nonbipartite matrix,
then an edge $e = (\firstdv, \seconddv)$
forms an edge gadget with the distinguished vertices $\firstdv, \seconddv$
such that 
%its signature 
$M_{e, \AA} = \AA$ is not \multblrkone/
by Theorem~\ref{t-step-1-ours}.

\subsection{Step 2: Reduction to discrete unitary matrix}

%In this step, we prove the following theorem:
\begin{theorem}[modification of Theorem 6.3 from~\cite{Cai-Chen-Lu-2013}]\label{t-step-2-ours}
Let $\AA\in \mathbb{C}^{m\times m}$ be a purified nonbipartite matrix.
Then either 
%1. $\eval(\AA)$ is tractable; or
  %1. we can construct (there exists) an edge gadget $\Gamma$
(I) there exists an edge gadget $\Gamma$
  such that $\gm_{\Gamma, \AA}$ is not \multblrkone/
  (a fortori, $\eval^{(\Delta)}(\AA)$ is \#P-hard for some $\Delta > 0$)
%  or 2. we can construct (there exists) a triple
or (II) there exists a triple
  $((M,N),$ $\FF,\fD)$ such that 
%3.1. for any an edge gadget $\Gamma$
\begin{enumerate}[wide]
\item
for any an edge gadget $\Gamma$
  such that $\gm_{\Gamma, \FF, \fD}$
  is not \multblrkone/, %(not \modblrkone/)
  we can construct an edge gadget $\Gamma'$ 
%(in polynomial time)
  so that $\gm_{\Gamma', \AA}$
  is not \multblrkone/;
\item
  %3.2. 
$\EVAL(\AA) \le \EVAL(\FF, \fD)$; and %{\color{red} (constructively (can be removed))}
  %  and 3.3.
\item $((M,N),\FF,\fD)$ satisfies $(\calU_1')$--$(\calU_4')$:
\begin{enumerate}[wide]
\item[$(\calU_1')$]
%$M$ and $N$ are positive integers that satisfy $2\hspace{0.04cm}|\hspace{0.04cm}N$
%  and . 
$\FF\in \mathbb{C}^{n\times n}$
  for some $n\ge 1$, and
  $\fD=(\DD^{[0]},\ldots,\DD^{[N-1]})$ is a sequence of $N$\\
  $n\times n$ diagonal matrices for some \emph{even} $N>1$.\label{CONDITIONUP}

\item[$(\calU_2')$]
$\FF$ is a {symmetric} $M$-discrete unitary matrix,
  where $M\ge 1$ and $M\hspace{0.04cm}|\hspace{0.04cm} N$.

%\item[$(\calU_3')$]
%For all $i\in [n]$, $D_{i}^{[0]}=1$, and for all $r\in [N-1]$ and $i\in [n]$,
%  $D_{i}^{[r]}$ is either zero or a power of $\omega_N$.

\item[$(\calU_3')$]
%For all $i\in [n]$, $D_{i}^{[0]}=1$.
$\DD^{[0]}$ is the identity matrix.
For each $r\in [N-1]$, either 
  $\DD^{[r]}=\00$ or $\DD^{[r]}$\\ has an entry equal to $1$. %\ \Longrightarrow\  %\exists\hspace{0.06cm} i\in [n],
%  \hspace{0.06cm}D_{i}^{[r]}=1.
%$

\item[$(\calU_4')$] For all $r \in [N-1]$ and $i\in [n]$,
  $D^{[r]}_i\in \mathbb{Q}(\oo_N)$ and $|D_i^{[r]}|\in \{0,1\}$.
%  either $D_{i}^{[r]}=0$ or $\big|D_i^{[r]}\big|=1$.
%If $\big|D_i^{[r]}\big|=1$, then it has the form $$D_{i}^{[r]}=\oo_N^k\cdot
%  \frac{\sum_{0\le j<N} a_j\cdot \oo_N^{j}}{\sum_{0\le j<N} b_j \cdot \oo_N^{j}},$$
%for some $k\in \mathbb{Z}$ and non-negative integers
%  $a_0,a_1,...,a_{N-1},b_0,b_1,...,b_{N-1}$.
  %or a power of $\omega_N$.
\end{enumerate}
\end{enumerate}
\end{theorem}
%{\color{red} (my addition)}
%We note that in Theorem~\ref{t-step-2-ours},
%we slightly abuse the notation for convenience
%as was done in~\cite{Cai-Chen-Lu-2013},
%and the pair $(\FF, \fD)$ in the statement of Theorem~\ref{t-step-2-ours}
%is not the pair also denoted by $(\FF, \fD)$ and used in the derivation of this theorem.

In addition to $(\FF, \fD)$,
we will also need to introduce a purified pair $(\pur \FF, \fD)$,
where $\pur \FF$ is a purification of $\FF$ 
to relate to the purification matrix $\pur \AA$.
%as an auxiliary tool
%
%that will be used to relate to the purification matrix $\pur \AA$.

%%%%%%%%%%%%%%%%%%%%%%%%%%%%%%%%%%%%%%%%%%%%%%%%%%%%%%%%%%%%%%%%%%%%
\begin{comment}
{\color{blue} (old text)
\subsection*{Step 3: Canonical Form of $\FF$ and $\fD$}

Now suppose we have a tuple $((M,N),\FF,\fD)$ that satisfies  $(\calU'_1)$--$(\calU'_4)$.
For convenience we still use $m$ to denote the number of rows
  and columns of $\FF$ and each $\DD^{[r]}$ in $\fD$,
  though it should be noted that this new $m$ is indeed the $n$ in Theorem \ref{t-step-2},
  which is different from the $m$ used in the first two steps.
Similar to the bipartite case, we adopt the following convention in
  this step: given an $n\times n$ matrix, we use $[0:n-1]$,
  instead of $[n]$, to index its rows and columns.

We start with the special case when $M=1$.
Since $\FF$ is $M$-discrete unitary, we must have $m=1$ and $\FF=(1)$.
In this case, it is clear that the problem $\eval(\CC,\fD)$ is tractable.
So in the rest of this section, we always assume that $M>1$.
}

{\color{red} (new text)
\end{comment}
%%%%%%%%%%%%%%%%%%%%%%%%%%%%%%%%%%%%%%%%%%%%%%%%%%%%%%%%%%%%%%%%%%%%
\subsection{Step 3: Canonical form of $\FF$ and $\fD$}

Now suppose we have a tuple $((M,N),\FF,\fD)$ that satisfies  $(\calU'_1)$--$(\calU'_4)$.
For convenience we still use $m$ to denote the number of rows
  and columns of $\FF$ and each $\DD^{[r]}$ in $\fD$,
  though it should be noted that this new $m$ is indeed the $n$ in Theorem \ref{t-step-2-ours},
  which is different from the $m$ used in the first two steps.
Similar to the bipartite case, we adopt the following convention in
  this step: given an $n\times n$ matrix, we use $[0:n-1]$,
  instead of $[n]$, to index its rows and columns.

We start with the special case when $M=1$.
Since $\FF$ is $M$-discrete unitary, we must have $m=1$ and $\FF=(1)$.
In this case, it is clear that the problem $\eval(\FF,\fD)$ is tractable.
So in the rest of this section, we always assume that $M>1$.

%%%%%%%%%%%%%%%%%%%%%%%%%%%%%%%%%%%%%%%%%%%%%%%%%%%%%%%%%%%%%%%%%%%%
\begin{comment}
{\color{blue} (old text)
\subsubsection*{Step 3.1: Entries of $\DD^{[r]}$ 
  are Either $0$ or Powers of $\oo_N$}\ 

\begin{theorem}\label{nonbipstep0}
Suppose $((M,N), \FF,\fD)$ satisfies $(\calU_1')$--$(\calU_4')$, and $M>1$.
Then~either $\eval(\FF,\fD)$ is \#P-hard or $((M,N),\FF,\fD)$
  satisfies the following condition $(\calU_5')$: 
\begin{enumerate}
\item[$(\calU_5')$] For all $r\in [N-1]$, %and $i\in [0:m-1]$,
  entries of $\DD^{[r]}$ are either zero or powers of $\oo_N$.\label{CONDITIONUP5}
\end{enumerate}
\end{theorem}
}

{\color{blue} (new text)}
\end{comment}
%%%%%%%%%%%%%%%%%%%%%%%%%%%%%%%%%%%%%%%%%%%%%%%%%%%%%%%%%%%%%%%%%%%%

\subsubsection{Step 3.1: Entries of $\DD^{[r]}$ are either $0$ or powers of $\oo_N$}%\ 

\begin{theorem}[modification of Theorem 6.4 from~\cite{Cai-Chen-Lu-2013}]\label{nonbipstep0-ours}
Suppose $((M,N), \FF,\fD)$ satisfies $(\calU_1')$--$(\calU_4')$, and $M>1$.
Then~either there exists an edge gadget $\Gamma$
such that $\gm_{\Gamma, \FF, \fD}$ is not \modblrkone/ %(and so not \multblrkone/ either)
or $((M,N),\FF,\fD)$
  satisfies the following condition $(\calU_5')$: 
\begin{enumerate}[wide]
\item[$(\calU_5')$] For all $r\in [N-1]$, %and $i\in [0:m-1]$,
  entries of $\DD^{[r]}$ are either zero or powers of $\oo_N$.\label{CONDITIONUP5} %{\color{red} (JYC: entries is from CCL, is each entry more correct because otherwise it might mean all together? If not critical then fine)}
\end{enumerate}
\end{theorem}

\subsubsection{Step 3.2: Fourier decomposition}\label{reference-for-sec-6.3.2-ours}

Let $q$ be a prime power.
We say $\WW$ is a \emph{nondegenerate} matrix in $\mathbb{Z}_{q}^{2\times 2}$
  if $\WW\xx\ne \00$
for all $\xx\ne \00\in \mathbb{Z}_q^2$.
%The following lemma gives some equivalent characterizations of non-degenerate
%  matrices.
%The proof is elementary, so we omit it here.
%We have the following lemma (Lemma 6.5) from~\cite{Cai-Chen-Lu-2013}.

\begin{lemma}[Lemma 6.5 from~\cite{Cai-Chen-Lu-2013}]\label{equivaequiva-ours}
Let $q$ be a prime power and $\WW \in \mathbb{Z}_q^{2\times 2}$.
The following statements are equivalent: 
(1) $\WW$ is nondegenerate; (2)
  $\xx\mapsto \WW\xx$ is a bijection from $\mathbb{Z}_{q}^2$ to
  itself; and 
(3) $\det(\WW)$ is invertible in $\mathbb{Z}_q$.
  %W_{1,1}W_{2,2}-W_{1,2}W_{2,1}\in \mathbb{Z}_q^*$, that is,
\end{lemma}
%\begin{proof}
%1.)\hspace{0.13cm}$\Leftrightarrow$\hspace{0.13cm}2.) is trivial.

%We prove 3.)\hspace{0.13cm}$\Rightarrow$\hspace{0.13cm}1.).
%Since $\det(\WW)\in\mathbb{Z}_q^*$, by Cramer's rule, there exists an
%  $\WW^{-1}\in \mathbb{Z}_{q}^{2\times 2}$ such that $\WW\WW^{-1}=\WW^{-1}\WW$
%  is the identity matrix.
%Suppose $\WW\xx=0$ for some $\xx\in \mathbb{Z}_q^2$. We have $\WW^{-1}\WW\xx=\xx=\00$ and
%thus, $\WW$ is non-degenerate.

%Finally, we prove 1.)\hspace{0.13cm}$\Rightarrow$\hspace{0.13cm}3.).
%For $W_{1,1}$ and $W_{2,1}$, either there is an $\alpha\in \mathbb{Z}_q$
%  such that $W_{1,1}=\alpha W_{2,1}$ or there is an $\alpha\in \mathbb{Z}_q$
%  such that $W_{2,1}=\alpha W_{1,1}$.
%Without loss of generality, we assume $W_{2,1}=\alpha W_{1,1}$. Let
%$$
%\WW'=\begin{pmatrix}1 & 0\\ -\alpha & 1\end{pmatrix}\WW,
%$$
%then $W'_{2,1}=0$ and $\det(\WW)=\det(\WW')=W_{1,1}'W_{2,2}'$.
%On the other hand,
%\end{proof}

\begin{definition}[generalized Fourier matrix, Definition 6.6 from~\cite{Cai-Chen-Lu-2013}]\label{GeneralizedFourierMatrix-ours}
Let $q$ be a prime power and $\WW=(W_{ij})$ be a symmetric nondegenerate matrix
  in $\mathbb{Z}_q^{2\times 2}$. 
We say a $q^2\times q^2$ matrix $\boldsymbol{\calF}_{q,\WW}$ is 
  a \emph{$(q,\WW)$-generalized Fourier matrix} if there exists
  a bijection $\rho$ from $[0:q^2-1]$ to $[0:q-1]^2$  such that 
\begin{equation*}
(\boldsymbol{\calF}_{q,\WW})_{i,j}=\omega_{q}^{W_{11}x_1y_1+ W_{12}x_1y_2
  +W_{21}x_2y_1+ W_{22}x_2y_2}\ \ \ \text{for all $i,j\in [0:q^2-1]$},
\end{equation*}
where %$(x_1,x_2)=\rho(i)$, $(y_1,y_2)=\rho(j)$, and $\rho:i\mapsto
  %(\lfloor i/q\rfloor, i-q\lfloor i/q\rfloor)$ is a map from
  %$[0:q^2-1]$ to $\mathbb{Z}_q\times \mathbb{Z}_q$.
  $\xx=(x_1,x_2)=\rho(i)$ and $\yy=(y_1,y_2)=\rho(j)$. 
\end{definition}

%For each $i\in [h]$, we let $\mathbb{Z}_{q_i}$
%  to denote the $i$th set in the Cartesian product.
%The first $2r$ components correspond to the $r$ $\HH_4$ matrices
%  in the decomposition (more exactly, $x_{2i+1}$ and $x_{2i+2}$
%  correspond to the $(i-1)$th $\HH_4$), so
%  $q_i=2$ for $i\in [2r]$.
%By Theorem \ref{step3-1}, we know there exists integers $k_i$
%For $i> 2r$, we use $\mathbb{Z}_{q_i}$ to denote the%
%  , $i\in [2r]$ to denote the first $2r$ components of
%  $\rho(x)$, and $x_{p_i,\alpha,k,}$

\begin{theorem}[modification of Theorem 6.7 from~\cite{Cai-Chen-Lu-2013}]\label{t-step-3-ours}
Suppose $((M,N),\FF,\fD)$ satisfies conditions $(\calU_1')$--$(\calU_5')$.
Then either there exists an edge gadget $\Gamma$
such that $\gm_{\Gamma, \FF, \fD}$ is not \modblrkone/ or there exist %(and so not \multblrkone/ either)
  a permutation $\Sigma$ of $[0:m-1]$
such that
\[
\FF_{\Sigma,\Sigma}=\left(\bigotimes_{i=1}^g {\boldsymbol{\calF}}_{d_i,\WW^{[i]}}\right)
\otimes\left(\bigotimes_{i=1}^\ell \boldsymbol{\calF}_{q_i,k_i}\right),
\]
where $\dd=(d_1,\ldots,d_g)$ and $\calW=(\WW^{[1]},\ldots,\WW^{[g]})$ 
  are two sequences,
  for some $g\ge 0$.
  (Note that the $g$ here can be $0$, in which case $\dd$ and
  $\calW$ are  empty.)
  For each $i\in [g]$, $d_i>1$ is a power of $2$ and $\WW^{[i]}$
  is a $2\times 2$ symmetric nondegenerate matrix over $\mathbb{Z}_{d_i}$;  
  $\qq=(q_1,\ldots,q_\ell)$ and
  $\kk=(k_1,\ldots,k_\ell)$ are two sequences for some
  $\ell\ge 0$ {(}again $\ell$ can be $0$). %, in which case
  %both $\qq$ and $\kk$ are empty\emph{)}, in which
  For each $i\in [\ell]$, $q_i$ is a
  prime power, $k_i\in \mathbb{Z}_{q_i}$, and $\gcd(q_i,k_i)=1$. 
\end{theorem}

%\begin{definition}
%Let $f(x_1,\ldots,x_t,y_1,\ldots,y_t)=\sum_{i,j\in [t]}a_{i,j}x_iy_j$ be a
%  $2t$-variable non-degenerate quadratic polynomial.
%We say $f$ is \emph{symmetric} if for all $i,j\in [t]$, $a_{i,j}=a_{j,i}$.
%\end{definition}

Assume there does exist a permutation $\Sigma$, together with
  the four sequences, such that
  $\FF_{\Sigma,\Sigma}$ satisfies the equation above; otherwise,  there exists an edge gadget $\Gamma$
  such that $\gm_{\Gamma, \FF, \fD}$ is not \modblrkone/. %(and so not \multblrkone/ either)
Then we apply $\Sigma$ to $\DD^{[r]}$, $r\in [0:N-1]$, to
  get a new sequence $\fD_{\Sigma}$ of $N$ diagonal matrices
  in~which the $r${th} matrix of $\fD_{\Sigma}$ is
\[
\left(\begin{matrix}
D^{[r]}_{\Sigma(0)}\\
%& D^t_{\Sigma(2)}\\
& \ddots\\
& & D^{[r]}_{\Sigma(m-1)}\\
\end{matrix}\right).
\]
%It is clear that %permuting the rows and columns of
%  $\FF$ and $\DD^{[r]}$ in $\fD$ by the same permutation $\Sigma$
%  does not affect the complexity of $\eval(\FF,\fD)$, so
  %$\eval(\FF_{\Sigma,\Sigma},\fD_{\Sigma})\equiv \eval(\FF,\fD)$.
Clearly $\eval(\FF_{\Sigma,\Sigma},\fD_{\Sigma})$ and $\eval(\FF,\fD)$
are equivalent. %{\color{red} (as problems? what about $\leftrightarrow$ stuff)}
%%% not only for unbounded degree. but for boudned degree too.
%We also have that 
%%%
%For any edge gadget $\Gamma$,
%the signature matrix
%$\gm_{\Gamma, \FF_{\Sigma,\Sigma},\fD_{\Sigma,\Sigma}}$ is \multblrkone/ (\modblrkone/)
%iff $\gm_{\Gamma, \FF, \fD}$ is \multblrkone/ (\modblrkone/). {\color{red} (here only \modblrkone/ is needed, maybe we don't need this sentence, we don't have a similar one in the biparite case, and we have discussed permutation issues before)}
%%%
From now on, we simply let $\FF$ and $\fD$ denote
  $\FF_{\Sigma,\Sigma}$ and $\fD_{\Sigma}$, respectively. Thus, we have
\begin{equation}\label{juju-ours}
\FF=\left(\bigotimes_{i=1}^g {\boldsymbol{\calF}}_{d_i,\WW^{[i]}}\right)
\otimes\left(\bigotimes_{i=1}^\ell \boldsymbol{\calF}_{q_i,k_i}\right).
\end{equation}

Before moving forward to Step 3.3, we rearrange the prime powers in $\dd$ and $\qq$ and
  divide them into groups according to different primes.

By (\ref{juju-ours}), there exist $\dd, \cal{W}, \pp,
  \bft, \mathcal{Q}$, and $\mathcal{K}$ such that
  the tuple $((M,N),\FF,\fD,(\dd, \calW, \pp,$ $ \bft,\cal{Q}, \cal{K}))$
  satisfies the following condition $(\calR')$:
\begin{enumerate}[wide]
\item[$(\calR_1')$]
$\dd=(d_1, \ldots,d_g)$ is a nonincreasing 
  sequence of powers of $2$ for some $g\ge 0$; 
  %, such that if $g>0$, then $d_1\ge \cdots$ $\ge d_g$;
$\calW=(\WW^{[1]}, \ldots,\WW^{[g]})$ is a sequence of  
   symmetric nondegenerate $2\times 2$ matrices over $\mathbb{Z}_{d_i}$
  (note that $\dd$ and $\calW$ can be empty); \label{CONDITIONRP}
$\pp=(p_1,\ldots,p_s)$ is a strictly increasing~sequence of $s$ primes 
for some $s\ge 1$, 
  starting with $p_1=2$;  
%  such that
%  $2=p_1<\cdots <p_s$; 
$\bft= (t_1,$ $\ldots,t_s)$ is a sequence of integers with $t_1\ge 0$
  and $t_i\ge 1$ for all $i>1$;
  $\mathcal{Q}=\{\qq_i: i\in [s]\}$ is
  a collection of sequences in which each $\qq_i=(q_{i,1},\ldots,q_{i,t_i})$
  is a nonincreasing 
  sequence of powers of $p_i$ (only
  $\qq_1$ can be empty as we always fix $p_1=2$
  even when no powers of $2$ occur in $\calQ$); 
$\mathcal{K}=\{\kk_i: i\in [s]\}$
  is a collection of sequences in which each $\kk_i=(k_{i,1},\ldots,k_{i,t_i})$
  is a sequence of length $t_i$.
Finally, for all $i\in [s]$ and $j\in [t_i]$, $k_{i,j}\in [0:q_{i,j}-1]$ and
  satisfies $\gcd(k_{i,j},q_{i,j})=\gcd(k_{i,j},p_i)=1$. %\vspace{0.016cm}

\item[$(\calR_2')$] $((M,N),\FF,\fD)$ satisfies conditions ($\calU'_{1}$)--($\calU'_5$),
  and %$\FF$ is an $m\times m$ matrix,
  %$$m=\prod_{i=1}^s q_i^{t_i},$$ and
\[
m=\prod_{i\in [g]} (d_i)^2\hspace{0.06cm} \times
  \prod_{i\in [s],j\in [t_i]}q_{i,j}\hspace{0.05cm}. 
\]

\item[$(\calR_3')$]
There is a bijection $\rho$ from
  $[0:m-1]$ to $\mathbb{Z}^2_{\dd}\times \mathbb{Z}_{\cal{Q}}$, where
\[
\mathbb{Z}^2_{\dd}=\prod_{i\in [g]} \left(\mathbb{Z}_{d_i}\right)^2\ \ \ \text{and}\ \ \
\mathbb{Z}_{\calQ}=\hspace{-0.16cm}\prod_{i\in [s],j\in [t_i]} \mathbb{Z}_{q_{i,j}},
\]
such that (for each $a\in [0:m-1]$, we use \[\big(x_{0,i,j}: i\in [g],j\in \{1,2\}\big)\in
  \mathbb{Z}^2_{\dd}\ \ \ \ \text{and}\
  \ \ \ \big(x_{1,i,j}:i\in [s],j\in [t_i]\big)\in \mathbb{Z}_{\calQ}
  \]
to denote the components of $\xx=\rho(a)$,
  where $x_{0,i,j}\in \mathbb{Z}_{d_i}$ and $x_{1,i,j}\in \mathbb{Z}_{q_{i,j}}$)
\[
F_{a,b}=\prod_{i\in [g]}\hspace{0.15cm}\oo_{d_i}^{(x_{0,i,1}\hspace{0.08cm}
  x_{0,i,2})\cdot \WW^{[i]} \cdot (y_{0,i,1}\hspace{0.08cm} y_{0,i,2})^{\text{T}}}
\hspace{-0.2cm}\prod_{i\in [s],j\in [t_i]}
  \omega_{q_{i,j}}^{k_{i,j}\cdot x_{1,i,j} y_{1,i,j}}
\]
for all $a,b\in [0:m-1]$, where $((x_{0,i,j}),(x_{1,i,j}))=\xx=\rho(a)$ and
  $\yy=\rho(b)$. 
\end{enumerate}

For convenience, from now on we will directly use $\xx\in \mathbb{Z}_{\dd}^2
  \times \mathbb{Z}_{\calQ}$ to 
  index the rows and columns of $\FF$, i.e.,
$F_{\xx,\yy}\equiv F_{\rho^{-1}(\xx),\rho^{-1}(\yy)}$.
\subsubsection{Step 3.3: Affine support for $\fD$}\label{sec:Affine-Support-ours}

Now we have a tuple $((M,N),\FF,\fD,$ $(\dd,\calW,\pp,\bft,\calQ,\calK))$
  that satisfies ($\calR'$).
In the next step, we show for every $r\in [N-1]$
  ($\DD^{[0]}$ is already known to be the identity matrix)
  the nonzero entries of $\DD^{[r]}$
  (in $\fD$) must have a coset structure; otherwise there exists an edge gadget $\Gamma$
    such that $\gm_{\Gamma, \FF, \fD}$ is not \modblrkone/. %(and so not \multblrkone/ either)

\def\hzqt{\tilde{\mathbb{Z}}_{\qq_i}}

For each $r\in [N-1]$, let $\Gamma_r\subseteq \mathbb{Z}^2_{\dd}\times \mathbb{Z}_{\calQ}$
   denote the set of $\xx$ such that the entry of $\DD^{[r]}$ 
   indexed by $\xx$ is nonzero.
We also use $\calZ$ to denote the set of $r\in [N-1]$ such that
  $\Gamma_r\ne \emptyset$. 
For convenience, we let $\tilde{\mathbb{Z}}_{\qq_i}$, $i\in [s]$,
  denote the following set (or group):
\[\hzqt=
\begin{cases}  \mathbb{Z}_{\qq_i}& \text{if $i>1$}\\[0.3ex]
  \mathbb{Z}^2_{\dd}\times \mathbb{Z}_{\qq_1} & \text{if $i=1$.}
\end{cases}\]
%When $i>1$, $$;
%and when $i=1$, $\tilde{\mathbb{Z}}_{\qq_1}=$.
This gives us a new way to denote the components of 
\[\xx\in \mathbb{Z}_{\dd}^2\times
  \mathbb{Z}_{\calQ}=\tilde{\mathbb{Z}}_{\qq_1}\times 
  \tilde{\mathbb{Z}}_{\qq_2} \times \cdots\times
  \tilde{\mathbb{Z}}_{\qq_s},\] 
%  \prod_{i\in [s]}\tilde{\mathbb{Z}}_{\qq_i},$$
i.e.,
  $\xx=(\xx_1,\ldots,\xx_s)$, where $\xx_i\in \tilde{\mathbb{Z}}_{\qq_i}$
  for each $i\in [s]$.% 

\begin{theorem}[modification of Theorem 6.8 from~\cite{Cai-Chen-Lu-2013}]\label{t-step-4-ours}
Assume that $((M,N),\FF,\fD,(\dd,\calW,\pp,\bft,\calQ,\calK))$ satisfies 
  condition $(\calR')$.
Then either there exists an edge gadget $\Gamma$
  such that $\gm_{\Gamma, \FF, \fD}$ is not \modblrkone/ or $\fD$ satisfies the %(and so not \multblrkone/ either)
  following condition: 
\begin{enumerate}[wide]
\item[$(\calL_1')$]
For every $r\in \calZ$, $\Gamma_r=\prod_{i=1}^{s} \Gamma_{r,i}$,
  where $\Gamma_{r,i}$ is a coset in $\tilde{\mathbb{Z}}_{\qq_i}$
  for all $i\in [s]$.\label{CONDITIONLP}
\end{enumerate}
\end{theorem} 

Suppose for any gadget $\Gamma$,
$\gm_{\Gamma, \FF, \fD}$ is \modblrkone/. Then % {\color{red} (and \multblrkone/) (remove)}
by Theorem \ref{t-step-4-ours},
the  tuple $((M,N),\FF,\fD,$ $(\dd,\calW,\pp,\bft,\calQ,\calK))$ satisfies
  not only ($\calR'$) but also ($\calL_1'$).
By ($\calU_3'$), $\fD$ also satisfies the following: 
\begin{enumerate}[wide]
\item[$(\calL_2')$]
For every $r\in \calZ$, there exists an
  $\fa^{[r]}\in \Gamma_r\subseteq\mathbb{Z}_{\dd}^2\times \mathbb{Z}_{\calQ}$
  %=\prod_{i\in [s]} \tilde{\mathbb{Z}}_{\qq_i}$ 
  such that the entry of $\DD^{[r]}$ indexed by
  $\fa^{[r]}$ is equal to $1$. 
\end{enumerate}

From now on, we refer to conditions
  $(\calL_{1}')$ and $(\calL_2')$ as condition $(\calL')$.

\subsubsection{Step 3.4: Quadratic structure}

In this final step within Step 3 for the nonbipartite case,
  we show that for any index $r\in [N-1]$, the nonzero entries of
  $\DD^{[r]}$ must have a quadratic
  structure; otherwise there exists an edge gadget $\Gamma$
    such that $\gm_{\Gamma, \FF, \fD}$ is not \modblrkone/. %(and so not \multblrkone/ either) 

We need the following notation.
Given $\xx$ in $\tilde{\mathbb{Z}}_{\qq_i}$ for some $i\in [s]$,
  we let $\ext_r(\xx)$, where $r\in \calZ$, denote the following unique vector: 
\[\left(\fa^{[r]}_1, \ldots,\fa^{[r]}_{i-1},\xx,\fa^{[r]}_{i+1},\ldots,
  \fa^{[r]}_s\right)
\in \prod_{j \in [s]} \tilde{\mathbb{Z}}_{\qq_j}.\]
%its $i^{th}$ component
%    $\xx_i'=\xx$, the vector given in  $\hat{\mathbb{Z}}_{\qq_i}$,
% \vspace{-0.1cm}and
%$$\xx_j'=\fa^{[r]}_j,\ \ \ \text{for all $j\ne i$}.\vspace{-0.1cm}$$
%Recall that $\fa^{[r]}$ is a vector we picked from $\Gamma_r$ in condition %$(\calL_2')$.
Given $\aa\in \tilde{\mathbb{Z}}_{\qq_i}$ for some $i\in [s]$,
  we let $\widetilde{\aa}=
  (\widetilde{\aa}_1,\ldots,\widetilde{\aa}_s) \in \prod_{j\in [s]}
  \tilde{\mathbb{Z}}_{\qq_j}$
  such that $\widetilde{\aa}_i=\aa$ and all other components are $\00$. 
%Also recall that we use $q_k$, where $k\in [s]$, to denote $q_{k,1}$.

\begin{theorem}[modification of Theorem 6.9 from~\cite{Cai-Chen-Lu-2013}]\label{nonbi-step-5-ours}
Suppose $((M,N),\FF,\fD,(\dd,\calW,\pp,\bft,\calQ,\calK))$ satisfies
  $(\calR')$ and $(\calL')$. Then either there exists an edge gadget $\Gamma$
    such that $\gm_{\Gamma, \FF, \fD}$ is not \modblrkone/ or %(and so not \multblrkone/ either)
  $\fD$ satisfies the following condition $(\calD')$:
\begin{enumerate}[wide]
\item[$(\calD_1')$] For all $r\in \calZ$
  and $\xx\in \Gamma_r$, we have\label{CONDITIONDP}
\begin{equation}\label{D1eqn-nonbi-step-5-ours}
D^{[r]}_{\xx}=D^{[r]}_{\extt_r({\xx_1})} D^{[r]}_{\extt_r({\xx_2})} \cdots
   D^{[r]}_{\extt_r({\xx_{s}})}.
\end{equation}

\item[$(\calD_2')$] For all $r\in \calZ$, $k\in [s]$, and $\aa\in
  {\Gamma_{r,k}^{\text{\emph{lin}}}}$, % \subseteq \tilde{\mathbb{Z}}_{\qq_k}$,
  there are $\bb\in \tilde{\mathbb{Z}}_{\qq_k}$ and $\alpha\in
  \mathbb{Z}_{N}$ such that
\begin{equation}\label{targetprove3-ours}
\oo_{N}^\alpha\cdot F_{\widetilde{\bb},\xx}=D^{[r]}_{\xx+\widetilde{\aa}}
  \cdot \overline{D^{[r]}_{\xx}}\ \ \ \text{for all\ $\xx\in\Gamma_r$.}
\end{equation}
\end{enumerate}
\end{theorem}

Note that in (\ref{targetprove3-ours}), 
  the expression on the left-hand side does not depend
  on other components of $\xx$ except the $k${th} component
  $\xx_k\in \tilde{\mathbb{Z}}_{\qq_k}$.
 % as all other components of $\widetilde{\bb}$ are $\00$.

%\item[$(\calD')$] for each $\qq_i$, there exists a qualified quadratic polynomial
%  $f_i(x_{i,1},\ldots,x_{i,h_i})$ such that
%$$
%D^t_{\rho(\xx)}=\prod_{i\in [s]}
%  \omega_{q_{i,1}}^{f_i(\xx_i)},\quad \forall\ \xx\in \Lambda^t;
%$$

%%%%%%%%%%%%%%%%%%%%%%%%%%%%%%%%%%%%%%%%%%%%%%%%%%%%%%%%%%%%%%%%%%%%
\begin{comment}
{\color{blue} (old text)
\subsection*{Tractability}\ 

\begin{theorem}\label{tractable-2}
Let $((M,N),\FF,\fD,(\dd,\calW,\pp,\bft,\calQ,\calK))$ be a tuple that
  satisfies all conditions $(\calR'), (\calL')$ and $(\calD')$.
Then $\eval(\FF,\fD)$ can be solved in polynomial time.
\end{theorem}
}

{\color{red} (new text)}
\end{comment}
%%%%%%%%%%%%%%%%%%%%%%%%%%%%%%%%%%%%%%%%%%%%%%%%%%%%%%%%%%%%%%%%%%%%
\subsection{Tractability} 

\begin{theorem}[modification of Theorem 6.10 from~\cite{Cai-Chen-Lu-2013}]\label{tractable-2-ours}
%Let $((M,N),\FF,\fD,(\dd,\calW,\pp,\bft,\calQ,\calK))$ be a tuple that
%satisfies all conditions $(\calR'), (\calL')$, and $(\calD')$.
%Then $\eval(\FF,\fD)$ can be solved in polynomial time.
%%%
%{\color{red} Let $((M,N),\FF,\fD,(\dd,\calW,\pp,\bft,\calQ,\calK))$ be a tuple that
%satisfies all conditions $(\calR'), (\calL')$, and $(\calD')$.
%Then $\eval(\FF,\fD)$ can be solved in polynomial time. (in red is the original in CCL, just asking why is was changed as it doesn't cause an overfull box, is it because the below one is shorter a bit?)}
%%%
If  $((M,N),\FF,\fD,(\dd,\calW,\pp,\bft,\calQ,\calK))$
  satisfies all conditions $(\calR'), (\calL')$, and $(\calD')$,
then $\eval(\FF,\fD)$ can be solved in polynomial time.
\end{theorem}

%% file: regularization-step.tex
\section{Proof of Theorem \ref{bi-step-1-ours} and Theorem \ref{t-step-1-ours}}\label{sec:Regularization}
%\section{Regularization step}
%{\color{red} (the name is like CCL section 7 it is relevant, since this is the goal of this section, the purification is moved back from this section, the subsection has step 1.2,
%this is in agreement with CCL)}
%
%\subsection{Step 1.2}\label{sec:b-t-step-1-proofs}
%
% as we moved purification definition earlier, and got rid of COUNT
% no need for subsection
%
%{\color{red} (lots of notations repeat)}

We prove Theorem \ref{bi-step-1-ours} and Theorem \ref{t-step-1-ours}
in this section.

%%%%%%%%%%%%%%%%%%%%%%%%%%%%%%%%%%%%%%%%%%%%%%%%%%%%%%%%%%%%%%%%%%%%
\begin{comment}

Finally we prove Theorem \ref{bi-step-1-ours} and Theorem \ref{t-step-1-ours}.
(We will work within the notation established in Section~\ref{sec:purification-ours}. {\color{red} (not sure about this sentence)})

{\color{red} (below is your proof for the bipartite case, you need to do the same for the nonbipartite one,
note that we need to do a bit more after the proof to show similar decomposition for $\pur \AA$,
read my previous proofs, and take into account the details because they will be passed to the shape conditions.
In other words, I formulated the statements of Theorems~\ref{bi-step-1-ours} and Theorem \ref{t-step-1-ours} so that
using these theorems it's possible to extract the additional conditions needed for the decomposition of $\pur \AA$ ($\calS_1$).
The statements don't mention $\pur \AA$, but had I included it, they would be much longer)}

{\color{red} (PS instead of doing this $N(i)$ neightbor thing, we can say it's obvious, it's just one step from the rectangularity for symmetric matrices (see the fragment in Section~\ref{sec:preliminaries-gcd}),
and we already use similar reasoning just to justify such a decomposition)}

\end{comment}
%%%%%%%%%%%%%%%%%%%%%%%%%%%%%%%%%%%%%%%%%%%%%%%%%%%%%%%%%%%%%%%%%%%%

\begin{proof}[Proof of Theorem \ref{bi-step-1-ours}]
Let $\AA \in \bbC^{m \times m}$ be a symmetric, connected, and bipartite matrix.
If $\AA$ is a regularized bipartite matrix, then
by taking a common multiplier of the orders of the roots of unity $\zeta_{i,j}$'s
in (\ref{eqn:regularized-form-bipartite}) it is easy to see that $\AA$ is \multblrkone/.
We prove the other direction, so assume $\AA$ is \multblrkone/.
In particular,  $\AA$ is rectangular.
Let $\AA$ be the bipartization of some $\BB \in \bbC^{k \times (m - k)}$ for some $k \in [m - 1]$. %{\color{red} (JYC: it's more correct to say $k \in [m]$, $m = 1$ is not excluded at this point)}
%Let $\{g_1, \ldots, g_d\}$, where $d \ge 0$, be a generating set of the set of nonzero entries of $\AA$.
%In particular $g_1, \ldots, g_d$ are multiplicatively independent.

We first show that all entries of $\BB$ are nonzero.
Let $N(i) = \{ j \mid B_{i,j} \ne 0 \}$ denote the set of neighbors of $i \in [k]$.
If $N(i) \not = N(i')$ for some $i, i' \in [k]$,
since $i$ and $i'$ are connected by a path, there are successive
vertices $i_1 =i, j_1, i_2, \ldots, i_\ell =i'$,
where $i_1, i_2, \ldots, i_\ell$ are on the LHS of the bipartite graph,
and for some $s \in [\ell]$
we have $N(i_s) \not = N(i_{s+1})$ but they intersect.
This violates $\AA$ being rectangular.

\begin{comment}
If $N(i) \not = N(i')$ for some $i, i' \in [k]$,
since $i$ and $i'$ are connected by a path, $i_1 =i, j_1, i_2, \ldots, i_\ell =i'$
{\color{red} (JYC: $i_1 =i, j_1, i_2, \ldots, j_{\ell - 1}, i_\ell =i'$ ($\ell \ge 2$), don't know which one is better)},
there must be
some $i_s, i_{s+1} \in [k]$ along the path
such that $N(i_s) \not = N(i_{s+1})$ but they intersect.
This violates $\AA$ being rectangular.
\end{comment}

So $N(i) = N(i')$ for  all $i, i' \in [k]$, and by $\AA$ being connected
it follows that $N(i) = [m - k]$ for all $i \in [k]$, i.e.,
all entries of $\BB$ are nonzero.

We are given some integer $N \ge 1$
such that 
%the Hadamard power 
$\AA^{\odot N}$ is block-rank-$1$.
%\blrkone/.
%%% /* because all entries of $\BB$ are nonzero. */
It follows that $\BB^{\odot N}$ has rank $1$.

\begin{comment}
Let $\{g_1, \ldots, g_d\}$ be a generating set for 
%{\color{red} of?}
%%%% FOR suggests non-uniqueness
the entries of $\BB$.
%In particular $g_1, \ldots, g_d$ are multiplicatively independent.
We can write $B_{i,j} = \widehat B_{i,j} \zeta_{i,j}$,
where $\widehat B_{i,j}$ all belong to the torsion-free
group generated by $\{g_1, \ldots, g_d\}$, and $\zeta_{i,j}$
are roots of unity for $i \in k$ and $j \in [m - k]$.
Let $N'$ be the least common multiple
of their orders. Then $\BB^{\odot N N'}$ has rank $1$.
% {\color{red} (seems to be concluded from the premise of the next sentence which is then inferred)}
Thus all $2$ by $2$ submatrices of $\BB^{\odot N N'}$
have determinant $0$,
which implies that all $2$ by $2$ submatrices
of the matrix $\widehat \BB = (\widehat B_{i,j})$
have determinant $0$,
since $g_1, \ldots, g_d$ are multiplicatively independent.
\end{comment}

%%%% The above proof is correct. The way to think about the passage
%%% above is: since it's given that $\BB^{\odot N}$ has rank $1$,
%% it is a column vector times a row vector, so  $\BB^{\odot N k}$ 
%%% is also rank 1.
%
% but your proof below is also fine, so we will just use yours below.

%{\color{red} 
Let $\{g_1, \ldots, g_d\}$ be a generating set
for
 the entries of $\BB$.
%In particular $g_1, \ldots, g_d$ are multiplicatively independent.
We can write $B_{i,j} = \widehat B_{i,j} \zeta_{i,j}$,
where $\widehat B_{i,j}$ all belong to the torsion-free
group generated by $\{g_1, \ldots, g_d\}$, and $\zeta_{i,j}$
are roots of unity for $i \in k$ and $j \in [m - k]$.
Let $N'$ be the least common multiple
of their orders. %Then $\BB^{\odot N N'}$ has rank $1$.
Since $\BB^{\odot N}$ has rank $1$, all $2$ by $2$ submatrices of $\BB^{\odot N}$ and thus of $\BB^{\odot N N'}$
have determinant $0$.
This implies that all $2$ by $2$ submatrices
of the matrix $\widehat \BB = (\widehat B_{i,j})$
have determinant $0$,
since $g_1, \ldots, g_d$ are multiplicatively independent. 
%(JYC: my alternative to the above paragraph)}

Thus $\widehat \BB = (\widehat B_{i,j})$ has rank $1$.
%So $\BB^{\odot N'}$ has rank $1$.
%%%
%%% you are right below.
%%% {\color{red} (Note that $\BB^{\odot N'}$ has rank $1$.) 
% (JYC: not used directly but good to mention so it's in ())}
%\sout{Therefore $\BB$, and hence $\AA$, are
%\multblrkone/.} {\color{red} (JYC: incorrect, we are given that $\AA$ is \multblrkone/, and concluded it for $\BB$ before)}
Writing $\widehat \BB$ as a product of
a column vector $(\mu_1, \ldots, \mu_k)^{T}$ and a row vector
$(\mu_{k+1}, \ldots, \mu_{m})$, we get the form
(\ref{eqn:regularized-form-bipartite}) for $\BB$.
If we choose $\mu_1=1$, then the factorization of $\BB$ 
is unique, and  all $\mu_i$ belong to
the multiplicative subgroup generated by  $\{g_1, \ldots, g_d\}$. 
\end{proof}

\begin{proof}[Proof of Theorem \ref{t-step-1-ours}]
Let $\AA \in \bbC^{m \times m}$ be a symmetric, connected, and nonbipartite matrix.
Again if  $\AA$ is regularized nonbipartite, then it is easy to see 
 that $\AA$ is \multblrkone/.
We prove the other direction, so assume $\AA$ is \multblrkone/.

%%%
%{\color{red}
%We first show that all entries of $\AA$ are nonzero.
%Let $N(i) = \{ j \mid A_{i,j} \ne 0 \}$ denote the set of neighbors of $i \in [k]$.
%If $N(i) \not = N(i')$ for some $i, i' \in [k]$,
%since $i$ and $i'$ are connected by a path, there are successive
%vertices $i_1 =i, i_2, \ldots, i_\ell =i'$,
%and for some $s \in [\ell]$
%we have $N(i_s) \not = N(i_{s+1})$ but they intersect.
%This violates $\AA$ being rectangular. (In the same way as in Theorem~\ref{bi-step-1-ours}, we can show that all entries of $\AA$ are nonzero.)
%
%From this point, reasoning similarly to
%the proof of Theorem~\ref{bi-step-1-ours}, we can take
%$\mu_1 =1$, then we can factor out $\mu_{k+1} \not =0$ (call it $\mu$),
%and rename the two sequences of diagonal values as 
%$\mu_1 =1, \mu_2, \ldots, \mu_m$ and $\nu_1=1, \nu_2, \ldots, \nu_m$.
%They are all nonzero and all belong to $\frak G$.
%So we have the following form:
%(JYC: alternative to the below fragment) 
%}
%%%

%%%
%Let 
%\[\AA'
%= \begin{pmatrix} 0 & \AA \\ \AA & 0 \end{pmatrix}
%\]
%be the bipartization of $\AA$, and 
%by Property~\ref{prop:equiv-direct-sum-mbrk1-perm},
%$\AA'$ is  \multblrkone/.
%%{\color{red} Since $\AA$ no connected and nonbipartite, it is easy to see that $\AA'$ is connected and bipartite. (JYC: we need it, in fact it's not that trivial.)}
%Applying Theorem~\ref{bi-step-1-ours} to $\AA'$, we get an expression
%for $\AA$ as in  (\ref{eqn:regularized-form-bipartite}),
%%%
Let
\[\AA'
= \begin{pmatrix} 0 & \AA \\ \AA & 0 \end{pmatrix}
\]
be the bipartization of $\AA$, and
by Property~\ref{prop:equiv-direct-sum-mbrk1-perm},
$\AA'$ is  \multblrkone/.
We show $\AA'$ is connected so that we can apply 
Theorem~\ref{bi-step-1-ours} to $\AA'$.
Since $\AA$ is nonbipartite, there is an odd cycle.
Since $\AA$ is connected, for every $i \in [m]$, there is a
closed walk of odd length.
%% walk means vertices may repeat
%%% by going to that cycle if need be, and take that odd cycle and then back
Then for every $i, j \in [m]$ there are both walks of odd length and even
length, $i=i_0, i_1, \ldots, i_s=j$.
This gives a walk in $\AA'$ between $i$ and $j$ of even length, %for {\color{red} some} odd $s$
and  a walk in $\AA'$ between $i$ and $m+j$ of odd length. %for {\color{red} some} even $s$.
Now we apply Theorem~\ref{bi-step-1-ours} to $\AA'$, and get an expression
for $\AA$ as in  (\ref{eqn:regularized-form-bipartite}),
%%%
where all $\mu_i$ in (\ref{eqn:regularized-form-bipartite}) are nonzero,
and all belong to the  multiplicative
group $\frak G$ generated by $\{g_1, \ldots, g_d\}$, a generating set
for
% {\color{red} of?}
%%% you are right
 the  entries of $\AA$ which must all be nonzero. As
in the proof of Theorem~\ref{bi-step-1-ours} we can take
$\mu_1 =1$, then we can factor out $\mu_{k+1} \not =0$ (call it $\mu$),
and rename the two sequences of diagonal values as 
$\mu_1 =1, \mu_2, \ldots, \mu_m$ and $\nu_1=1, \nu_2, \ldots, \nu_m$.
They are all nonzero and all belong to $\frak G$.
So we have the following form:
\[
\AA = \mu \left(\begin{matrix}
1 \\
& \mu_2 \\
& & \ddots\\
& & & \mu_m
\end{matrix}\right)
\left( \begin{matrix}
\zeta_{1,1} & \zeta_{1,2} & \ldots & \zeta_{1,m} \\
\zeta_{2,1} & \zeta_{2,2} & \ldots & \zeta_{2,m} \\
\vdots      & \vdots      & \ddots & \vdots \\
\zeta_{m,1} & \zeta_{m,2} & \ldots & \zeta_{m,m}
\end{matrix}\right)
\left(\begin{matrix}
1 \\
& \nu_{2} \\
& & \ddots\\
& & & \nu_{m}
\end{matrix}\right),
\]
where $\mu, \mu_2, \ldots, \mu_m, \nu_{2}, \ldots, \nu_{m} \in \frak G$,
and $\zeta_{i,j}$ are roots of unity.
Since $\AA$ is symmetric, we 
have $\mu \mu_i \nu_j \zeta_{i,j} = \mu \mu_j \nu_i \zeta_{j, i}$,
for all $i,j \in [m]$.
Since   $g_1, \ldots, g_d$ are multiplicatively independent,
and $(\mu_i \nu_j)/(\mu_j \nu_i) \in \frak G$,
we get $\zeta_{i,j} = \zeta_{j, i}$ for all $i, j \in [m]$.
%%%
%%% yes, below you rs is more succint.
%\sout{Let $N$ be a common multiplier of the orders of all $\zeta_{i,j}$, ($i,j \in [m]$),
%then $\AA^{\odot N}$ is symmetric and has rank 1, and
%is $\mu^N$ times the product of the column 
%vector $((1, \mu_2, \ldots, \mu_m)^{T})^{\odot N}$ and the row vector
%$(1, \nu_{2},  \ldots, \nu_{m})^{\odot N}$.
%Since $g_1, \ldots, g_d$ are multiplicatively independent,
%the product of the
 %column vector $(1, \mu_2, \ldots, \mu_m)^{T}$ and the row vector
%$(1, \nu_{2},  \ldots, \nu_{m})$ is also a symmetric matrix, and it follows
%that $\mu_i = \nu_i$ for all $i \in [m]$.} {\color{red} (JYC: the whole part is not needed)}
%{\color{red} 
Then $\mu_i \nu_j = \mu_j \nu_i$ for all $i, j \in [m]$.
Putting $j = 1$, we get $\mu_i = \nu_i$ for all $i \in [m]$. 

%(JYC: way shorter)}  yes you are right.

\end{proof}

%% file: gadget-cyclotomic-transformation-inside-meta.tex
\section{Proof of Theorem \ref{bi-step-2-ours}}\label{sec:proof-bi-step-2-ours}

%\subsection*{Cyclotomic gadget transformation}

In this section we adapt the proof from Section 8.1 of~\cite{Cai-Chen-Lu-2013}
to establish a connection between the signature matrices
$\gm_{\Gamma, \AA}$ and $\gm_{\Gamma, \CC, \fD}$,
for any edge gadget $\Gamma$,
where $\AA$ and $(\CC, \fD)$ are related in a precise way
to be described in Definition~\ref{def:ACR-gen}.
Essentially we will show that all the results
from Section 8.1 of~\cite{Cai-Chen-Lu-2013} 
can be carried over to our setting.

%In this section we extend a technique
%given from Section 8.1~\cite{Cai-Chen-Lu-2013} to establish
%a connection between the signature matrices $\gm_{\Gamma, \AA}$ and $\gm_{\Gamma, \CC, \fD}$,
%where $\AA$ and $(\CC, \fD)$ are related in a precise way
%to be described in Definition~\ref{def:ACR-gen}.
%%provided there is a nontrivial dependency between $\AA$ and $(\CC, \fD)$.
%
%We first recall some definitions from~\cite{Cai-Chen-Lu-2013}, Section 8.1
%We note that all results from Section 8.1 actually holds for connected $\AA$.
%
%{\color{red} (I'm not sure if it was already assumed in Section 8.1 from CCL
%but there is no usage or mention of the word connectedness there)}
%
%{\color{red} (copied from CCL, needed, but not necessarily bipartite or connected)}

Let $\AA$ be an $m\times m$  symmetric (but not necessarily bipartite) complex matrix,
and let $(\CC,\fD)$ be a pair that satisfies the
  following condition $(\calT)$:\label{CONDITIONT}
\begin{enumerate}[wide]
\item[($\calT_1$)]
  $\CC$ is an $n\times n$ symmetric complex matrix. 
\item[($\calT_2$)]
$\fD=(\DD^{[0]},\ldots,
  \DD^{[N-1]})$ is a sequence
  of $N$ $n\times n$ diagonal complex matrices for some $N\ge 1$. 
\item[($\calT_3$)]
Every diagonal entry in $\DD^{[0]}$ is a positive integer.
Moreover, for each $a\in [n]$, there exist nonnegative integers
  $\alpha_{a,0},\ldots,\alpha_{a,N-1}$ such that
\[
D_a^{[0]}=\sum_{b=0}^{N-1} \alpha_{a,b}\ \ \ \text{and}\ \ \
D_a^{[r]}=\sum_{b=0}^{N-1} \alpha_{a,b}\cdot\oo_N^{b\hspace{0.02cm}r}\ \ \ \  \text{for all $r\in [N-1]$.}
\]
In particular, we say that the tuple $(\alpha_{a,0},\ldots,\alpha_{a,N-1})$
  \emph{generates} the $a${th} entries of $\fD$.
\end{enumerate}

%We need the following definition.

\begin{definition}[Definition 8.1 from~\cite{Cai-Chen-Lu-2013}]\label{def:ACR-gen}
Let ${\mathscr R}=\{R_{a,b}:a\in [n],b\in [0:N-1]\hspace{0.02cm}\}$
  be a partition of $[m]$ \emph{(}note that any $R_{a,b}$
  here may be empty\emph{)} such that for every $a\in [n]$,
\[\label{eq:nonempty-R-union}
\bigcup^{N-1}_{b=0} R_{a,b}\ne \emptyset.
\]
We say $\AA$ can be \emph{generated} by $\CC$ using $\mathscr R$ if
  for all $i,j\in [m]$,
\begin{equation}\label{weneedit-ours}
A_{i,j}=C_{a,a'}\cdot \oo_{N}^{b+b'},\ \ \ \ \text{where
  $i\in R_{a,b}$ and $j\in R_{a',b'}$}. 
\end{equation}
\end{definition}

%%cur
%{\color{red} (old text)
%
%Given any pair $(\CC,\fD)$ that satisfies $(\calT)$, we prove the following lemma:
%}
%
%In~\cite{Cai-Chen-Lu-2013},
%a cyclotomic reduction lemma (Lemma 8.2) was shown.
%In fact, it was actually shown that $Z_\AA(\cdot) \equiv Z_{\CC, \fD}(\cdot)$
%in that proof and from this it was trivially concluded that $\EVAL(\AA) \equiv \EVAL(\CC, \fD)$.
%We will extend this observation to edge gadgets in Lemma~\ref{lem:cycl-gadget-trans}.
%However, we will need the former equality
%rather than the latter equivalence
%so we formulate this statement accordingly.
%

%In~\cite{Cai-Chen-Lu-2013},
Assuming  $(\CC,\fD)$ satisfies $(\calT)$
and  $\AA$ is \emph{generated} by $\CC$ using $\mathscr R$,
it was shown in~\cite{Cai-Chen-Lu-2013}
  that  $Z_\AA(G) = Z_{\CC, \fD}(G)$ for any undirected graph $G$.
From this, $\EVAL(\AA) \equiv \EVAL(\CC, \fD)$.
This is the following lemma, called the cyclotomic reduction lemma,
in~\cite{Cai-Chen-Lu-2013}.
%(Lemma 8.2, called the cyclotomic reduction lemma, in~\cite{Cai-Chen-Lu-2013}.)
We will extend this to edge gadgets in Lemma~\ref{lem:cycl-gadget-trans}.

\begin{lemma}[cyclotomic reduction lemma (Lemma 8.2 from~\cite{Cai-Chen-Lu-2013})]\label{twinreduction-ours} % / extended, rephrased
Assume that $(\CC,\fD)$ satisfies $(\calT)$ with
  nonnegative integers $\alpha_{a,b}$.
Let ${\mathscr R}=\{R_{a,b}\}$ be a partition of $[m]$ satisfying
\[
 |R_{a,b} | = \alpha_{a,b}\ \ \ \text{and}\ \ \ 
  m = \sum_{a=1}^n \sum_{b=0}^{N-1} \alpha_{a,b} \ge n,
\]
and let $\AA$ denote the matrix
  generated by $\CC$ using $\mathscr R$.
%Suppose matrix $\AA$ can be \emph{generated} by $\CC$ using partition
%  $\mathscr R=\{R_{a,b} \}$, and for any $a\in [n]$, the $a^{th}$ entries of $\fD$ can be
%  \emph{generated} by $(|R_{a,0}|,\ldots,|R_{a,N-1}|)$,
Then $\eval(\AA)\equiv \eval(\CC,\fD).$
\end{lemma}

Given any pair $(\CC, \fD)$ that satisfies $(\calT)$, we prove the following lemma.
%expanding the \emph{cyclotomic reduction lemma}, i.e., Lemma 8.2 from~\cite{Cai-Chen-Lu-2013}:

\begin{lemma}[cyclotomic transfer lemma for edge gadgets]\label{lem:cycl-gadget-trans}
Assume that $(\CC,\fD)$ satisfies $(\calT)$ with
  nonnegative integers $\alpha_{a,b}$.
Let ${\mathscr R}=\{R_{a,b}\}$ be a partition of $[m]$ satisfying
\[
 |R_{a,b} | = \alpha_{a,b}\ \ \ \text{and}\ \ \ 
  m = \sum_{a=1}^n \sum_{b=0}^{N-1} \alpha_{a,b} \ge n,
\]
and let $\AA$ denote the matrix
  generated by $\CC$ using $\mathscr R$.
%Next, 
Let $\Gamma = (V, E)$ be an edge gadget
with two distinguished vertices $\firstdv$ and $\seconddv$ (in this order),
and let $r  = \deg(\firstdv)$ and $r'  = \deg(\seconddv)$.
Then for any $i, j \in [m]$,
 $\gm_{\Gamma, \AA}(i, j) = \omega_N^{rb + r' b'}  \gm_{\Gamma, \CC, \fD} (a, a')$,
where $a, a' \in [n]$,
 $R_{a,b} \ni i$, $R_{a',b'} \ni j$ are uniquely determined by $(i,j)$.
%In particular, $\gm_{\Gamma, \AA}$ is \multblrkone/ (\modblrkone/)
%iff $\gm_{\Gamma, \CC, \fD}$ is \multblrkone/ (\modblrkone/). 
%(or just say it here instead of having a corollary, not sure)}
\end{lemma}
\begin{proof}
We define a surjective map $\rho$ from
$\{\xi \mid \xi: V \rightarrow [m]\}$ 
to $\{\eta \mid \eta: V \rightarrow [n]\}$. 
Let $\xi: V \rightarrow [m]$.
 Then $\eta = \rho(\xi)$ is the following vertex assignment from $V$ to $[n]$:
For any $v \in V$, since $\mathscr R$ is a partition of $[m]$,
there is a unique pair $(a(v), b(v))$
such that $\xi(v) \in R_{a(v), b(v)}$.
Then let $\rho(\xi)(v) = a(v)$, and we also let $\xi_2(v) = b(v)$.
It is easy to check that $\rho$ is surjective.
We can write $\twt_{\Gamma, \AA}(\xi)$ as
\[
\twt_{\Gamma, \AA}(\xi) = \prod_{u v \in E} A_{\xi(u), \xi(v)}
= \prod_{u v \in E} C_{\eta(u), \eta(v)} \cdot \omega_N^{\xi_2(u) + \xi_2(v)}
= \prod_{u v \in E} C_{\eta(u), \eta(v)} \cdot \omega_N^{\xi_2(u)} \cdot \omega_N^{\xi_2(v)}.
\]

Fix $i, j \in [m]$ and let $i \in R_{a,b}$ and $j \in R_{a',b'}$.
%We are ready to proceed to the proof.
The  $(i, j)${th} entry of the matrix $\gm_{\Gamma, \AA}$ is
\[
\gm_{\Gamma, \AA}(i, j) =
\sum_{\substack{\xi \colon V \to [m] \\ \xi(\firstdv) = i,~ \xi(\seconddv) = j}} \twt_{\Gamma, \AA}(\xi)
%= \sum_{\substack{\eta \colon V \to [n] \\ 
%\eta(\firstdv) = a,~ \eta(\seconddv) = a'}} \sum_{\substack{~~\xi \colon V \to [m]: ~
%\rho(\xi) = \eta \\ \xi(\firstdv) = i,~ \xi (\seconddv) = j}} \twt_{\Gamma, \AA}(\xi)
= \sum_{\substack{\eta \colon V \to [n] \\ \eta(\firstdv) = a,~ \eta(\seconddv) = a'}} \sum_{\substack{~~\xi \colon V \to [m],~  \rho(\xi) = \eta \\ ~~\xi(\firstdv) = i,~ \xi (\seconddv) = j}} \twt_{\Gamma, \AA}(\xi).
\]

%where the last equality is true because
%for any $\xi \colon V \to [n]$ and $\eta \colon V \to [m]$
%with $\xi(\firstdv) = i_1$, $\xi(\seconddv) = i_2$ and $\rho(\xi) = \eta$,
%we must have $\eta(\firstdv) = a_1$ and $\eta(\seconddv) = a_2$.

Now for any given $\eta \colon V \to [n]$
with $\eta(\firstdv) = a$ and $\eta(\seconddv) = a'$,
we have
%\begingroup
%\allowdisplaybreaks
%\begin{align*}
\begin{eqnarray*}
& & \sum_{\substack{\xi \colon V \to [m],~ \rho(\xi) = \eta \\ \xi(\firstdv) = i,~ \xi(\seconddv) = j}} \twt_{\Gamma, \AA}(\xi)
= \prod_{u v \in E} C_{\eta(u), \eta(v)}
\times \sum_{\substack{\xi \colon V \to [m],~ \rho(\xi) = \eta \\ \xi(\firstdv) = i,~ \xi(\seconddv) = j}}
\prod_{~w \in V} \omega_N^{\xi_2(w) \cdot \deg(w)} \\
&=& \prod_{u v \in E} C_{\eta(u), \eta(v)}
\times \sum_{\substack{\xi \colon V \to [m],~ \rho(\xi) = \eta \\ \xi(\firstdv) = i,~ \xi (\seconddv) = j}}
\left( \prod_{w \in V \setminus \{\firstdv, \seconddv\}} \omega_N^{\xi_2(w) \cdot \deg(w)} \right)
\omega_N^{\xi_2(\firstdv) \cdot \deg(\firstdv)} \omega_N^{\xi_2(\seconddv) \cdot \deg(\seconddv)} \\
&=& \prod_{u v \in E} C_{\eta(u), \eta(v)}
\times \left( \prod_{w \in V \setminus \{\firstdv, \seconddv\}}
\left( \sum_{k = 0}^{N - 1} \big| R_{\eta(w), k} \big| \omega_N^{k \cdot \deg(w)} \right) \right)
\omega_N^{b \cdot r} \omega_N^{b' \cdot r'} \\
&=& \prod_{u v \in E} C_{\eta(u), \eta(v)}
\times \left( \prod_{w \in V \setminus \{\firstdv, \seconddv\}} D_{\eta(w)}^{[\deg(w) \bmod N]} \right)
\times \omega_N^{b r +b' r'}.
%\omega_N^{\xi_2(\firstdv) \cdot \deg(\firstdv)} \omega_N^{\xi_2(\seconddv) \cdot \deg(\seconddv)}
\end{eqnarray*}
%\end{align*}
%\endgroup
Summing the above equality over all $\eta \colon V \to [n]$
with $\eta(\firstdv) = a$ and $\eta(\seconddv) = a'$,
we obtain
\begin{align*}
(\gm_{\Gamma, \AA})(i,  j) &= \sum_{\substack{\eta \colon V \to [n] \\ \eta(\firstdv) = a,~ \eta(\seconddv) = a'}} \prod_{u v \in E} C_{\eta(u), \eta(v)}
\times \left( \prod_{w \in V \setminus \{\firstdv, \seconddv\}} D_{\eta(w)}^{[\deg(w) \bmod N]} \right) \times \omega_N^{b r +b' r'} \\
&= \omega_N^{b r +b' r'} \gm_{\Gamma, \CC, \fD}(a, a'),
\end{align*}
and the lemma is proved.
%Note that
%\[
%((S^{\odot r_1})^T \gm_{\Gamma, \CC, \fD)} S^{\odot r_2})_{i_1, i_2} = (\gm_{\Gamma, \CC, \fD)})_{a_1, a_2} \omega_N^{b_1 r_1} \omega_N^{b_2 r_2}.
%\]
\end{proof}

%\begin{remark}
%An alternative more straightforward proof can be obtained
%by a technique called holographic transformation~\cite{Cai-Chen-Book}.
%{\color{red} (just a note not sure if needed)}
%\end{remark}

%{\color{red} JYC: note that from this lemma,
%it's easy to conclude that $Z_\AA(G) = Z_{\CC, \fD}(G)$ for any graph $G$
%so the cyclotomic reduction lemma from the CCL paper follows.
%Maybe we should mention it.}

%In the proof of Lemma~\ref{lem:cycl-gadget-trans},
%condition~(\ref{eq:nonempty-R-union}) was not actually used.
%However, it is crucial for the following corollary.
%(which is a modification of Corollary~8.3
%(the inverse cyclotomic reduction lemma) from~\cite{Cai-Chen-Lu-2013} for edge gadgets)
%{\color{red} (but for both directions)}
Given  $\AA$ and $(\CC, \fD)$ as above, let 
 $\pur \AA$, ${\pur \CC}$ be the purifications of
$\AA$, $\CC$, respectively.
Then $\pur \AA$ is generated by ${\pur \CC}$ using the same $\mathscr R$.
If $(\CC, \fD)$ satisfies condition $(\calT)$, then so does
the pair  $({\pur \CC},\fD)$.

\begin{corollary}\label{cor:cycl-blr1-equiv}
Under the conditions of Lemma~\ref{lem:cycl-gadget-trans},
 $\gm_{\Gamma, \AA}$ is \multblrkone/ (\modblrkone/)
iff $\gm_{\Gamma, \CC, \fD}$ is \multblrkone/ (\modblrkone/).
The same conclusion holds 
for $\gm_{\Gamma, \pur \AA}$ and $\gm_{\Gamma, \pur \CC, \fD}$.
%we have $\gm_{\Gamma, \AA} = (S^{\odot r_1})^T \gm_{\Gamma, \CC, \fD} S^{\odot r_2}$,
%so $\gm_{\Gamma, \AA}$ is \multblrkone/ (\modblrkone/)
%iff $\gm_{\Gamma, \CC, \fD}$ is \multblrkone/ (\modblrkone/).
\end{corollary}
%\begin{proof}
%Follows from Lemma~\ref{lem:cycl-gadget-trans}.
%It can be checked directly.
%%%
%We note that for the forward implication in Corollary~\ref{cor:cycl-blr1-equiv},
%%$\Rightarrow$,
%condition~(\ref{eq:nonempty-R-union}) is crucial:
%we use the fact that for any $a \in [n]$,
%$\bigcup_{k=0}^{N-1} R_{a, k} \ne \emptyset$. {\color{red} (JYC: wrong equation, am I right, but If I label that is just a repetition)}
%%%
We note that for the forward implicaiton in Corollary~\ref{cor:cycl-blr1-equiv},
the condition $\bigcup_{b=0}^{N-1} R_{a, b} \ne \emptyset$ is crucial.
% so
%there exists $i \in [m]$ such that $i \in R_{a, k}$ 
%for some $k \in [0 : N - 1]$.
%(namely, $i \in R_{a, b}$ for some $b \in [0 : N - 1]$)
%such that $S_{a, i} = \omega_N^b \ne 0$.
%\end{proof}

%%%
%{\color{red} Given  $\AA$ and $(\CC, \fD)$ as above, let 
% $\pur \AA$, ${\pur \CC}$ be the purifications of
%$\AA$, $\CC$, respectively.
%Then $\pur \AA$ is generated by ${\pur \CC}$ using the same $\mathscr R$.
%If $(\CC, \fD)$ satisfies condition $(\calT)$, then so does
%the pair  $({\pur \CC},\fD)$. (JYC: used in the derivation of the above corollary, so moved up)} 
%%%
%{\color{red} (JYC: leave it, but make the references to it parenthetical)}
From Corollary~\ref{cor:cycl-blr1-equiv} 
and Theorem~\ref{thm:gadget-thickening-non-mult-bl-rk-1}, we have the following corollary.
\begin{corollary}[inverse cyclotomic transfer lemma for edge gadgets]\label{inversetwin-ours}
Let $\AA$ and $(\CC, \fD)$ be as above, 
satisfying condition $(\calT)$.
Let $\pur \AA$, ${\pur \CC}$ be the purifications of
$\AA$, $\CC$, respectively. 
If  $\Gamma$ is an edge gadget, such that 
$\gm_{\Gamma, {\pur \CC}, \fD}$ has a $2$ by $2$
submatrix $\left( \begin{matrix}
M_{i,k} & M_{i, \ell}\\
M_{j,k} & M_{j, \ell}
\end{matrix}\right)
$ that has no zero entries and $|M_{i,k}M_{j, \ell}| \not = |M_{i, \ell} M_{j,k}|$,
then  $\gm_{\Gamma, {\pur \CC}, \fD}$ is not \modblrkone/,
and for some $p \ge 1$, there is an edge gadget $\Gamma' = T_p(\Gamma)$,
such that $\gm_{\Gamma', \AA}$ is not \multblrkone/. %{\color{red} (JYC: this corollary is used peripherally in meta2, so I'll remove it later)}
\end{corollary}

%$$
%\left(\begin{matrix} C_{i,k} & C_{i,\ell}\\  C_{j,k} & C_{j,\ell}
%\end{matrix}\right)$$  such that all four entries are non-zero and
%$|C_{i,k}C_{j,\ell} |\ne |C_{i,\ell}C_{j,k}|$,
%  then $\eval(\CC,\fD)$ is \#P-hard.
%\end{corollary}

%% file: meta-argument-bounded-degree-graphs.tex
\subsection{Step 2.1}\label{sec:step-2.1-bip}
%{\color{red} (make sure Theorem~\ref{bi-step-1-ours} applies here, i.e. what is stated in there is enough, which is why I expanded on it)}
%
%{\color{red} (mention $(g_1, \ldots, g_d)$ and $(p_1, \ldots, p_d)$ here or inside $\calS$, not sure, if here, there are subtle issues)}
%
Let $\AA$ be a regularized bipartite matrix
and let $\pur \AA$ be the purification of $\AA$
obtained by going from a generating set
 $(g_1, \ldots, g_d)$ of nonzero entries of $\AA$
 to the $d$ smallest primes $(p_1, \ldots, p_d)$, $d \ge 0$.
There exist a positive integer $N$ and
six sequences $\boldsymbol{\mu}$, $\boldsymbol{\nu}$,
$\pur{\boldsymbol{\mu}}$, $\pur{\boldsymbol{\nu}}$, $\mm$, and $\nn$
such that $(\AA,(N,\boldsymbol{\mu},\boldsymbol{\nu},\mm,\nn))$,
$(\pur \AA,(N,\pur{\boldsymbol{\mu}},\pur{\boldsymbol{\nu}},\mm,\nn))$
satisfy the following condition:\label{CONDITIONS1}
%By a permutation of rows and columns so that 
%$\pur{\boldsymbol{\mu}} = (\pur{\mu_1}, \ldots, 
% $\boldsymbol{\mu} = (\mu_1, \mu_2, $, $\boldsymbol{\nu}$  that puts  
%By a simultaneous row and column permutation 
%Up to a simultaneous row and column permutation by the same permutation
%of both $\AA$ and $\pur \AA$
%(this permutation is obtained by collecting the entries of $\pur \AA$
%of equal norm in decreasing order),
\begin{enumerate}[wide]
\item[($\calS_1$)] $\AA$ is the bipartization of an $m\times n$ matrix $\BB$,
  so $\AA$ is $(m+n)\times (m+n)$.
$\boldsymbol{\mu}$ $=(\mu_1,\ldots,\mu_s)$ and
  $\boldsymbol{\nu}=(\nu_1,\ldots,\nu_t)$ are two sequences,
each consisting of pairwise distinct nonzero complex numbers where $s\ge 1$ and $t\ge 1$
such that $\mu_1, \ldots, \mu_s, \nu_1, \ldots, \nu_t$ generate
a torsion-free multiplicative group.
$\mm=(m_1,\ldots,m_s)$ and $\nn=(n_1,\ldots,n_t)$ are two sequences of positive
  integers such that $m=\sum m_i$ and $n=\sum n_i$.
The rows of $\BB$ are indexed by $\xx=(x_1,x_2)$, where $x_1\in [s]$
  and $x_2\in [m_{x_1}]$; the columns of $\BB$ are indexed by
  $\yy=(y_1,y_2)$, where $y_1\in [t]$ and $y_2\in [n_{y_1}]$.
We have, for all $\xx,\yy$,
\[
B_{\xx,\yy}=B_{(x_1,x_2),(y_1,y_2)}=\mu_{x_1}\nu_{y_1} S_{\xx,\yy},
\] 
where $\SS=\{S_{\xx,\yy}\}$ is an $m\times n$ matrix in which
  every entry is a power of $\oo_N$:
\[
\BB = \left(\begin{matrix}
\mu_1\II_{m_1}\\
& \hspace{-0.5cm}\mu_2\II_{m_2}\hspace{-0.5cm} \\
& & \hspace{-0.5cm}\ddots\hspace{-0.5cm}\\
& & & \mu_s\II_{m_s}
\end{matrix}\right)\hspace{-0.1cm}
\left( \begin{matrix}
\SS_{(1,*),(1,*)} & \hspace{-0.25cm}\SS_{(1,*),(2,*)} & \hspace{-0.25cm}\ldots & \hspace{-0.25cm}\SS_{(1,*),(t,*)} \\
\SS_{(2,*),(1,*)} & \hspace{-0.25cm}\SS_{(2,*),(2,*)} & \hspace{-0.25cm}\ldots & \hspace{-0.25cm}\SS_{(2,*),(t,*)} \\
\vdots & \hspace{-0.25cm}\vdots & \hspace{-0.25cm}\ddots & \hspace{-0.25cm}\vdots \\
\SS_{(s,*),(1,*)} & \hspace{-0.25cm}\SS_{(s,*),(2,*)} & \hspace{-0.25cm}\ldots & \hspace{-0.25cm}\SS_{(s,*),(t,*)}
\end{matrix}\right)\hspace{-0.1cm}
\left(\begin{matrix}
\nu_1\II_{n_1} \\
& \hspace{-0.4cm}\nu_2\II_{n_2}\hspace{-0.4cm} \\
& & \hspace{-0.4cm}\ddots\hspace{-0.4cm}\\
& & & \nu_t\II_{n_t}
\end{matrix}\right), 
\]
where $\II_{k}$ denotes the $k\times k$ identity matrix.

$\pur \BB$ is the purification of $\BB$, 
$\pur \AA$ is the purification of $\AA$, and
$\pur \AA$ is also the bipartization of $\pur \BB$.
  %so $\pur \AA$ is $(m+n)\times (m+n)$.
We may assume 
$\pur{\boldsymbol{\mu}}$ $=(\pur \mu_1,\ldots, \pur \mu_s)$ and
  $\pur{\boldsymbol{\nu}}=(\pur \nu_1,\ldots,\pur \nu_t)$, the purifications of $\boldsymbol{\mu}$ and $\boldsymbol{\nu}$, respectively,
  are strictly
decreasing sequences of
  positive rational numbers,
by a simultaneous row and column permutation by the same permutation
%%% old
%of both $\AA$ and $\pur \AA$
%%% new
applied to both $\AA$ and $\pur \AA$.
%The rows and columns of $\pur \BB$ are also indexed
%in the same way as $\BB$.
%%% this is implied by purification
% by $\xx=(x_1,x_2)$ where $x_1\in [s]$
%  and $x_2\in [m_{x_1}]$; the columns of $\pur \BB$ are indexed by
%  $\yy=(y_1,y_2)$ where $y_1\in [t]$ and $y_2\in [n_{y_1}]$.
We have, for all $\xx,\yy$,
\[
\pur B_{\xx,\yy}=\pur B_{(x_1,x_2),(y_1,y_2)}=\pur \mu_{x_1}\pur \nu_{y_1} S_{\xx,\yy},
\] 
%where $\SS=\{S_{\xx,\yy}\}$ is an $m\times n$ matrix in which
%  every entry is a power of $\oo_N$:
so that
\[
\pur \BB = \left(\begin{matrix}
\pur \mu_1\II_{m_1}\\
& \hspace{-0.5cm}\pur \mu_2\II_{m_2}\hspace{-0.5cm} \\
& & \hspace{-0.5cm}\ddots\hspace{-0.5cm}\\
& & & \pur \mu_s\II_{m_s}
\end{matrix}\right)\hspace{-0.1cm}
\left( \begin{matrix}
\SS_{(1,*),(1,*)} & \hspace{-0.25cm}\SS_{(1,*),(2,*)} & \hspace{-0.25cm}\ldots & \hspace{-0.25cm}\SS_{(1,*),(t,*)} \\
\SS_{(2,*),(1,*)} & \hspace{-0.25cm}\SS_{(2,*),(2,*)} & \hspace{-0.25cm}\ldots & \hspace{-0.25cm}\SS_{(2,*),(t,*)} \\
\vdots & \hspace{-0.25cm}\vdots & \hspace{-0.25cm}\ddots & \hspace{-0.25cm}\vdots \\
\SS_{(s,*),(1,*)} & \hspace{-0.25cm}\SS_{(s,*),(2,*)} & \hspace{-0.25cm}\ldots & \hspace{-0.25cm}\SS_{(s,*),(t,*)}
\end{matrix}\right)\hspace{-0.1cm}
\left(\begin{matrix}
\pur \nu_1\II_{n_1} \\
& \hspace{-0.4cm}\pur \nu_2\II_{n_2}\hspace{-0.4cm} \\
& & \hspace{-0.4cm}\ddots\hspace{-0.4cm}\\
& & & \pur \nu_t\II_{n_t}
\end{matrix}\right). 
\]
Note that the matrix $\SS$ consisting of roots of unity is
the same for $\BB$ and $\pur \BB$.

%%% JYC: below: this is implied. by purification from A to \pur A.
%
%Furthermore, we can additionally assume that $(\pur \mu_1, \ldots, \pur \mu_s)$ and $(\pur \nu_1, \ldots, \pur \nu_t)$
%are the purifications of $(\mu_1, \ldots, \mu_s)$ and $(\nu_1, \ldots, \nu_t)$, respectively,
%each obtained by going from $(g_1, \ldots, g_d)$ to $(p_1, \ldots, p_d)$.
%,
%where $\{g_1, \ldots, g_d\}$ is a fixed generating set for the set of nonzero entries of $\AA$,
%and $p_1 \le \ldots \le p_d$ are the first $d$ primes.
\end{enumerate}
%{\color{red} (added, can be removed)
%
%The condition \emph($\calS_1$\emph) corresponds to (is similar)
%to the condition \emph($\calS_1$\emph) from~\cite{Cai-Chen-Lu-2013}, p. 955.
%}

We let 
\[I=\bigcup_{i\in [s]} \big\{ (i,j):
j\in [m_i]\hspace{0.03cm}\big\}\ \ \ \text{and}\ \ \ J
  =\bigcup_{i\in [t]} \big\{ (i,j):
  j\in [n_i]\hspace{0.03cm}\big\},\]respectively.
We use $\{0\}\times I$ to index the first $m$ rows (or columns) of $\AA$ (and $\pur \AA$)
  and $\{1\}\times J$ to index the last $n$ rows (or columns) of $\AA$ (and $\pur \AA$).
%We also view $\BB$ as an $s\times t$ block matrix:
%Its $(i,j)^{th}$ block is an $m_i\times n_j$ matrix $c_id_j\SS_{(i,*),(j,*)}$.
Given $\xx\in I$ and $j\in [t]$, we let 
  \[\SS_{\xx,(j,*)}=\big(S_{\xx,(j,1)},\ldots,S_{\xx,(j,n_j)}\big)\in \mathbb{C}^{n_j}\]
denote the $j${th} block of the $\xx${th} row vector of $\SS$.
Similarly, given $\yy\in J$ and $i\in [s]$, \[\SS_{(i,*),\yy}
= \big(S_{(i,1),\yy},\ldots,S_{(i,m_i),\yy}\big)\in \mathbb{C}^{m_i}\]
denotes the $i${th} block of the $\yy${th} column vector of $\SS$.

%\begin{figure}
%\center
%\includegraphics[height=3.1cm]{figure_2}
%\caption{Gadget for constructing graph $G^{[p]}$, $p\ge 1$.}\label{figure_2}
%\end{figure}

%{\color{red} (added, can be omitted, up to you)
%
%It is important to note that $\pur \AA$ can be obtained from $\AA$
%by the following symbolic substitution:
%$\mu_i \to \pur \mu_i$ for $i \in [s]$ and $\nu_i \to \pur \nu_i$ for $i \in [t]$.
%We also note that $\AA$ and $\pur \AA$ share the ``common'' matrix $\SS$.
%}
%
\begin{figure}
\center
\includegraphics[height=3.1cm]{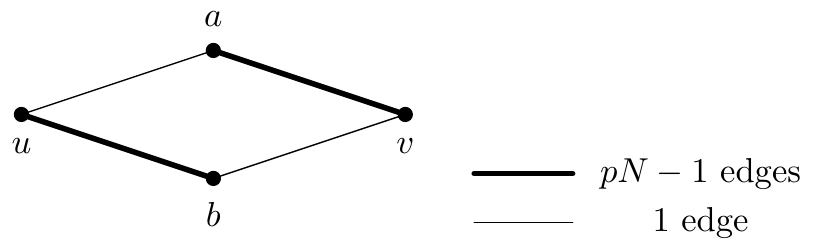}%{figure_2}
\caption{Gadget for constructing graph $G^{[p]}$, $p\ge 1$.
% (old). 
%Gadget $\Gamma^{[p]}$, $p \ge 1$, corresponding to Figure 8.1 from~\cite{Cai-Chen-Lu-2013}. {\color{red} (new, the correspondence is not exact, we need to use different letters and remove the superscript so it need editing)}
}\label{figure_2-ours}
\end{figure}

%The following is a modification of Lemma 8.5 from~\cite{Cai-Chen-Lu-2013}.
\begin{lemma}[modification of Lemma 8.5 from~\cite{Cai-Chen-Lu-2013}]\label{hahajaja-ours}
Suppose $(\AA,(N,\boldsymbol{\mu},\boldsymbol{\nu},\mm,\nn))$,
$(\pur \AA,(N,\pur{\boldsymbol{\mu}},\pur{\boldsymbol{\nu}},\mm,\nn))$ satisfy $(\calS_1)$.
Then either there exists an edge gadget $\Gamma$
such that $\gm_{\Gamma, \AA}$ is not \multblrkone/, %(a fortiori, $\EVAL^{(\Delta)}(\AA)$ is \#P-hard for some $\Delta > 0$)
or $(\AA,(N,\boldsymbol{\mu},\boldsymbol{\nu},\mm,\nn))$,
  $(\pur \AA,(N,\pur{\boldsymbol{\mu}},\pur{\boldsymbol{\nu}},\mm,\nn))$ satisfy
  the following two conditions:\label{CONDITIONS23}
\begin{enumerate}[wide]
\item[$(\calS_2)$] For all $\xx,\xx'\in I$, either there exists
  an integer $k$ such that $\SS_{\xx,*}=\oo_N^k\cdot \SS_{\xx',*}$ or
  for every $j\in [t]$,
$\langle \SS_{\xx,(j,*)},\SS_{\xx',(j,*)}\rangle =0.$
\item[$(\calS_3)$]
For all $\yy,\yy'\in J$, either there exists an integer $k$ such that
  $\SS_{*,\yy}=\oo_N^k\cdot \SS_{*,\yy'}$ or for every
  $i\in [s]$, $\langle \SS_{(i,*),\yy},\SS_{(i,*),\yy'}\rangle =0$.
\end{enumerate}
%{\color{red} (added, can be removed)
%
%The conditions ($\calS_2$), ($\calS_3$) correspond (are similar)
%to the conditions ($\calS_2$), ($\calS_3$) from~\cite{Cai-Chen-Lu-2013}, p. 955.
%}
\end{lemma}

\begin{proof}
%%%%%%%%%%%%%%%%%%%%%%%%%%%%%%%%%%%%%%%%%%%%%%%%%%%%%%%%%%%%%%%%%%%%
\begin{comment}
{\color{red} (old text)

Assume $\eval(\AA)$ is not \#P-hard. Assume for every edge gadget $\Gamma$, $\gm_{\Gamma, \AA}$ is \multblrkone/. We prove ($\calS_2$) here.
($\calS_3$) is similar.

Let $G=(V,E)$ be an undirected graph. We construct
  a new graph $G^{[p]}$ for each $p\ge 1$ by replacing every edge $uv$ in $E$ with
  a gadget shown in Figure \ref{figure_2}.
Formally we define graph $G^{[p]}=(V^{[p]},E^{[p]})$ as follows:
$$
V^{[p]}=V\cup\big\{a_e,b_e:e\in E\big\}
$$
and $E^{[p]}$ contains the following edges: For each $e=uv\in E$, add
\begin{enumerate}
\item one edge $(u,a_e)$ and $(b_e,v)$; and 
\item $(pN-1)$ parallel edges $(a_e,v)$ and $(u,b_e)$.
\end{enumerate}
The construction of $G^{[p]}$ gives us an $(m+n)\times (m+n)$ matrix
  $\AA^{[p]}$ such that
$$
Z_{\AA^{[p]}}(G)=Z_\AA(G^{[p]}),\ \ \ \text{for all undirected graphs $G$.}
$$
Thus, we have $\eval(\AA^{[p]})\le \eval(\AA)$,
  and $\eval(\AA^{[p]})$ is also not \#P-hard. 
}

{\color{red} (new text)}

{\color{red} (here I'm using $V^{[p]},E^{[p]}$ to denote the vertex and edge sets of $\Gamma^{[p]}$, in CCL it's for $G^{[p]}$,
maybe we should use a bit different letters to avoid confusion when comparing to CCL)}

{\color{red} Assume for every edge gadget $\Gamma$, $\gm_{\Gamma, \AA}$ is \multblrkone/. (can remove this sentence, or use an argument by contradiction futher)} We prove ($\calS_2$) here.
($\calS_3$) is similar.

\end{comment}
%%%%%%%%%%%%%%%%%%%%%%%%%%%%%%%%%%%%%%%%%%%%%%%%%%%%%%%%%%%%%%%%%%%%
We adapt the proof for Lemma~8.5 from~\cite{Cai-Chen-Lu-2013}.
We prove ($\calS_2$) here;
the proof of ($\calS_3$) is similar.
Consider the edge gadget $\Gamma^{[p]}$ for each $p \ge 1$ as shown 
in Figure~\ref{figure_2-ours}.
Formally, $\Gamma^{[p]}=(V^{[p]},E^{[p]})$ is defined as follows:
\[
V^{[p]}=\big\{u, v, a, b\big\},
\]
where $u, v$ are the distinguished vertices of $\Gamma^{[p]}$ (in this order)
% (in this order) no ned for order. it's obvious symmeric.
and $E^{[p]}$ contains the following edges:
\begin{enumerate}
\item one edge $(u,a)$ and $(b,v)$ and 
\item $(pN-1)$ parallel edges $(a,v)$ and $(u,b)$.
\end{enumerate}
The construction of $\Gamma^{[p]}$ gives us an $(m+n)\times (m+n)$ matrix
  $\pur \AA^{[p]} = \gm_{\Gamma^{[p]}, \pur \AA}$.  

If there exists $p \ge 1$, such that $\gm_{\Gamma^{[p]}, \pur \AA}$
is not  \modblrkone/, then by 
Theorem~\ref{thm:gadget-thickening-non-mult-bl-rk-1},  there exists
$\ell \ge 1$,  such that 
$\gm_{T_\ell(\Gamma^{[p]}), \AA}$ is non-\multblrkone/
and we are done by taking $\Gamma = T_\ell(\Gamma^{[p]})$.
So we may assume that for all  $p \ge 1$, 
 $\pur \AA^{[p]} = \gm_{\Gamma^{[p]}, \pur \AA}$ is \modblrkone/.

The entries of $\pur \AA^{[p]}$ are as follows. First,
\[
\pur A^{[p]}_{(0,\uu),(1,\vv)}=\pur A^{[p]}_{(1,\vv),(0,\uu)}=0,\ \ \ \text{for
  all $\uu \in I$ and $\vv \in J$.}
\]
So $\pur \AA^{[p]}$ is a block diagonal matrix with two blocks of
  $m\times m$ and $n\times n$, respectively.
The entries in the upper-left $m\times m$ block are
\begin{align*}
\pur A^{[p]}_{(0,\uu),(0,\vv)}&= \left(\sum_{\aa\in J}
  \pur A_{(0,\uu),(1,\aa)}(\pur A_{(0,\vv),(1,\aa)})^{pN-1}\right)
  \left(\sum_{\bb\in J}
  (\pur A_{(0,\uu),(1,\bb)})^{pN-1}\pur A_{(0,\vv),(1,\bb)}\right)\\[0.3ex]
  &= \left(\sum_{\aa\in J}
  \pur B_{\uu,\aa}(\pur B_{\vv,\aa})^{pN-1}\right)
  \left(\sum_{\bb\in J}
  (\pur B_{\uu,\bb})^{pN-1}\pur B_{\vv,\bb}\right)
\end{align*}
for all $\uu,\vv\in I$. The first factor of the last expression is
\[
\sum_{\aa\in J} \pur \mu_{u_1}\pur \nu_{a_1}S_{\uu,\aa} (\pur \mu_{v_1}\pur \nu_{a_1})^{pN-1}\overline{S_{\vv,\aa}}
%=\mu_{u_1}\mu_{v_1}^{pN-1} \sum_{\aa\in J} \nu_{a_1}^{pN}
%  S_{\uu,\aa}\overline{S_{\vv,\aa}}
=\pur \mu_{u_1}\pur \mu_{v_1}^{pN-1} \sum_{i\in [t]} \pur \nu_{i}^{pN} \langle
  \SS_{\uu,(i,*)},\SS_{\vv,(i,*)}\rangle.
\]
Similarly, we have for the second factor
\[ \sum_{\bb\in J}
  (\pur B_{\uu,\bb})^{pN-1}\pur B_{\vv,\bb}
=\pur \mu_{u_1}^{pN-1}\pur \mu_{v_1} \sum_{i\in [t]}\pur \nu_i^{pN}
  \overline{\langle
  \SS_{\uu,(i,*)},\SS_{\vv,(i,*)}\rangle}.
\]
As a result, we have
\[
\pur A^{[p]}_{(0,\uu),(0,\vv)}=(\pur \mu_{u_1}\pur \mu_{v_1})^{pN}\left|\hspace{0.05cm}
  \sum_{i\in [t]} \pur \nu_i^{pN} \langle
  \SS_{\uu,(i,*)},\SS_{\vv,(i,*)}\rangle\hspace{0.05cm} \right|^2.
\]
It is clear that the upper-left $m\times m$ block of $\pur \AA^{[p]}$
  is nonnegative.
This holds for its
  lower-right $n\times n$ block as well, so $\pur \AA^{[p]}$ is a nonnegative
   matrix. 

Now let $\uu\ne \vv$ be two arbitrary indices in $I$ (if $|I|=1$,
  $(\calS_2)$ is trivially true); then we have
\[\pur A^{[p]}_{(0,\uu),(0,\uu)}\pur A^{[p]}_{(0,\vv),(0,\vv)}=(\pur \mu_{u_1}\pur \mu_{v_1})^{2pN}\left(
\sum_{i\in [t]} n_i\cdot \pur \nu_i^{pN} \right)^4,\]
which is positive, and
\[
\pur A^{[p]}_{(0,\uu),(0,\vv)}\pur A^{[p]}_{(0,\vv),(0,\uu)}=
  (\pur \mu_{u_1}\pur \mu_{v_1})^{2pN} \left|\hspace{0.05cm}\sum_{i\in [t]} \pur \nu_i^{pN} \langle
  \SS_{\uu,(i,*)},\SS_{\vv,(i,*)}\rangle \hspace{0.05cm}\right|^4.
\]
Since $\pur \AA^{[p]}$ is \modblrkone/, 
%{\color{red} (actually, even \blrkone/, this is what we use here)}
\begin{equation}\label{temptemptem-ours}
\left|\hspace{0.05cm}\sum_{i\in [t]} \pur \nu_i^{pN} \langle
  \SS_{\uu,(i,*)},\SS_{\vv,(i,*)}\rangle\hspace{0.05cm}\right|
\in \left\{0, \sum_{i\in [t]}n_i\cdot \pur \nu_i^{pN}\right\}. 
\end{equation}

On the other hand, the following inequality always holds:
For any $p\ge 1$,
\begin{equation}\label{temptemp-ours}
\left|\hspace{0.05cm}\sum_{i\in [t]} \pur \nu_i^{pN}\cdot \langle
  \SS_{\uu,(i,*)},\SS_{\vv,(i,*)}\rangle
  \hspace{0.05cm}\right|\le \sum_{i\in [t]}n_i\cdot \pur \nu_i^{pN}.
\end{equation}
%since $|\langle \SS_{\uu,(i,*)},\SS_{\vv,(i,*)}\rangle|\le n_i$.
If there exists $p \ge 1$ such that the equality in 
(\ref{temptemp-ours}) holds, then $\SS$ must satisfy
  $|\langle \SS_{\uu,(i,*)},\SS_{\vv,(i,*)}\rangle|=n_i$
  for all $i\in [t]$ and thus
  $\SS_{\uu,(i,*)}=(\oo_N)^{k_i}\cdot \SS_{\vv,(i,*)}$ for some $k_i\in [0:N-1]$.
Furthermore, for equality to hold
these $k_i$'s must be the same. %since we assumed (\ref{temptemp}) is
%  an equality:
%  $$\left|\sum_{i\in [t]} \nu_i^{pN}n_i\cdot \oo_N^{k_i}\right|=\sum_{i\in [t]}n_i\cdot
%  \nu_i^{pN}.$$
%%%%As a result, $\SS_{\uu,*}$ and $\SS_{\vv,*}$ are linearly dependent, which
%%%%  contradicts with our assumption.
%%%
%This is one of the two alternatives in  $(\calS_2)$. {\color{red} (This is the first alternative in $(\calS_2)$. JYC: more correct?)}
%%%
This is the first alternative in $(\calS_2)$.

%Now suppose vectors $\SS_{\uu,*}$ and $\SS_{\vv,*}$
%  are linearly dependent. Then
%because entries of $\SS$ are all powers of $\oo_N$,
%  there must exist an integer $k\in [0:N-1]$ such that
%$\SS_{\uu,*}=\oo_N^k\cdot \SS_{\vv,*}$,
%and we are done.%
%Otherwise, assuming $\SS_{\uu,*}$ and
%  $\SS_{\vv,*}$ are linearly independent,
%  we have
%This is because,\vspace{-0.062cm} if the left-hand side is equal to the right-hand side,
%  then $|\langle \SS_{\uu,(i,*)},\SS_{\vv,(i,*)}\rangle|=n_i$
%It then follows from (\ref{temptemptem-ours}) that 

Suppose (\ref{temptemp-ours}) is a strict inequality  for all $p \ge 1$.
Then from (\ref{temptemptem-ours}),
\[
\sum_{i\in [t]} \pur \nu_i^{pN} \langle
  \SS_{\uu,(i,*)},\SS_{\vv,(i,*)}\rangle=0\ \ \ \text{for all $p\ge 1$.}
\]
As $(\pur \nu_1,  \ldots,\pur \nu_t)$ is strictly decreasing, these equations form a 
  Vandermonde system. It follows that
$ 
\langle
  \SS_{\uu,(i,*)},\SS_{\vv,(i,*)}\rangle=0$ {for all $i\in [t]$}. 
This is the second alternative in $(\calS_2)$, which proves
$(\calS_2)$.
%This finishes the proof of $(\calS_2)$.
%
%{\color{red} (JYC: in this proof we get a nonnegative \blrkone/ signature)}
\end{proof}

%%%%%%%%%%%%%%%%%%%%%%%%%%%%%%%%%%%%%%%%

In what follows, we assume $(\AA,(N,\boldsymbol{\mu},\boldsymbol{\nu},\mm,\nn))$
and
$(\pur \AA,(N,\pur{\boldsymbol{\mu}},\pur{\boldsymbol{\nu}},\mm,\nn))$ 
satisfy $(\calS_1), (\calS_2)$, and $(\calS_3)$.

%{\color{red} (moved Meta1 over here)}

The proof of Lemma~\ref{hahajaja-ours} was adapted from that of 
Lemma 8.5 in~\cite{Cai-Chen-Lu-2013}.
In fact, we will need to adapt many proofs from~\cite{Cai-Chen-Lu-2013} 
in a similar fashion.
In order to only highlight the essential point, we 
introduce the following \emph{meta-argument}.
%by introducing the so-called \emph{meta-arguments}.

%{\color{red} (meta-remark)}
%
%We need to introduce the following crucial meta-argument.
\begin{argument}[{\sl Meta}$_1$]\label{meta1} %\label{meta:sec-8-A-ccl}
We have the following:
\begin{enumerate}
\item\label{item:meta1-thickening} Either $\gm_{\Gamma, \pur \AA}$ is \modblrkone/
for every edge gadget $\Gamma$,
or there is an edge gadget $\Gamma$
such that $\gm_{\Gamma, \pur \AA}$ is not \modblrkone/,
so by Theorem~\ref{thm:gadget-thickening-non-mult-bl-rk-1},
for some $p \ge 1$,
$\gm_{T_p(\Gamma), \AA}$ is not \multblrkone/;

\item\label{item:meta1-assumption} Referring to the matrix $\AA$
from Section 8.2 in~\cite{Cai-Chen-Lu-2013},
all statements (including theorems, lemmas, corollaries, properties, etc.
 numbered from 8.5 to 8.7) from~\cite{Cai-Chen-Lu-2013}
can be proved under the assumption that every signature $\gm_{\Gamma, \AA}$ %($\AA$ is from~\cite{Cai-Chen-Lu-2013}) %edge
is \modblrkone/,
and whenever it is concluded that, as a possible scenario, $\EVAL(\AA)$ 
%($\AA$ is from~\cite{Cai-Chen-Lu-2013}) 
is $\#$P-hard,
this is because an edge gadget $\Gamma$ has been constructed
such that $\gm_{\Gamma, \AA}$ is not \modblrkone/. 
This statement can be checked directly;
%
%
%in particular, all the statements (theorems, lemmas, corollaries, properties, etc.) numbered from 8.5 to 8.7,
%can be proved under the assumption that every edge signature $\gm_{\Gamma, \AA}$ ($\AA$ is from~\cite{Cai-Chen-Lu-2013}) is \modblrkone/
%and whenever it is concluded that, as a possible scenario, $\EVAL(\AA)$ ($\AA$ is from~\cite{Cai-Chen-Lu-2013}) is $\#$P-hard,
%this is because a gadget, say, $\Gamma$ has been constructed
%such that $\gm_{\Gamma, \AA}$ ($\AA$ is from~\cite{Cai-Chen-Lu-2013}) is not \modblrkone/ (by Lemma 7.5 and then Corollary 2.6 from~\cite{Cai-Chen-Lu-2013}).
%This statement can be checked directly;
%
\item\label{item:meta1-application} Thus, in the notation of this paper, provided all the corresponding conditions for $\pur \AA$
(in place of $\AA$ from~\cite{Cai-Chen-Lu-2013})
are satisfied, we can apply the corresponding reasoning from~\cite{Cai-Chen-Lu-2013} for $\pur \AA$
and make the corresponding conclusions for $\pur \AA$.
%
%%%%%%%%%%%%%%%%%%%%%%%%%%%%%%%%%%%%%%%%%%%%%%%%%%%%%%%%%%%%%%%%%%%%
\begin{comment}
\item\label{item:meta1-constructivity} In fact, the assumption in~\ref{item:meta1-assumption},
can be trivially weakened to range over the edge gadgets $\Gamma$
constructed in the corresponding part from~\cite{Cai-Chen-Lu-2013}
(described in~\ref{item:meta1-assumption}).
Moreover, if this weakened assumption is violated,
then it can be checked such an edge gadget $\Gamma$
can always be constructed explicitly.
Thus, if in~\ref{item:meta1-application} the corresponding
conclusions are not true for $\pur \AA$,
then we can explicitly get an edge gadget $\Gamma$ such that
$\gm_{\Gamma, \pur \AA}$ is not \modblrkone/
and therefore by (the proof of) Theorem~\ref{thm:gadget-thickening-non-mult-bl-rk-1},
for some explicit (we can explicitly find) $p \ge 1$,
$\gm_{T_p(\Gamma), \AA}$ is not \multblrkone/,
and the edge gadget $T_p(\Gamma)$ is constructed explicitly.
\end{comment}
%%%%%%%%%%%%%%%%%%%%%%%%%%%%%%%%%%%%%%%%%%%%%%%%%%%%%%%%%%%%%%%%%%%%
\end{enumerate}
%In the above, $\AA$ from~\cite{Cai-Chen-Lu-2013} corresponds to our $\pur \AA$ in this paper.
\end{argument}
We have the following corollary, which is the same as Corollary 8.6 from~\cite{Cai-Chen-Lu-2013}. % which repeats Corollary 8.6 from~\cite{Cai-Chen-Lu-2013}.
%{\color{red} ,whose proof can be repeated verbatim (JYC: ours and CCL's have the same formulation and the proof but they are in different frameworks, are we allowed to say they are the same and cite it?)}
% corollary. %For all
  %$i\in [s]$ and $j\in [t]$, the $(i,j)${th} block
  %$\SS_{(i,*),(j,*)}$ of $\SS$ has exactly the same rank as $\SS$.
  
%(which is Corollary 8.6~from~\cite{Cai-Chen-Lu-2013})
\begin{corollary}\label{fullrank-ours}
%[cf. Corollary 8.6 from~\cite{Cai-Chen-Lu-2013}]\label{fullrank-ours}
For all $i\in [s]$ and $j\in [t]$,
% {\color{red} (the rank of) (remove this text, it's a typo in CCL)} 
the $(i,j)${th} 
  block matrix $\SS_{(i,*),(j,*)}$ of $\SS$ has the same rank as $\SS$.
\end{corollary}
\begin{proof}
The proof is the same as that of Corollary 8.6 from~\cite{Cai-Chen-Lu-2013}.
\end{proof}
%{\color{red} (the proof is exactly the same as in CCL, include it for convenience)}
%{\color{red} (the proof is the same as in CCL and therefore can omitted here)
%\begin{proof}
%Without loss of generality, we prove $\rank(\SS_{(1,*),(1,*)})=\rank(\SS)$.
%
%First, we use Lemma \ref{hahajaja-ours} to show that
%$$
%\text{rank}\left(\begin{matrix} \SS_{(1,*),(1,*)}
%%\\\SS_{(2,*),(1,*)}
%\\ \vdots
%  \\ \SS_{(s,*),(1,*)}\end{matrix}\right)=\text{rank}(\SS).
%$$
%To see this, we take any $h=\text{rank}(\SS)$ rows of $\SS$
%  which are linearly independent.
%Since any two of them $\SS_{\xx,(*,*)}$ and $\SS_{\yy,(*,*)}$
%  are linearly independent, by condition $(\calS_2)$, the two subvectors
%  $\SS_{\xx,(1,*)}$ and $\SS_{\yy,(1,*)}$ are orthogonal.
%Therefore, the corresponding $h$ rows of the matrix on the left-hand side
%  are pairwise orthogonal, and the left-hand side is at least $h$.
%Of course it cannot be larger than $h$, so it is equal to $h$.
%
%By using condition ($\calS_3$), we can similarly show that
%$$
%\text{rank}(\SS_{(1,*),(1,*)})=\text{rank}\left(\begin{matrix} \SS_{(1,*),(1,*)}
%  %\\\SS_{(2,*),(1,*)}
%  \\ \vdots \\ \SS_{(s,*),(1,*)}\end{matrix}\right).
%$$
%As a result, we have $\text{rank}(\SS_{(1,*),(1,*)})=\text{rank}(\SS)$.
%\end{proof}
%}
%
%{\color{red} (same as in CCL)}

Now suppose $h=\text{rank}(\SS)$. Then by Corollary \ref{fullrank-ours},
  there must exist indices $1\le i_1<\ldots<i_h\le m_1$
  and $1\le j_1<\ldots<j_h\le n_1$ such that the $\{(1,i_1),\ldots,(1,i_h)\}\times
  \{(1,j_1),\ldots,(1,j_h)\}$ submatrix of $\SS$ has full rank $h$.
Without loss of generality
we assume $i_k=k$ and $j_k=k$
  for all $k\in [h]$ (if this is not true, we can apply an
  appropriate permutation $\Pi$ to the rows and columns of $\AA$
  so that the new $\SS$ has this property; this permutation is within
the first block and so it does not affect
the monotonicity of 
$\pur{\boldsymbol{\mu}}$ and $\pur{\boldsymbol{\nu}}$).
%%% because this further permutation happens withn each block (\mu_i I)
We use $\HH$ to denote this $h\times h$ matrix: $H_{i,j}=S_{(1,i),(1,j)}$.\vspace{0.007cm}

By Corollary \ref{fullrank-ours} and Lemma \ref{hahajaja-ours},
  for every index $\xx\in I$, there exists
  a unique pair of integers $j\in [h]$ and $k\in [0:N-1]$ \vspace{-0.1cm}such that
\begin{equation}\label{case1-ours}
\SS_{\xx,*} = \oo_N^k \cdot \SS_{(1,j),*}.
\end{equation}
This gives us a partition of the index set $\{0\}\times I$:
\[
{\mathscr R}_0 =\big\{R_{(0,i,j),k}:
  i\in [s],j\in [h],k\in [0:N-1]\hspace{0.03cm}\big\}.
\]
For every $\xx\in I$, $(0,\xx)\in R_{(0,i,j),k}$ if
  $i=x_1$ and $\xx,j,k$ satisfy (\ref{case1-ours}).
By Corollary \ref{fullrank-ours}, 
  \[\bigcup_{k\in [0:N-1]} R_{(0,i,j),k}\ne \emptyset\ \ \
  \text{for all $i\in [s]$ and $j\in [h]$.}\]

Similarly, for every index $\yy\in J$ there exists a unique pair of integers $j\in [h]$
  and $k\in [0:N-1]$ such that
\begin{equation}\label{case2-ours}
\SS_{*,\yy}= \oo_N^k \cdot \SS_{*,(1,j)},
\end{equation}
and we partition $\{1\}\times J$ into
\[
{\mathscr R}_1 =\big\{R_{(1,i,j),k}:i\in [t],j\in [h],k\in [0:N-1]
  \hspace{0.03cm}\big\}.
\]
For every $\yy\in J$, $(1,\yy)\in R_{(1,i,j),k}$ if
  $i=y_1$ and $\yy,j,k$ satisfy (\ref{case2-ours}).
By Corollary \ref{fullrank-ours},
\[ \bigcup_{k\in [0:N-1]} R_{(1,i,j),k}\ne \emptyset\ \ \
  \text{for all $ i \in [t]$ and $j\in [h]$.}\]

Now we define $(\CC, \fD)$ and $(\pur \CC, \fD)$, and use the 
cyclotomic reduction lemma (Lemma \ref{twinreduction-ours})
to show that $\EVAL(\AA) \equiv \EVAL(\CC, \fD)$ (and $\EVAL(\pur \AA) \equiv \EVAL(\pur \CC, \fD)$),
and use the cyclotomic transfer lemma for edge gadgets 
(Lemma~\ref{lem:cycl-gadget-trans})
to show that for every edge gadget $\Gamma$, $\gm_{\Gamma, \AA}$ (resp., $\gm_{\Gamma, \pur \AA}$)
is \multblrkone/
% (\modblrkone/)
iff $\gm_{\Gamma, \CC, \fD}$ (resp., $\gm_{\Gamma, \pur \CC, \fD}$)
is \multblrkone/.
% (\modblrkone/).
The same is true replacing \multblrkone/ by \modblrkone/.
This will allow us to move between the frameworks $\eval(\AA)$ (resp., $\eval(\pur \AA)$)
 and $\eval(\CC, \fD)$ (resp., $\eval(\pur \CC, \fD)$).

First, $\CC$ is an $(s+t)h\times (s+t)h$ matrix which is the bipartization
  of an $sh\times th$ matrix $\FF$.
We use the set $I'\equiv [s]\times [h]$ to index the rows of $\FF$ and $J'\equiv[t]\times [h]$
  to index the columns of $\FF$.
We have
\[
F_{\xx,\yy}=\mu_{x_1}\nu_{y_1}H_{x_2,y_2}=\mu_{x_1}\nu_{y_1}S_{(1,x_2),(1,y_2)}
\ \ \ \text{for all $\xx\in I'$, $\yy\in J'$},\]
or equivalently, 
\[
\FF = \left(\begin{matrix}
\mu_1\II \\
& \mu_2\II \\
& & \ddots\\
& & & \mu_s\II
\end{matrix}\right)
\left( \begin{matrix}
\HH & \HH & \ldots & \HH \\
\HH & \HH & \ldots & \HH \\
\vdots & \vdots & \ddots & \vdots \\
\HH & \HH & \ldots & \HH
\end{matrix}\right)
\left(\begin{matrix}
\nu_1\II \\
& \nu_2\II \\
& & \ddots\\
& & & \nu_t\II
\end{matrix}\right),
\]
where $\II$ is the $h\times h$ identity matrix.
We use $(\{0\}\times I')\cup (\{1\}\times J')$ to
  index the rows and columns of $\CC$. %a $2\times 2$ block matrix,

Second, $\fD=(\DD^{[0]},\ldots,\DD^{[N-1]})$ is a sequence of $N$ diagonal
  matrices of the same size as $\CC$.
We use $\{0\}\times I'$ to index the first $sh$ entries and
  $\{1\}\times J'$ to index the last $th$ entries. The
  $(0,\xx)${th} entries of $\fD$ are generated by % ${\mathscr R}_0$,
  $(|R_{(0,x_1,x_2),0}|, \ldots,
   |R_{(0,x_1,x_2),N-1}|)$,
  and the $(1,\yy)${th} entries of $\fD$ are
  generated by  $(|R_{(1,y_1,y_2),0}|,\ldots, |R_{(1,y_1,y_2),N-1}|)$:
\begin{eqnarray*}
D^{[r]}_{(0,\xx)}=\sum_{k=0}^{N-1} \big|R_{(0,x_1,x_2),k}\big|\cdot \oo_N^{kr}\ \  \ \text{
  and}\ \ \  
D^{[r]}_{(1,\yy)}=\sum_{k=0}^{N-1} \big|R_{(1,y_1,y_2),k}\big|\cdot \oo_N^{kr},
\end{eqnarray*}
for all $r\in [0:N-1],\xx=(x_1,x_2)\in I'$, and $\yy=(y_1,y_2)\in J'$.

The same purification producing the
 substitutions
$\mu_i \to \pur \mu_i$ for $i \in [s]$ and 
$\nu_i \to \pur \nu_i$ for $i \in [t]$
goes from $\FF$ to $\pur \FF$,
and from $\CC$ to $\pur \CC$, respectively
the bipartizations of $\FF$ and $\pur \FF$.

We have
\[
\pur F_{\xx,\yy}=\pur \mu_{x_1}\pur \nu_{y_1}H_{x_2,y_2}=\pur \mu_{x_1}\pur \nu_{y_1}S_{(1,x_2),(1,y_2)}
\ \ \ \text{for all $\xx\in I'$, $\yy\in J'$},\]
or equivalently, 
\[
\pur \FF = \left(\begin{matrix}
\pur \mu_1\II \\
& \pur \mu_2\II \\
& & \ddots\\
& & & \pur \mu_s\II
\end{matrix}\right)
\left( \begin{matrix}
\HH & \HH & \ldots & \HH \\
\HH & \HH & \ldots & \HH \\
\vdots & \vdots & \ddots & \vdots \\
\HH & \HH & \ldots & \HH
\end{matrix}\right)
\left(\begin{matrix}
\pur \nu_1\II \\
& \pur \nu_2\II \\
& & \ddots\\
& & & \pur \nu_t\II
\end{matrix}\right).
\]
%where $\II$ is the $h\times h$ identity matrix.
%We also use $(\{0\}\times I')\cup (\{1\}\times J')$ to
%  index the rows and columns of $\pur \CC$. %a $2\times 2$ block matrix,
%  
%It is easy to see that $\pur \CC$ is the purification of $\CC$
%obtained by going
%from $(g_1, \ldots, g_d)$ to $(p_1, \ldots, p_d)$
%(in the same way as $\pur \AA$ is the purification of $\AA$
%obtained by going from $(g_1, \ldots, g_d)$ to $(p_1, \ldots, p_d)$),
%although we will not use this fact (explicitly).
%
%{\color{red} (added, can be omitted, up to you)
%
%It is also important to note that $\pur \CC$ can be obtained from $\CC$
%in the same way as $\pur \AA$ is obtained from $\AA$
%which is by the following symbolic substitution:
%$\mu_i \to \pur \mu_i$ for $i \in [s]$ and $\nu_i \to \pur \nu_i$ for $i \in [t]$.
%}
%
%
%{\color{red} (new text, not sure about the statement of the lemma yet)}

This finishes the construction of $(\CC, \fD)$ and $(\pur \CC, \fD)$.
We prove the following lemma.
\begin{lemma}[modification of Lemma 8.7 from~\cite{Cai-Chen-Lu-2013}]\label{lem:A-CD-cycl-gen}
%{\color{red} (my preferred option)}
%
The matrix $\AA$ is generated from $\CC$ using $\mathscr R_0 \cup \mathscr R_1$.
As a consequence, $\EVAL(\AA) \equiv \EVAL(\CC, \fD)$
and for any edge gadget $\Gamma$, $\gm_{\Gamma, \AA}$
is \multblrkone/ (\modblrkone/)
iff $\gm_{\Gamma, \CC, \fD}$
is \multblrkone/ (\modblrkone/).
The same statements hold with $\AA$ and $\CC$ replaced by $\pur \AA$ and $\pur \CC$, respectively.
\end{lemma}
\begin{proof}
We only show that $\AA$ can be generated from $\CC$ using ${\mathscr R}_0\cup {\mathscr R}_1$;
that $\pur \AA$ can be generated from $\pur \CC$ using ${\mathscr R}_0\cup {\mathscr R}_1$
can be shown similarly.
%To get the first statement, we can then apply
%the cyclotomic edge gadget reduction lemma.
%
%{\color{red} 
%First we show that $\AA$ can be generated from $\CC$ using
%  ${\mathscr R}_0\cup {\mathscr R}_1$. (old text)
%}

Let $\xx,\xx'\in I$, $(0,\xx)\in R_{(0,x_1,j),k}$, and $(0,\xx')\in$
  $R_{(0,x_1',j'),k'}$. Then
we have \[A_{(0,\xx),(0,\xx')}= C_{(0,x_1,j),(0,x_1',j')}=0,\] since
  $\AA$ and $\CC$ are the bipartizations of $\BB$ and $\FF$, respectively,
and the upper-left block is $0$ in a bipartization matrix.
%%%
%\sout{and the upper-left $n \times n$ block is $0$
%in a $2n \times 2n$ bipartization matrix}. %{\color{red} (JYC: not true, it's $m \times m$ and $(m + n) \times (m + n)$ respectively, but only for $\AA$, remove or suggest replacement}
%%%
Therefore, it holds trivially that
\[
A_{(0,\xx),(0,\xx')}=C_{(0,x_1,j),(0,x_1',j')}\cdot \oo_N^{k+k'}.
\]
%holds trivially. 
Clearly, this 
  also holds for the lower-right $n\times n$ block of $\AA$.

Let $\xx\in I$, $(0,\xx)\in R_{(0,x_1,j),k}$, $\yy\in J$, and
  $(1,\yy)\in R_{(1,y_1,j'),k'}$ for some $j,k,j',k'$. By (\ref{case1-ours}) and (\ref{case2-ours}), we have
\begin{align*}
A_{(0,\xx),(1,\yy)}&=\mu_{x_1}\nu_{y_1} S_{\xx,\yy }
 =\mu_{x_1}\nu_{y_1} S_{(1,j), \yy }\cdot \oo_N^{k}\\
  &=\mu_{x_1}\nu_{y_1} S_{( 1,j),( 1,j')}\cdot \oo_N^{k+k'}
  =C_{(0,x_1,j),(1,y_1,j')}\cdot \oo_N^{k+k'}.
\end{align*}
A similar equation also holds for the lower-left block.
Thus,
  $\AA$ can be generated from $\CC$ using ${\mathscr R}_0\cup {\mathscr R}_1$.
Moreover, the construction of $\fD$ implies that
  $\fD$ can be generated from
  the partition ${\mathscr R}_0\cup {\mathscr R}_1$.
%{\color{red} The lemma then follows directly from the Cyclotomic reduction lemma. (old text)}

Now to obtain the remaining statements,
we apply Lemma~\ref{twinreduction-ours} 
%and~\ref{lem:cycl-gadget-trans}.
and Corollary~\ref{cor:cycl-blr1-equiv} to Lemma~\ref{lem:cycl-gadget-trans}. %{\color{red} (JYC: don't understand "to Lemma~\ref{lem:cycl-gadget-trans}", just cross it out)}.
\end{proof}

Before moving forward to the next step, we 
  summarize our progress so far.
%{\color{red} We showed that $\eval(\AA)$ is either \#P-hard or equivalent
%  to $\eval(\CC,\fD)$, where the pair
%  $(\CC,\fD)$ satisfies the following conditions
%  ({\sl Shape}$_1$)--({\sl Shape}$_3$): (old text)%\label{SHAPECONDITION}
%}
We showed that either there is an edge gadget $\Gamma$
such that $\gm_{\Gamma, \AA}$ is not \multblrkone/ %(a fortiori, $\EVAL^{(\Delta)}(A)$ is \#P-hard for some $\Delta > 0$)
or we can construct a pair $(\CC, \fD)$ along with a pair $(\pur \CC, \fD)$
such that $\EVAL(\AA) \equiv \EVAL(\CC, \fD)$ and $\EVAL(\pur \AA) \equiv \EVAL(\pur \CC, \fD)$.
Furthermore, 
for any edge gadget $\Gamma$, $\gm_{\Gamma, \AA}$ is \multblrkone/ (\modblrkone/)
iff $\gm_{\Gamma, \CC, \fD}$ is \multblrkone/ (\modblrkone/),
and the same
is true for $\pur \AA$ and $\pur \CC$ in place of $\AA$ and $\CC$, respectively,
and $(\CC, \fD)$, $(\pur \CC, \fD)$ also satisfy  the following conditions
  ({\sl Shape}$_1$)--({\sl Shape}$_3$): \label{SHAPECONDITION}
\begin{enumerate}[wide]
%\item[]\hspace{-0.25cm}({\sl Shape}$_1$)\ \ $\CC\in \mathbb{C}^{m\times m}$ 
\item[({\sl Shape}$_1$)] $\CC\in \mathbb{C}^{m\times m}$ 
%(note that the $m$ here
%  is different from the $m$ used at the beginning of Step 2.1) is
%  the bipartization of an $sh\times th$ matrix $\FF$ (so $m=(s+t)h$).
is the bipartization of an $sh\times th$ matrix $\FF$ (so $m=(s+t)h$;
this $m$ is different from the $m$ used at the beginning of Step 2.1).  
%$\FF$ is an $s\times t$ block matrix.
$\FF$  has  $s\times t$ blocks of $h \times h$ each, 
and we use $I=[s]\times [h]$
  and $J=[t]\times [h]$
  to index the rows and columns of $\FF$, respectively.
$\pur \FF$  and $\pur \CC$ are purifications of $\FF$ and $\CC$,
respectively.
  
%%$\pur \CC\in \mathbb{C}^{m\times m}$ 
%(again, note that the $m$ here
%    is different from the $m$ used at the beginning of Step 2.1) is also
%    the bipartization of an $sh\times th$ matrix $\pur \FF$ (so $m=(s+t)h$).
%  $\pur \FF$ is an $s\times t$ block matrix. We also use $I=[s]\times [h]$
%    and $J=[t]\times [h]$
%    to index the rows and columns of $\pur \FF$, respectively.
  
%\item[]\hspace{-0.25cm}({\sl Shape}$_2$)\ \ There are two 
\item[({\sl Shape}$_2$)] There are two 
  sequences $\boldsymbol{\mu}
  =(\mu_1,\ldots,\mu_s)$ and
  $\boldsymbol{\nu}$ $=(\nu_1,\ldots,\nu_t)$,
each consisting of pairwise distinct nonzero complex numbers. %,
%Next, $\mu_1, \ldots, \mu_s, \nu_1, \ldots, \nu_t$ are in the multiplicative subgroup
%generated by $\{g_1, \ldots, g_d\}$.
%In particular, 
$\mu_1, \ldots, \mu_s, \nu_1, \ldots, \nu_t$ generate
a torsion-free multiplicative group.
There is also 
  an $h\times h$ full-rank matrix $\HH$ whose entries are all powers of $\oo_N$ for
  some positive integer $N$.
The entries of $\FF$ can be expressed using $\boldsymbol{\mu},\boldsymbol{\nu}$,
  and $\HH$ explicitly as follows:  
  \[F_{\xx,\yy}=\mu_{x_1}\nu_{y_1}H_{x_2,y_2}\ \ \ \text{ 
  for all $\xx\in I$ and $\yy\in J$.}\]
Moreover, the purification
substitutes $\mu_i \to \pur \mu_i$ for $i \in [s]$ and
$\nu_i \to \pur \nu_i$ for $i \in [t]$, and 
goes from $\FF$ to $\pur \FF$,
and from $\CC$ to $\pur \CC$, respectively.
$\pur{\boldsymbol{\mu}}
  =(\pur \mu_1,\ldots,\pur \mu_s)$ and
  $\pur{\boldsymbol{\nu}}$ $=(\pur \nu_1,\ldots,\pur \nu_t)$
are strictly decreasing sequences of positive rational numbers.
The entries of $\pur \FF$ can be expressed using $\pur{\boldsymbol{\mu}},\pur{\boldsymbol{\nu}}$,
  and $\HH$ explicitly as follows:  
  \[\pur F_{\xx,\yy}=\pur \mu_{x_1}\pur \nu_{y_1}H_{x_2,y_2}\ \ \ \text{ 
  for all $\xx\in I$ and $\yy\in J$.}\]

%\item[]\hspace{-0.25cm}({\sl Shape}$_3$)\ \ $\fD=(\DD^{[0]},\ldots,\DD^{[N-1]})$ is a sequence of
\item[({\sl Shape}$_3$)] $\fD=(\DD^{[0]},\ldots,\DD^{[N-1]})$ is a sequence of
  $m\times m$ diagonal matrices.
We use $(\{0\}\times I)\cup(\{1\}\times J)$ to index the rows and columns
  of the matrices $\CC$ (and $\pur \CC$) and $\DD^{[r]}$.
  $\fD$ satisfies $(\calT_3)$, so for all $r\in [N-1]$,
    $\xx\in [s]\times [h]$, and $\yy\in [t]\times [h]$, 
  \[
  D^{[r]}_{(0,\xx)}=\overline{D^{[N-r]}_{(0,\xx)}}\ \ \text{and}\ \
  D^{[r]}_{(1,\yy)}=\overline{D^{[N-r]}_{(1,\yy)}}. 
  \]
\end{enumerate}
%{\color{red} (added, can be removed)
%
%The conditions ({\sl Shape}$_1$)--({\sl Shape}$_3$) correspond to (are similar) to
%the conditions ({\sl Shape}$_1$)--({\sl Shape}$_3$) from~\cite{Cai-Chen-Lu-2013}, p. 959.
%}

%%%%%%%%%%%%%%%%%%%%%%%%%%%%%%%%%%%%%%%%%%%%%%

%%%%%%%%%%%%%%%%%%%%%%%%%%%%%%%%%%%%%%%%%%%%%%

%{\color{red} (meta-remark)}

%%% go back remark
%Clearly, if $(\CC, \fD)$, $(\CC, \fD)$ are pairs that satisfy ({\sl Shape}$_1$)--({\sl Shape}$_3$),
%we can restore the corresponding matrices $\AA$, $\pur \AA$ which satisfy ($\calS_1$)--($\calS_3$).

\subsection{Step 2.2}

We introduce the following second meta-argument.
%crucial meta-argument.
\begin{argument}[{\sl Meta}$_2$]\label{meta2} %\label{meta:sec-8-CD-ccl}
We have the following: 
\begin{enumerate}
\item\label{item:meta2-thickening} Either $\gm_{\Gamma, \pur \CC, \fD}$ (equiv., $\gm_{\Gamma, \pur \AA}$) is \modblrkone/
for every edge gadget $\Gamma$,
or there is an edge gadget $\Gamma$
such that $\gm_{\Gamma, \pur \CC, \fD}$ (equiv., $\gm_{\Gamma, \pur \AA}$ by Lemma~\ref{lem:A-CD-cycl-gen}) is not \modblrkone/,
%{\color{red}((or say) and then by Lemma~\ref{lem:A-CD-cycl-gen}, $\gm_{\Gamma, \pur \AA}$ is not \modblrkone/)},
so by Theorem~\ref{thm:gadget-thickening-non-mult-bl-rk-1},
for some $p \ge 1$,
$\gm_{T_p(\Gamma), \AA}$ is not \multblrkone/;

\item\label{item:meta2-assumption} 
Referring to the pair $(\CC, \fD)$
from Section 8.3 in~\cite{Cai-Chen-Lu-2013},
all statements (including theorems, lemmas, corollaries, properties, etc.
 numbered from 8.8 to 8.23) from~\cite{Cai-Chen-Lu-2013}
can be proved under the assumption that every signature $\gm_{\Gamma, \CC, \fD}$  is \modblrkone/, %edge
and whenever it is concluded that, as a possible scenario, $\EVAL(\CC, \fD)$
is $\#$P-hard,
this is because an edge gadget, say, $\Gamma$ has been constructed
such that $M_{\Gamma, \CC, \fD}$  is not \modblrkone/.
This statement can be checked directly;
\item\label{item:meta2-application} Thus, in the notation of this paper, provided all the corresponding conditions for $(\pur \CC, \fD)$
(in place of $(\CC, \fD)$ from~\cite{Cai-Chen-Lu-2013})
are satisfied, we can apply the corresponding reasoning from~\cite{Cai-Chen-Lu-2013} for $(\pur \CC, \fD)$
and make the corresponding conclusions for $(\pur \CC, \fD)$.
%
%%%%%%%%%%%%%%%%%%%%%%%%%%%%%%%%%%%%%%%%%%%%%%%%%%%%%%%%%%%%%%%%%%%%
\begin{comment}

\item\label{item:meta2-constructivity} In fact, the assumption in~\ref{item:meta2-assumption},
can be trivially weakened to range over the edge gadgets $\Gamma$
constructed in the corresponding part from~\cite{Cai-Chen-Lu-2013}
(described in~\ref{item:meta2-assumption}).
Moreover, if this weakened assumption is violated,
then it can be checked such an edge gadget $\Gamma$
can always be constructed explicitly.
Thus, if in~\ref{item:meta2-application} the corresponding
conclusions are not true for $(\pur \CC, \fD)$,
then we can explicitly get an edge gadget $\Gamma$ such that
$\gm_{\Gamma, \pur \CC, \fD}$ (equiv. $\gm_{\Gamma, \pur \AA}$) is not \modblrkone/
%{\color{red}((or say) and then by Lemma~\ref{lem:A-CD-cycl-gen}, $\gm_{\Gamma, \pur \AA}$ is not \modblrkone/)}
and therefore by (the proof of) Theorem~\ref{thm:gadget-thickening-non-mult-bl-rk-1},
for some explicit (we can explicitly find) $p \ge 1$,
$\gm_{T_p(\Gamma), \AA}$ is not \multblrkone/,
and the edge gadget $T_p(\Gamma)$ is constructed explicitly.
\end{comment}
%%%%%%%%%%%%%%%%%%%%%%%%%%%%%%%%%%%%%%%%%%%%%%%%%%%%%%%%%%%%%%%%%%%%
\end{enumerate}
%In the above, $\AA$, $(\CC, \fD)$ from~\cite{Cai-Chen-Lu-2013} respectively correspond to our $\pur \AA$, $(\pur \CC, \fD)$ in this paper.
\end{argument}

In Step 2.2, we prove % since $(\CC,\fD)$ is constructed from $\AA$
  the following lemma.
%%(a modification of Lemma 8.8 from~\cite{Cai-Chen-Lu-2013}):%  using the cyclotomic reduction lemma,

\begin{lemma}[modification of Lemma 8.8 from~\cite{Cai-Chen-Lu-2013}]\label{shapecondition-ours}
%Either $\eval(\CC,\fD)$ is \#P-hard; or
%  $\HH$ and $\DD^{[0]}$ satisfy the following two conditions: 
Either there is an edge gadget $\Gamma$
such that $\gm_{\Gamma, \AA}$ is not \multblrkone/ or
 $\HH$ and $\DD^{[0]}$ satisfy the following two conditions:
\begin{enumerate}[wide]
%\item[]\hspace{-0.25cm}\emph{({\sl Shape}$_4$)}\ \ 
\item[\emph{({\sl Shape}$_4$)}]
  $({1}/{\sqrt{h}})\cdot \HH$ is a
%%% JYC corrected a typo above. n--> h, the order of H.
  unitary matrix, i.e., 
\[\langle \HH_{i,*},\HH_{j,*}\rangle
  =\langle \HH_{*,i},\HH_{*,j}\rangle =0,\ \ \ \text{for all $i\ne j\in [h]$.}\]

%\item[]\hspace{-0.25cm}\emph{({\sl Shape}$_5$)}\ \ $\DD^{[0]}$ satisfies,
\item[\emph{({\sl Shape}$_5$)}] $\DD^{[0]}$ satisfies,
for all $\xx\in I$ and for all $\yy\in J$,
\[D^{[0]}_{(0,\xx)}=
  D^{[0]}_{(0,(x_1,1))}\ \ \ \text{and}\ \ \ 
  D^{[0]}_{(1,\yy)}=D^{[0]}_{(1,(y_1,1))}.\] 
\end{enumerate}
\end{lemma}
%{\color{red} proof taken from~\cite{Cai-Chen-Lu-2013} as it introduces four equations
%for the next sublemma}
\begin{proof}
We rearrange the entries of $\DD^{[0]}$ indexed by $\{1\}\times J$
  into a $t\times h$ matrix 
\begin{equation}\label{eqn:def-X}
X_{i,j}=D^{[0]}_{(1,(i,j))}\ \ \ \text{for all $i\in [t]$ and $j\in [h]$}
\end{equation}
and rearrange its entries indexed by $\{0\}\times I$ into an $s\times h$ matrix 
\begin{equation}\label{eqn:def-Y}
Y_{i,j}=D^{[0]}_{(0,(i,j))}\ \ \ \text{for all $i\in [s]$ and $j\in [h]$.}
\end{equation}
Note that by condition $(\calT_3)$, all entries of $\XX$ and $\YY$
  are positive integers.

The proof has two stages. First, we show in Lemma \ref{stage1-ours}
  that either we can construct an edge gadget $\Gamma$ such that $\gm_{\Gamma, \AA}$ is not \multblrkone/ or $\HH,\XX$ and $\YY$ must satisfy
\begin{align}\label{assumption11-ours}
&\langle \HH_{i,*}\circ\overline{\HH_{j,*}},\XX_{k,*}\rangle=0\ \ \ \text{for all
  $k\in [t]$ and $i\ne j\in [h]$}\ \ \ \ \text{and}\\[0.6ex]
\label{assumption21-ours}
&\langle \HH_{*,i}\circ\overline{\HH_{*,j}},\YY_{k,*}\rangle=0\ \ \
\text{for all $k\in [s]$ and $i\ne j\in [h]$.} 
\end{align}
We use $U$ to denote the set of $h$-dimensional
  vectors that are orthogonal to
\[\HH_{1,*}\circ \overline{\HH_{2,*}},\ \HH_{1,*}\circ
  \overline{\HH_{3,*}},\ \ldots,\ \HH_{1,*}\circ \overline{\HH_{h,*}}.\]
The above set of $h-1$ vectors is linearly independent.
This is because
\[\sum_{i=2}^{h} a_i\big(\HH_{1,*}\circ \overline{\HH_{i,*}}\big)=\HH_{1,*}\circ
  \left(\sum_{i=2}^h a_i\overline{\HH_{i,*}}\right),\]
and\vspace{0.003cm} if $\sum_{i=2}^{h} a_i (\HH_{1,*}\circ \overline{\HH_{i,*}})=\00$,
  then $\sum_{i=2}^h a_i\overline{\HH_{i,*}}=\00$
  since all entries of $\HH_{1,*}$ are nonzero.
Because $\HH$ has full rank, we have $a_i=0$, $i=2,\ldots,h$.
As a result, $U$ is a linear space of dimension $1$ over $\mathbb{C}$.

Second, we show in Lemma \ref{stage2-ours} that, assuming (\ref{assumption11-ours}) and
  (\ref{assumption21-ours}), either
\begin{align}\label{assumption12-ours}
&\langle \HH_{i,*}\circ\overline{\HH_{j,*}},(\XX_{k,*})^2\rangle =0\ \ \ \text{for
  all $k\in [t]$ and $i\ne j\in [h]$}\ \ \ \ \text{and}\\[0.6ex]
\label{assumption22-ours}
&\langle \HH_{*,i}\circ\overline{\HH_{*,j}},(\YY_{k,*})^2\rangle =0\ \ \ \text{for all
  $k\in [s]$ and $i\ne j\in [h]$,} 
\end{align}
or we can construct an edge gadget $\Gamma$ such that $\gm_{\Gamma, \AA}$ is not \multblrkone/. Here we use $(\XX_{k,*})^2$ 
  to denote $\XX_{k,*}\circ\XX_{k,*}$.%\vspace{0.005cm} %{\color{red} (JYC: like in CCL, not $\odot 2$ hehehe, of course I prefer $2$ here, ignore this remark)}

Equations (\ref{assumption11-ours}) and (\ref{assumption12-ours}) then
  imply that both $\XX_{k,*}$ and $(\XX_{k,*})^2$ are in $U$
  and thus they are linearly dependent (since the dimension of $U$ is $1$).
On the other hand, by $(\calT_3)$, every entry in $\XX_{k,*}$ is a positive
  integer. Therefore, $\XX_{k,*}$ must have the form $u\cdot \11$,
  for some positive integer $u$.
The same argument works for $\YY_{k,*}$ and the latter must also
  have the form $u'\cdot \11$.
By (\ref{assumption11-ours}) and (\ref{assumption21-ours}), this further implies that
\[
\langle \HH_{i,*},\HH_{j,*}\rangle =0\ \ \text{and}\ \
\langle \HH_{*,i},\HH_{*,j}\rangle =0\ \ \ \text{for all $i\ne j\in [h]$.}
\]
This finishes the proof of Lemma \ref{shapecondition-ours}.\end{proof}

%{\color{red} (new text)
%
%Notice that equations~(\ref{assumption11-ours}), (\ref{assumption12-ours}),
%(\ref{assumption21-ours}, (\ref{assumption22-ours})
%are precisely equations (8.8), (8.9), (8.10) and (8.11) on p. 960 from~\cite{Cai-Chen-Lu-2013}, correspondingly.
%}

Now we proceed to the two stages of the proof.
In the first stage, we prove the following lemma.
Given $(\CC,\fD)$ and $({\pur \CC}, \fD)$ that satisfy 
conditions ({\sl Shape}$_1$)--({\sl Shape}$_3$), let 
$\XX$ and $\YY$ be defined in (\ref{eqn:def-X}) and (\ref{eqn:def-Y}).
%modification of Lemma 8.9 from~\cite{Cai-Chen-Lu-2013}:
\begin{lemma}[modification of Lemma 8.9 from~\cite{Cai-Chen-Lu-2013}]\label{stage1-ours}
%Either $\HH,\XX,\YY$ satisfy \emph{(\ref{assumption11-ours})} and
%  \emph{(\ref{assumption21-ours})}, or $\eval(\CC,\fD)$ is \#P-hard.
Either $\HH,\XX,\YY$ satisfy \emph{(\ref{assumption11-ours})} and
  \emph{(\ref{assumption21-ours})}, or there exists
  an edge gadget $\Gamma$ such that $\gm_{\Gamma, \AA}$ is
  not \multblrkone/.
\end{lemma}
\begin{proof}
We adapt  the proof of Lemma 8.9 from~\cite{Cai-Chen-Lu-2013}
but with  $(\pur \CC, \fD)$
in place of $(\CC, \fD)$ in~\cite{Cai-Chen-Lu-2013}.
It is easy to see that $(\pur \CC, \fD)$
satisfies the conditions needed for the proof of 
Lemma 8.9 from~\cite{Cai-Chen-Lu-2013},
and so do our $\HH, \XX, \YY$ derived from $(\pur \CC, \fD)$.
We use the same edge gadget $\Gamma = \Gamma^{[p]}$, for some $p \ge 1$,
 in Figure~\ref{figure_2-ours}, and get a signature  matrix
$\gm_{\Gamma, {\pur \CC}, \fD}$.
%%% old
By Corollary~\ref{inversetwin-ours},
either we get an edge gadget $\Gamma$ such that $\gm_{\Gamma, {\pur \CC}, \fD}$
is not \modblrkone/, or the same equation (8.13) in~\cite{Cai-Chen-Lu-2013}
holds for our matrix $\gm_{\Gamma, {\pur \CC}, \fD}$.
%%% new %%% correction
%Then
%either we get an edge gadget $\Gamma$ such that $\gm_{\Gamma, {\pur \CC}, \fD}$
%is not \modblrkone/, or the same equation (8.13) in~\cite{Cai-Chen-Lu-2013}
%holds for our matrix $\gm_{\Gamma, {\pur \CC}, \fD}$.
%More precisely, one can check that the conditions needed to apply Corollary~\ref{inversetwin-ours}
%are satisfied each time, which are slightly stronger than the property of non-\modblrkone/,
%although in our reasoning we only need to have the latter.
%%%
Thus (and this is essentially our Argument~\ref{meta2} ({\sl Meta}$_2$) in action),
we get that 
\begin{enumerate}
\item either our $\HH, \XX, \YY$ 
satisfy equations (\ref{assumption11-ours}) and (\ref{assumption21-ours})
(which are  (8.8) and (8.9) from~\cite{Cai-Chen-Lu-2013}),
\item or there exists an edge gadget $\Gamma' = T_\ell(\Gamma)$, for some $\ell \ge 1$,
such that $\gm_{\Gamma', \AA}$ is not  \multblrkone/.
\end{enumerate}
\end{proof}

All subsequent arguments invoking Argument~\ref{meta2} ({\sl Meta}$_2$) follow
a similar vein.
% remarks can be formulated in all subsequent arguments involving Argument~\ref{meta2} ({\sl Meta}$_2$).

In the second stage, we prove the following lemma.
%which is a modification of Lemma 8.10 from~\cite{Cai-Chen-Lu-2013}:
\begin{lemma}[modification of Lemma 8.10 from~\cite{Cai-Chen-Lu-2013}]\label{stage2-ours}
Suppose matrices $\HH$, $\XX$, and $\YY$ satisfy both
  \emph{(\ref{assumption11-ours})} and \emph{(\ref{assumption21-ours})}. Then
  either they also satisfy \emph{(\ref{assumption12-ours})}
  and \emph{(\ref{assumption22-ours})} 
or there exists an edge gadget $\Gamma$
such that $\gm_{\Gamma, \AA}$ is not \multblrkone/.
\end{lemma}
%{\color{red} (meta-proved, new)}
\begin{proof}
We adapt the proof of Lemma 8.10 from~\cite{Cai-Chen-Lu-2013}
and apply it to 
 $(\pur \CC, \fD)$
in place of $(\CC, \fD)$ in~\cite{Cai-Chen-Lu-2013}.
% (in the notation of~\cite{Cai-Chen-Lu-2013}).
It is easy to see that $(\pur \CC, \fD)$
satisfies the conditions needed for the proof of Lemma 8.10 
from~\cite{Cai-Chen-Lu-2013},
and so do our $\HH, \XX, \YY$ derived from $(\pur \CC, \fD)$.
We use the same edge gadget $\Gamma = \Gamma^{[p]}$, for some $p \ge 1$,
from~\cite{Cai-Chen-Lu-2013} as depicted in Figure~8.2 (p.~962)
with distinguished vertices $u$ and $v$ (in this order), and
get a signature matrix
$\RR^{(p)} = \gm_{\Gamma, {\pur \CC}, \fD}$.
%%% old
By Corollary~\ref{inversetwin-ours},
either we get an edge gadget $\Gamma$ such that $\RR^{(p)}$
is not \modblrkone/, or the same equations concerning $\RR^{(p)}$
in~\cite{Cai-Chen-Lu-2013} (p.~963)
hold for our matrix as well.
%%% new %%% correction
%Then
%either we get an edge gadget $\Gamma$ such that $\RR^{(p)}$
%is not \modblrkone/, or the same equations concerning $\RR^{(p)}$
%in~\cite{Cai-Chen-Lu-2013} (p.~963)
%hold for our matrix as well.
%More precisely, one can check that the conditions needed to apply Corollary~\ref{inversetwin-ours}
%are satisfied each time, which are slightly stronger than the property of non-\modblrkone/,
%although in our reasoning we only need to have the latter.
%%%
Thus by Argument~\ref{meta2} ({\sl Meta}$_2$),
we get that
\begin{enumerate}
\item either our $\HH, \XX, \YY$
satisfy equations  (\ref{assumption12-ours}) and (\ref{assumption22-ours})
(which are   (8.10) and (8.11) from~\cite{Cai-Chen-Lu-2013}),
\item or there exists an edge gadget $\Gamma' = T_\ell(\Gamma)$, for some $\ell \ge 1$,
such that $\gm_{\Gamma', \AA}$ is not  \multblrkone/.
\end{enumerate}
\end{proof}

\subsection{Step 2.3}\label{D-is-rank-one-ours}

%{\color{red} (old text)
%
%Now we get a pair $(\CC,\fD)$ that satisfies ({\sl Shape}$_1
  %$)--({\sl Shape}$_5$) since otherwise, by Lemma \ref{shapecondition},
  %$\eval(\CC,\fD)$ is \#P-hard and we are done.
%}

Now we get pairs $(\CC,\fD)$, $(\pur \CC, \fD)$ that satisfy ({\sl Shape}$_1
  $)--({\sl Shape}$_5$).
We can use ({\sl Shape}$_5$) to express $\DD^{[0]}$
in  a tensor product form.
% since otherwise, by Lemma \ref{shapecondition-ours},
%we can construct an edge gadget $\Gamma$ such that
%$\gm_{\Gamma, \AA}$ is not \multblrkone/ and we are done.
%
%In particular, by using ({\sl Shape}$_5$), 
We define two diagonal matrices $\KK^{[0]}$
  and $\LL^{[0]}$ as follows.
$\KK^{[0]}$ is an $(s+t)\times (s+t)$ diagonal matrix.
We use $(0,i)$, $i\in [s]$, to index its first $s$ rows
  and $(1,j)$, $j\in [t]$, to index its last $t$ rows.
Its diagonal entries are
\[
K^{[0]}_{(0,i)}=D^{[0]}_{(0,(i,1))}\ \ \ \text{and}\ \ \
K^{[0]}_{(1,j)}=D^{[0]}_{(1,(j,1))},\ \ \ \text{for all $i\in [s]$ and $j\in [t]$.}
\]
$\LL^{[0]}$ is the $2h\times 2h$ identity matrix.
We use $(0,i)$, $i\in [h]$, to index its first $h$ rows 
  and $(1,j)$, $j\in [h]$, to index its last $h$ rows.
By ({\sl Shape}$_5$), we have
\begin{equation}\label{conditionbb-ours}
D^{[0]}_{(0,\xx)}=K^{[0]}_{(0,x_1)}\cdot L^{[0]}_{(0,x_2)}\ \ \ \text{and}\ \ \
D^{[0]}_{(1,\yy)}=K^{[0]}_{(1,y_1)}\cdot L^{[0]}_{(1,y_2)}
\end{equation} 
{for all $\xx\in I$ and $\yy\in J$,}
or equivalently,
\begin{equation}\label{fufufufu-ours}
\DD^{[0]}=\left(\begin{matrix}
\DD^{[0]}_{(0,*)}\\ & \DD^{[0]}_{(1,*)}
\end{matrix}\right)=\left(\begin{matrix}
\KK^{[0]}_{(0,*)}\otimes \LL^{[0]}_{(0,*)}\\
& \KK^{[0]}_{(1,*)}\otimes \LL^{[0]}_{(1,*)}
\end{matrix}\right).
\end{equation}
The goal of Step 2.3 is to prove a similar
  statement for $\DD^{[r]}$, $r\in [N-1]$,
and these equations will allow us in Step 2.4 to decompose
  $\eval(\CC,\fD)$ into two subproblems.

In the proof of Lemma \ref{shapecondition-ours}, we crucially used the property (from $(\calT_3)$)
  that all the diagonal entries of $\DD^{[0]}$ are positive integers.
However, for $r\ge 1$, $(\calT_3)$ only gives us some very weak properties about $\DD^{[r]}$.
For example, the entries are not guaranteed to be real numbers.
So the proof in~\cite{Cai-Chen-Lu-2013} for Step~2.3 is difficult.
 %that we are going to present here is more difficult.
However, our proof here simply follows Step~2.3 in~\cite{Cai-Chen-Lu-2013}
and use  Argument~\ref{meta2} ({\sl Meta}$_2$).
We prove the following lemma.
%(corresponding to Lemma 8.11 from~\cite{Cai-Chen-Lu-2013}):
%
%{\color{red} (The only difference is in the first sentence, so they are actually the same)}
\begin{lemma}[modification of Lemma 8.11 from~\cite{Cai-Chen-Lu-2013}]\label{horrible-ours}
%Suppose $(\CC,\fD)$, $(\pur \CC, \fD)$ are pairs that satisfy
%Let $(\CC,\fD)$, $(\pur \CC, \fD)$ be pairs that satisfy %%% correction
Let  $(\CC,\fD)$ be a pair that satisfies
  \emph{({\sl Shape}$_1$)--({\sl Shape}$_5$)}.
Then either there exists an edge gadget $\Gamma$
such that $\gm_{\Gamma, \AA}$ is not \multblrkone/ or
 we have the following additional condition:
\begin{enumerate}[wide]
%\item[]\hspace{-0.45cm}\emph{({\sl Shape}$_6$)}\ \ There exist diagonal matrices $\KK^{[0]}$
\item[\emph{({\sl Shape}$_6$)}] There exist diagonal matrices $\KK^{[0]}$
  and $\LL^{[0]}$ such that $\DD^{[0]},\KK^{[0]}$, and $\LL^{[0]}$ satisfy \emph{(\ref{fufufufu-ours})}.
Every entry of $\KK^{[0]}$ is a positive integer, and $\LL^{[0]}$ is the
  $2h\times 2h$ identity matrix.
For each $r\in [N-1]$, there
  exist two diagonal matrices $\KK^{[r]}$ and $\LL^{[r]}$. $\KK^{[r]}$
  is an $(s+t)\times (s+t)$ matrix, and $\LL^{[r]}$ is a $2h\times 2h$
  matrix. We index $\KK^{[r]}$ and $\LL^{[r]}$ in the same way we index
  $\KK^{[0]}$ and $\LL^{[0]}$, respectively. Then \label{SHAPECONDITION6} %So $\KK^{[r]}_{(0,*)}$ and $\KK^{[r]}_{(1,*)}$
  %are $s\times s$ \vspace{-0.16cm}and $t\times t$, respectively,
  %while both $\LL^{[r]}_{(0,*)}$ and $\LL^{[r]}_{(1,*)}$ are $h\times h$.
\[
\DD^{[r]}=\left(\begin{matrix}
\DD^{[r]}_{(0,*)}\\ & \DD^{[r]}_{(1,*)}
\end{matrix}\right)=\left(\begin{matrix}
\KK^{[r]}_{(0,*)}\otimes \LL^{[r]}_{(0,*)}\\
& \KK^{[r]}_{(1,*)}\otimes \LL^{[r]}_{(1,*)}
\end{matrix}\right).
\]
Moreover, the norm of every entry in $\LL^{[r]}$ is
  either $0$ or $1$, and for any $r\in [N-1]$, 
\begin{align*}
\KK^{[r]}_{(0,*)}=\00 \ \Longleftrightarrow \ \LL^{[r]}_{(0,*)}=\00\ \ \ & {and}\ \ \ 
  \KK^{[r]}_{(1,*)}=\00 \ \Longleftrightarrow \ \LL^{[r]}_{(1,*)}=\00;\\[0.12ex]
\LL^{[r]}_{(0,*)}\ne \00 \ \Longrightarrow\ \exists\hspace{0.05cm}i\in [h],\ L^{[r]}_{(0,
  i)}=1\ \ \ &\text{and}\ \ \ \LL^{[r]}_{(1,*)}\ne \00 \
  \Longrightarrow\ \exists\hspace{0.05cm}i\in [h],\ L^{[r]}_{(1,i)}=1. 
\end{align*}
\end{enumerate}
\end{lemma}

We now present the proof of Lemma \ref{horrible-ours}.
Fix an $r\in [N-1]$ to be any index.
We use the following notation.
Consider the diagonal matrix $\DD^{[r]}$. %which corresponds to vertices of degree $r \bmod N$.
 It has two parts: \[\DD^{[r]}_{(0,*)}\in \mathbb{C}^{
  sh\times sh}\ \ \ \text{and}\ \ \ \DD^{[r]}_{(1,*)}
  \in \mathbb{C}^{th\times th}.\]
The first part has $s$ blocks, where each
  block is a diagonal matrix with $h$ entries. We will rearrange
  the entries indexed by $(0,*)$ into another 
  $s\times h$ matrix, which we denote as
$\DD$ (just as we did with $\DD^{[0]}$ in the proof of
  Lemma \ref{shapecondition-ours}), where %its $i$-th row $\DD_{i,*}$, for $i\in [s]$,
%denotes the values of the $i$-th block and the $j$-th entry of
%the $i$-th row $D_{i,j}$, for $j\in [h]$, denotes the
%  $j$-th entry of that $i$-th block. % of the diagonal matrix $\DD^{[r]}_{(0,*)}$.
%More exactly,
\[
D_{i,j}=D^{[r]}_{(0,(i,j))}\ \ \ \text{for all $i\in [s]$ and $j\in [h]$.}
\]
%{\color{red} (old text)
%We prove the following lemma in Section \ref{proofofrank1-ours}:
%}

We first prove the following lemma by
% Argument~\ref{meta2} 
({\sl Meta}$_2$),
and then use it to prove Lemma~\ref{horrible-ours}.
%We prove the following lemma by Argument~\ref{meta2} ({\sl Meta}$_2$) soon.
  
%A similar lemma can be proved for $\DD^{[r]}_{(1,*)}$.
  %and Section \ref{proofroot}, respectively.

\begin{lemma}[modification of Lemma 8.12 from~\cite{Cai-Chen-Lu-2013}]\label{rank1-ours}
Either there exists an edge gadget $\Gamma$
such that $\gm_{\Gamma, \AA}$ is not \multblrkone/, or we have 
\begin{enumerate}[label=(\arabic*),ref=\arabic*]
\item\label{item-1-in-lm-for-old-8-12} $\text{\rm rank}(\DD)\le 1$ and 
\item\label{item-2-in-lm-for-old-8-12} for each $i\in [s]$, all nonzero entries of $\DD_{i,*}$ have the same norm.
%for any $i,j,j'\in [h]$, if
%  $D_{i,j}\ne 0$ and $D_{i,j'}\ne 0$, then $|D_{i,j}|=|D_{i,j'}|$.
\end{enumerate}
\end{lemma}

\begin{proof}[Proof of Lemma~\ref{rank1-ours}]
In~\cite{Cai-Chen-Lu-2013}, Lemma~8.13 (vanishing lemma) was
first proved in Section~8.4.1. This is a general lemma not 
%related to $\eval$-problems,
specific to $\eval$-problems,
and remains valid in our setting. Then the proof of Lemma~8.12 is given in Section 8.4.2
in~\cite{Cai-Chen-Lu-2013}. 
Let $\pur \CC$ be the purification of $\CC$ which is 
obtained corresponding to 
the  purification going from $\AA$ to $\pur \AA$.  
We apply the proof of Lemma~8.12 from~\cite{Cai-Chen-Lu-2013} to
 $(\pur \CC, \fD)$
in place of $(\CC, \fD)$ from~\cite{Cai-Chen-Lu-2013}.
It is easy to see that $(\pur \CC, \fD)$
satisfies the conditions needed for the proof of Lemma~8.12 
from~\cite{Cai-Chen-Lu-2013}.
We use the same gadget sequence $\Gamma^{[n]}$
as depicted in Figure~8.3 (p.~967 of~\cite{Cai-Chen-Lu-2013})
with distinguished vertices $u$ and $v$ (in this order), and
get a sequence of
 signature matrices $\RR^{[n]} = \gm_{\Gamma^{[n]}, {\pur \CC}, \fD}$.
%%% old
By going through the argument in~\cite{Cai-Chen-Lu-2013},
e.g., for Property~8.14 (p.~969) we use Corollary~\ref{inversetwin-ours}
to reach the same conclusion. It follows that
either we get an edge gadget $\Gamma^{[n]}$ for some $n \ge 1$, such that 
$\RR^{[n]}$
is not \modblrkone/, or the two items (\ref{item-1-in-lm-for-old-8-12})
and (\ref{item-2-in-lm-for-old-8-12}) in the statement of
Lemma~\ref{rank1-ours} hold.
%%% new %%% correction
%By going through the argument in~\cite{Cai-Chen-Lu-2013},
%e.g., for Property~8.14 (p.~969) we use the property of non-\modblrkone/
%to reach the same conclusion. It follows that
%either we get an edge gadget $\Gamma^{[n]}$ for some $n \ge 1$, such that 
%$\RR^{[n]}$
%is not \modblrkone/, or the two items (\ref{item-1-in-lm-for-old-8-12})
%and (\ref{item-2-in-lm-for-old-8-12}) in the statement of
%Lemma~\ref{rank1-ours} hold.
%More precisely, one can check that the conditions needed to apply Corollary~\ref{inversetwin-ours}
%are satisfied each time, which are slightly stronger than the property of non-\modblrkone/,
%although in our reasoning we only need to have the latter.
%%%

Thus by Argument~\ref{meta2} ({\sl Meta}$_2$),
either 1. there exists an edge gadget $\Gamma' = T_\ell(\Gamma^{[n]})$
for some $n, \ell \ge 1$,
%(where $\Gamma$ is an edge gadget and $p \ge 1$)
such that $\gm_{\Gamma', \AA}$ is not \multblrkone/,
or 2.  $\text{\rm rank}(\DD)\le 1$ and,
for each $i\in [s]$, all nonzero entries of our $\DD_{i,*}$ have the same norm.
\end{proof}
%(which is precisely the second part of the statement of Lemma~\ref{rank1-ours})
%and we are done.
%\end{lemma}
%\begin{lemma}\label{rootofunity}
%If $\fD$ satisfies condition $(\calT_3)$, then for any $r\in [N-1]$
%  and $\xx,\yy\in I$,
%$$
%\left|D^{[r]}_{(0,\xx)}\right|=\left|D^{[r]}_{(0,\yy)}\right|\ne 0\ \
%\Longrightarrow\ \ D^{[r]}_{(0,\xx)}\Big/D^{[r]}_{(0,\yy)}\ \text{is a power of $\oo_N$.}
%$$
%\end{lemma}

\begin{proof}[Proof of Lemma~\ref{horrible-ours}]
Lemma~\ref{horrible-ours} follows from Lemma~\ref{rank1-ours}
in the same way as Lemma 8.11 follows from Lemma 8.12 (see p. 965 
in~\cite{Cai-Chen-Lu-2013}).
% Lemma 8.11 does from Lemma 8.12 (see p. 965 from~\cite{Cai-Chen-Lu-2013}).
\end{proof}

\input{technical-modifications-bounded-degree-gadget-based-inside-meta}

\section{Proofs of Theorem~\ref{step30-ours} and Theorem~\ref{bi-step-3-ours}}\label{sec:bip-case-inner-part}

Let $((M,N),\CC,\fD)$ be a tuple that satisfies $(\calU_1)$--$(\calU_4)$
  and let $\FF\in \mathbb{C}^{m\times m}$ be the upper-right block of $\CC$.
In this section,
  we index the rows and columns of an $n\times n$ matrix with $[0:n-1]$.
  
%{\color{red} (In other words, we denote by $(\CC, \fD)$
%the previously obtained $(\XX, \fY)$ at the end of Section~\ref{step25-ours}.
%The same was implicitly done in~\cite{Cai-Chen-Lu-2013} at the beginning of Section 9, p. 980.
%Note that $(\XX, \fY)$ is already ``purified''
%so we will not have an associated ``purified'' pair (added, can be turned into a footnote))}

%{\color{red} (first introduce a meta argument)}

We introduce the following third meta-argument ({\sl Meta}$_3$).
This ({\sl Meta}$_3$) resembles ({\sl Meta}$_2$) at a superficial level;
however remember that past Step 2 the pair $(\CC,\fD)$
that satisfies Theorem~\ref{bi-step-2-ours} is the inner pair from Step 2, which is
distinct from the pairs $(\CC,\fD)$ and $(\pur \CC,\fD)$ %{\color{red} (JYC: pairs or just (and $(\pur \CC,\fD)$), which one?)}
referred to
in ({\sl Meta}$_2$). In particular, the $\CC$ of the inner pair is the
bipartization of a square matrix $\HH$ (now redenoted as $\FF$) consisting of roots of unity
only, and clearly  already purified (so here there is no separate $\pur \CC$).

\begin{argument}[{\sl Meta}$_3$]\label{meta3}
We have the following:
\begin{enumerate}
\item\label{item:meta3-thickening} Either $\gm_{\Gamma, \CC, \fD}$ is \modblrkone/
for every edge gadget $\Gamma$,
or there is an edge gadget $\Gamma$
such that $\gm_{\Gamma, \CC, \fD}$ is not \modblrkone/;

\item\label{item:meta3-assumption} Referring to the pair $(\CC, \fD)$
from Section 9 in~\cite{Cai-Chen-Lu-2013},
all statements from Section 9 
%(proofs of Theorems 5.4 and 5.6) 
through Section 11 
%(proof of Theorem 5.9) 
in~\cite{Cai-Chen-Lu-2013} (including theorems, lemmas, corollaries, properties, etc.
numbered from 9.1 to 11.2)
can be proved under the assumption that every signature $M_{\Gamma, \CC, \fD}$  %edge 
%($(\CC, \fD)$ is from~\cite{Cai-Chen-Lu-2013}) 
is \modblrkone/
and whenever it is concluded that, as a possible scenario, $\EVAL(\CC, \fD)$ 
%($(\CC, \fD)$ is from~\cite{Cai-Chen-Lu-2013}) 
is $\#$P-hard,
this is because an edge gadget, say, $\Gamma$ has been constructed
such that $M_{\Gamma, \CC, \fD}$ 
%($(\CC, \fD)$ is from~\cite{Cai-Chen-Lu-2013}) i
is not \modblrkone/.
% (by Corollary 8.3 from~\cite{Cai-Chen-Lu-2013}).
This statement can be checked directly;

\item\label{item:meta3-application} Thus, 
%in the notation of this paper, 
provided our $(\CC, \fD)$
satisfies all the corresponding conditions for $(\CC, \fD)$ in~\cite{Cai-Chen-Lu-2013},
we can apply the corresponding reasoning from~\cite{Cai-Chen-Lu-2013} 
to our $(\CC, \fD)$
and make the corresponding conclusions.
% for $(\CC, \fD)$.
\end{enumerate}
%%%%%%%%%%%%%%%%%%%%%%%%%%%%%%%%%%%%%%%%%%%%%%%%%%%%%%%%%%%%%%%%%%%%
\begin{comment}

\item\label{item:meta3-constructivity} In fact, the assumption in~\ref{item:meta3-assumption},
can be trivially weakened to range over the edge gadgets $\Gamma$
constructed in the corresponding part from~\cite{Cai-Chen-Lu-2013}
(described in~\ref{item:meta3-assumption}).
Moreover, if this weakened assumption is violated,
then it can be checked such an edge gadget $\Gamma$
can always be constructed explicitly.
Thus, if in~\ref{item:meta3-application} the corresponding
conclusions are not true for $(\CC, \fD)$,
then we can explicitly get an edge gadget $\Gamma$ such that
$\gm_{\Gamma, \CC, \fD}$ is not \modblrkone/.

\end{enumerate}
In the above, $(\CC, \fD)$ from~\cite{Cai-Chen-Lu-2013} means $(\CC, \fD)$ introduced at the beginning of Section 9, p. 980 in~\cite{Cai-Chen-Lu-2013}
and corresponds to our $(\CC, \fD)$ in this paper introduced in this section.
{\color{red}
(In other words, $(\CC, \fD)$ from~\cite{Cai-Chen-Lu-2013}
here corresponds to the redenoted $(\XX, \YY)$ from Section 8.6 (Step 2.5) in~\cite{Cai-Chen-Lu-2013}
where the current $(\CC, \fD)$ corresponds to the redenoted $(\XX, \YY)$ from Section~\ref{step25-ours}. (this is not quite right, must be removed to avoid mess))
}
\end{comment}
%%%%%%%%%%%%%%%%%%%%%%%%%%%%%%%%%%%%%%%%%%%%%%%%%%%%%%%%%%%%%%%%%%%%
\end{argument}

\subsection{The group condition}

We first show that either $\FF$ satisfies the following
  condition or there exists an edge gadget $\Gamma$
such that $\gm_{\Gamma, \CC, \fD}$ is not \modblrkone/:

%\begin{figure}
%\center
%\includegraphics[height=6cm]{figure_5}
%\caption{The gadget for $p=1$ (Note that the subscript $e$ is suppressed).}\label{figure_5}
%\end{figure}

\begin{lemma}[modification of Lemma 9.1 from~\cite{Cai-Chen-Lu-2013}]\label{groupcondition1-ours}
Let $((M,N),\CC,\fD)$ be a tuple that satisfies $(\calU_1)$--$(\calU_4)$.
Then either~$\FF$ satisfies the group condition \emph{(\GC)},
%%%
%\begin{enumerate}
%\item[] \hspace{-0.6cm}{\emph{(row-\GC)}}\ \ \ $\forall\hspace{0.06cm} i,j\in [0:m-1]$,
%  $\exists\hspace{0.06cm} k\in [0:m-1]$ such that
%  $\FF_{k,*}=\FF_{i,*}\circ \FF_{j,*}$; \label{CONDITIONGC}
%
%\item[] \hspace{-0.6cm}{\emph{(column-\GC)}}\ \ \ $\forall\hspace{0.06cm} i,j\in [0:m-1]$,
%  $\exists\hspace{0.06cm} k\in [0:m-1]$ such that
%  $\FF_{*,k}=\FF_{*,i}\circ \FF_{*,j}$, 
%\end{enumerate}
%%%
\begin{enumerate}[wide]
\item[\emph{(row-\GC)}] for all $i,j\in [0:m-1]$, %$\forall\hspace{0.06cm} i,j\in [0:m-1]$,
  $\exists\hspace{0.06cm} k\in [0:m-1]$ such that
  $\FF_{k,*}=\FF_{i,*}\circ \FF_{j,*}$; \label{CONDITIONGC}

\item[\emph{(column-\GC)}] for all $i,j\in [0:m-1]$, %$\forall\hspace{0.06cm} i,j\in [0:m-1]$,
  $\exists\hspace{0.06cm} k\in [0:m-1]$ such that
  $\FF_{*,k}=\FF_{*,i}\circ \FF_{*,j}$, 
\end{enumerate}
or  there exists an edge gadget $\Gamma$
such that $\gm_{\Gamma, \CC, \fD}$ is not \modblrkone/.
%
%{\color{red} (added, can be removed)
%
%The group conditions (\emph{(\GC)}) correspond (are similar)
%to the group conditions (\emph{(\GC)}) from~\cite{Cai-Chen-Lu-2013}, p. 981.
%}
\end{lemma}
%{\color{red} (meta-proved)}
\begin{proof}
We prove  this lemma by Argument~\ref{meta3} ({\sl Meta}$_3$)
adapting the proof of Lemma 9.1 from~\cite{Cai-Chen-Lu-2013}.
The following details can be noted.
%we would like to emphasize a few details.
%\begin{quote}
The proof of Lemma 9.1 in~\cite{Cai-Chen-Lu-2013} uses a family of gadgets 
parameterized by $p \ge 1$; the gadget for $p=1$ is
depicted in Figure~9.1 (p. 981 in~\cite{Cai-Chen-Lu-2013}).
We use the same gadgets.
Let $\Gamma^{[p]}$, where $p \ge 1$, denote these edge gadgets.
Then in~\cite{Cai-Chen-Lu-2013} for the obtained matrices $\AA^{[p]}$, where $p \ge 1$, it was noted that $Z_{\AA^{[p]}}(G) = Z_{\CC, \fD}(G^{[p]})$
for all $G$, where $G^{[p]}$ is the graph obtained by replacing every edge of $G$ by $\Gamma^{[p]}$.
This uses the fact that in $\Gamma^{[p]}$ the distinguished vertices
are each of degree divisible by $N$, and $D^{[0]} = \II_{2 m}$. %, although in our reasoning the property of non-\modblrkone/ alone is sufficient. %%% correction
Thus in terms of signature matrices, $\AA^{[p]} = \gm_{\Gamma^{[p]}, \CC, \fD}$ 
for our $(\CC, \fD)$,
for all $p \ge 1$.

%%%%%%%%%%%%%%%%%%%%%%%%%%%%%%%%%%%%%%%%%%%%%%%%%%%%%%%%%%%%%%%%%%%%
\begin{comment}

{\color{red} (or say the following)

The reason this is true is because
\begin{enumerate}
\item $\AA^{[p]} = \gm_{\Gamma^{[p]}, \CC, \fD}$ for $p \ge 1$ (in our notation);
\item In $\Gamma^{[p]}$ the distinguished vertices
are each of degree divisible by $N$;
\item $D^{[0]} = \II_{2 m}$.
\end{enumerate}
In our argument, we only need the first property.}
\end{comment}
%%%%%%%%%%%%%%%%%%%%%%%%%%%%%%%%%%%%%%%%%%%%%%%%%%%%%%%%%%%%%%%%%%%%

Then just as in~\cite{Cai-Chen-Lu-2013}, we either get the desired properties
\emph{(\GC)},
or we can exhibit an edge gadget $\Gamma^{[p]}$ for some $p \ge 1$,
for which the signature matrix $\AA^{[p]} = \gm_{\Gamma^{[p]}, \CC, \fD}$ is not \modblrkone/.
%\end{quote}
\end{proof}

Similar remarks can be formulated in all subsequent arguments involving Argument~\ref{meta3} ({\sl Meta}$_3$).

%%%%%%%%%%%%%%%%%%%%%%%%%%%%%%%%%%%%%%%%%%%%%%%%%%%%%%%%%%%%%%%%%%%%
\begin{comment}

{\color{red} (JYC: in this proof we get a nonnegative \blrkone/ signature)}

{\color{red} (JYC: in this proof we use the same gadget construction as in the proof of Lemma 9.1 (described on p. 981) and Figure 9.1 from~\cite{Cai-Chen-Lu-2013} (p. 981, shows only for $p = 1$), can incorporate some of this info into the above proof if needed)}
\end{proof}

{\color{red} (old, to be removed)
\begin{proof}
Note that the conditions $(\calU_1)$--$(\calU_4)$ are identical
to the conditions $(\calU_1)$--$(\calU_4)$ from~\cite{Cai-Chen-Lu-2013}.
Next, we note that proof of Lemma 9.1 to the current $((M,N),\CC,\fD)$
is edge gadget based and it actually yields the desired result.
\end{proof}
}

{\color{red} (we may need to formulate this property but only for understanding the group structure)}

\end{comment}
%%%%%%%%%%%%%%%%%%%%%%%%%%%%%%%%%%%%%%%%%%%%%%%%%%%%%%%%%%%%%%%%%%%%

Next we prove a property concerning
  discrete unitary matrices that satisfy (\GC).
Given an $n\times n$ matrix $\AA$,  let $A^{\text{R}}$ 
  denote the set of its row vectors
  $\{\AA_{i,*}\}$ and $A^{\text{C}}$ denote the set of its column
  vectors $\{\AA_{*,j}\}$. For general matrices, it is possible
  that $|A^{\text{R}}|, |A^{\text{C}}|<n$, 
  since $\AA$ may have duplicate rows or columns.
But if $\AA$ is $M$-discrete unitary, then it is clear that 
  $|A^{\text{R}}|=|A^{\text{C}}|=n$. 

\begin{property}[Property 9.2 from~\cite{Cai-Chen-Lu-2013}]\label{gcproperty1-ours}
If $\AA\in \mathbb{C}^{n\times n}$ is an $M$-discrete unitary matrix
  that satisfies \emph{(}\GC\emph{)},
then $A^{\text{R}}$ and $A^{\text{C}}$ are finite Abelian groups \emph{(}of order $n$\emph{)}
  under the Hadamard product.
\end{property}
\begin{proof}
The proof is the same as that of Property 9.2 from~\cite{Cai-Chen-Lu-2013}.
\end{proof}

%{\color{red} (new text)}
\subsection{Proof of Theorem~\ref{step30-ours}}

In this section, we prove Theorem \ref{step30-ours} by showing that 
  $(\calU_1)$--$(\calU_4)$ indeed imply $(\calU_5)$.
  
%Suppose $\eval(\CC,\fD)$ is not \#P-hard; 
%  otherwise we can construct (there exists) an edge gadget $\Gamma$
%  such that $\gm_{\Gamma, \CC, \fD}$ is not \modblrkone/.
Assume for any edge gadget $\Gamma$,
   $\gm_{\Gamma, \CC, \fD}$ is \modblrkone/;
   otherwise we are done.
By Lemma \ref{groupcondition1-ours}, $((M,N),\CC,\fD)$ 
  satisfies   (\GC).
  %only $(\calU_1)$--$(\calU_4)$, but also.
Fixing $r$ to be any index in $[N-1]$,
  we will prove $(\calU_5)$ for the $(i,i)$th
  entries of $\DD^{[r]}$, where $i\in [m:2m-1]$.
The proof for the first half of $\DD^{[r]}$ is similar.
For simplicity, let $\DD$ be the $m$-dimensional vector such that
\[
D_i=D^{[r]}_{m+i}\ \ \ \text{for all $i\in [0:m-1]$}.
\]
Also let $K=\{i\in [0:m-1]:
  D_i\ne 0\}$.
If $|K|=0$, then there is nothing to prove; if $|K|=1$, then
  by $(\calU_3)$, the only nonzero entry in $\DD$ must be $1$.
So we assume $|K|\ge 2$.

%We start with a useful lemma. 
The following lemma is Lemma 9.3 from~\cite{Cai-Chen-Lu-2013};
the same proof works here.
It implies that to prove Theorem \ref{step30-ours}, i.e.,  
  ($\calU_5$),~it suffices to prove that $D_i$ is a root of unity for
  every $i\in K$. 
%We claim that $D_i$, for every $i\in K$, must be a root of unity;
%  otherwise $\eval(\CC,\fD)$ is \#P-hard, which contradicts the assumption.
%Actually the lemma below shows that such a claim is all we need
%  to prove Theorem :

\begin{lemma}[Lemma 9.3 from~\cite{Cai-Chen-Lu-2013}]\label{fieldlemma-ours}
If $D\in \mathbb{Q}(\oo_N)$
  is a root of unity, %and there exists a positive even integer $N$ such that
%$$
%D=\oo_N^k\cdot \frac{\sum_{i=1}^{N-1}a_i\cdot \oo_N^i}{
%  \sum_{i=1}^{N-1}b_i\cdot \oo_N^i},
%$$
%where $k\in \mathbb{Z}$ and $a_0,...,a_{N-1},b_0,...,b_{N-1}$ are non-negative integers,
%then $D$ must be a power of $\oo_N$.
%%%JYC added $N$ is even...
then $D$ must be a power of $\oo_N$. %\emph{(}$N$ is \emph{even} by %\emph{($\calU_1$)}.\emph{)}
\end{lemma}

%\begin{proof} %[Proof of Lemma \ref{fieldlemma}]
%The proof is the same as that of Lemma 9.3 from~\cite{Cai-Chen-Lu-2013}.
%\end{proof}

%\def\ZZ{\mathbf{Z}}

%We delay the proof to the end of the section.
Next we show that every $D_i$, $i\in K$, is a root of unity.
Suppose for a contradiction that this is not true.
The next lemma is Lemma~9.4 from~\cite{Cai-Chen-Lu-2013},
and the same proof works.
Define $\ZZ=(Z_0,\ldots,Z_{m-1})$,
  where $Z_i=(D_i)^N$.
%  There exists an infinite sequence $\{P_n\}_{n\rightarrow \infty}$
%  such that $$(D_{i}^{NP_n}:i\in K)
%  $$ approaches to the all-one vector (of dimension $|K|$), but yet
%  never equals to it, when $n \rightarrow \infty$.

\begin{lemma}[Lemma 9.4 from~\cite{Cai-Chen-Lu-2013}]\label{lemma-approach-1-ours}
Suppose there is a $k\in K$ such that $Z_k$ is not a root of unity.
Then there exists an
  infinite integer sequence $\{P_n\} $
  such that when $n\rightarrow \infty$, the vector sequence $((Z_k)^{P_n} :k
\in K)$ approaches, but never reaches, the all-one vector $\11_{|K |}$.
 %of
%dimension $|K|$.
\end{lemma}
%\begin{proof}
%The proof is the same as that of Lemma 9.4 from~\cite{Cai-Chen-Lu-2013}.
%\end{proof}

%%%%%%%%%%%%%%%%%%%%%%%%%%%%%%%%%%%%%%%%%%%%%%%%%%%%%%%%%%%%%%%%%%%%
\begin{comment}

%%%
{\color{red} (alternative start, without explaining the notation like above)}

In this section, we prove Theorem \ref{step30-ours} by showing that 
  $(\calU_1)$--$(\calU_4)$ indeed imply $(\calU_5)$.
%In this section we explain to how prove Theorem~\ref{step30-ours}.
First, we can start by introducing the same notation as was done in Section 9.2 from~\cite{Cai-Chen-Lu-2013}.
After that, we repeat the proofs of (general) Lemmas 9.3 and 9.4 from Section 9.2 from~\cite{Cai-Chen-Lu-2013}.
(These lemmas are general and not related to $\EVAL$ problems.)

{\color{red} (to the proof of Theorem~Theorem~\ref{step30-ours})}
\end{comment}
%%%%%%%%%%%%%%%%%%%%%%%%%%%%%%%%%%%%%%%%%%%%%%%%%%%%%%%%%%%%%%%%%%%%

Now we apply Argument~\ref{meta3} ({\sl Meta}$_3$)
adapting %and adapt %and adapting
the reasoning after the proof of Lemma~9.4 from Section 9.2 in~\cite{Cai-Chen-Lu-2013} to prove Theorem~\ref{step30-ours}.
%
%In particular, we would like to emphasize a few details.
The following details can be noted.
%\begin{quote}
The proof in~\cite{Cai-Chen-Lu-2013} uses a family of gadgets parameterized 
by $p \ge 1$; this  gadget family is depicted in Figure~9.2 (p.~985
in~\cite{Cai-Chen-Lu-2013}). We use
the same gadgets. 
Let $\Gamma^{[p]}$, where $p \ge 1$, denote these edge gadgets.
Then in~\cite{Cai-Chen-Lu-2013} for the obtained matrices $\AA^{[p]}$, 
where $p \ge 1$, it was noted that $Z_{\AA^{[p]}}(G) = Z_{\CC, \fD}(G^{[p]})$
for all $G$, where $G^{[p]}$ is the graph obtained by replacing 
every edge of $G$ by $\Gamma^{[p]}$.
This uses the fact that in $\Gamma^{[p]}$ the distinguished vertices
are each of degree divisible by $N$, and $\DD^{[0]} = \II_{2 m}$. %, although in our reasoning the property of non-\modblrkone/ alone is sufficient.  %%% correction
Thus in terms of signature matrices,
 $\AA^{[p]} = \gm_{\Gamma^{[p]}, \CC, \fD}$ for our $(\CC, \fD)$, for all $p \ge 1$. 

%%%%%%%%%%%%%%%%%%%%%%%%%%%%%%%%%%%%%%%%%%%%%%%%%%%%%%%%%%%%%%%%%%%%
\begin{comment}

{\color{red} (or say the following)

The reason this is true is because
\begin{enumerate}
\item $\AA^{[p]} = \gm_{\Gamma^{[p]}, \CC, \fD}$ for $p \ge 1$ (in our notation);
\item In $\Gamma^{[p]}$ the distinguished vertices
are each of degree divisible by $N$;
\item $D^{[0]} = \II_{2 m}$.
\end{enumerate}
In our argument, we only need the first property.}

\end{comment}
%%%%%%%%%%%%%%%%%%%%%%%%%%%%%%%%%%%%%%%%%%%%%%%%%%%%%%%%%%%%%%%%%%%%

Then just as in~\cite{Cai-Chen-Lu-2013}, we either get the desired properties,
or we can exhibit an edge gadget $\Gamma^{[p]}$ for some $p \ge 1$,
for which the signature matrix $\AA^{[p]} = \gm_{\Gamma^{[p]}, \CC, \fD}$ is not \modblrkone/.
%\end{quote}

%Theorem \ref{step30-ours} is proved.

%%%%%%%%%%%%%%%%%%%%%%%%%%%%%%%%%%%%%%%%%%%%%%%%%%%%%%%%%%%%%%%%%%%%
\begin{comment}

{\color{red} (JYC: in this proof we use the same gadget construction as in the proof fragment starting after the proof of Lemma 9.4 (described on pp. 984 -- 985) and Figure 9.1 from~\cite{Cai-Chen-Lu-2013} (p. 985, shows only for $p = 1$, the caption in~\cite{Cai-Chen-Lu-2013} is incorrect, it works for all $p \ge 1$), can incorporate some of this info into the above proof if needed)}

{\color{red} (JYC: in this proof we get a \modblrkone/ signature, not just \blrkone/)}

%{\color{red} (meta-proved, empty)}
%\begin{proof}[Proof of Theorem~\ref{step30-ours}]
%
%\end{proof}
%%Theorem \ref{step30-ours} is proved.

{\color{red} (new text)}
\end{comment}
%%%%%%%%%%%%%%%%%%%%%%%%%%%%%%%%%%%%%%%%%%%%%%%%%%%%%%%%%%%%%%%%%%%%

\subsection{Decomposing $\FF$ into Fourier matrices}

The same sequence of Lemma~9.5, Property~9.6, and 
Lemma~9.7 in~\cite{Cai-Chen-Lu-2013} (with the same proof) now implies
our Theorem~\ref{bi-step-3-ours}.
For the reader's convenience we restate these here.

Suppose $((M,N),\CC,\fD)$ satisfies   
  $(\calU_1)$--$(\calU_5)$ and (\GC); 
  otherwise there exists an edge gadget $\Gamma$
  such that $\gm_{\Gamma, \CC, \fD}$ is not \modblrkone/.
%We prove Theorem~\ref{bi-step-3-ours}.
%We start with the following simple observation.
%To decompose $\FF$ into Fourier matrices (recall that $\FF$ is the
%  upper-right $m\times m$ block matrix of $\CC$),
% we first show that if $M=pq$ and $\gcd(p,q)=1$,
%  then up to a permutation of rows and columns, $\FF$ is
%  the tensor product of two smaller matrices,
%both of which are discrete unitary and satisfy (\GC).
%Note that $p$ and $q$ here are not necessarily primes or prime powers.
%
\begin{lemma}[Lemma 9.5 from~\cite{Cai-Chen-Lu-2013}]\label{decomp1-ours}
Let $\FF\in \mathbb{C}^{m\times m}$ be an $M$-discrete unitary matrix that
  satisfies \emph{(}\GC\emph{)},
  where $M=pq$, $p,q>1$, and $\gcd(p,q)=1$.
Then there exist two permutations $\Pi$ and $\Sigma$ over $[0:m-1]$
  such that
$ 
\FF_{\Pi,\Sigma}=\FF'\otimes \FF'',
$ 
where $\FF'$ is a $p$-discrete unitary matrix, $\FF''$ is a $q$-discrete
  unitary matrix, and both of them satisfy \emph{(}\GC\emph{)}.
\end{lemma}
%\begin{proof}
%The proof is the same as that of Lemma 9.5 from~\cite{Cai-Chen-Lu-2013}.
%\end{proof}
%
%Next we prove Lemma~\ref{decomp2-ours} which deals with the case 
%when $M$ is a prime power. %The following property is easy to prove.
%
\begin{property}[Property 9.6 from~\cite{Cai-Chen-Lu-2013}]\label{gcproperty2-ours}
Let $\AA$ be an $M$-discrete unitary matrix that satisfies the group condition
  \emph{(}\GC\emph{)}. If $M$ is a prime power, then 
  one of its entries is equal to $\oo_M$.
\end{property}
%\begin{proof}
%The proof is the same as that of Property 9.6 from~\cite{Cai-Chen-Lu-2013}.
%\end{proof}
%
\begin{lemma}[Lemma 9.7 from~\cite{Cai-Chen-Lu-2013}]\label{decomp2-ours}
Let $\FF\in \mathbb{C}^{m\times m}$ be an $M$-discrete unitary matrix that
  satisfies \emph{(}\GC\emph{)}.
Moreover, $M=p^k$ is a prime power for some $k\ge 1$.
Then there exist two permutations $\Pi$ and $\Sigma$ such that
$ 
\FF_{\Pi,\Sigma}=\boldsymbol{\mathcal{F}}_{M}\otimes\hspace{0.04cm} \FF',
$ 
where $\FF'$ is an $M'$-discrete unitary matrix, $M'=p^{k'}$ for some
  $k'\le k$, and $\FF'$ satisfies \emph{(}\GC\emph{)}.
\end{lemma}
%\begin{proof}
%The proof is the same as that of Lemma 9.7 from~\cite{Cai-Chen-2017}.
%\end{proof}
%
Theorem \ref{bi-step-3-ours} then follows.
% from Lemma \ref{decomp1-ours}
%  and Lemma \ref{decomp2-ours}.

%{\color{red} (new text)}
%
%{\color{red} (here we outline what was done in CCL, Section 10)}

\section{Proof of Theorem~\ref{bi-step-4-ours}}

As described in Section~\ref{subsec:Fourier-decomposition}, after
Theorem~\ref{bi-step-3-ours} is proved, we may assume that we have a 
$4$-tuple $((M,N),\CC,\fD,(\pp,\bft,\fq))$ %%%\sout{$((M,N),\CC,\fD,(\qq,\bft,\fq))$} {\color{red} $((M,N),\CC,\fD,(\pp,\bft,\fq))$ (JYC: same typo in CCL, p. 991, $\qq_i$ already used for powers of $p_{i j}$, see p. 942, for the explanation)}
 that satisfies
  condition ($\calR$).
Also we may assume that $\gm_{\Gamma, \CC, \fD}$ is \modblrkone/
for every edge gadget $\Gamma$;
otherwise,
  we are done.
%For every $r$ in $\cT$ (recall that $\cT$ is the set of $r\in [N-1]$ such
  %that $\Delta_r\ne \emptyset$), we show that
%$\Lambda_r$ and $\Delta_r$ must be cosets in $\zqt$.

The purpose of Section~10 in~\cite{Cai-Chen-Lu-2013}
is to prove Theorem 5.8 in~\cite{Cai-Chen-Lu-2013}.
We have the corresponding statement, Theorem~\ref{bi-step-4-ours}.
(Note that starting from Section~\ref{sec:bip-case-inner-part},
$(\CC, \fD)$ is the same in this paper as the $(\CC, \fD)$ in~\cite{Cai-Chen-Lu-2013}
starting from Section~9.)
Lemma~10.1 from~\cite{Cai-Chen-Lu-2013} is still valid without change,
which gives a direct product form for $\Lambda_r$ and $\Delta_r$,
once it is proved that they are indeed cosets in $\zqt$.
This is stated in Conditions ($\calL_1$) and  ($\calL_2$), respectively.

The main content of Theorem 5.8 in~\cite{Cai-Chen-Lu-2013}
is to prove that $\Lambda_r$ and $\Delta_r$ are indeed cosets in $\zqt$,
for all $r \in [N-1]$, unless $\EVAL(\CC, \fD)$ is \#P-hard. In our setting,
Theorem~\ref{bi-step-4-ours} does the same under the
condition that $\gm_{\Gamma, \CC, \fD}$ is \modblrkone/
for every edge gadget $\Gamma$.
To this end, we apply Argument~\ref{meta3} ({\sl Meta}$_3$)
to adapt the proof  of Theorem~5.8  in Section 10 of~\cite{Cai-Chen-Lu-2013},
now for Theorem~\ref{bi-step-4-ours}.

%{\color{red} (normally, we could stop here, but we can go further)}
%
%However, we would like to emphasize a few details.

The following details can be noted.
The proof in Section 10 of~\cite{Cai-Chen-Lu-2013}
uses an edge gadget, which we will denote by $\Gamma = \Gamma_r$,
one for each fixed $r \in  \cT$;
%[N-1]$; 
this  gadget is depicted in Figure~10.1 (p.~992
in~\cite{Cai-Chen-Lu-2013}).
We use
the same gadget $\Gamma$. 
In~\cite{Cai-Chen-Lu-2013}
 for the obtained matrix $\AA$ corresponding to $\Gamma$
(this  $\AA$ is not to be confused with the beginning matrix  $\AA$ 
that gives rise to the pair $(\CC, \fD)$), 
it was noted that $Z_{\AA}(G) = Z_{\CC, \fD}(G')$,
for all $G$, where $G'$ is the graph obtained by replacing every edge of $G$ by $\Gamma$.
This uses the fact that in $\Gamma$ the distinguished vertices
are each of degree divisible by $N$, and
 $\DD^{[0]} = \II_{2 m}$. %, although in our reasoning the property of non-\modblrkone/ alone is sufficient.  %%% correction
Thus in terms of signature matrices,
this new matrix $\AA = \gm_{\Gamma, \CC, \fD}$ for our $(\CC, \fD)$.

%%%%%%%%%%%%%%%%%%%%%%%%%%%%%%%%%%%%%%%%%%%%%%%%%%%%%%%%%%%%%%%%%%%%
\begin{comment}

The reason this is true is because
\begin{enumerate}
\item $\AA = \gm_{\Gamma, \CC, \fD}$ (in our notation);
\item In $\Gamma$ the distinguished vertices
are each of degree divisible by $N$;
\item $D^{[0]} = \II_{2 m}$.
\end{enumerate}
In our argument, we only need the first property.}

\end{comment}
%%%%%%%%%%%%%%%%%%%%%%%%%%%%%%%%%%%%%%%%%%%%%%%%%%%%%%%%%%%%%%%%%%%%

%We note that this $\AA$ is local for Section 10 in~\cite{Cai-Chen-Lu-2013}
%and not related to the initial $\AA$ in~\cite{Cai-Chen-Lu-2013}.

Then  just as  in~\cite{Cai-Chen-Lu-2013},  we either get the desired properties,
or for the edge gadget $\Gamma$, the signature matrix $\gm_{\Gamma, \CC, \fD}$ is not \modblrkone/.
Similarly we can prove for $\Delta_r$, for $r \in \cS$.

This allows the rest of the proof on pp. 992--994
from~\cite{Cai-Chen-Lu-2013} to go through.
In particular, we may repeat the proofs of Lemmas 10.2 and 10.3 from~\cite{Cai-Chen-Lu-2013} (these have no direct relation to edge gadgets)
%%% old
%and %%%%%%%%%%%%then %added then
%finally proving Theorem~\ref{bi-step-4-ours},
%and showing that $(\calL_1)$, $(\calL_2)$ hold.
%%% new %%% correction
and finally finish the proof of Theorem~\ref{bi-step-4-ours},
showing that $(\calL_1)$, $(\calL_2)$ hold.
%%%
%{\color{red} 
That $(\calL_3)$ holds is concluded at the end of Section~\ref{subsec:Step-3.3}. %{\color{red} (is vs was, both are fine but which one reads better?)}

\subsection{A Corollary of Theorem~\ref{bi-step-4-ours}}

Now that we have proved Theorem~\ref{bi-step-4-ours}, we know that
either there exists an edge gadget $\Gamma$ such that $\gm_{\Gamma, \CC, \fD}$ 
is not \modblrkone/, or we may assume
that $(\calL)$ holds. Thus, $\Lambda_r$ and
$\Delta_r$ are cosets.
The next is Corollary~10.4 from~\cite{Cai-Chen-Lu-2013};
the same proof works.

\begin{corollary}[Corollary 10.4 from~\cite{Cai-Chen-Lu-2013}]\label{lemma8-4-ours}
Let $\HH$ be the $m \times |\Delta_r|$  submatrix
obtained from  $\FF$ by restricting to the columns indexed by $\Delta_r$.
Then for any two rows $\HH_{\uu,*}$ and $\HH_{\vv,*}$, where $\uu, \vv \in \zqt$,
either there exists some $\alpha \in \mathbb{Z}_M$ such that
$\HH_{\uu,*} = \omega_M^\alpha\cdot \HH_{\vv,*}$ or
$\langle \HH_{\uu,*}, \HH_{\vv,*} \rangle = 0$. 

Similarly we denote by $\GG$ the $|\Lambda_r| \times m$ submatrix
obtained from  $\FF$ by restricting to the rows indexed by $\Lambda_r$.
Then for any two columns $\GG_{*,\uu}$ and $\GG_{*,\vv}$, where $\uu, \vv \in \zqt$,
either there exists an $\alpha \in \mathbb{Z}_M$ such that
$\GG_{*,\uu} = \omega_M^\alpha\cdot \GG_{*,\vv}$ or
$\langle \GG_{*,\uu}, \GG_{*,\vv} \rangle = 0$. 
\end{corollary}
%\begin{proof}
%The proof is the same as that of Corollary 10.4 from~\cite{Cai-Chen-Lu-2013}.
%\end{proof}
%%%%%%%%%%%%%%%%%%%%%%%%%%%%%%%%%%%%%%%%%%%%%%%%%%%%%%%%%%%%%%%%%%%%
\begin{comment}

{\color{red} (JYC: the proof of Corollary 10.4 uses Lemma 10.3 from CCL whose statements we didin't include, however, this corollary is used in CCL in Section 11)}
%\begin{proof}
%The rows of $\HH$ are restrictions of $\FF$.
%Any two rows $\HH_{\uu,*}, \HH_{\vv,*}$  satisfy
%$$\HH_{\uu,*} \circ \overline{\HH_{\vv,*}} = \FF_{\uu-\vv,*}\hspace{0.03cm} |_{\Delta_r}
%\hspace{-0.1cm}
%= \HH_{\uu-\vv,*},$$ which is a row in $\HH$.
%If this $\HH_{\uu-\vv,*}$ is a constant, namely $\omega_M^\alpha$
%for some  $\alpha \in \mathbb{Z}_M$, then
%$\HH_{\uu,*} = \omega_M^\alpha\hspace{0.05cm} \HH_{\vv,*}$;
%otherwise, Lemma~\ref{kkkbb-ours} says that
%$\langle \HH_{\uu,*}, \HH_{\vv,*} \rangle = 0$. 
%
%The proof for $\GG$ is exactly the same.
%\end{proof}

{\color{red} (some explanations from CCL following this proof)}

\end{comment}
%%%%%%%%%%%%%%%%%%%%%%%%%%%%%%%%%%%%%%%%%%%%%%%%%%%%%%%%%%%%%%%%%%%%

As part of a discrete unitary matrix $\FF$,
all columns $\{\HH_{*,\uu}\hspace{0.06cm}|\hspace{0.06cm} \uu \in \Delta_r\}$ of
$\HH$ must be orthogonal and thus $\text{rank}(\HH)=|\Delta_r|$.
We denote by $n$ the cardinality $|\Delta_r|$.
There must be $n$ linearly independent rows in $\HH$.
We may start with $\fb_0=\00$ and assume the
$n$ vectors
$\fb_0=\00,\fb_1,\ldots,\fb_{n-1}\in \zqt$ are the
indices of  a set of linearly independent rows.
By Corollary~\ref{lemma8-4-ours}, these must be orthogonal as row vectors
(over $\mathbb{C}$).
Since the rank of the matrix $\HH$ is exactly $n$,
it is clear that all other rows must be a multiple of
these rows, since the only alternative is to be orthogonal to them all,
by Corollary~\ref{lemma8-4-ours} again,
which is~absurd.
A symmetric statement for $\GG$ also holds.

%%%%%%%%%%%%%%%%%%%%%%%%%%%%%%%%%%%%%%%%%%%%%%%%%%%%%%%%%%%%%%%%%%%%
\begin{comment}

{\color{red} (new text)}

{\color{red} (here we outline what was done in CCL, Section 11)}

\end{comment}
%%%%%%%%%%%%%%%%%%%%%%%%%%%%%%%%%%%%%%%%%%%%%%%%%%%%%%%%%%%%%%%%%%%%

\section{Proof of Theorem~\ref{bi-step-5-ours}}
%{\color{red} (copied and modified fragment)}

Let $((M,N),\CC,\fD,(\pp,\bft,\fq))$ be a tuple that
  satisfies $(\calR)$ and $(\calL)$ including $(\calL_3)$.
%We also assume that there is no edge gadget $\Gamma$ such that
%$M_{\CC,\fD}$ is not \multblrkone/.
%$M_{\Gamma,\CC,\fD}$ is not \modblrkone/.
We also assume that $M_{\Gamma,\CC,\fD}$ is  \modblrkone/
for every edge gadget $\Gamma$.
%%% JYC
%(Note that every nonzero entries of $\CC$ is a root of unity,
%\CC is \multblrkone/ iff \modblrkone/. but thst doesn't seem to extend to 
%to gadgets $M_{\CC,\fD}$ . ???
By $(\calL)$, we have
$\Lambda_r=\prod_{i=1}^s \Lambda_{r,i}$ {for every $r\in \cS$ and} 
$\Delta_r=\prod_{i=1}^s \Delta_{r,i}$ {for every $r\in \cT$,}
where both $\Lambda_{r,i}$ and $\Delta_{r,i}$ are cosets
  in $\mathbb{Z}_{\qq_i}$.

The purpose of Section 11 in~\cite{Cai-Chen-Lu-2013}
is to prove Theorem 5.9 in~\cite{Cai-Chen-Lu-2013}.
We have the corresponding statement, Theorem~\ref{bi-step-5-ours}, which is
to establish a quadratic structure of
the nonzero entries of the diagonal matrices $\DD^{[r]}$,
more specifically, for the first  half entries of
$\DD^{[r]}$ when $r \in \cS$, and the second half entries of
$\DD^{[r]}$ when $r \in \cT$.
Below we prove ($\calD_1$) and $(\calD_3)$.
The other parts, i.e., $(\calD_2)$ and
  $(\calD_4)$,  of Theorem~\ref{bi-step-5-ours}  are proved similarly.

Let $\GG$ denote the $|\Lambda_r|\times m$ submatrix
  of $\FF$ whose row set is $\Lambda_r\subseteq \zqt$.
The following lemma is Lemma~11.1 from~\cite{Cai-Chen-Lu-2013},
and remains valid in our setting with the same proof.
Let $n=|\Lambda_r|\ge 1$.
\begin{lemma}[Lemma 11.1 from~\cite{Cai-Chen-Lu-2013}]\label{useful1-ours}
There exist
  vectors $\fb_0=\00,\fb_1,\ldots,\fb_{n-1}\in \zqt$ such that
\begin{enumerate}[wide]
%\item $\{\GG_{*,\fb_i},i\in [0:n-1]\}$ are linearly independent;
%%% this is implied by the next item
\item $\{\GG_{*,\fb_i}:i\in [0:n-1]\hspace{0.02cm}\}$ forms an orthogonal basis;
\item for all $\bb\in \zqt$, 
$\exists\hspace{0.05cm}i\in [0:n-1]$ and $\alpha\in \mathbb{Z}_M$
  such that $\GG_{*,\bb}=\omega_M^\alpha\cdot \GG_{*,\fb_i}$; 
\item let $A_i$ be the set of $\bb\in \zqt$ s.t.
  $\GG_{*,\bb}$ is linearly dependent on $\GG_{*,\fb_i}$; 
  then \[|A_0|=|A_1|=\ldots=|A_{n-1}|= {m}/{n}. \]
\end{enumerate}
\end{lemma}
A symmetric statement also holds for the $m \times |\Delta_r|$
  submatrix of $\FF$ whose column set is $\Delta_r$, where we replace
  $n = |\Lambda_r|$ by $|\Delta_r|$.
% which could be different.

\begin{figure}
\center
\includegraphics[height=12cm]{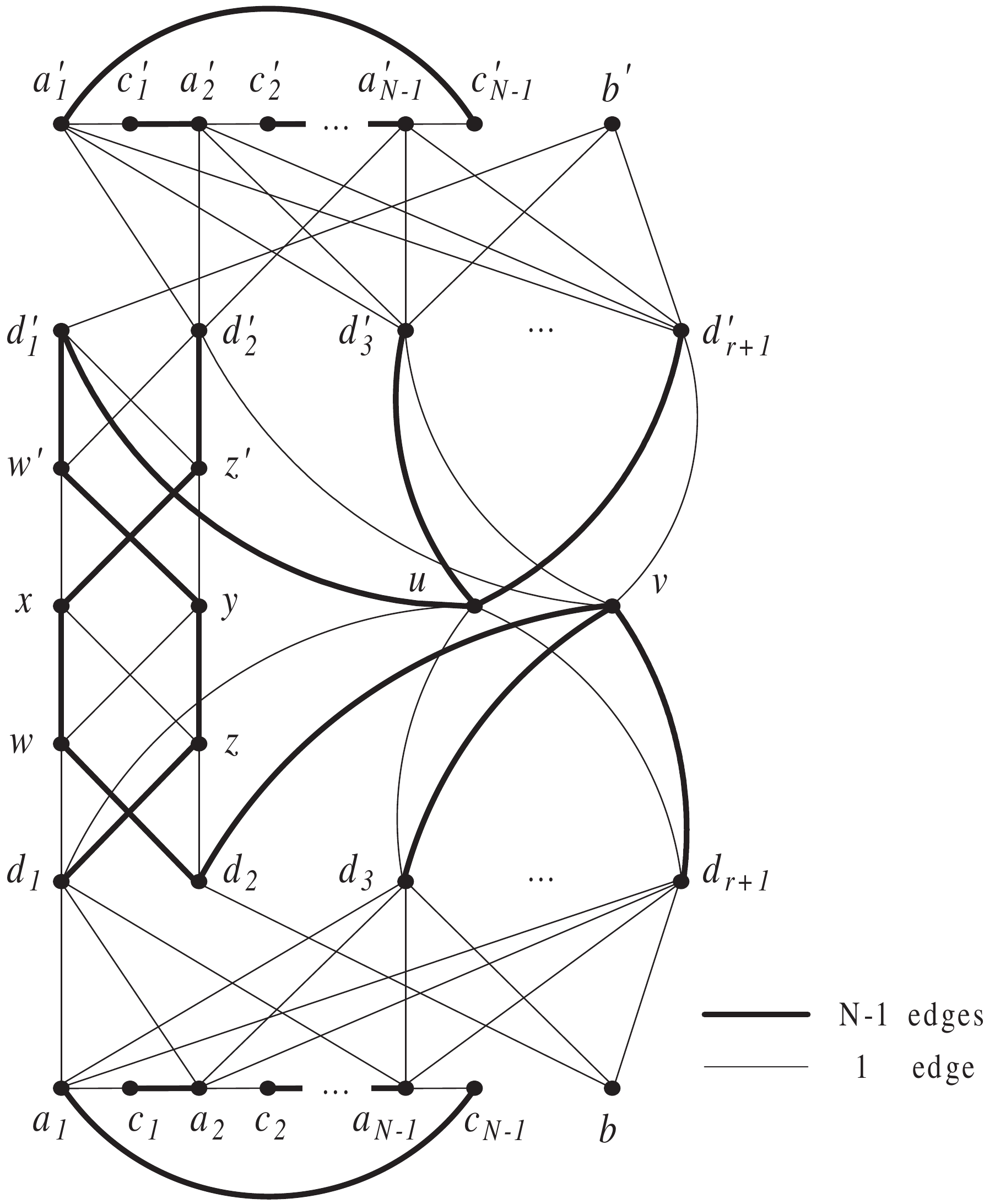}%\vspace{0.25cm}
\caption{The gadget $\Gamma^{[1]}$ with distinguished vertices
$u$ and $v$.}\label{figure_8}
\end{figure}

%{\color{red} (altervative start)}

We use Argument~\ref{meta3} ({\sl Meta}$_3$)
adapting %and adapt
the proof in Section 11 in~\cite{Cai-Chen-Lu-2013}
to prove Theorem~\ref{bi-step-5-ours}. 
The following details can be noted.
The proof in Section~11 of~\cite{Cai-Chen-Lu-2013}
uses an edge gadget construction, one for each $r\in \cS$,
which is described on pp. 996--997.
This  gadget  $\Gamma^{[1]}$ with  distinguished vertices
$u$ and $v$ (in this order), reproduced here in Figure~\ref{figure_8},
was Figure~11.1 in~\cite{Cai-Chen-Lu-2013}.
The actual gadget used in the proof is to take two copies of $\Gamma^{[1]}$,
and identify the respective copies of the vertices $u, v, x, y$.
We will call it $\Gamma = \Gamma^{[2]}$, and will use it in the
proof of Theorem~\ref{bi-step-5-ours}.
% as well.
In~\cite{Cai-Chen-Lu-2013} for the obtained matrix $\AA$ 
corresponding to $\Gamma$
(again this  $\AA$ is not to be confused with the beginning matrix  $\AA$
that gives rise to the pair $(\CC, \fD)$),
it was noted that $Z_{\AA}(G) = Z_{\CC, \fD}(G^{[2]})$,
for all $G$, where $G^{[2]}$ is the graph obtained by replacing every 
edge of $G$ by $\Gamma$.
This uses the fact that in $\Gamma$ the distinguished vertices
are each of degree divisible by $N$, and
 $\DD^{[0]} = \II_{2 m}$. %, although in our reasoning the property of non-\modblrkone/ alone is sufficient.  %%% correction
Thus in terms of signature matrices,
this new $\AA = \gm_{\Gamma, \CC, \fD}$ for  our $(\CC, \fD)$.
Then  just as  in~\cite{Cai-Chen-Lu-2013}   we either get the desired properties,
or for the edge gadget $\Gamma$, the signature matrix $\gm_{\Gamma, \CC, \fD}$ 
is not \modblrkone/.
The key facts in this derivation in~\cite{Cai-Chen-Lu-2013}  are the expression for
$A_{(0, \uu), (0, \vv)}$ in equation (11.3) and
the assertion in (11.5) on page 999.
Using Argument~\ref{meta3} ({\sl Meta}$_3$), we derive the same
equations here.

The  rest of the proof (on pp. 997--1001)
from~\cite{Cai-Chen-Lu-2013} all go through.
% This allows the rest of the proof (on pp. 997--1001)
%from~\cite{Cai-Chen-Lu-2013} to go through.
In particular, we may repeat the proofs of Lemma 11.2 from~\cite{Cai-Chen-Lu-2013} 
and finally prove Theorem~\ref{bi-step-5-ours}.

%% file: technical-modifications-bounded-degree-gadget-based-inside-meta.tex
\subsection{Step 2.4}%\label{sec:decomposition}

After Step 2.3, we obtain pairs $(\CC,\fD)$ and $(\pur \CC,\fD)$ that satisfy conditions
  ({\sl Shape}$_1$)--({\sl Shape}$_6$).
By ({\sl Shape}$_2$), we have
\[
\CC=\left(\begin{matrix}\hspace{-0.15cm}\00& \FF\\ \FF^T&\00\end{matrix}\right)=
\left(\begin{matrix}\hspace{-0.15cm}\00&\MM\otimes
  \HH\\ (\MM\otimes \HH)^T&\00\end{matrix}\right),
\]
where $\MM$ is an $s\times t$ matrix of rank $1$, $M_{i,j}=\mu_i\nu_j$,
  and $\HH$ is the $h\times h$ matrix defined in ({\sl Shape}$_2$).
By ({\sl Shape}$_2$), we also have
\[
\pur \CC=\left(\begin{matrix}\hspace{-0.15cm}\00& \pur \FF\\ \pur \FF^T&\00\end{matrix}\right)=
\left(\begin{matrix}\hspace{-0.15cm}\00&\pur \MM\otimes
  \HH\\ (\pur \MM\otimes \HH)^T&\00\end{matrix}\right),
\]
where $\pur \MM$, the purification of $\MM$
obtained by respectively going from   $\boldsymbol{\mu}$ and $\boldsymbol{\nu}$
 to $\boldsymbol{\pur \mu}$ and $\boldsymbol{\pur \nu}$
 is also an $s\times t$ matrix of rank $1$, $\pur M_{i,j}=\pur \mu_i\pur \nu_j$.
By ({\sl Shape}$_5$) and ({\sl Shape}$_6$), we have
  {for every $r\in [0:N-1]$}
\[
\DD^{[r]}=\left(\begin{matrix}
\DD^{[r]}_{(0,*)} &  \\ & \DD^{[r]}_{(1,*)}
\end{matrix}\right)=\left(\begin{matrix}
\KK^{[r]}_{(0,*)}\otimes \LL^{[r]}_{(0,*)}& \\
 & \KK^{[r]}_{(1,*)}\otimes \LL^{[r]}_{(1,*)}
\end{matrix}\right).
\]
Every entry in $\LL^{[r]}$ either is $0$ or has norm $1$
  and $\LL^{[0]}$ is the $2h\times 2h$ identity matrix.

Thus $\CC$, $\pur \CC$ and $\DD^{[r]}$ are all expressed
in a tensor product form. We now define
%Using these matrices, we define 
three new $\eval$ problems by defining
three pairs $(\CC',{\frak K})$, $(\pur \CC',{\frak K})$ and
  $(\CC'',{\frak L})$.
These give a decomposition of
both  problems $\eval(\CC,\fD)$ and $\eval(\pur \CC, \fD)$
as a respective tensor product of
two problems, an \emph{outer} problem  $\eval(\CC',{\frak K})$
for  $\eval(\CC,\fD)$,
and  $\eval(\pur \CC',{\frak K})$ for  $\eval(\pur \CC, \fD)$,
% respectively,
and a common \emph{inner} problem $\eval(\CC'',{\frak L})$. %{\color{red} (JYC: I hope problem are understood as frameworks?)}

More specifically,
%in a  which give rise to three problems
we define
  $\eval(\CC',{\frak K})$, $\eval(\pur \CC',{\frak K})$ and $\eval(\CC'',{\frak L})$
as follows.
First, $\CC'$ is the bipartization of $\MM$, so it is $(s+t)\times (s+t)$,
  and $\frak K$ is a sequence of $N$ diagonal matrices 
  also of this size: $(\KK^{[0]},\ldots,\KK^{[N-1]})$.
Also, $\pur \CC'$ is the bipartization of $\pur \MM$, so it is $(s+t)\times (s+t)$.
Second, $\CC''$ is the bipartization of $\HH$, and it is $2h\times 2h$,
  and $\frak L$ is the sequence of $N$ diagonal matrices: $(\LL^{[0]},\ldots,
  \LL^{[N-1]})$.
Next, we prove a lemma that essentially reduces 
 $\eval(\CC, \fD)$
to its inner problem  $\eval(\CC'', \fL)$.
This is a weakened version of  Lemma~8.24 in~\cite{Cai-Chen-Lu-2013},
and it will only be used in our tractability proof.
(The proof for \#P-hardness will use a gadget version, 
Corollary~\ref{cor:hahaclaim-ours-edge-gadget-bip-same-different-pur} 
following Claim~\ref{hahaclaim-ours-edge-gadget-bip-same-pur}.)
The proof of Lemma~8.24 in~\cite{Cai-Chen-Lu-2013} uses the so-called
second pinning lemma (Lemma~4.3 of~\cite{Cai-Chen-Lu-2013}, p.~938).
That proof is nonconstructive, and 
we do not have a constructive proof.
It also uses the third  pinning lemma 
(Corollary 8.4 from~\cite{Cai-Chen-Lu-2013}, p.~954). To maintain constructivity
in this paper, we devise an alternative route, without using these
pinning lemmas of~\cite{Cai-Chen-Lu-2013}.
%
%%%The following lemma will be used in the tractability part. {\color{red} (JYC: we already said: "This is a weakened ..." above, so I'd remove this sentence)}
%
%which is a weakened version of Claim~8.24 in~\cite{Cai-Chen-Lu-2013}
%and which will be used in the tractability part.
\begin{lemma}[weakened form of Lemma 8.24 from~\cite{Cai-Chen-Lu-2013}]\label{jajajaja-ours}
$\EVAL(\CC, \fD) \le \EVAL^\leftrightarrow(\CC'', \fL)$.
Moreover,
$\EVAL^\rightarrow(\CC, \fD) \le \EVAL^\rightarrow(\CC'', \fL)$ and
$\EVAL^\leftarrow(\CC, \fD) \le \EVAL^\leftarrow(\CC'', \fL)$.
%$Z^\rightarrow_{\CC', \fK}$ and $Z^\leftarrow_{\CC', \fK}$
The same is true for $(\pur \CC, \fD)$ replacing  $(\CC, \fD)$.
% $(\CC', \fL)$ are replaced by $(\pur \CC, \fD)$, $(\pur \CC', \fL)$, respectively.
%are computible in polynomial time.
%Next,
%$\EVAL^\rightarrow(\CC, \fD) \le \EVAL^\rightarrow(\CC'', \fL)$ and
%$\EVAL^\leftarrow(\CC, \fD) \le \EVAL^\leftarrow(\CC'', \fL)$.
%As a result, $\EVAL(\CC, \fD) \le \EVAL^\leftrightarrow(\CC'', \fL)$.
\end{lemma}
%\begin{lemma}[weakening of Claim~8.25 from~\cite{Cai-Chen-Lu-2013}]\label{jajajaja-ours}
%$Z^\rightarrow_{\CC', \fK}$ and $Z^\leftarrow_{\CC', \fK}$
%are computible in polynomial time.
%As a result, the problems $Z^\rightarrow_{\CC, \fD}$ and $Z^\leftarrow_{\CC, \fD}$
%are correspondingly reducible to $Z^\rightarrow_{\CC'', \fL}$ and $Z^\leftarrow_{\CC'', \fL}$
%in polynomial time
%and so $\eval(\CC, \fD)$ can be computed
%from the union of $Z^\rightarrow_{\CC'', \fL}$ and $Z^\leftarrow_{\CC'', \fL}$ in polynomial time.
%%(using a single oracle query on the input pair $(G, u^*)$
%%where $G$ is a connected graph and $u^* \in V(G)$ (not needed))
%\end{lemma}
\begin{proof}
We only prove for $(\CC,\fD)$; the proof for $(\pur \CC,\fD)$ is the same.
Let $G$ be a \emph{connected} undirected graph and let $u^*$ be one of its vertices.
Then by Property~\ref{verytrivial-ours} and Lemma~\ref{tensorproduct-ours}, we have
\begin{align*}
&Z_{\CC,\fD}(G)=Z_{\CC,\fD}^{\rightarrow}(G,u^*)+Z_{\CC,\fD}^{\leftarrow}(G,u^*),\\
&Z_{\CC,\fD}^{\rightarrow}(G,u^*)=
  Z_{\CC',{\frak K}}^{\rightarrow}(G,u^*)\cdot Z_{\CC'',
  {\frak L}}^{\rightarrow}(G,u^*),\ \ \ \ \text{and}\\
&Z_{\CC ,\fD }^{\leftarrow}(G,u^*)=
  Z_{\CC',{\frak K}}^{\leftarrow}(G,u^*)\cdot 
  Z_{\CC'',{\frak L}}^{\leftarrow}(G,u^*).
\end{align*}
As $\MM$ has rank $1$, both $Z_{\CC',{\frak K}}^{\rightarrow}$
  and $Z_{\CC',{\frak K}}^{\leftarrow}$ can be computed in polynomial time.
We only prove for $Z_{\CC',{\frak K}}^{\rightarrow}$ here.
If $G$ is not bipartite, $Z_{\CC',{\frak K}}^{\rightarrow}(G,u^*)$
  is trivially $0$;
otherwise let $U\cup V$ be the vertex set of $G$, $u^*\in U$, and every edge $uv\in E$ has
  one vertex $u$ from $U$ and one vertex $v$ from $V$.
Let $\Xi$ denote the set of assignments $\xi$ which map $U$ to $[s]$ and
  $V$ to $[t]$. Then (note that we use
  $\KK^{[r]}$ to denote $\KK^{[r\bmod N]}$ for any $r\ge N$)
\begin{align*}
Z_{\CC',{\frak K}}^{\rightarrow}(G,u^*) &= \sum_{\xi\in \Xi}
  \left(\prod_{uv\in E} \mu_{\xi(u)}\cdot \nu_{\xi(v)}\right) \left(\prod_{u\in U}
  K^{[\text{deg}(u) ]}_{(0,\xi(u))}\right)\left(
  \prod_{v\in V} K^{[\text{deg}(v) ]}_{(1,\xi(v))}\right)\\[0.1ex]
  &= 
  \prod_{u\in U} \left(\sum_{i\in [s]} (\mu_i)^{\text{deg}(u)}
  \cdot K^{[\text{deg}(u) ]}_{(0,i)}\right)
  \times \prod_{v\in V} \left(\sum_{j\in [t]} (\nu_j)^{\text{deg}(v)}
  \cdot K^{[\text{deg}(v) ]}_{(1,j)}\right) ,
\end{align*}
which can be computed in polynomial time.
The lemma then follows.
\end{proof}
In~\cite{Cai-Chen-Lu-2013}, the proof of Lemma~8.24
uses Claim 8.25. We now prove a version of Claim 8.25
for edge gadgets, whereby we can circumvent Lemma~8.24 of~\cite{Cai-Chen-Lu-2013}.
%We also need to prove a version
%of Claim 8.25 (used in the proof of Claim 8.24)
%for edge gadgets that will be used
%in the \#P-hardness part.
%This way we will not need
%to prove a version of Claim 8.24 from~\cite{Cai-Chen-Lu-2013}
%for the \#P-hardness part.
Our version of Claim 8.25 is about the purified $(\pur \CC', \fD)$.
%However, we will prove a version
%corresponding to the $(\pur \CC', \fL')$
%rather than $(\CC', \fL)$:
%in this case
%the proof from~\cite{Cai-Chen-Lu-2013}
%can be easily adapted because
%$(\CC', \fL)$ (in the notation of~\cite{Cai-Chen-Lu-2013})
%is already purified.
%{\color{red} (basically, say this resonates with CCL)}
%%%
%%%% what ?  there is no meaning of $(\CC', \fL)$ , $(\pur \CC', \fL')$...

%{\color{red} back to the hardness part}

%Our goal is to avoid the reductions from
%the proof of Claim~8.24 in~\cite{Cai-Chen-Lu-2013}.
%For this we will formulate and prove a version
%of Claim~8.25 from~\cite{Cai-Chen-Lu-2013}
%in the nonpurified setting for edge-gadgets.

%{\color{red} (purified version, adapted with minimal changes)}
%
%For convenience, in Claims~\ref{hahaclaim-ours-edge-gadget-bip-same-pur} and~\ref{hahaclaim-ours-edge-gadget-bip-different-pur} below, we identify an edge gadget with its underlying graph. {\color{red} (not sure if we need to say it)}
%
%
Let  $\Gamma = (U \cup V, E)$
be a  connected, bipartite edge gadget
with distinguished vertices $\firstdv, \seconddv$ (in this order).
In Claim~\ref{hahaclaim-ours-edge-gadget-bip-same-pur} we assume
$\firstdv, \seconddv$ belong to the same bipartite part,
say, $\firstdv, \seconddv \in U$. Thus, $\gm_{\Gamma, \pur \CC, \fD}$ has the form
\[
\gm_{\Gamma, \pur \CC, \fD} = \begin{pmatrix} \subgm^{[0]}_{\Gamma, \pur \CC, \fD} &  {\bf 0}\\
{\bf 0} & \subgm^{[1]}_{\Gamma, \pur \CC, \fD} \end{pmatrix}.
\]
The proof is easily adapted to the case if
$\firstdv \in U$ and $\seconddv \in V$.
\begin{claim}[modification of Claim 8.25 from~\cite{Cai-Chen-Lu-2013}]\label{hahaclaim-ours-edge-gadget-bip-same-pur} %\label{clm:8.25-for-bd-egde-gadget}
For any $\Gamma$ as above,
there is a connected, bipartite edge gadget $\Gamma' = (U' \cup V', E')$
with distinguished vertices $\firstdv, \seconddv$ (in this order)
such that $\firstdv, \seconddv \in U \subset U'$ and $V \subset V'$,
and 
\begin{equation}\label{gjgi-edge-gadget-bip-same-pur}
\gm_{\Gamma', \CC'', \fL} = h^{|U \cup V| - 2} \cdot \gm_{\Gamma, \CC'', \fL}.
\end{equation}
For any $i \in \{0, 1\}$,
if $\subgm^{[i]}_{\Gamma,  \CC'', \fL} \ne \00$,
then
we can choose $\Gamma'$ so that
$\subgm^{[i]}_{\Gamma', {\pur \CC'}, \fK}$ has no zero entries.
%In particular, $\gm_{\Gamma', {\pur \CC'}, \fK} \ne \00$.
%(so $\subgm^{[i]}_{\Gamma', \pur \CC', \fK} \ne 0$),
%or we can show that $\subgm^{[i]}_{\Gamma, \CC'', \fL} = 0$.
%
%{\color{red}
%The additional notation is specified as follows.
%We have
%\[
%\gm_{\Gamma, \CC'', \fL} = \begin{pmatrix} \subgm_{\Gamma, \CC'', \fL}^{[1, 1]} \\ & \subgm_{\Gamma, \CC'', \fL''}^{[2, 2]} \end{pmatrix}
%\]
%where $\subgm_{\Gamma, \CC'', \fL}^{[1, 1]}$, $\subgm_{\Gamma, \CC'', \fL}^{[2, 2]}$ are $h \times h$ matrices,
%and if $G'$ is constructed, then
%\[
%\gm_{\Gamma', \CC'', \fL} = \begin{pmatrix} \subgm_{\Gamma', \CC'', \fL)}^{[1, 1]} \\ & \subgm_{\Gamma', \CC'', \fL}^{[2, 2]} \end{pmatrix} 
%\]
%where $\subgm_{\Gamma', \CC'', \fL}^{[1, 1]}$, $\subgm_{\Gamma', \CC'', \fL}^{[2, 2]}$
%are $h \times h$ matrices, and
%\[
%\gm_{\Gamma', \pur \CC', \fK} = \begin{pmatrix} \subgm_{\Gamma', \pur \CC', \fK}^{[1, 1]} \\ & \subgm_{\Gamma', \pur \CC', \fK}^{[2, 2]} \end{pmatrix}
%\]
%where $\subgm_{\Gamma', \pur \CC', \fK}^{[1, 1]}$, $\subgm_{\Gamma', \pur \CC', \fK}^{[2, 2]}$
%are $s \times s$ and $t \times t$ matrices, respectively.
%(must be removed, defined it long ago)}
\end{claim}
\begin{proof}
%We prove the statement only for $i = 0$. The case $i = 1$ is the same.
%can be dealt with similarly.
%
Let $U_0 = U \setminus \{\firstdv, \seconddv\}$.
We construct a bipartite  edge gadget  $\Gamma'=(U'\cup V',E')$
with parameters $\ell_u\in [s], \ell_v\in [t]$ to be determined shortly,
for all $u \in U_0, v \in V$. First,
$U'=U\cup \widehat{V}$ {and} $V'=V\cup \widehat{U_0}$, {where
  $\widehat{V}= \{\widehat{v}:v\in V \}$  and
  $\widehat{U_0}= \{\widehat{u}:u\in U_0 \}$.}
We make $\firstdv, \seconddv$ the first and second distinguished vertices
in $\Gamma'$, respectively.
The edge set $E'$ contains $E$ over $U\cup V$, and the following edges:
$\ell_uN$ parallel edges between $u$ and $\widehat{u}$, for every $u\in U_0$,
  and $\ell_vN$ parallel edges between $v$ and $\widehat{v}$, for every $v\in V$.
Clearly, $\Gamma'$ is a connected and bipartite edge gadget. %%%added this sentence
For each $u\in U$ (and $v\in V$), we use $r_u$ (and $r_v$)
  to denote its degree in $\Gamma$.
Then in $\Gamma'$ the degrees of $u\in U$ (and $v \in V$) are
congruent to $r_u$ (and $r_v$) $\bmod\, N$, while
the degrees of $\widehat{u} \in \widehat{U_0}$ and $\widehat{v} \in \widehat{V}$
are all congruent to $0$ $\bmod\, N$.

To prove (\ref{gjgi-edge-gadget-bip-same-pur}),
we take any $x, y \in [h]$. Let $\eta$ be an assignment that maps
  $U$ and $V$ to $[h]$ such that $\eta(\firstdv) = x$ and $\eta(\seconddv) = y$.
Given $\eta$, let $\Xi$ denote the set of assignments $\xi$ over $U' \cup V'$
that map $U'$ and $V'$ to $[h]$ and that satisfy $\xi(u) = \eta(u)$ for all $u \in U$
(so $\xi(\firstdv) = x$ and $\xi(\seconddv) = y$) and $\xi(v) = \eta(v)$ for all $v \in V$.
Recall that in the definition of $\text{wt}_{\Gamma', \CC'',{\frak L}}(\xi)$
(see Defintion~\ref{def:M(Gamma,C,D)})
 the vertex weights of $\firstdv$ and $\seconddv$ are excluded. %,
We have %we
\begin{align*}
\sum_{\xi\in \Xi}\text{wt}_{\Gamma', \CC'',{\frak L}}(\xi) ={}&%&=
  \sum_{\xi\in \Xi}\hspace{0.1cm}\left(\prod_{uv\in E}H_{\eta(u),\eta(v)}
  \prod_{u\in U_0} (H_{\eta(u),\xi(\widehat{u})})^{\ell_uN}
  \prod_{v\in V} (H_{\xi(\widehat{v}),\eta(v)})^{\ell_vN}\right)\\
  &\times %&\hspace{4em}
  \left(\prod_{u\in U_0} L^{[r_u ]}_{(0,\eta(u))}
  L^{[0]}_{(1,\xi(\widehat{u}))}\right)
  \left(\prod_{v\in V} L^{[r_v ]}_{(1,\eta(v))}
  L^{[0]}_{(0,\xi(\widehat{v}))}\right)\\
={}&\sum_{\xi\in \Xi}\text{wt}_{\Gamma, \CC'',{\frak L}}(\eta) %&=
  \hspace{0.06cm}=\hspace{0.06cm}h^{|\widehat{U_0}\cup
  \widehat{V}|}\cdot \text{wt}_{\Gamma, \CC'',{\frak L}}(\eta).
\end{align*}
The second equation uses the fact that the entries of $\HH$ are powers of
  $\oo_N$ (thus $(H_{i,j})^N=1$) and $\LL^{[0]}$ is the identity matrix.
It follows that the $(x,y)${th} entry
of the signature matrix is
 \[\subgm_{\Gamma', \CC'', \fL}^{[0]}(x, y) 
= h^{|U \cup V| -2} \cdot \subgm_{\Gamma, \CC'', \fL}^{[0]}(x, y)\]
for each $x, y \in [h]$.
The same proof works for
$\subgm_{\Gamma', \CC'', \fL}^{[1]} 
= h^{|U \cup V| -2}  \cdot \subgm_{\Gamma, \CC'', \fL}^{[1]}$
by exchanging $\LL^{[r]}_{(0,*)}$ with $\LL^{[r]}_{(1,*)}$
in the above derivation.
This proves (\ref{gjgi-edge-gadget-bip-same-pur}). %{\color{red} (JYC: not enough, we also need to flip the subscript in the entries of $\HH$, what do I add?)}

Now let  $\subgm^{[i]}_{\Gamma,  \CC'', \fL} \ne \00$
for some $i \in \{0, 1\}$.
%assume $\gm_{\Gamma, {\pur \CC}, \fD} \ne \00$,
%then $\subgm^{[i]}_{\Gamma, {\pur \CC}, \fD} \ne \00$
% for either $i=0$ or $i=1$.
Without loss of generality assume $\subgm^{[0]}_{\Gamma,  \CC'', \fL} \ne \00$.
%%%
%Suppose $U_0\ne \emptyset$. {\color{red} (JYC: I want to cross it out because later we don't suppose $V \ne \emptyset$ later)}
%%%
For any $u \in U_0$,  
if $\KK^{[r_u]}_{(0,*)} = \00$, then by condition ({\sl Shape}$_6$),
$\LL^{[r_u]}_{(0,*)} = \00$ as well. Then
 $\subgm^{[0]}_{\Gamma,  \CC'', \fL} =\00$, a contradiction.
So  $\KK^{[r_u]}_{(0,*)} \ne \00$ for all  $u \in U_0$.
Similarly $\KK^{[r_v]}_{(1,*)} \not = \00$, for all $v \in V$.
(Note that these statements are also vacuously true if $U_0 = \emptyset$
or $V = \emptyset$.)
Next we wish to choose $\ell_u \in [s]$ for each $u \in U_0$,
such that
\begin{equation}\label{opop-ours-for-uinU_0}
\sum_{i\in [s]} \pur \mu_i^{\ell_uN+r_u}\cdot  K^{[r_u ]}_{(0,i)}\ne 0.
\end{equation}
If for all $\ell_u \in [s]$ an equality holds in (\ref{opop-ours-for-uinU_0}),
this is a full-ranked Vandermonde system 
 since $\pur \mu_1>\ldots>\pur \mu_s>0$, which would force
 $\KK^{[r_u]}_{(0,*)} = \00$, a contradiction.
Thus for each $u \in U_0$, some  $\ell_u \in [s]$ can be chosen so that
(\ref{opop-ours-for-uinU_0}) holds.
(It is also vacuously true if $U_0 = \emptyset$.)
Similarly, for each $v \in V$, we can  choose $\ell_v \in [t]$
such that
\begin{equation}\label{opop-ours-for-vinV}
\sum_{i\in [t]} \pur \nu_i^{\ell_vN+r_v}\cdot  K^{[r_v ]}_{(1,i)}\ne 0.
\end{equation}
(Again, (\ref{opop-ours-for-vinV}) is vacuously true if $V = \emptyset$.)

We use these parameters $\ell_u, \ell_v$  in the construction of
$\Gamma'$.

%Clearly, $\Gamma'$ is a connected and bipartite edge gadget with 
%$\firstdv, \seconddv \in U \subset U'$ and $V \subset V'$.
%The degrees of $\firstdv, \seconddv$ are respectively $r_{\firstdv}, r_{\seconddv}$,
%the degree of $u \in U_0$ (or $v \in V$) is $r_u + \ell_u N$ (or $r_v + \ell_v N$),
%and the degree of $\widehat u$ (or $\widehat v$) is $\ell_u N$ (or $\ell_v N$).
%We now use $\Gamma'$ to prove Claim \ref{hahaclaim-ours-edge-gadget-bip-same-pur}. %\vspace{0.01cm}

Next we prove $\subgm^{[0]}_{\Gamma', {\pur \CC'}, \fK}$ has no zero entries.
For each $x, y \in [s]$, we have
(the sum is over all $\xi$ that map $U'$ to $[s]$, $V'$ to $[t]$
and satisfy $\xi(\firstdv) = x$, $\xi(\seconddv) = y$)
\begin{align*}
\subgm_{\Gamma', \pur \CC', \fK}^{[0]}(x, y)={}& %&=
  \sum_{\xi}
\left(\prod_{uv\in E} \pur M_{\xi(u),\xi(v)}
\prod_{u\in U_0} \pur M_{\xi(u),\xi(\widehat{u})}^{\ell_u N}\prod_{v\in V}
  \pur M_{\xi(\widehat{v}),\xi({v})}^{\ell_vN} \right)\\ 
  &\times \left( %&\hspace{4em}
\prod_{u\in U_0} K^{[r_u ]}_{(0,\xi(u))}
  K^{[0]}_{(1,\xi(\widehat{u}))}\right)\left(
\prod_{v\in V} K^{[r_v ]}_{(1,\xi(v))}
  K^{[0]}_{(0,\xi(\widehat{v}))}\right)\\ 
={}&%& = 
 \pur \mu_x^{r_{\firstdv}} \pur \mu_y^{r_{\seconddv}} \prod_{u\in U_0}\left(\sum_{i\in [s]} \pur \mu_i^{\ell_uN+r_u}\cdot
  K^{[r_u ]}_{(0,i)} \right)
  \prod_{v\in V}\left(\sum_{i\in [t]} \pur \nu_i^{\ell_vN+r_v}\cdot
  K^{[r_v ]}_{(1,i)}\right)\\ &\times %&\hspace{4em}
\prod_{\widehat{u}\in \widehat{U_0}}\left(\sum_{i\in [t]} \pur \nu_i^{\ell_uN} \cdot K^{[0]}_{(1,i)}
\right) \prod_{\widehat{v}\in \widehat{V}}\left(\sum_{i\in [s]}
  \pur \mu_i^{\ell_vN} \cdot K^{[0]}_{(0,i)}\right). 
\end{align*}
It is nonzero 
for each $x, y \in [s]$: The first two factors are nonzero because $\pur \mu_x, \pur \mu_y > 0$;
the middle two factors are nonzero because of
  the way we picked $\ell_u$ and $\ell_v$;
the latter two factors are nonzero because $\pur \mu_i,\pur \nu_i>0$,
  and by ({\sl Shape}$_6$), every entry of $\KK^{[0]}$ is a positive integer. 
(The middle two factors are also nonzero even in the case
 $U_0 = \emptyset$
or $V = \emptyset$, as a product over an empty index set is $1$. The same is true for the last two factors.)

If  $\subgm^{[1]}_{\Gamma,  \CC'', \fL} \ne \00$,
then 
$\subgm^{[1]}_{\Gamma', {\pur \CC'}, \fK}$ has no zero entries.
This can be proved by exchanging 
$\KK^{[r]}_{(0,*)}$ with $\KK^{[r]}_{(1,*)}$
in (\ref{opop-ours-for-uinU_0}) and in (\ref{opop-ours-for-vinV}). %{\color{red} (JYC: not enough, we also need to swap $\mu_i$ and $\nu_j$, what do I add?)}
\end{proof}

\begin{corollary}\label{cor:hahaclaim-ours-edge-gadget-bip-same-different-pur}
Let $\Gamma$ be an edge gadget such that $\gm_{\Gamma, \CC'', \fL}$
is not \multblrkone/ (not \modblrkone/).
Then 
%we can construct
there is a connected edge gadget $\Gamma'$ %changed from an edge gadget
% in polynomial time 
such that $\gm_{\Gamma', \pur \CC, \fD}$
is not \multblrkone/ (not \modblrkone/).
\end{corollary}
\begin{proof}
Let $\firstdv, \seconddv$ be the first
and second distinguished vertices of $\Gamma$, respectively.
% correspondingly.
By Lemma~\ref{lem:edge-gadget-component-nmrk1},
$\firstdv, \seconddv$ lie in the same connected component of $\Gamma$, call it $\Gamma_0$,
and $\gm_{\Gamma_0, \CC'', \fL}$ is not \multblrkone/ (\modblrkone/),
and so without loss of generality we may assume $\Gamma = \Gamma_0$.
%(Here we treat $\Gamma_0$ as an edge gadget with distinguished vertices $\firstdv, \seconddv$ (in this same order).)
It is also immediate that $\Gamma$ is a bipartite edge gadget
for otherwise $\gm_{\Gamma, \CC'', \fL} = 0$
which cannot be  non-\multblrkone/ (non-\modblrkone/).
We consider the case when $\firstdv, \seconddv$ are
in the same bipartite component of $\Gamma$;
the other case is similar.
Since $\gm_{\Gamma, \CC'', \fL}$
is not \multblrkone/ (not \modblrkone/),
by Property~\ref{prop:equiv-direct-sum-mbrk1},
for some $i \in \{0, 1\}$,
$\subgm^{[i]}_{\Gamma, \CC'', \fL}$ 
is not  \multblrkone/ (not  \modblrkone/), which is certainly nonzero.
By Claim~\ref{hahaclaim-ours-edge-gadget-bip-same-pur},
there is a connected bipartite edge gadget $\Gamma'$
%, and for the same $i  \in \{0, 1\}$, $\subgm^{[i]}_{\Gamma', {\CC''}, \fL}$ is not \multblrkone/ (not \modblrkone/)
such that $\subgm^{[i]}_{\Gamma', {\CC''}, \fL}$ is not \multblrkone/ (not \modblrkone/) for the same $i  \in \{0, 1\}$, 
as demonstrated by (\ref{gjgi-edge-gadget-bip-same-pur}).
%In any case, 
Also by Claim~\ref{hahaclaim-ours-edge-gadget-bip-same-pur}, for this $i \in \{0, 1\}$,
$\subgm^{[i]}_{\Gamma', {\pur \CC'}, \fK} \ne \00$.
We have
\[
\subgm^{[i]}_{\Gamma', {\pur \CC}, \fD} =
\subgm^{[i]}_{\Gamma', {\pur \CC'}, \fK} \otimes
\subgm^{[i]}_{\Gamma', {\CC''}, \fL}. % \quad \text{where } i \in \{0, 1\}.
\]
%{\color{red} (JYC: can't use $i$, it's already fixed, do $i \to j$ any alternatives?)
%\[
%\subgm^{[j]}_{\Gamma', {\pur \CC}, \fD} =
%\subgm^{[j]}_{\Gamma', {\pur \CC'}, \fK} \otimes
%\subgm^{[j]}_{\Gamma', {\CC''}, \fL} \quad \text{where/for } j \in \{0, 1\}.
%\]
%}
By Property~\ref{prop:equiv-tensor-product-mbrk1},
$\subgm^{[i]}_{\Gamma', {\pur \CC}, \fD}$ is
not \multblrkone/ (not \modblrkone/).
\end{proof}

\subsection{Step 2.5}\label{step25-ours}

We are almost done with Step 2.
It is (a normalized version of) the inner pair $(\CC'',{\frak L})$
that will be renamed $(\CC, {\frak D})$ that fulfills the requirements
in~Theorem~\ref{bi-step-2-ours}.
The only conditions $(\calU_i)$ that are possibly violated by $(\CC'',{\frak L})$
  are $(\calU_1)$ ($N$ might be odd) and $(\calU_2)$ ($H_{i,1}$ and $H_{1,j}$
  might not be $1$). We deal with $(\calU_2)$ first.

What we will do below is to normalize $\HH$ (in $\CC''$), so that
it becomes a discrete unitary matrix for some positive integer $M$ that divides $N$,
while preserving the following: %{\color{red} (JYC: not sure (don't think) bullets go along with Items plus if we extend, the space is stretched, better use 1), 2), way easier even to cite, see below)}
\begin{itemize}
%\begin{enumerate}[label={$\bullet$ Item \arabic*.}, ref=\arabic*, wide]%[\label={}]
\item\mbox{Item 1.\,}
%\label{item:norm-dir-eval-comp} 
The complexity of $\EVAL^\rightarrow(\CC'', \fL)$, $\EVAL^\leftarrow(\CC'', \fL)$ (and thus of $\EVAL^\leftrightarrow(\CC'', \fL)$);
\item\mbox{Item 2.\,}
%\label{item:norm-mbr1-edge-gadget} 
The validity of the condition $\gm_{\Gamma, \CC'', \fL}$
is or is not \multblrkone/ (\modblrkone/)
where $\Gamma$ is any connected edge gadget.
%\end{enumerate}
\end{itemize}

First, without loss of generality, we may assume $\HH$ satisfies
$H_{1,1}=1$ since otherwise we can divide $\HH$ by $H_{1,1}$, which does
%{\color{red}  not affect the complexity of $\eval(\CC'',{\frak L})$. (old)}
  not affect the desired requirements.
Then we construct the following pair: $(\XX,{\frak Y})$.
  $\XX$ is the bipartization of an $h\times h$ matrix over $\mathbb{C}$,
  whose $(i,j)${th} entry is
%(note that this pair $(\CC,\fD)$ is different from the one that appeared earlier
%  in this section, but is the one we need to prove Theorem \ref{bi-step-2})
$H_{i,j} \overline{H_{1,j}H_{i,1}} = H_{i,j}/(H_{1,j}H_{i,1})$;
  ${\frak Y}$ is a sequence $(\YY^{[0]},\ldots,\YY^{[N-1]})$
  of $2h\times 2h$ diagonal matrices;
$\YY^{[0]}$ is the identity matrix.
  Let %$\calS$ and $\calT$ denote
\[
\calS=\{r\in [0:N-1]:\LL^{[r]}_{(0,*)}\ne \00\}\ \ \
\text{and}\ \ \
\calT=\{r\in [0:N-1]:\LL^{[r]}_{(1,*)}\ne \00\};
\]
then we have
\[\YY^{[r]}_{(0,*)}=\00\ \ \text{for all $r\notin \calS$\ \ \ and}\ \ \
  \YY^{[r]}_{(1,*)}=\00\ \ \text{for all $r\notin \calT$.}\vspace{0.04cm}\]
For each $r\in \calS$ (or $r\in \calT$), by
  ({\sl Shape}$_6$) there must be an
  $a_r\in [h]$ (or $b_r\in [h]$, resp.) such that
the $(0,a_r)$th entry of $\LL^{[r]}$ is $1$
  (or the $(1,b_r)$th entry of $\LL^{[r]}$ is $1$, resp.).
%$$L^{[r]}_{(0,a_r)}=1\ \ \ \ \Big(\text{and $L^{[r]}_{(1,b_r)}=1$, resp.}\Big) $$
Set 
\[
Y^{[r]}_{(0,i)}=L^{[r]}_{(0,i)} \left(\frac{H_{i,1}}{H_{a_r,1}}\right)^r\    \text{for
  all $i\in [h]$;} \ \ 
Y^{[r]}_{(1,j)}=L^{[r]}_{(1,j)} \left(\frac{H_{1,j}}{ H_{1,b_r}}\right)^r\ 
  \text{for all $j\in [h]$.} 
\]
%{\color{red} We show that $\eval(\CC'',{\frak L})\equiv \eval(\XX,{\frak Y})$. 
%For $\eval(\XX,{\frak Y})\le \eval(\CC'',{\frak L})$,
%  we let $G=(U\cup V, E)$ be a connected undirected graph and $u^*$ be a vertex in $U$. (old)}
For the purpose of Item~1,
%~\ref{item:norm-dir-eval-comp},
we show that $\EVAL^\rightarrow(\CC'', \fL) \equiv \EVAL^\rightarrow(\XX, \fY)$.
Let $G=(U\cup V, E)$ be a connected bipartite
 undirected graph and let $u^*$ be a vertex in $U$.
For every $r \in \calS$ (and $r\in \calT$), we use $U_r\subseteq U$
  (and $V_r\subseteq V$, resp.) to denote the set of
  vertices with degree $r\bmod N$.
It is clear that if $U_r\ne \emptyset$ for some $r\notin \calS$
  or if $V_r\ne$  $\emptyset$ for some $r\notin \calT$,
  both $Z_{\CC'',{\frak L}}^\rightarrow(G,u^*)$ and
  $Z_{\XX,{\frak Y}}^\rightarrow(G,u^*)$ are trivially zero.
Otherwise, we have
\[\label{xaxaabove-ours}
Z^{\rightarrow}_{\CC'',{\frak L}}(G,u^*)=\left(\prod_{r\in \calS }
(H_{a_r,1})^{r|U_r|}\right)\left(\prod_{r\in \calT }
(H_{1,b_r})^{r|V_r|} \right)\cdot Z^{\rightarrow}_{\XX,{\frak Y}}(G,u^*).%.&&\text{and}\\
%Z^{\leftarrow}_{\CC'',{\frak L}}(G,u^*)=\left(\prod_{r\in \calS }
%(H_{a_r,1})^{r|V_r|}\right)\left(\prod_{r\in \calT }
%(H_{1,b_r})^{r|U_r|} \right)\cdot Z^{\leftarrow}_{\XX,{\frak Y}}(G)&&
\]
%{\color{red} So the problem of computing $Z_{\XX,{\frak Y}}^\rightarrow$
%  is reducible to computing $Z_{\CC'',{\frak L}}^\rightarrow$.
%By combining it with the Second Pinning Lemma (Lemma \ref{pinning2}),
%  we know that computing $Z^{\rightarrow}_{\XX,{\frak Y}}$ is reducible
%  to $\eval(\CC'',{\frak L})$.
%A similar statement can be proved for $Z_{\XX,{\frak Y}}^\leftarrow$,
%  and it follows that $\eval(\XX,{\frak Y})\le \eval(\CC'',{\frak L}).$ 
%The other direction, $\eval(\CC'',{\frak L})\le \eval(\XX,{\frak Y})$,
%  can be proved similarly. (old)}
So the problem of computing $Z^\rightarrow_{\XX,{\frak Y}}$
  is reducible to computing $Z^\rightarrow_{\CC'',{\frak L}}$ and vice versa.
(Here we use the fact the entries of $\HH$ being powers of $\oo_N$ are all nonzero.)
In other words, $\EVAL^\rightarrow(\CC'', \fL)  \equiv  
\EVAL^\rightarrow(\XX, \fY)$.
Similarly, we can prove $\EVAL^\leftarrow(\CC'', \fL) \equiv \EVAL^\leftarrow(\XX, \fY)$.
We conclude that $\EVAL^\leftrightarrow(\CC'', \fL) \equiv \EVAL^\leftrightarrow(\XX, \fY)$.
This finishes Item~1.

%  for both pairs $(\XX,{\frak Y})$ and $(\CC'',{\frak L})$, computing $Z^\rightarrow$
%  (and $Z^{\leftarrow}$) is reducible to computing  $Z$.
%It then follows that $\eval(\XX,{\frak Y})\equiv \eval(\CC'',{\frak L})$.

Now, for the purpose of Item~2,
%~\ref{item:norm-mbr1-edge-gadget},
 we show that for every connected edge gadget $\Gamma$,
$\gm_{\Gamma, \CC'', \fL}$ is \multblrkone/ (\modblrkone/)
iff $\gm_{\Gamma, \XX, \fY}$ is \multblrkone/ (\modblrkone/).
Clearly, if $\Gamma$ is not bipartite,
then both $\gm_{\Gamma, \CC'', \fL}$ and $\gm_{\Gamma, \XX, \fY}$
are zero matrices and we are done.
Thus we may assume that $\Gamma$ is bipartite.
As in Claim~\ref{hahaclaim-ours-edge-gadget-bip-same-pur},
we identify an edge gadget with its underlying graph.
Let $\Gamma = (U \cup V, E)$ be a connected bipartite edge gadget
with distinguished vertices $\firstdv, \seconddv$ (in this order).
We consider the case when $\firstdv, \seconddv$ lie
in the same bipartite part of $\Gamma$,
say, $\firstdv, \seconddv \in U$.
The case when $\firstdv, \seconddv$ lie in different bipartite parts of $\Gamma$
can be done similarly.

%We establish a relation between $R^{[1, 1]}_{\Gamma(G, \CC'', \fL)}$
%and $R^{[1, 1]}_{\Gamma(G, \XX, \fY)}$.

First, we show that $\subgm^{[0]}_{\Gamma, \CC'', \fL}$ is \multblrkone/ (\modblrkone/)
iff $\subgm^{[0]}_{\Gamma, \XX, \fY}$ is \multblrkone/ (\modblrkone/).
%{\color{red} (JYC: inserted this sentence, because "otherwise we are done" below doesn't make sense, but then we have a little repetition below "It follows that ...")}
%%%
Let $r_1 = \deg(\firstdv)$ and $r_2 = \deg(\seconddv)$.
For every $r \in \calS$ (and $r\in \calT$), we use $U_r\subseteq U \setminus \{\firstdv, \seconddv\}$
  (and $V_r\subseteq V$, resp.) to denote the set of
  vertices with degree $r\bmod N$.
It is clear that if $U_r\ne \emptyset$ for some $r\notin \calS$
  or if $V_r\ne$  $\emptyset$ for some $r\notin \calT$,
  both $\subgm^{[0]}_{\Gamma, \CC'', \fL}$ and
  $\subgm^{[0]}_{\Gamma, \XX, \fY}$ are trivially zero matrices and we are done.
Otherwise, for each $i, j \in [h]$, we have
\[
\subgm^{[0]}_{\Gamma, \CC'', \fL}(i, j) = (H_{i, 1})^{r_1} (H_{j, 1})^{r_2}
\left(\prod_{r \in \calS} (H_{a_r, 1})^{r |U_r|}\right)
\left(\prod_{r \in \calT} (H_{1, b_r})^{r |V_r|}\right) \cdot
\subgm^{[0]}_{\Gamma, \XX, \fY}(i, j),
\]
where $(H_{i, 1})^{r_1}, (H_{j, 1})^{r_2} \ne 0$ (these two extra factors
are due to the fact that vertex weights of $\firstdv, \seconddv$ are not
included in the signature matrices) and %{\color{red} (JYC: which one, $\subgm^{[0]}_{\Gamma, \CC'', \fL}$ or $\subgm^{[0]}_{\Gamma, \XX, \fY}$, we have two, or just say "any of the signature matrices")}
\[
\left(\prod_{r \in \calS} (H_{a_r, 1})^{r |U_r|}\right)
\left(\prod_{r \in \calT} (H_{1, b_r})^{r |V_r|}\right) \ne 0.
\]
It follows that $\subgm^{[0]}_{\Gamma, \CC'', \fL}$ is \multblrkone/ (\modblrkone/)
iff $\subgm^{[0]}_{\Gamma, \XX, \fY}$ is \multblrkone/ (\modblrkone/).
Similarly, $\subgm^{[1]}_{\Gamma, \CC'', \fL}$ is \multblrkone/ (\modblrkone/)
iff $\subgm^{[1]}_{\Gamma, \XX, \fY}$ is \multblrkone/ (\modblrkone/).
From this, by Property~\ref{prop:equiv-direct-sum-mbrk1}, we conclude that $\gm_{\Gamma, \CC'', \fL}$ is \multblrkone/ (\modblrkone/)
iff $\gm_{\Gamma, \XX, \fY}$ is \multblrkone/ (\modblrkone/).
This finishes Item~2.

One can verify that $(\XX,{\frak Y})$ satisfies $(\calU_1)$--$(\calU_4)$,
  except that $N$ might be odd.
In particular, the upper-right $h\times h$ block of $\XX$ is an $M$-discrete
  unitary matrix for some positive integer $M\hspace{0.06cm}|\hspace{0.06cm}N$,
and ${\frak Y}$ satisfies both $(\calU_3)$ and $(\calU_4)$ (which uses the
  fact that every entry of $\HH$ is a power of $\oo_N$).

If $N$ is even, then we are done with Step 2; otherwise we extend
  ${\frak Y}$ to be \[{\frak Y}'=\{\YY^{[0]},
  \ldots,\YY^{[N-1]},\YY^{[N]},\ldots,\YY^{[2N-1]}\}, \]
  where $\YY^{[r]}=\YY^{[r-N]}$, for all $r\in [N:2N-1]$.
%{\color{red} (We have $\eval(\XX,{\frak Y})\equiv \eval(\XX,{\frak Y}')$, since
%$Z_{\XX,{\frak Y}}(G)=Z_{\XX,{\frak Y}'}(G)$, {for all undirected $G$,}
%and the new tuple $((M,2N),\XX,{\frak Y}')$ now satisfies conditions $(\calU_1)$--$(\calU_4)$. (old))}
We have $\eval^\rightarrow(\XX,{\frak Y})\equiv \eval^\rightarrow(\XX,{\frak Y}')$, since
$Z_{\XX,{\frak Y}}(G)=Z_{\XX,{\frak Y}'}(G)$, {for all undirected $G$,}
and similarly 
$\eval^\leftarrow(\XX,{\frak Y})\equiv \eval^\leftarrow(\XX,{\frak Y}')$.
This satisfies Item~1.
%Item~\ref{item:norm-dir-eval-comp}.
We also have $\gm_{\Gamma, \XX, \fY} = \gm_{\Gamma, \XX, \fY'}$
for all edge gadgets $\Gamma$.
This satisfies Item~2.
%.,Item~\ref{item:norm-mbr1-edge-gadget}.
So the new tuple $((M,2N),\XX,{\frak Y}')$ satisfies conditions $(\calU_1)$--$(\calU_4)$.

We state the above results for $(\XX, \fY)$, which is a normalization of
$(\CC'', \fL)$.
\begin{lemma}\label{lem:norm-dir-eval-equiv}
$\EVAL^\rightarrow(\CC'', \fL) \equiv \EVAL^\rightarrow(\XX, \fY)$ and
$\EVAL^\leftarrow(\CC'', \fL) \equiv \EVAL^\leftarrow(\XX, \fY)$.
As a consequence, $\EVAL^\leftrightarrow(\CC'', \fL) \equiv \EVAL^\leftrightarrow(\XX, \fL)$.
\end{lemma}

\begin{lemma}\label{lem:norm-mbr1-edge-gadget}
For any connected edge gadget $\Gamma$, 
we have $\gm_{\Gamma, \CC'', \fL}$ is \multblrkone/ (\modblrkone/)
iff $\gm_{\Gamma, \XX, \fY}$ is \multblrkone/ (\modblrkone/).
\end{lemma}

%%% commented out
As a corollary, we have the following.
\begin{corollary}\label{cor:norm-mbr1-edge-gadget-all}
Let $\Gamma$ be an edge gadget (not necessarily connected).
If $\gm_{\Gamma, \XX, \fY}$ is not \multblrkone/ (not \modblrkone/),
then there is a connected edge gadget $\Gamma'$
such that $\gm_{\Gamma', \CC'', \fL}$ is not \multblrkone/ (not \modblrkone/).
\end{corollary}
\begin{proof}
Let $\firstdv, \seconddv$ be the first
and second distinguished vertices of $\Gamma$, correspondingly.
By Lemma~\ref{lem:edge-gadget-component-nmrk1},
$\firstdv, \seconddv$ lie in the same connected component $\Gamma_0$ of $\Gamma$
and $\gm_{\Gamma_0, \XX, \fY}$ is not \multblrkone/ (\modblrkone/).
Taking $\Gamma' = \Gamma_0$, we are done by Lemma~\ref{lem:norm-mbr1-edge-gadget}.
\end{proof}
%%%
%\begin{proof}
%%By Lemma~\ref{foo},
%Clearly, the distinguished vertices of $\Gamma$, say $\firstdv, \seconddv$ (in this same order)
%belong to the same connected component
%for otherwise $\rank \gm_{\Gamma, \XX, \fY} \le 1$
%and so $\gm_{\Gamma, \XX, \fY}$ is not \multblrkone/ (and \modblrkone/).
%Let $\Gamma_0$ be this connected component.
%We view $\Gamma_0$ as an edge gadget with the same distinguished vertices $\firstdv, \seconddv$ (in this order).
%Note that other connected components must constribute a nonzero global factor to $\gm_{\Gamma, \XX, \fY}$
%for otherwise $\gm_{\Gamma, \XX, \fY} = 0$ is not \multblrkone/ (and \modblrkone/).
%Then $\gm_{\Gamma_0, \XX, \fY)}$ is clearly not \multblrkone/ (not \modblrkone/)
%so taking $\Gamma' = \Gamma_0$, we are done by Lemma~\ref{lem:norm-mbr1-edge-gadget}.
%
%%{\color{red} (I don't justify the rank argument, by now any able reader must figure it out, right?
%%I don't say $\Gamma'$ can be constructed in polynomial time as it's obvious, also is it better to start with $\Gamma_0$ instead of $\Gamma'$,
%%I guess the following justification is unnecessary: as $\gm_{\Gamma, \XX, \fY} = Z_{\XX, \fY}(\Gamma \setminus \Gamma_0) \gm_{\Gamma_0, \XX, \fY}$ where $\Gamma \setminus \Gamma_0$ as is a union of the connected components of $\Gamma$ except $\Gamma_0$, $\gm_{\Gamma_0, \XX, \fY)}$ is clearly not \multblrkone/ (not \modblrkone/))}
%\end{proof} 

Combining Lemmas~\ref{lem:A-CD-cycl-gen}, \ref{jajajaja-ours} and~\ref{lem:norm-dir-eval-equiv},
we obtain the following corollary.
%Notice that we use the trivial fact $\EVAL(\CC) \le \EVAL^\leftrightarrow(\CC, \fD)$.
\begin{corollary}\label{cor:bi-step-2-reduction}
$\EVAL(\AA) \le \EVAL^\leftrightarrow(\XX, \fY)$.
\end{corollary}
%Because all the reductions have been obtained constructively,
%the final reduction in Corollary~\ref{cor:bi-step-2-reduction} is also constructive.

Next, combining Theorem~\ref{thm:gadget-thickening-non-mult-bl-rk-1},
 Lemma~\ref{lem:A-CD-cycl-gen},
Corollaries~\ref{cor:hahaclaim-ours-edge-gadget-bip-same-different-pur}, \ref{cor:norm-mbr1-edge-gadget-all},
we obtain the following corollary.
Notice that we use the fact that if $\Gamma$ is connected,
then $T_p(\Gamma)$, where $p \ge 1$, is also connected.
\begin{corollary}\label{cor:bi-step-2-edge-gadget-transformation}
Let $\Gamma$ be an edge gadget (not necessarily connected).
If $\gm_{\Gamma, \XX, \fY}$ is not \multblrkone/ (not \modblrkone/),
%then we can construct a connected edge gadget $\Gamma'$ in polynomial time
then there is a connected edge gadget $\Gamma'$
such that $\gm_{\Gamma', \AA}$ is not \multblrkone/.
\end{corollary}

Since $(\XX,\fY)$ already satisfies $(\calU_1)$--$(\calU_4)$,
Corollaries~\ref{cor:bi-step-2-reduction} and~\ref{cor:bi-step-2-edge-gadget-transformation}
finalize the proof of Theorem~\ref{bi-step-2-ours}.

We now rename  $(\CC, \fD)$ to be the
pair $(\XX, \fY)$, which satisfies  
 Theorem~\ref{bi-step-2-ours}.

%% file: tractability.tex
\section{Tractability: Proof of Theorem \ref{tractable-1-ours}}\label{tractabilitysec-ours}

In this section we describe the proof of
Theorem~\ref{tractable-1-ours}.
identical to the tractability proof of Section~12 
in~\cite{Cai-Chen-Lu-2013}. This technical point is our avoidance
of using the so-called pinning lemmas in~\cite{Cai-Chen-Lu-2013}
(especially for the  so-called second pinning lemma (Lemma 4.3 from~\cite{Cai-Chen-Lu-2013}), which
we painstakingly avoided, because we could not find a constructive
proof of it \emph{and}, later in the paper we will claim our
dichotomy in this paper is effective, which requires this
constructivity). Thus, if a reader is not concerned with
this effectiveness, one can safely skip this section, and simply
use the proof in~\cite{Cai-Chen-Lu-2013}.

Let $((M,N),\CC,\fD,(\pp,\bft,\fq))$ be a tuple that satisfies
  $(\calR)$,\hspace{0.07cm}$(\calL)$,\hspace{0.07cm}$(\calD)$.
We  show that $\eval^\leftrightarrow(\CC,\fD)$ 
  is tractable by reducing it to the following problem.
Let $q=p^k$ be a prime power for some prime $p$ and
  positive integer $k$. The input of $\eval(q)$ is a
  quadratic polynomial $f(x_1,x_2,\ldots,x_n)=\sum_{i,j\in [n]}a_{i,j}x_ix_j$,
  where $a_{i,j}\in \mathbb{Z}_q$ for all $i,j$, and
  the output is
\[
Z_q(f)=\sum_{x_1,\ldots,x_n\in \mathbb{Z}_q} \oo_q^{f(x_1,\ldots,x_n)}.\label{ZQF}
\] 

%{\color{red} We postpone the proof of the following theorem to the end of this section. (old text)}

In~\cite{Cai-Chen-Lu-2013} the following theorem is shown.
%We will make use of the following theorem from~\cite{Cai-Chen-Lu-2013}
%whose proof is given in section 12.3 of~\cite{Cai-Chen-Lu-2013}
%(and is partitioned into Lemmas 12.7 and 12.9). {\color{red} (new text)}

\begin{theorem}[Theorem 12.1 from~\cite{Cai-Chen-Lu-2013}]\label{pinyan-ours}
Let $q$ be a prime power. Then $\eval(q)$ can be solved in 
  polynomial time \emph{(}in $n$, the number
  of variables\emph{)}.
\end{theorem}

The reduction goes as follows.
First, we use conditions $(\calR),(\calL)$ and $(\calD)$ to show that
  $\eval^\leftrightarrow(\CC,\fD)$ can be decomposed into $s$ smaller problems,
  where $s$ is the number of primes in the sequence $\pp$:
  $\eval^\leftrightarrow(\CC^{[1]},\fD^{[1]}),\ldots,\eval^\leftrightarrow(\CC^{[s]},\fD^{[s]}).$
If each of these $s$ problems is tractable, then so is $\eval^\leftrightarrow(\CC,\fD)$.
Second, we reduce each $\eval^\leftrightarrow(\CC^{[i]},\fD^{[i]})$ to $\eval(q)$ for some   
  appropriate prime power $q$
  that will become clear later. 
It follows from Theorem \ref{pinyan-ours} that all $\eval^\leftrightarrow(\CC^{[i]},\fD^{[i]})$'s
  can be solved in polynomial time.

%{\color{red} (new text)}
\subsection{Step 1}

For each integer $i\in [s]$, we define a $2m_i\times 2m_i$ matrix $
  \CC^{[i]}$ where $m_i=|\mathbb{Z}_{\qq_i}|$:
  $\CC^{[i]}$ is the bipartization of the following $m_i\times m_i$
  matrix $\FF^{[i]}$, where
\begin{equation}\label{step1-1-ours}
F^{[i]}_{\xx,\yy}=\prod_{j\in [t_i]} \oo_{q_{i,j}}^{
  x_{j}y_{j}}\ \ \ \text{for all $\xx=(x_1,\ldots,x_{t_i}),\yy
  =(y_1,\ldots,y_{t_i})\in \mathbb{Z}_{\qq_i}$}.
\end{equation}
We index the rows and columns of $\FF^{[i]}$ by
  $\xx\in \mathbb{Z}_{\qq_i}$ and index the rows and columns
  of $\CC^{[i]}$ by $\{0,1\}\times
  \mathbb{Z}_{\qq_i}$.
We let $x_{j}$, $j\in [t_i]$, 
  denote the $j${th} entry of $\xx\in \mathbb{Z}_{q_{i,j}}$.
By $(\calR_3)$,
\begin{equation}\label{tensor-prod-F-diff-primes-ours}
F_{\xx,\yy}=F^{[1]}_{\xx_1,\yy_1}\cdot F^{[2]}_{\xx_2,\yy_2} \cdots
   F^{[s]}_{\xx_s,\yy_s}\ \ \ \text{for all $\xx,\yy\in \zqt$}.
\end{equation}

For each integer $i\in [s]$, we define a sequence
  of $N$ $2m_i\times 2m_i$ diagonal matrices \[\fD^{[i]}=\big(\DD^{[i,0]},\ldots,
  \DD^{[i,N-1]}\big).\] $\DD^{[i,0]}$ is the $2m_i\times 2m_i$ identity matrix; for every
  $r\in [N-1]$, we set 
\begin{eqnarray*}
&\DD^{[i,r]}_{(0,*)}=\00\ \text{if $r\notin \calS$}\ \ \ \text{and}\ \ \
D^{[i,r]}_{(0,\xx)}=D^{[r]}_{(0,\ext_r(\xx))}\ \text{for all $\xx\in \mathbb{Z}_{\qq_i}$
  if $r\in \calS$;}& \\[0.3ex]
&\DD^{[i,r]}_{(1,*)}=\00\ \text{if $r\notin \calT$}\ \ \ \text{and}\ \ \
D^{[i,r]}_{(1,\xx)}=D^{[r]}_{(1,\ext_r'(\xx))}\ \text{for all
  $\xx\in \mathbb{Z}_{\qq_i}$\ if $r\in \calT$}.& 
\end{eqnarray*}
By conditions $(\calD_1)$ and $(\calD_2)$, we have
\begin{equation}\label{step1-2-ours}
D^{[r]}_{(b,\xx)}=D^{[1,r]}_{(b,\xx_1)}\cdots  D^{[s,r]}_{(b,\xx_s)}\ \ \
\text{for all $b\in \{0,1\}$ and  $\xx\in \zqt$.}
\end{equation}
Equation (\ref{step1-2-ours}) is valid for all $\xx\in \zqt$.
For example, for $b=0$ and $\xx\in\mathbb{Z}_{\fq}-\Lambda_r$,
  the left-hand side is $0$ because $\xx\notin \Lambda_r$.
The right-hand side is also $0$, because there exists an index $i\in [s]$
  such that $\xx_i\notin \Lambda_{r,i}$ and thus $\ext_r(\xx_i)\notin \Lambda_r$.
  %$D^{[i,r]}_{(0,\xx_i)}=0$.
It then follows from (\ref{step1-1-ours}),\hspace{0.05cm}(\ref{step1-2-ours}), and the following lemma that 
  if $\eval^\leftrightarrow(\CC^{[i]},\fD^{[i]})$ {is in polynomial time for all $i\in [s]$},
  then $\eval^\leftrightarrow(\CC,\fD)$ {is also in polynomial time}. 
 
%For each $i\in \{0,1,2\}$, $\FF^{[i]}$ is an $m_i\times n_i$ complex matrix,
%  for some positive integers $m_i$ and $n_i$, and
%  $\CC^{[i]}$ is the bipartization of $\FF^{[i]}$;
%${\frak D}^{[i]}=\{\DD^{[i,0]},\ldots,\DD^{[i,N-1]}\}$ is
%  a sequence of $N$ $(m_i+n_i)\times (m_i+n_i)$ diagonal matrices, where
%$$
%\DD^{[i,r]}=\left(\begin{matrix} \PP^{[i,r]} &  \00 \\
%%$$
%and $\PP^{[i,r]}$ and $\QQ^{[i,r]}$ are $m_i\times m_i$ and $n_i\times n_i$
%   diagonal matrices, respectively.

\begin{lemma}[modification of Lemma 12.2 from~\cite{Cai-Chen-Lu-2013}]\label{tensorproduct2-ours}
%%%JYC edited
Suppose we have the following matrices:
for each $i\in \{0,1,2\}$, 
$\CC^{[i]}$ is the~bipartization of an $m_i \times m_i$ complex matrix $\FF^{[i]}$;
${\frak D}^{[i]}=(\DD^{[i,0]},\ldots,$ $\DD^{[i,N-1]})$ is
  a sequence of $N$ $2m_i\times 2m_i$ diagonal matrices for
  some $N\ge 1$, where 
\[
\DD^{[i,r]}=\left(\begin{matrix} \PP^{[i,r]} & \\
  & \QQ^{[i,r]}\end{matrix}\right)\vspace{-0.1cm}
\]
and $\PP^{[i,r]}$ and $\QQ^{[i,r]}$ are $m_i\times m_i$
   diagonal matrices;
%For each $i\in \{0,1,2\}$, $(\CC^{[i]},\fD^{[i]})$ satisfies
%  condition \emph{(Pinning)}.
%%% no longer needed
% $(\CC^{[i]},\fD^{[i]})$ satisfies
%   \emph{({\sl Pinning})};
\[
\FF^{[0]}=\FF^{[1]}\otimes \FF^{[2]},\ \ \PP^{[0,r]}=
  \PP^{[1,r]}\otimes \PP^{[2,r]} \ \ \text{and}\ \ 
  \QQ^{[0,r]}=\QQ^{[1,r]}\otimes \QQ^{[2,r]}
\] 
for all $r\in [0:N-1]$ \emph{(}so  
  $m_0=m_1 m_2$\emph{)}.
%Then for any \emph{connected} graph $G=(V,E)$ and vertex $u^*$ in $G$, we have
%\begin{eqnarray}
%&Z_{\CC^{[0]},\fD^{[0]}}^\rightarrow (G,u^*)=Z_{\CC^{[1]},\fD^{[1]}}^\rightarrow (G,u^*)
%  \cdot Z_{\CC^{[2]},\fD^{[2]}}^\rightarrow (G,u^*)&\text{and}\label{juju}\\
%&Z_{\CC^{[0]},\fD^{[0]}}^\leftarrow (G,u^*)=Z_{\CC^{[1]},\fD^{[1]}}^\leftarrow (G,u^*)
%  \cdot Z_{\CC^{[2]},\fD^{[2]}}^\leftarrow (G,u^*)&\hspace{-0.3cm}\nonumber.
If $\eval^\leftrightarrow(\CC^{[1]},\fD^{[1]})$ and $\eval^\leftrightarrow(\CC^{[2]},\fD^{[2]})$
  are tractable, then $\eval^\leftrightarrow(\CC^{[0]},\fD^{[0]})$ is also tractable.
%Then if both $\eval(\CC^{[1]},\fD^{[1]})$ and $\eval(\CC^{[2]},\fD^{[2]})$
%  are tractable, $\eval(\CC^{[0]},\fD^{[0]})$ is also tractable.
\end{lemma}
%\begin{lemma}\label{tensorproduct2}
% Moreover, $m_0=m_1\cdot m_2$,
%$$\FF^{[0]}=\FF^{[1]}\otimes \FF^{[2]},\ \ \PP^{[0,r]}=
%  \PP^{[1,r]}\otimes \PP^{[2,r]},\ \ \text{and}\ \ \
%  \QQ^{[0,r]}=\QQ^{[1,r]}\otimes \QQ^{[2,r]},\ \
%  \ \text{for all $r\in [0:N-1]$.}$$
%\end{lemma}
\begin{proof}
As shown earlier, it suffices to restrict the inputs to
$(\leftarrow, G, u^*)$ and $(\rightarrow, G, u^*)$
where $u^* \in V(G)$ and $G$ is connected.
Now we can simply apply Lemma~\ref{tensorproduct-ours}.
%Now it suffices to apply Lemma~\ref{tensorproduct-ours}. %{\color{red} (JYC: "it suffices" repeats; also as shown earlier now refers to that added paragraph after Def 2.7, or should it be "As noted earlier")}
\end{proof}
%{\color{red} (my addition)}
\begin{remark}
We note that unlike the proof of Lemma 12.2 from~\cite{Cai-Chen-Lu-2013},
in the proof of Lemma~\ref{tensorproduct2-ours} we do not invoke 
the so-called second pinning lemma (Lemma 4.3 from~\cite{Cai-Chen-Lu-2013})
with the goal of giving a constructive proof. %{\color{red} (JYC: feels like "with the goal" is also negated when reading this?)}
See Section~\ref{sec:constructivity} for further discussions on the topic of constructivity.
\end{remark}

We now use condition ($\calD_4$) to prove the following lemma.
%  about $\DD^{[i,r]}_{(1,*)}$, $r\in \calT$.
 %(recall $\calS$ is the set of $r\in [N-1]$
  %such that $\DD^{[r]}\ne \00$, and $\Gamma_{r,i}$ is a coset in $\hat{\mathbb{Z}}_{\qq_i}$ for every $i\in [s]$,
  %such that, $\Gamma_r=\prod_{i\in [s]}\Gamma_{r,i}$):
\begin{lemma}[Lemma 12.3 from~\cite{Cai-Chen-Lu-2013}]\label{useitonce2-ours} %{\color{red} (same but not the same)}
Given $r\in \calT,i\in [s]$ and $\aa\in \Delta_{r,i}^{\text{\rm lin}}$,
  there exist $\bb\in {\mathbb{Z}}_{\qq_i}$ and $\alpha\in \mathbb{Z}_N$ such that
    the following equation holds for all $\xx\in \Delta_{r,i}$:
\[
D^{[i,r]}_{(1,\xx+\aa)}\cdot \overline{D^{[i,r]}_{(1,\xx)}}=
  \oo_N^\alpha\cdot F^{[i]}_{\bb,\xx}.
\]
\end{lemma}
\begin{proof}
The proof is the same as that of Lemma 12.3 from~\cite{Cai-Chen-Lu-2013}. %{\color{red} (added this one)}
\end{proof}
%%% gone
%{\color{red} (JYC: can remove the below proof or are we supposed to keep it to show it's constructive, also suggestions for the above? again the issue is that the formulations are the same but the frameworks are different)}
%\begin{proof}
%By the definition of $\DD^{[i,r]}$, we have
%\[
%D^{[i,r]}_{(1,\xx+\aa)}\cdot \overline{D^{[i,r]}_{(1,\xx)}}=
%  D^{[r]}_{(1,\ext_r'(\xx+\aa))}\cdot \overline{
%  D^{[r]}_{(1,\ext_r'(\xx))}}=D^{[r]}_{(1,\ext_r'(\xx)+\widetilde{\aa})}\cdot \overline{
%  D^{[r]}_{(1,\ext_r'(\xx))}}.
%\]
%Recall that $\widetilde{\aa}$ is the vector in $\mathbb{Z}_\calQ$
%  such that $\widetilde{\aa}_i=\aa$ and $\widetilde{\aa}_j=\00$ for all other $j\ne i$.
%
%Then by condition ($\calD_4$), we know there exist $\bb\in {\mathbb{Z}}_{\qq_i}$ and
%  $\alpha\in \mathbb{Z}_N$ such that
%\begin{equation*}
%D^{[i,r]}_{(1,\xx+\aa)}\cdot \overline{D^{[i,r]}_{(1,\xx)}}=\oo_N^\alpha\cdot
%  F_{\widetilde{\bb},\ext_r'(\xx)}=\oo_N^{\alpha}\cdot  F^{[i]}_{\bb,\xx}
%  \ \ \ \text{for all $\xx\in \Delta_{r,i}$,}%\qedhere
%\end{equation*}
%and the lemma is proved.
%\end{proof}

One can also prove a similar lemma for the other block
  of $\DD^{[i,r]}$, 
  using ($\calD_3$).

%{\color{red} (new text)}
\subsection{Step 2}

For convenience, in this step  we abuse the notation slightly and
  use $\eval^\leftrightarrow(\CC,\fD)$ to
 denote one of the subproblems $\eval^\leftrightarrow(\CC^{[i]},\fD^{[i]})$, $i\in [s]$, 
 defined in the last step.
Then by using conditions $(\calR),(\calL)$, and $(\calD)$, we summarize
  the properties of this new pair $(\CC,\fD)$ that we need in the reduction as
  follows:
\begin{enumerate}[wide]
\item[($\calF_1$)] There is a prime $p$ and a nonincreasing
  sequence $\boldsymbol{\pi}=
  (\pi_1, \ldots,\pi_h)$ of powers of the same $p$.
%In particular, we let $q$ denote $q_1$.
$\FF$ is an $m\times m$ complex matrix, where $m=\pi_1\pi_2\cdots \pi_h$,
  and $\CC$ is the bipartization of $\FF$.
We let $\pi$ denote $\pi_1$.
We also use $\mathbb{Z}_{\boldsymbol{\pi}}\equiv \mathbb{Z}_{\pi_1}\times \cdots
  \times \mathbb{Z}_{\pi_h}$ to index the rows and columns of $\FF$.
Then $\FF$ satisfies
\[
F_{\xx,\yy}=\prod_{i\in [h]} \oo_{\pi_i}^{x_{i}y_{i}}
\ \ \ \text{for all $\xx=(x_1,\ldots,x_h)$ and $\yy
  =(y_1,\ldots,y_h)\in \mathbb{Z}_{\boldsymbol{\pi}}$,} 
\]
where we use $x_{i}\in \mathbb{Z}_{\pi_i}$ to denote the $i${th} entry of $\xx$, $i\in [h]$.\label{CONDITIONF14}

\item[($\calF_2$)] $\fD=(\DD^{[0]},\ldots,\DD^{[N-1]})$ is a sequence
  of $N$ $2m\times 2m$ diagonal matrices for some $N\ge 1$
  with $ \pi \hspace{0.06cm}|\hspace{0.06cm}N$.
$\DD^{[0]}$ is the identity matrix, and every diagonal entry of $\DD^{[r]}$,
   $r\in [N-1]$, is either $0$ or a power of $\oo_N$.
We use $\{0,1\}\times \mathbb{Z}_{\boldsymbol{\pi}}$ to index the
  rows and columns of matrices $\CC$ and $\DD^{[r]}$.
(The condition $\pi\hspace{0.05cm}|\hspace{0.05cm}N$
  is from the condition $M\hspace{0.05cm}|\hspace{0.05cm}N$
in $(\calU_2)$, and the expression of $M$ in terms of the prime %\sout{$(\calU_1)$}  (JYC: miscited in CCL, p 1003 as well)}
powers, stated after $(\calR_3)$. The $\pi$ here is one of the $q_{i}
= q_{i,1}$ there.)

\item[($\calF_3$)] For each $r\in [0:N-1]$, we use $\Lambda_r$ and $\Delta_r$ to denote
\[
\Lambda_r=\{\xx\in \mathbb{Z}_{\boldsymbol{\pi}}\hspace{0.07cm}\big|
  \hspace{0.07cm} D^{[r]}_{(0,\xx)}\ne 0\}\ \ \ \text{and}\ \ \
\Delta_r=\{\xx\in \mathbb{Z}_{\boldsymbol{\pi}}\hspace{0.07cm}\big|
  \hspace{0.07cm} D^{[r]}_{(1,\xx)}\ne 0\}.
\]
We use $\cS$ to denote the set of $r$ such that $\Lambda_r\ne \emptyset$
  and $\cT$ to denote the set of $r$ such that $\Delta_r\ne \emptyset$.
Then for every $r\in \cS$, $\Lambda_r$ is a coset in $\mathbb{Z}_{\boldsymbol{\pi}}$;
  for every $r\in \cT$, $\Delta_r$ is a coset in $\mathbb{Z}_{\boldsymbol{\pi}}$.
For each $r\in \cS$ (and $r\in \cT$),
  there is an $\fa^{[r]}\in \Lambda_r$ ($\fb^{[r]}\in
  \Delta_r$, resp.) such that
\[D_{(0,\fa^{[r]})}^{[r]}=1\ \ \text{$\Big($and
  $D_{(1,\fb^{[r]})}^{[r]}=1\text{,\ resp.}\Big)$}.\]

\item[$(\calF_4)$] For all $r\in \cS$ and $\aa\in \lin{\Lambda_r}$,
  there exist $\bb\in \mathbb{Z}_{\boldsymbol{\pi}}$ and $\alpha\in \mathbb{Z}_N$ such that
\[
D^{[r]}_{(0,\xx+\aa)}\overline{D^{[r]}_{(0,\xx)}}=\oo_{N}^\alpha\cdot
  \FF_{\xx,\bb} \ \ \ \text{for all $\xx\in \Lambda_r$};
\]
for all $r\in \cT$ and $\aa\in \lin{\Delta_r}$,
  there exist $\bb\in \mathbb{Z}_{\boldsymbol{\pi}}$ and $\alpha\in \mathbb{Z}_N$ such that
\[
D^{[r]}_{(1,\xx+\aa)}\overline{D^{[r]}_{(1,\xx)}}=\oo_N^\alpha\cdot
  \FF_{\bb,\xx}\ \ \ \text{for all $\xx\in \Delta_r$}. 
\]
%(We may need a few remarks to derive them from $(\calD_2)$ and $(\calD_4)$, respectively)
\end{enumerate}

%Now let $G$ be a connected graph and $u^* \in V(G)$.
To consider $Z^\rightarrow_{\CC,\fD}$, let
$G$ be a  connected graph and $u^* \in V (G)$. Below we reduce the computation
  of $Z^\rightarrow_{\CC,\fD}(G, u^*)$ to $\eval(\hq)$,
where $\hq=\pi$ if $p\ne 2$ and $\hq=2 \pi$ if $p=2$. The  $Z^\leftarrow$ part can be dealt with similarly. %{\color{red} (moved from below, we need it)}

Given $a\in \mathbb{Z}_{\pi_i}$ for some $i\in [h]$,
  let $\widehat{a}$ denote an element in $\mathbb{Z}_{\hq}$
  such that $\widehat{a}\equiv a\pmod{\pi_i}$.
As $\pi_i\hspace{0.08cm}| \hspace{0.08cm}\pi_1 = \pi \hspace{0.08cm}|\hspace{0.08cm} \hq$,
this lifting of $a$ is certainly feasible.
For defi\-niteness, we can choose $a$ itself if we consider
$a$ to be an integer between $0$ and $\pi_i-1$.

First, if $G$ is not bipartite, then $Z^\rightarrow_{\CC,\fD}(G)$ is trivially $0$.
From now on 
we assume $G=(U\cup V, E)$ to be bipartite: every
  edge has one vertex in $U$ and one vertex in $V$. We can assume $u^* \in U$. %{\color{red}, and $u^* \in U$}. {\color{red} (probably shortest)} {\color{red} / We can assume $u^* \in U$. / Without loss of generality, let $u^* \in U$.) (JYC: added but too long, can we say it shorter, we have to say it because it was also assumed in CCL)}

%{\color{red} Let $u^*$ be a vertex in $U$. Then we can decompose $Z_{\CC,\fD}(G)$ into
%\[
%Z_{\CC,\fD}(G)=Z_{\CC,\fD}^\rightarrow(G,u^*)+Z_{\CC,\fD}^\leftarrow(G,u^*).
%\]
%We will reduce
%  $Z_{\CC,\fD}^\rightarrow(G,u^*)$ to $\eval(\hq)$.
%The  $Z^\leftarrow $ part
% can be dealt with similarly. (JYC: only belongs to CCL, I'll remove it)}
 
%{\color{red} (my addition)}
Now we can proceed exactly as in Section 12.2 (Step 2) of~\cite{Cai-Chen-Lu-2013} starting from line -4, p.~1003.
This is because in~\cite{Cai-Chen-Lu-2013}, a reduction from $Z_{\CC,\fD}^\rightarrow(G,u^*)$ to $\eval(\hq)$
was also obtained. This way we can simply repeat the rest of Section 12.2 from~\cite{Cai-Chen-Lu-2013}.
After that we can repeat the material of Section 12.3 from~\cite{Cai-Chen-Lu-2013}
which will give us the proof of Theorem~\ref{pinyan-ours}.
%{\color{red} (this is Theorem 12.3 from~\cite{Cai-Chen-Lu-2013} whose proof is
%partitioned into Lemma 12.7 (for the case $q = p^k$ where $p$ is an odd prime and $k$ is a positive integer)
%and Lemma 12.9 (for the case $q = 2^k$ where $k$ is a positive integer) from Section 12.3 in~\cite{Cai-Chen-Lu-2013}).
%(not sure if it's needed at this point, since we already said that the proof of Theorem~\ref{pinyan-ours} is given there).}

%% file: meta-argument-bounded-degree-graphs-unipairite.tex
%{\color{red} (new text)}
\section{Proof of Theorem \ref{t-step-2-ours}}
%{\color{red} (make sure Theorem~\ref{t-step-1-ours} applies here, i.e. what is stated in there is enough, which is why I expanded on it)}
%
%{\color{red} (mention $(g_1, \ldots, g_d)$ and $(p_1, \ldots, p_d)$ here or inside $\calS$, not sure, if here, there are subtle issues)}

Let $\AA$ be a regularized nonbipartite matrix
and let $\pur \AA$ be the purification of $\AA$
obtained by going from a  generating set  $(g_1, \ldots, g_d)$ 
 of nonzero entries of $\AA$
  to the $d$ smallest primes $(p_1, \ldots, p_d)$, ($d \ge 0$).
%Up to a simultaneous row and column permutation by the same permutation
%of both $\AA$ and $\pur \AA$
Up to a simultaneous row and column permutation by the same permutation,
which is applied to both $\AA$ and $\pur \AA$
(this permutation is obtained by collecting the entries of $\pur \AA$
of equal norm in decreasing order),
we may assume that there exist a positive integer $N$, and three
  sequences $\boldsymbol{\kappa}$, $\pur{\boldsymbol{\kappa}}$, and $\mm$ such that
  $(\AA,(N,\boldsymbol{\kappa},\mm))$ and $(\pur \AA,(N,\pur{\boldsymbol{\kappa}},\mm))$ satisfy the following condition:
\begin{enumerate}[wide]
\item[($\calS_1'$)] $\AA$ is an $m\times m$ symmetric matrix.\label{CONDITIONSP}
$\boldsymbol{\kappa}=(\kappa_1, \ldots,\kappa_s)$ is a sequence
of pairwise distinct nonzero complex numbers, where $s\ge 1$. % such that
%  $\kappa_1>\kappa_2>\ldots>\kappa_s>0$.
$\mm=(m_1,\ldots,m_s)$ is a sequence of positive
  integers such that $m=\sum m_i$.
The rows and columns of $\AA$ are indexed by $\xx=(x_1,x_2)$, where $x_1\in [s]$
  and $x_2\in [m_{x_1}]$.
For all $\xx,\yy$, $\AA$ satisfies
\[
A_{\xx,\yy}=A_{(x_1,x_2),(y_1,y_2)}=\kappa_{x_1}\kappa_{y_1} S_{\xx,\yy},
\]
where $\SS=\{S_{\xx,\yy}\}$ is a symmetric matrix in which
  every entry is a power of $\oo_N$: 
\begin{equation*}
\hspace{-0.2cm}\AA= \begin{pmatrix}
\kappa_1\II_{m_1}\\
& \hspace{-1cm}\kappa_2\II_{m_2}\hspace{-1cm} \\
& & \hspace{-1cm}\ddots\hspace{-1cm}\\
& & & \kappa_s\II_{m_s}
\end{pmatrix} \hspace{-0.12cm}
\left( \begin{matrix}
\SS_{(1,*),(1,*)}\hspace{-0.26cm} & \SS_{(1,*),(2,*)}\hspace{-0.25cm} & \ldots\hspace{-0.25cm} & \SS_{(1,*),(s,*)} \\
\SS_{(2,*),(1,*)}\hspace{-0.26cm} & \SS_{(2,*),(2,*)}\hspace{-0.25cm} & \ldots\hspace{-0.25cm} & \SS_{(2,*),(s,*)} \\
\vdots\hspace{-0.26cm} & \vdots\hspace{-0.25cm} & \ddots\hspace{-0.25cm} & \vdots \\
\SS_{(s,*),(1,*)}\hspace{-0.26cm} & \SS_{(s,*),(2,*)}\hspace{-0.25cm} & \ldots\hspace{-0.25cm} & \SS_{(s,*),(s,*)}
\end{matrix}\right)\hspace{-0.12cm}
\left(\begin{matrix}
\kappa_1\II_{m_1} \\
& \hspace{-1cm}\kappa_2\II_{m_2}\hspace{-1cm} \\
& & \hspace{-1cm}\ddots\hspace{-1cm}\\
& & & \kappa_s\II_{m_s}
\end{matrix}\right)  
\end{equation*}
where $\II_{m_i}$ is the $m_i\times m_i$ identity matrix.

We can also assume that $\kappa_1, \ldots, \kappa_s$
are in the multiplicative subgroup generated by $\{g_1, \ldots, g_d\}$.  %{\color{red} moved here}

The purification matrix $\pur \AA$ is also 
$m \times m$ symmetric.
$\pur{\boldsymbol{\kappa}}=(\pur \kappa_1, \ldots,\pur \kappa_s)$ 
is the purification of $\boldsymbol{\kappa}=(\kappa_1, \ldots,\kappa_s)$ obtained by going from $(g_1, \ldots, g_d)$ to the $d$ smallest primes $(p_1, \ldots, p_d)$, and
is a strictly
  decreasing sequence of
  positive rational numbers, where $s\ge 1$. % such that
%  $\kappa_1>\kappa_2>\ldots>\kappa_s>0$.
%$\mm=(m_1,\ldots,m_s)$ is a sequence of positive
%  integers such that $m=\sum m_i$.
The rows and columns of $\pur \AA$ are also indexed by $\xx=(x_1,x_2)$, where $x_1\in [s]$
  and $x_2\in [m_{x_1}]$.
For all $\xx,\yy$, $\pur \AA$ satisfies
\[
\pur A_{\xx,\yy}=\pur A_{(x_1,x_2),(y_1,y_2)}=\pur \kappa_{x_1}\pur \kappa_{y_1} S_{\xx,\yy},
\]
%where $\SS=\{S_{\xx,\yy}\}$ is a symmetric matrix in which
%  every entry is a power of $\oo_N$: 
so that
\begin{equation*}
\hspace{-0.2cm}\pur \AA= \begin{pmatrix}
\pur \kappa_1\II_{m_1}\\
& \hspace{-1cm}\pur \kappa_2\II_{m_2}\hspace{-1cm} \\
& & \hspace{-1cm}\ddots\hspace{-1cm}\\
& & & \pur \kappa_s\II_{m_s}
\end{pmatrix} \hspace{-0.12cm}
\left( \begin{matrix}
\SS_{(1,*),(1,*)}\hspace{-0.26cm} & \SS_{(1,*),(2,*)}\hspace{-0.25cm} & \ldots\hspace{-0.25cm} & \SS_{(1,*),(s,*)} \\
\SS_{(2,*),(1,*)}\hspace{-0.26cm} & \SS_{(2,*),(2,*)}\hspace{-0.25cm} & \ldots\hspace{-0.25cm} & \SS_{(2,*),(s,*)} \\
\vdots\hspace{-0.26cm} & \vdots\hspace{-0.25cm} & \ddots\hspace{-0.25cm} & \vdots \\
\SS_{(s,*),(1,*)}\hspace{-0.26cm} & \SS_{(s,*),(2,*)}\hspace{-0.25cm} & \ldots\hspace{-0.25cm} & \SS_{(s,*),(s,*)}
\end{matrix}\right)\hspace{-0.12cm}
\left(\begin{matrix}
\pur \kappa_1\II_{m_1} \\
& \hspace{-1cm}\pur \kappa_2\II_{m_2}\hspace{-1cm} \\
& & \hspace{-1cm}\ddots\hspace{-1cm}\\
& & & \pur \kappa_s\II_{m_s}
\end{matrix}\right).  
\end{equation*}

We let
$I= \{(i,j):i\in [s],j\in [m_i]\hspace{0.02cm}\}.$
%where $\II_{m_i}$ is the $m_i\times m_i$ identity matrix.
%%%
%{\color{red} (the generators to be defined here, to be moved from above)}
%
%We have a generating set $\{g_1, \ldots, g_d\}$ 
%of the set of nonzero entries of $\AA$ {\color{red} (all entries of $\AA$ are nonzero)},
%and the $d$ smallest primes $p_1 < \ldots < p_d$.
%(In particular, $g_1, \ldots, g_d$ are multiplicatively independent.)
%Here $d \ge 0$.
%Next, $\pur \AA$ is the purification of $\AA$ obtained
%by going from $(g_1, \ldots, g_d)$ to $(p_1, \ldots, p_d)$.

%%%
%{\color{red} We can also assume that $\kappa_1, \ldots, \kappa_s$
%are in the multiplicative subgroup generated by $g_1, \ldots, g_d$. (JYC: can't say it here because we already said that $\pur \kappa$ is a purification $\kappa$)}
%%%
%(whereas $\pur \kappa_1, \ldots \pur \kappa_s$
%are in the multiplicative subgroup generated by $p_1, \ldots, p_d$).
%
%{\color{red} (to make $\kappa_i$ and $\pur \kappa_i$ related)}
%
%Furthermore, we can additionally assume that $(\pur \kappa_1, \ldots, \pur \kappa_s)$
%is the purification of $(\kappa_1, \ldots, \kappa_s)$.
%each obtained by going from $(g_1, \ldots, g_d)$ to $(p_1, \ldots, p_d)$.
%
%Let %We let
%$I= \{(i,j):i\in [s],j\in [m_i]\hspace{0.02cm}\}.$
\end{enumerate}

The proof of Theorem \ref{t-step-2-ours}, just like the proof
 of Theorem \ref{bi-step-2-ours},
  consists of five steps.
%All the proofs use the following strategy:
It uses the  following strategy.
We construct from  $\AA$ its bipartization $\AA'$,
  a $2m\times 2m$ symmetric matrix.
We also construct from  $\pur \AA$ its bipartization $\pur \AA'$.
  %a $2m\times 2m$ symmetric matrix as well.
Then we just apply the lemmas for the bipartite case to $\AA'$ and $\pur \AA'$,
  and show that either there is an edge gadget $\Gamma$ such that
  $\gm_{\Gamma, \AA'}$ is not \multblrkone/ or $\AA'$ has certain properties.
These properties are then transferred to $\AA$.
%Finally, we use these properties of $\AA'$ to derive properties of $\AA$. 

To this end, we need the following lemma.

\begin{lemma}[modification of Lemma 13.1 from~\cite{Cai-Chen-Lu-2013}]\label{biparbipar-ours}
Let $\AA$ be a symmetric matrix, and let $\AA'$ be its bipartization.
Let $\Gamma$ be an edge gadget such that $\gm_{\Gamma, \AA'}$
is not \multblrkone/ (not \modblrkone/).
Then there is an edge gadget $\Gamma'$ such that $\gm_{\Gamma', \AA}$
is not \multblrkone/ (not \modblrkone/).
\end{lemma}
\begin{proof}
Let $\firstdv, \seconddv$ be the first
and second distinguished vertices of $\Gamma$, correspondingly.
By Lemma~\ref{lem:edge-gadget-component-nmrk1},
$\firstdv, \seconddv$ lie in the same connected component of $\Gamma$, call it $\Gamma_0$,
and $\gm_{\Gamma_0, \AA'}$ is not \multblrkone/ (not \modblrkone/).
%and so without loss of generality we may assume $\Gamma = \Gamma_0$.
%(Here we treat $\Gamma_0$ as an edge gadget with distinguished vertices $\firstdv, \seconddv$ (in this same order).)
It is also immediate that $\Gamma_0$ is a bipartite edge gadget
for otherwise $\gm_{\Gamma, \AA'} = 0$, which contradicts $\gm_{\Gamma, \AA'}$
being non-\multblrkone/ (non-\modblrkone/).
Let $\Gamma_0 = (U \cup V, E)$ be a bipartization of $\Gamma_0$.
It is easy to check that the following hold.
\begin{enumerate}
\item If $\firstdv, \seconddv$ are in the same bipartite part of $\Gamma_0$, say $\firstdv, \seconddv \in U$, then
\[
\gm_{\Gamma_0, \AA'} = \begin{pmatrix} \gm_{\Gamma_0, \AA} \\
 & \gm_{\Gamma_0, \AA} \end{pmatrix};
\]
\item If $\firstdv, \seconddv$ are in different bipartite parts of $\Gamma_0$, say $\firstdv \in U$ and $\seconddv \in V$, then
\[
\gm_{\Gamma_0, \AA'} = \begin{pmatrix} & \gm_{\Gamma_0, \AA} \\
\gm_{\Gamma_0, \AA} \end{pmatrix}.
\]
\end{enumerate}
Now by 
% haha, quote itself. Lemma~\ref{biparbipar-ours} {\color{red} and 
Properties~\ref{prop:equiv-direct-sum-mbrk1} 
and~\ref{prop:equiv-direct-sum-mbrk1-perm},
we get that $\gm_{\Gamma_0, \AA}$ is not \multblrkone/ (not \modblrkone/)
so taking $\Gamma' = \Gamma_0$ we are done.
\end{proof}
%{\color{red} (no need for this remark, remove it)
%\begin{remark}
%It can be checked that Lemma~\ref{biparbipar-ours} also holds when $\AA$ is nonsymmetric
%but in this case we have to consider directed edge gadgets.
%\end{remark}
%}

%{\color{red} (new text)}
\subsection{Step 2.1}

\begin{lemma}[modification of Lemma 13.2 from~\cite{Cai-Chen-Lu-2013}]\label{step221-nonbip-ours}
Suppose that $(\AA,(N,\boldsymbol{\kappa},\mm))$ and $(\pur \AA,(N,\pur{\boldsymbol{\kappa}},\mm))$ satisfy $(\calS_1')$. Then either
  there exists an edge gadget $\Gamma$
  such that $\gm_{\Gamma, \AA}$ is not \multblrkone/
  or $(\AA,(N,\boldsymbol{\kappa},\mm))$ and $(\pur \AA,(N,\pur{\boldsymbol{\kappa}},\mm))$ satisfy
  the following condition:
\begin{enumerate}[wide]
\item[$(\calS_2')$] For all $\xx,\xx'\in I$, either there exists
  an integer $k$ such that $\SS_{\xx,*}=\oo_N^k\cdot \SS_{\xx',*}$, or
  for every $j\in [s]$, 
$ 
\langle \SS_{\xx,(j,*)},\SS_{\xx',(j,*)}\rangle =0. 
$ 
%\item[$(\calS_3)$]
%For all $\yy,\yy'\in J$, either there exists an integer $k$ such that
%  $\SS_{*,\yy}=\oo_N^k\cdot \SS_{*,\yy'}$ or for every
%  $i\in [s]$,
%$$
%\langle \SS_{(i,*),\yy},\SS_{(i,*),\yy'}\rangle =0.
%$$
\end{enumerate}\end{lemma}\begin{proof}
Let $\AA'$ and $\pur \AA'$
 be the bipartizations of $\AA$ and $\pur \AA$, respectively. 
Suppose that for any edge gadget $\Gamma$,
$\gm_{\Gamma, \AA}$ is \multblrkone/. 
From Lemma \ref{biparbipar-ours}, for any edge gadget $\Gamma$,
$\gm_{\Gamma, \AA'}$ is \multblrkone/.
We apply Lemma~\ref{hahajaja-ours} to the
sequences $(\AA',(N,\boldsymbol{\kappa},\boldsymbol{\kappa},
  \mm,\mm))$ and $(\pur \AA',(N,\pur{\boldsymbol{\kappa}},\pur{\boldsymbol{\kappa}},\mm,\mm))$.
The  $\SS$ matrices of $\AA'$ and $\pur \AA'$ are the same; it is
also the same   $\SS$ from $\AA$ and $\pur \AA$ here.
%Note that the $\SS$ matrix in Lemma \ref{hahajaja-ours}
  %is exactly the same $\SS$ here.
%Also $(\AA',(N,\boldsymbol{\kappa},\boldsymbol{\kappa},
  %\mm,\mm))$ and $(\pur \AA',(N,\pur{\boldsymbol{\kappa}},\pur{\boldsymbol{\kappa}},\mm,\mm))$ 
The condition $(\calS_1)$ is satisfied.
So by Lemma \ref{hahajaja-ours} together with the assumption that for any edge gadget $\Gamma$,
$\gm_{\Gamma, \AA'}$ is \multblrkone/, $\SS$ satisfies $(\calS_2)$
  which is exactly the same as $(\calS_2')$ here. (For Lemma \ref{hahajaja-ours},
  $\SS$ also needs to satisfy $(\calS_3)$, but since $\SS$ is symmetric here,
  $(\calS_3)$ is the same as $(\calS_2)$.)
\end{proof}

We have the following corollary, which is the 
same as Corollary~13.3 from~\cite{Cai-Chen-Lu-2013}.
The proof is the same as that of Corollary~\ref{fullrank-ours}
(see Corollary~8.6 in~\cite{Cai-Chen-Lu-2013}). %{\color{red} (JYC: but there's no proof of Corollary~\ref{fullrank-ours} either, just a comment)}

\begin{corollary}
%[cf. Corollary 13.3 from~\cite{Cai-Chen-Lu-2013}]
\label{fullrank-nonbip-ours}
For all $i,j\in [s]$,  $\SS_{(i,*),(j,*)}$ has the same rank as $\SS$. 
\end{corollary}

Next  we build pairs $(\FF,\fD)$ and $(\pur \FF,\fD)$, and
  apply the cyclotomic reduction lemma (Lemma~\ref{twinreduction-ours}) and
% the cyclotomic edge gadget transformation lemma 
the cyclotomic transfer lemma for edge gadgets (Lemma~\ref{lem:cycl-gadget-trans})
  on $\AA$ and $\pur \AA$.% such that 
%$\eval(\AA)\equiv\eval(\FF,\fD).$

Let $h=\text{rank}(\SS)$. By Corollary \ref{fullrank-nonbip-ours}
and condition
$(\calS_2')$,
 there exist $1\le i_1<\ldots<i_h\le m_1$
such that the $h$ 
columns
 $\{\SS_{(1,*), (1, i_1)}, \ldots, \SS_{(1,*), (1, i_h)}\}$
are pairwise orthogonal, and 
these columns form a submatrix of rank $h$.
Without loss of generality, we may assume $i_k=k$
  for all $k\in [h]$ (if this is not the case, we can apply an
  appropriate permutation $\Pi$ to the rows and columns of $\AA$
  so that the new $\SS$ has this property; this does not affect
the monotonicity of
$\pur{\boldsymbol{\kappa}}$ since the permutation
  $\Pi$ is within the first block).

Since $\SS$ is symmetric, the $h$
rows $\{\SS_{(1, 1), (1,*)}, \ldots, \SS_{(1, h), (1,*)}\}$
are pairwise orthogonal. 
 By condition 
$(\calS_2')$, for every $h < i \le m_1$, the $i$th 
 row $\SS_{(1, i), (1,*)}$
% $(S_{(1, i), (1,1)}, S_{(1, i), (1,2)},
%\ldots, S_{(1, i), (1,m_1)})$ 
%for $h < i \le m_1$
is a multiple of one of the $h$ rows
$\{\SS_{(1, 1), (1,*)}, \ldots, \SS_{(1, h), (1,*)}\}$, otherwise
together they would form a submatrix of rank $h+1 > \text{rank}(\SS)$.
Thus each truncated $h$-dimensional row vector
 $(S_{(1, i), (1,1)}, \ldots, S_{(1, i), (1,h)})$ for $h < i \le m_1$
is also a multiple of one of the truncated $h$ row vectors
$(S_{(1,i'), (1,1)}, \ldots, S_{(1, i'), (1,h)})$ for some $1 \le i' \le h$.
Therefore the row rank of
the upper-left-most $h \times h$ submatrix of 
 $\SS_{(1,*),(1,*)}$ is $h$. %{\color{red} (JYC: overall, thank you for explaining this step better than CCL did)}

Let $\HH$ denote this $h\times h$
  symmetric matrix: $H_{i,j}=S_{(1,i),(1,j)}$, $i,j \in [h]$,
and it has rank $h$.
%%%%%%
  %and $1\le j_1<...<j_h\le n_1$,
  %such that the $\{(1,i_1),\ldots,(1,i_h)\}\times
  %\{(1,i_1),\ldots,(1,i_h)\}$ submatrix of $\SS$ has full rank $h$ (using
  %the fact that $\SS$ is symmetric).
%Without loss of generality (if this is not the case, we can apply an
  %appropriate permutation $\Pi$ to the rows and columns of $\AA$
  %so that the new $\SS$ has this property; this does not affect
%the monotonicity of
%$\pur{\boldsymbol{\kappa}}$), assume $i_k=k$
  %for all $k\in [h]$.
%Let $\HH$ denote this $h\times h$
  %symmetric matrix: $H_{i,j}=S_{(1,i),(1,j)}$.
From Corollary \ref{fullrank-nonbip-ours} and Lemma \ref{step221-nonbip-ours},
  for every index $\xx\in I$, there exist
  two unique integers $j\in [h]$ and $k\in [0:N-1]$ such that 
\begin{equation}\label{t-case1-bipa-ours}
\SS_{\xx,*} = \oo_N^k \cdot \SS_{(1,j),*}\ \ \ \text{and}\ \ \
  \SS_{*,\xx}=\oo_N^k \cdot \SS_{*,(1,j)}.
\end{equation}
This gives us a partition of the index set $I$
\[
{\mathscr R} =\big\{R_{(i,j),k}:
  i\in [s],j\in [h],k\in [0:N-1]\hspace{0.02cm}\big\}.
\]
For every $\xx\in I$, $\xx\in R_{(i,j),k}$ iff
  $i=x_1$ and $\xx,j,k$ satisfy (\ref{t-case1-bipa-ours}).
By Corollary \ref{fullrank-nonbip-ours}, 
  \[\bigcup_{k\in [0:N-1]} R_{(i,j),k}\ne \emptyset\ \ \
  \text{for all $i\in [s]$ and $j\in [h]$.}\]

Now we define $(\FF,\fD)$ and $(\pur \FF,\fD)$, and use
Lemmas~\ref{twinreduction-ours}, \ref{lem:cycl-gadget-trans},
   and $\mathscr{R} $ to show that
$\EVAL(\AA) \equiv \EVAL(\FF, \fD)$,
and $\EVAL(\pur \AA) \equiv \EVAL(\pur \FF, \fD)$,
and that for every edge gadget $\Gamma$, $\gm_{\Gamma, \AA}$ (resp., $\gm_{\Gamma, \pur \AA}$)
is \multblrkone/ (\modblrkone/)
iff $\gm_{\Gamma, \FF, \fD}$ (resp., $\gm_{\Gamma, \pur \FF, \fD}$)
is \multblrkone/ (\modblrkone/).
This will allow us to move between the frameworks $\eval(\AA)$ (resp., $\eval(\pur \AA)$)
 and $\eval(\FF, \fD)$ (resp., $\eval(\pur \FF, \fD)$).
First, $\FF$ is an $sh\times sh$ matrix.
We use $I'= [s]\times [h]$ to index the rows and columns of $\FF$.
Then
\[
F_{\xx,\yy}=\kappa_{x_1}\kappa_{y_1}H_{x_2,y_2}=\kappa_{x_1}\kappa_{y_1}S_{(1,x_2),(1,y_2)}
  \ \ \ \text{ for all $\xx,\yy\in I'$,}
\]
or equivalently, 
\[
\FF = \left(\begin{matrix}
\kappa_1\II \\
& \kappa_2\II \\
& & \ddots\\
& & & \kappa_s\II
\end{matrix}\right)
\left( \begin{matrix}
\HH & \HH & \ldots & \HH \\
\HH & \HH & \ldots & \HH \\
\vdots & \vdots & \ddots & \vdots \\
\HH & \HH & \ldots & \HH
\end{matrix}\right)
\left(\begin{matrix}
\kappa_1\II \\
& \kappa_2\II \\
& & \ddots\\
& & & \kappa_s\II
\end{matrix}\right),
\]
where $\II$ is the $h\times h$ identity matrix.\vspace{0.003cm}
%To index the rows (and columns) of $\CC$, a $2\times 2$ block matrix,
%  we use $(\{0\}\times I')\times (\{1\}\times J')$.

Second, $\fD=(\DD^{[0]},\ldots,\DD^{[N-1]})$\vspace{-0.004cm} is a sequence of $N$ diagonal
  matrices of the same size as $\FF$.
We use $I'$ to index its diagonal entries.
The $\xx${th} entries are 
% $(|R_{(x_1,x_2),0}|,\ldots,
%  |R_{(x_1,x_2),N-1}|)$:
\[
D^{[r]}_{\xx}=\sum_{k=0}^{N-1} \big|R_{(x_1,x_2),k}\big|\cdot \oo_N^{kr}
\ \ \ \text{for all $r\in [0:N-1],\xx\in I'$.}
\] 

Finally, $\pur \FF$ is an $sh\times sh$ matrix.
We also use $I'= [s]\times [h]$ to index the rows and columns of $\pur \FF$.
Then
\[
\pur F_{\xx,\yy}=\pur \kappa_{x_1}\pur \kappa_{y_1}H_{x_2,y_2}=\pur \kappa_{x_1}\pur \kappa_{y_1}S_{(1,x_2),(1,y_2)}
  \ \ \ \text{ for all $\xx,\yy\in I'$,}
\]
or equivalently, 
\[
\pur \FF = \left(\begin{matrix}
\pur \kappa_1\II \\
& \pur \kappa_2\II \\
& & \ddots\\
& & & \pur \kappa_s\II
\end{matrix}\right)
\left( \begin{matrix}
\HH & \HH & \ldots & \HH \\
\HH & \HH & \ldots & \HH \\
\vdots & \vdots & \ddots & \vdots \\
\HH & \HH & \ldots & \HH
\end{matrix}\right)
\left(\begin{matrix}
\pur \kappa_1\II \\
& \pur \kappa_2\II \\
& & \ddots\\
& & & \pur \kappa_s\II
\end{matrix}\right).
\]
%where $\II$ is the $h\times h$ identity matrix.\vspace{0.003cm}
%To index the rows (and columns) of $\pur \CC$, a $2\times 2$ block matrix,
%  we use $(\{0\}\times I')\times (\{1\}\times J')$.

We use 
%the cyclotomic reduction lemma (
Lemmas~\ref{twinreduction-ours} and \ref{lem:cycl-gadget-trans}
%)
%and the cyclotomic edge gadget transformation lemma (
%and Lemma \ref{lem:cycl-gadget-trans}
%) 
to prove the next lemma.

\begin{lemma}[modification of Lemma 13.4 from~\cite{Cai-Chen-Lu-2013}]\label{lem:A-FD-cycl-gen}
%{\color{red} (my preferred option)}
%
The matrix
$\AA$ is generated from $\FF$ using $\mathscr R$.
As a consequence, $\EVAL(\AA) \equiv \EVAL(\FF, \fD)$
and for any edge gadget $\Gamma$, $\gm_{\Gamma, \AA}$
is \multblrkone/ (\modblrkone/)
iff $\gm_{\Gamma, \FF, \fD}$
is \multblrkone/ (\modblrkone/).
The same statements hold with $\AA$ and $\FF$ replaced by $\pur \AA$ and $\pur \FF$, respectively.
\end{lemma}
\begin{proof}
%We show that $\AA$ can be generated from $\CC$ using $\mathscr R$;
We show that $\AA$ can be generated from $\FF$ using $\mathscr R$;
%that $\pur \AA$ can be generated from $\pur \CC$ using $\mathscr R$ 
that $\pur \AA$ can be generated from $\pur \FF$ using $\mathscr R$
is shown similarly.
%can be shown similarly.
%First we 
Let $\xx,\yy\in I$, $\xx \in R_{(x_1,j),k}$ and
  $\yy\in R_{(y_1,j'),k'}$ for some $j,k,j',k'$. By (\ref{t-case1-bipa-ours}),
\begin{equation*}
A_{\xx,\yy}=\kappa_{x_1}\kappa_{y_1} S_{\xx,\yy }
 % =\kappa_{x_1}\kappa_{y_1} S_{(1,j), \yy } \oo_N^{k}
  =\kappa_{x_1}\kappa_{y_1} S_{( 1,j),( 1,j')}\cdot  \oo_N^{k+k'}
  =F_{(x_1,j),(y_1,j')} \cdot \oo_N^{k+k'}.
\end{equation*}
So $\AA$ can be generated from $\FF$ using ${\mathscr R}$.
The construction of $\fD$ implies that $\fD$ can be generated from 
  ${\mathscr R}$.
Then the lemma follows from 
Lemmas~\ref{twinreduction-ours} and \ref{lem:cycl-gadget-trans}.
%
% the cyclotomic reduction lemma
%and the cyclotomic edge gadget transformation lemma.
\end{proof}

%{\color{red} (if choosing Meta2 to refer only to $(\pur \CC, \fD)$ but not to $\AA$ in the bipartite part, then need to add it)
%
\begin{corollary}\label{cor:A-FD-mbr1-transition}
%We have $\EVAL(\AA) \equiv \EVAL(\FF, \fD)$.
%Next, 
If $\Gamma$ is an edge gadget such that $\gm_{\Gamma, \pur \FF, \fD}$ is not \multblrkone/
(which is true in particular if $\gm_{\Gamma, \pur \FF, \fD}$ is not \modblrkone/),
then for some $p \ge 1$, $\gm_{T_p(\Gamma), \AA}$ is not \multblrkone/.
\end{corollary}
\begin{proof}
%The first part follows from Lemma~\ref{lem:A-FD-cycl-gen}.
%We prove the second part. 
By Lemma~\ref{lem:A-FD-cycl-gen}, $\gm_{\Gamma, \pur \AA}$ is not \multblrkone/.
Then by Theorem~\ref{thm:gadget-thickening-non-mult-bl-rk-1}, for some $p \ge 1$, $\gm_{T_p(\Gamma), \AA}$ is not \multblrkone/.
%
%((or just say) The corollary follows from Lemma~\ref{lem:A-FD-cycl-gen} and Theorem~\ref{thm:gadget-thickening-non-mult-bl-rk-1}.)
\end{proof}

In light of Corollary~\ref{cor:A-FD-mbr1-transition}, our goal will be either to construct an edge gadget $\Gamma$ such that $\gm_{\Gamma, \pur \FF, \fD}$ is not \multblrkone/ (in fact, not \modblrkone/) or impose additional conditions on $(\FF, \fD)$ (and $(\pur \FF, \fD)$), by adapting the corresponding proofs from~\cite{Cai-Chen-Lu-2013}. 
% with the help of~\cite{Cai-Chen-Lu-2013}. (JYC: we do both, we get non-\multblrkone/ matrices when transitioning from the inner to the original part)
%%(We recall that every matrix that is not \modblrkone/ is also not \multblrkone/.)
%}

%{\color{red} (new text)}
\subsection{Steps 2.2 and 2.3}

Now we have pairs $(\FF,\fD)$ and $(\pur \FF,\fD)$ that satisfy the following condition ({\sl Shape}$'$):
\begin{enumerate}[wide]
%\item[]\hspace{-0.4cm}({\sl Shape}$_1'$)\ \ $\FF\in \mathbb{C}^{m\times m}$ 
\item[({\sl Shape}$_1'$)] $\FF\in \mathbb{C}^{m\times m}$ 
  is a symmetric $s\times s$ block matrix. (Note that the $m$ here
  is different from the $m$ used in ($\calS_1'$) and Step 2.1.)
   We use $I=[s]\times [h]$
  to index its rows and columns, where $m = sh$.\label{SHAPECONDITIONP}

Similarly, the purification matrix $\pur \FF\in \mathbb{C}^{m\times m}$ 
  is also a symmetric $s\times s$ block matrix.
% (Again, the $m$ here
%  is different from the $m$ used in Step 2.1.)
%   We also use $I=[s]\times [h]$
%  to index its rows and columns.

%\item[]\hspace{-0.4cm}({\sl Shape}$_2'$)\ \  There is a 
\item[({\sl Shape}$_2'$)] There is a 
  sequence  $\boldsymbol{\kappa}
  =(\kappa_1,\ldots,\kappa_s)$ of pairwise distinct nonzero complex numbers
  %such that $\mu_1, \ldots, \mu_s$ are in the multiplicative subgroup
such that $\kappa_1, \ldots, \kappa_s$ are in the multiplicative subgroup
  generated by $\{g_1, \ldots, g_d\}$. In particular, $\kappa_1, \ldots, \kappa_s$ generate
  a torsion-free multiplicative group.
  There is also an $h\times h$ matrix $\HH$ of full rank, whose entries are all powers of $\oo_N$, for
  some $N\ge 1$.
We have
\[
F_{\xx,\yy}=\kappa_{x_1}\kappa_{y_1}H_{x_2,y_2}\ \ \ \text{for all $\xx,\yy\in I$}.
\]
%Next, $\kappa_1, \ldots, \kappa_s$ are in the multiplicative subgroup
%generated by $\{g_1, \ldots, g_d\}$.
%In particular, $\kappa_1, \ldots, \kappa_s$ generate
%a torsion-free multiplicative group.

Moreover, there is a 
  strictly decreasing sequence  $\pur{\boldsymbol{\kappa}}
  =(\pur \kappa_1,\ldots,\pur \kappa_s)$,
the purification of $\boldsymbol{\kappa}$ obtained by going from $(g_1, \ldots, g_d)$ to the $d$ smallest primes $(p_1, \ldots, p_d)$,
consisting of positive rational numbers.
We have
\[
\pur F_{\xx,\yy}=\pur \kappa_{x_1} \pur \kappa_{y_1}H_{x_2,y_2}\ \ \ \text{for all $\xx,\yy\in I$}.
\]

%%%%%%%%%%%%%%%%%%%%%%%%%%%%%%%%%%%%%%%%%%%%%%%%%%%%%%%%%%%%%%%%%%%%
\begin{comment}

%{\color{red} (defined generators here)}

We have a generating set $\{g_1, \ldots, g_d\}$
of nonzero entries of $\FF$ {\color{red} (all entries of $\FF$ are nonzero)},
and the $d$ smallest primes $p_1 < \ldots < p_d$.
(In particular, $g_1, \ldots, g_d$ are multiplicatively independent.)
Here $d \ge 0$.
{\color{red} (Next, $\pur \FF$ is the purification of $\FF$ obtained
by going from $(g_1, \ldots, g_d)$ to $(p_1, \ldots, p_d)$.) (not needed at this point but can say it)}
\end{comment}
%%%%%%%%%%%%%%%%%%%%%%%%%%%%%%%%%%%%%%%%%%%%%%%%%%%%%%%%%%%%%%%%%%%%

%We can also assume that $\kappa_1, \ldots, \kappa_s$
%are in the multiplicative subgroup generated by $g_1, \ldots, g_d$.
%(whereas $\pur \kappa_1, \ldots \pur \kappa_s$
%are in the multiplicative subgroup generated by $p_1, \ldots, p_d$).

%%%%%%%%%%%%%%%%%%%%%%%%%%%%%%%%%%%%%%%%%%%%%%%%%%%%%%%%%%%%%%%%%%%%
\begin{comment}

{\color{red} (to make $\kappa_i$ and $\pur \kappa_i$ related)}

Furthermore, we can additionally assume that $(\pur \kappa_1, \ldots, \pur \kappa_s)$
is the purification of $(\kappa_1, \ldots, \kappa_s)$
obtained by going from $(g_1, \ldots, g_d)$ to $(p_1, \ldots, p_d)$.

\end{comment}
%%%%%%%%%%%%%%%%%%%%%%%%%%%%%%%%%%%%%%%%%%%%%%%%%%%%%%%%%%%%%%%%%%%%

%\item[]\hspace{-0.4cm}({\sl Shape}$_3'$)\ \ $\fD=(\DD^{[0]},\ldots,\DD^{[N-1]})$ is a sequence of $N$
\item[({\sl Shape}$_3'$)] $\fD=(\DD^{[0]},\ldots,\DD^{[N-1]})$ is a sequence of $N$
  $m\times m$ diagonal matrices.
  $\fD$ satisfies $(\calT_3)$, so for all $r\in [N-1]$ 
    and $\xx\in I$, we have
  \[
  D^{[r]}_{\xx}=\overline{D^{[N-r]}_{\xx}}.
  \]
\end{enumerate}

Now suppose for any edge gadget $\Gamma$, $\gm_{\Gamma, \pur \FF, \fD}$ is \multblrkone/.
%{\color{red} (otherwise, we are done)}

We define $(\CC,{\fD}')$ and $(\pur \CC,\fD')$:
$\CC$ is the bipartization of $\FF$; $\pur \CC$ is the bipartization of $\pur \FF$; ${\fD}'$
  is a sequence of $N$ copies of 
%=\{\widehat{\DD}^{[0]},\ldots,\widehat{\DD}^{[N-1]}\}$,
%  where
\[
%\widehat{\DD}^{[r]}=
\left(\begin{matrix}
\DD^{[r]}\\ & \DD^{[r]}
\end{matrix}\right). %\ \ \ \text{for all $r\in [0:N-1]$.}
\]
The proof of the following lemma is the same as that of Lemma \ref{biparbipar-ours}.

\begin{lemma}[modification of Lemma 13.5 from~\cite{Cai-Chen-Lu-2013}]\label{checkcheck2-ours}
Let $\Gamma$ be an edge gadget such that $\gm_{\Gamma, \CC, \fD'}$
is not \multblrkone/ (not \modblrkone/).
Then 
%we can construct
there is an edge gadget $\Gamma'$ 
%in polynomial time 
such that $\gm_{\Gamma', \FF, \fD}$
is not \multblrkone/ (not \modblrkone/).
The same conclusion holds for $\gm_{\Gamma, \pur \CC, \fD'}$ and $\gm_{\Gamma', \pur \FF, \fD}$.
%{\color{red} (JYC: added, need to use namely the purified part because the meta for the bip case does; however, this brings about a little issue later on)}
\end{lemma}

By the contrapositive of Lemma \ref{checkcheck2-ours}, 
and our assumption that 
for  any edge gadget $\Gamma$, $\gm_{\Gamma, \pur \FF, \fD}$ is \multblrkone/,
we get that  for any edge gadget $\Gamma$, $\gm_{\Gamma, \CC, \fD'}$ is \multblrkone/.
By ({\sl Shape}$_1'$)--({\sl Shape}$_3'$),  
  $(\CC, {\fD}')$ and $(\pur \CC, {\fD}')$ also satisfy ({\sl Shape}$_1$)--({\sl Shape}$_3$). %$(\CC, {\fD}')$ also satisfies 
It then follows from Lemma \ref{shapecondition-ours} and Lemma \ref{horrible-ours} that
  $(\CC, {\fD}')$ and $(\pur \CC, {\fD}')$ also satisfy ({\sl Shape}$_4$)--({\sl Shape}$_6$). %$(\CC, {\fD}')$ also satisfies 
%\sout{By the way $(\CC, {\fD}')$ is built from $(\FF,\fD)$, the latter
%  must satisfy the following conditions:} 
By the way $(\CC, {\fD}')$ and $(\pur \CC, {\fD}')$ are built from $(\FF,\fD)$ and $(\pur \FF,\fD)$, the latter two
  must satisfy the following conditions:
\begin{enumerate}[wide]
%\item[]\hspace{-0.4cm}{({\sl Shape}$_4'$)}\ \ $\HH/\sqrt{h}$ is unitary:
\item[({\sl Shape}$_4'$)] $\HH/\sqrt{h}$ is unitary:
  $\langle \HH_{i,*},\HH_{j,*}\rangle
  =\langle \HH_{*,i},\HH_{*,j}\rangle =0$ for all $i\ne j\in [h]$.  

%\item[]\hspace{-0.4cm}{({\sl Shape}$_5'$)}\ \  For all $\xx\in I$, \vspace{-0.18cm}
\item[({\sl Shape}$_5'$)]  For all $\xx\in I$,
\[D^{[0]}_{\xx}=
  D^{[0]}_{(x_1,1)}.\] 

%\item[]\hspace{-0.4cm}{({\sl Shape}$_6'$)}\ \  For each $r\in [N-1]$, there
\item[({\sl Shape}$_6'$)] For each $r\in [N-1]$, there
  are diagonal matrices $\KK^{[r]}\in \mathbb{C}^{s\times s},
  \LL^{[r]}\in \mathbb{C}^{h\times h}$. 
The norm of every diagonal entry in $\LL^{[r]}$ is
  either $0$ or $1$. We have
\[
\DD^{[r]} =\KK^{[r]} \otimes \LL^{[r]},\ \ \ \text{for all $r\in [N-1]$.}
\]
For all $r\in [N-1]$, 
$\KK^{[r]} =\00$ implies $\LL^{[r]} =\00$; 
$\LL^{[r]} \ne \00$ implies 
  one of its entries is $1$.
%that there exists an $i\in [h]$ such that $L^{[r]}_i =1$.
%%%
%{\color{red} ("implies" is what is used in CCL, p. 1015, line -3, but for the bipartite case in CCL and our paper, we use "iff" which one is more correct, see p 964 CCL, Lemma 8.11; and Lemma~\ref{horrible-ours}, basically do we correct it or no and keep it as in CCL)} 
\end{enumerate} 

In particular, ({\sl Shape}$_5'$) means that by setting 
\[ K^{[0]}_i=D^{[0]}_{(i,1)}\ \ \text{and}\  \ L^{[0]}_j=1\ \ \ \text{for all
  $i\in [s]$ and $j\in [h]$,}\]
we have $\DD^{[0]}=\KK^{[0]}\otimes \LL^{[0]}$.
%, where $\LL^{[0]}$ is
%%  the  identity matrix.
By $(\calT_3)$ in ({\sl Shape}$_3'$), the entries of $\KK^{[0]}$ are positive integers.

%{\color{red} (new text)}
\subsection{Step 2.4}

Suppose $(\FF,\fD)$ and $(\pur \FF,\fD)$ satisfy ({\sl Shape}$_1'$)--({\sl Shape}$_6'$).
From ({\sl Shape}$_2'$) we have $\FF=\MM\otimes \HH$, where $\MM$
  is an $s\times s$ matrix of rank $1$:
  $M_{i,j}=\kappa_i\kappa_j$ for all $i,j\in [s]$.
From ({\sl Shape}$_2'$) we also have $\pur \FF=\pur \MM\otimes \HH$, where $\pur \MM$
  is also an $s\times s$ matrix of rank $1$:
  $\pur M_{i,j}=\pur \kappa_i \pur \kappa_j$ for all $i,j\in [s]$.

We reduce
  $\eval(\FF,\fD)$ to two problems $\eval(\MM,{\frak K})$ and $\eval(\HH,{\frak L})$,
  where
\[
{\frak K}=\big(\KK^{[0]},\ldots,\KK^{[N-1]}\big)\ \ \ \text{and}\ \ \
{\frak L}=\big(\LL^{[0]},\ldots,\LL^{[N-1]}\big).
\]
By the tensor product expression of $\eval(\FF,\fD)$ in  terms of
the two problems $\eval(\MM,\fK)$ (the outer problem) and 
$\eval(\HH,\fL)$ (the inner problem), and the fact that
$\eval(\MM,\fK)$ is tractable, we have a reduction
from $\eval(\FF,\fD)$ to $\eval(\HH,\fL)$.
The proof of the following lemma is essentially the same as that of Lemma \ref{jajajaja-ours}.

\begin{lemma}[weakened form of Lemma 13.6 from~\cite{Cai-Chen-Lu-2013}]
%$\eval(\FF,\fD)\equiv \eval(\HH,\calL)$.
$\eval(\FF,\fD)\le \eval(\HH,\fL)$.
%$\EVAL(\FF, \fD) \le \EVAL^\leftrightarrow(\HH, \fL)$.
%Moreover,
%$\EVAL^\rightarrow(\FF, \fD) \le \EVAL^\rightarrow(\HH, \fL)$ and
%$\EVAL^\leftarrow(\FF, \fD) \le \EVAL^\leftarrow(\HH, \fL)$.
\end{lemma}

Next,
by the tensor product expression of $\eval(\pur \FF,\fD)$ in  terms of
 $\eval(\pur \MM,\fK)$  and
$\eval(\HH,\fL)$, we have the following relation
\[
\gm_{\Gamma, \pur \FF, \fD}
= \gm_{\Gamma,  \pur \MM, \fK}
\otimes
\gm_{\Gamma, \HH, \fL},\]
for any edge gadget $\Gamma$ (not necessarily connected).
%
%
%Next, 
This relation will allow us to transfer an edge gadget that is
not \multblrkone/ (not \modblrkone/) for $(\HH, \fL)$ to an edge gadget that is
not \multblrkone/ (not \modblrkone/) for $(\pur \FF, \fD)$, provided that
$\gm_{\Gamma,  \pur \MM, \fK}$ is nonzero.
The proof of the following claim is essentially the same as that of Claim~\ref{hahaclaim-ours-edge-gadget-bip-same-pur}.

\begin{claim}[modification of Claim 8.25 from~\cite{Cai-Chen-Lu-2013}]\label{hahaclaim-ours-edge-gadget-nonbip-pur}
%{\color{red} (JYC: not sure we need to say it: it's for the bipartite case and for the nonbipartite case we don't have an analog formulated in CCL, ususaly we place modifications for CCL's claims)}
For any connected edge gadget $\Gamma = (V, E)$
with distinguished vertices $\firstdv, \seconddv$ (in this order),
there is a connected edge gadget $\Gamma' = (V', E')$
with distinguished vertices $\firstdv, \seconddv$ (in this order)
such that 
 $\firstdv, \seconddv \in V \subset V'$,
and
\begin{equation}\label{gjgi-edge-gadget-nonbip-pur}
\gm_{\Gamma', \HH, \fL} = h^{|V| - 2} \cdot \gm_{\Gamma, \HH, \fL}.
\end{equation}
Furthermore, 
if  $\gm_{\Gamma, \HH, \fL} \ne \00$, then we can choose 
$\Gamma'$ such that  $\gm_{\Gamma', \pur \MM, \fK}$ 
has no zero
entries.
%
%Given a connected edge gadget $\Gamma = (V, E)$
%with distinguished vertices $\firstdv, \seconddv$ (in this order), %($u_1 \ne u_2$)
%either we can construct a new connected, bipartite edge gadget $\Gamma' = (V', E')$
%with distinguished vertices $\firstdv, \seconddv$ (in this order)
%in polynomial time such that $\firstdv, \seconddv \in V \subset V'$,
%\begin{equation}\label{gjgi-edge-gadget-nonbip-pur}
%\gm_{\Gamma', \HH, \fL} = h^{|V| - 2} \cdot \gm_{\Gamma, \HH, \fL},
%\end{equation}
%and all entries of $\gm_{\Gamma', \pur \MM, \fK}$ are nonzero
%(so $\gm_{\Gamma', \pur \MM, \fK} \ne 0$),
%or we can show that $\gm_{\Gamma, \HH, \fL} = 0$.
\end{claim}

From Claim~\ref{hahaclaim-ours-edge-gadget-nonbip-pur}, we have the following corollary whose proof is essentially the same as that of Corollary~\ref{cor:hahaclaim-ours-edge-gadget-bip-same-different-pur}. %{\color{red} (JYC: added for clarity, suggest rewording of the sentence or just remove it)}
\begin{corollary}\label{cor:hahaclaim-ours-edge-gadget-nonbip-pur}
Let $\Gamma$ be an edge gadget such that $\gm_{\Gamma, \HH, \fL}$
is not \multblrkone/ (not \modblrkone/).
Then 
%we can construct
there is a connected edge gadget $\Gamma'$ %changed from an edge gadget
% in polynomial time 
such that $\gm_{\Gamma', \pur \FF, \fD}$
is not \multblrkone/ (not \modblrkone/).
\end{corollary}

%{\color{red} (new text)}
\subsection{Step 2.5}
%{\color{red} (if choosing Meta2 to refer only to $(\pur \CC, \fD)$ but not to $\AA$ in the bipartite part, then need change the below fragment to refer to $(\pur \FF, \fD)$ instead of $\AA$)}
%
Finally we normalize the matrix $\HH$ in the same way we did for the bipartite case
  and obtain a new pair $(\ZZ, \fY)$ such that
\begin{enumerate}
\item $(\ZZ, \fY)$ satisfies conditions $(\calU_1')$--$(\calU_4')$;
%\item $\EVAL(\AA) \le \EVAL^\leftrightarrow(\ZZ, \fY)$;
\item $\EVAL(\AA) \le \EVAL(\ZZ, \fY)$;
\item for any edge gadget (not necessarily connected) $\Gamma$,
if $\gm_{\Gamma, \ZZ, \fY}$ is not \multblrkone/ (not \modblrkone/),
then  there is  a connected edge gadget $\Gamma'$
%we can construct a connected edge gadget $\Gamma'$ in polynomial time
such that $\gm_{\Gamma', \AA}$ is not \multblrkone/.
\end{enumerate}
%%%moved here
Below we rename the pair $(\ZZ, \fY)$
(which is the normalized inner pair $(\HH, \fL)$)
as $(\FF,\fD)$.

%Next, combining Lemma~\ref{lem:A-FD-cycl-gen},
%Corollaries~\ref{cor:hahaclaim-ours-edge-gadget-bip-same-different-pur}, \ref{cor:norm-mbr1-edge-gadget-all}
%and Theorem~\ref{thm:gadget-thickening-non-mult-bl-rk-1} (and Corollary~\ref{cor:gadget-thickening-non-mod-bl-rk-1}),
%we obtain the following corollary.
%(Notice that we use the fact that if $\Gamma$ is connected,
%then $T_p(\Gamma)$ where $p \ge 1$ is also connected.)

%Since $(\ZZ, \fY)$ already satisfies $(\calU_1')$--$(\calU_4')$,
%Corollaries~\ref{cor:t-step-2-reduction} and~\ref{cor:t-step-2-edge-gadget-transformation}
%finalize the proof of Theorem~\ref{t-step-2}.

%\noindent
%Below we rename the pair $(\ZZ, \fY)$
%(which is the normalized inner pair $(\HH, \fL)$)
%as $(\FF,\fD)$.

%{\color{red} (new text)}
\section{Proofs of Theorem \ref{nonbipstep0-ours} and Theorem \ref{t-step-3-ours}}

Suppose $((M,N),\FF,\fD)$ satisfies ($\calU_1'$)--($\calU_4'$).
We prove Theorem \ref{nonbipstep0-ours} and \ref{t-step-3-ours} in this section. 
We first prove that if $\FF$ does not satisfy the group condition (\GC),
  then there exists an edge gadget $\Gamma$ such that $\gm_{\Gamma, \FF, \fD}$ is not \modblrkone/.
This is done by applying Lemma \ref{groupcondition1-ours} (for the bipartite
  case) to the bipartization $\CC$ of $\FF$. 

%Then we use (\GC), together with ($\calU'$), to show that $\FF$
%  can be decomposed into (generalized) Fourier matrices.

\begin{lemma}[modification of Lemma 14.1 from~\cite{Cai-Chen-Lu-2013}]\label{groupcondition2-ours}
Suppose $((M,N),\FF,\fD)$ satisfies conditions $(\calU_1')$--$(\calU_4')$.
 Then either the matrix $\FF$ satisfies the group condition \emph{(\GC)}
  or there is an edge gadget $\Gamma$ such that $\gm_{\Gamma, \FF, \fD}$ is not \modblrkone/. 
\end{lemma}
\begin{proof}
Assume for every edge gadget $\Gamma$, $\gm_{\Gamma, \FF, \fD}$ is \modblrkone/. 
Let $\CC$ and ${\frak E}=(\EE^{[0]}, \ldots,\EE^{[N-1]})$ be %the
%  bipartizations of $\FF$ and $\fD$:
\[
\CC=\left( \begin{matrix}\00 & \FF \\ \FF &\00\end{matrix}\right)\ \ \text{and}\ \
  \EE^{[r]} =\left( \begin{matrix}\DD^{[r]}& \00 \\ \00 & \DD^{[r]}\end{matrix}\right)\ \ \
\text{for all $r\in [0:N-1]$.}\]
By $(\calU_1')$--$(\calU_4')$, $((M,N),\CC,{\frak E})$
  satisfies $(\calU_1)$--$(\calU_4)$.
%%% can't use it
%Furthermore, by the contrapositive of Lemma~\ref{checkcheck2-ours} {\color{red} Furthermore, by the contrapositive of the analog of Lemma~\ref{checkcheck2-ours} but stated for the new (inner) $(\FF, \fD)$ (JYC: This is simply incorrect because we use it for the inner $(\FF, \fD)$, in CCL, p. 1016. Lemma 14.1 suffers from the same problem, in the referred-to Lemma 13.5 there, $(\FF, \fD)$, $(\CC, \fD')$ are clearly from the outer world so it's technically an error in CCL; I can either add another lemma, or just replace the faulty sentence in which case suggest an alternative)}, 
%%%
Arguing similarly to Lemma~\ref{checkcheck2-ours}, from the contrapositive, we infer that
%for every edge gadget $\Gamma$ such that $\gm_{\Gamma, \CC, \fE}$
%
%is not \modblrkone/,
%we can construct an edge gadget $\Gamma'$ in polynomial time such that $\gm_{\Gamma', \FF, \fD}$
%is not \modblrkone/ and, thus,
   for any edge gadget $\Gamma$, $\gm_{\Gamma, \CC, \fE}$ is \modblrkone/.
It follows from Lemma \ref{groupcondition1-ours} that
  $\FF$ satisfies the group condition (\GC).
\end{proof}

%{\color{red} (new text)}
\subsection{Proof of Theorem \ref{nonbipstep0-ours}}

We prove Theorem \ref{nonbipstep0-ours} again, using $\CC$ and ${\frak E}$
defined above. 
%{\color{red} (Recall that ${\frak E}=(\EE^{[0]}, \ldots,\EE^{[N-1]})$. (JYC: added for you, actually, it's defined right before, it's in close proximity, ignore the blue text above.)}
 %:  the bipartizations
 % of $\FF$ and $\fD$, respectively. 

We are given  that
$((M,N), \FF,\fD)$ satisfies $(\calU_1')$--$(\calU_4')$, and $M>1$.
So $((M,N), \CC,{\frak E})$ satisfies conditions $(\calU_1)$--$(\calU_4)$.
Suppose for every edge gadget $\Gamma$, 
 $\gm_{\Gamma, \FF,\fD}$ is \modblrkone/.
By Lemma~\ref{checkcheck2-ours},
for every edge gadget $\Gamma$,
$\gm_{\Gamma, \CC,{\frak E}}$ is \modblrkone/.
By Lemma~\ref{groupcondition2-ours},
$\FF$ satisfies  the group condition \emph{(\GC)}.
Now by Theorem~\ref{step30-ours}, 
the tuple  $((M,N), \CC,{\frak E})$ satisfies
$(\calU_5)$.
Thus, for all $r \in [N-1]$,
 every entry of $\DD^{[r]}$
  is either $0$ or a power of $\oo_N$.
This establishes $(\calU'_5)$. Theorem~\ref{nonbipstep0-ours}
is proved.

%%%%%%%%%%%%%%%%%%%%%%%%%%%%%%%%%%%%%%%%%%%%%%%%%%%%%%%%%%%%%%%%%%%%
\begin{comment}

Suppose for every edge gadget $\Gamma$, $\gm_{\Gamma, \FF, \fD}$ is \modblrkone/.
On the one hand, for every edge gadget $\Gamma$ such that $\gm_{\Gamma, \CC, \fE}$
is not \modblrkone/,
we can construct an edge gadget $\Gamma'$ in polynomial time such that $\gm_{\Gamma', \FF, \fD}$
is not \modblrkone/ and so
   for any edge gadget $\Gamma$, $\gm_{\Gamma, \CC, \fE}$ is \modblrkone/. 
On the other hand, $((M,N),\CC,{\frak E})$ satisfies conditions $(\calU_1)$--$(\calU_4)$.
Thus, using Theorem \ref{step30-ours}, ${\frak E}$ must satisfy ($\calU_5$):
Every entry of $\EE^{[r]}$, $r\in [N-1]$, is either $0$ or a power of $\oo_N$.
It then follows directly that every entry of $\DD^{[r]}$, $r\in [N-1]$,
  is either $0$ or a power of $\oo_N$.

\end{comment}
%%%%%%%%%%%%%%%%%%%%%%%%%%%%%%%%%%%%%%%%%%%%%%%%%%%%%%%%%%%%%%%%%%%%

%{\color{red} (new text)}
\subsection{Proof of Theorem \ref{t-step-3-ours}}

In this section we prove Theorem \ref{t-step-3-ours}.
However, here we cannot simply reduce it, using $(\CC,{\frak E})$, to the bipartite
  case (Theorem \ref{bi-step-3-ours}),
  because in Theorem \ref{t-step-3-ours},
  we are only allowed to permute the rows and columns symmetrically,
  while in Theorem \ref{bi-step-3-ours}, one can use two different permutations
  to permute the rows and columns.
But as we will see below,
  for most of the lemmas we need here, their proofs are exactly the same
  as those for the bipartite case.
The only exception is the counterpart of Lemma \ref{decomp2-ours},
  in which we have to bring in the generalized Fourier matrices
  %(see Section 4.3 for the definition).
%%% JYC I don't know why it is called Section 4.3
(see Definitions~\ref{FourierMatrix-ours} and \ref{GeneralizedFourierMatrix-ours}).

Suppose $\FF$ satisfies (\GC). % (otherwise we already
%  know that $\eval(\FF,\fD)$ is \#P-hard). 
Let $F^{\text{R}}$ denote the set of row vectors $\{\FF_{i,*} \}$ of $\FF$ and
  $F^{\text{C}}$ denote the set of column vectors $\{\FF_{*,j}\}$ of $\FF$.
Since $\FF$ satisfies (\GC), by Property \ref{gcproperty1-ours},
  both $F^{\text{R}}$ and $F^{\text{C}}$ are finite Abelian groups
  of order $m$, under the Hadamard product.

%%% if we had stated the full assumption that $\FF$ is symmetric
%%% there would be no need for both $F^{\text{R}}$ and $F^{\text{C}}$ 
%%% they are the same

We start by proving a symmetric version of Lemma \ref{decomp1-ours},
  stating that
  when $M=pq$~and $\gcd(p,q)=1$ (note that $p$ and $q$ are not necessarily primes),
  a permutation of $\FF$ 
  is the tensor pro\-duct of two smaller discrete unitary matrices, both
  of which satisfy the group condition.

\begin{lemma}[Lemma 14.2 from~\cite{Cai-Chen-Lu-2013}]\label{sys-decomp1-ours}
Suppose $\FF\in \mathbb{C}^{m\times m}$ is symmetric and $M$-discrete
  unitary and satisfies \emph{(}\GC\emph{)}. 
  Moreover, $M=pq$, $p,q>1$, and $\gcd(p,q)=1$.
Then there is a permutation $\Pi$ of $[0:m-1]$
  such that
$ 
\FF_{\Pi,\Pi}=\FF'\otimes \FF'',
$ 
where $\FF'$ is a symmetric $p$-discrete unitary matrix,
  $\FF''$ is a symmetric $q$-discrete
  unitary matrix, and both of them satisfy \emph{(}\GC\emph{)}.
\end{lemma}
\begin{proof}
The proof is the same as that of Lemma 14.2 from~\cite{Cai-Chen-Lu-2013}.
\end{proof}

As a result, we only need to deal with the case when $M=p^\beta$ is a prime power.
%In the bipartite case (Lemma \ref{decomp2}), we can always permute
%  the rows and columns so that there is a diagonal entry $F_{i,i}$
%  satisfies $F_{i,i}=\oo_N^{\alpha_{i,i}}$ and $p\nmid \alpha_{i,i}$.
%This is not the case when we are only allowed to permute $\FF$
%  symmetrically, in which it is
%  possible that all $\alpha_{i,i}$s' along the diagonal are multiples of $p$.
%However, as shown by the following lemma, this cannot be true if
%  $p$ is odd.

\begin{lemma}[Lemma 14.3 from~\cite{Cai-Chen-Lu-2013}]\label{t-decomp1-ours}
Suppose $\FF\in \mathbb{C}^{m\times m}$ is symmetric and $M$-discrete unitary and
  satisfies \emph{(}\GC\emph{)}. Moreover, $M=p^{\beta}$ is
  a prime power, $p\ne 2$, and $\beta\ge 1$.
Then there must exist an integer $k\in [0:m-1]$ such that
  $p\nmid \alpha_{k,k}$, where
  $F_{k,k}=\oo_M^{\alpha_{k,k}}$.
\end{lemma}
%\begin{proof}
%The proof is the same as that of Lemma 14.3 from~\cite{Cai-Chen-Lu-2013}.
%\end{proof}
%%% NOT quite the same... in CCL there was a wolog, to set i,j =1,2
%%% but in M-discrtete unitary, we had normalized, so that 1 is not general
%
%{\color{red} (added the corrected proof from CCL, kept i, j}
\begin{proof}
For $i,j\in [0:m-1]$, we let $\alpha_{i,j}$ denote the integer in $[0:M-1]$
  such that $F_{i,j}=\oo_M^{\alpha_{i,j}}$. 
Assume the lemma is not true, that is, $p\hspace{0.06cm}|\hspace{0.06cm}
  \alpha_{k,k}$ for all $k$.
Then because $\FF$ is $M$-discrete unitary, and $M = p^{\beta}$, there must exist
  $i\ne j\in [0:m-1]$ such that $p\nmid \alpha_{i,j}$.
%Without loss of generality, we assume $p\nmid \alpha_{3,2}=\alpha_{2,3}$.

By (\GC), there exists a $k\in [0:m-1]$
  such that $\FF_{k,*}=\FF_{i,*}\circ \FF_{j,*}$.
However,
\[
\omega_M^{\alpha_{k,k}}=F_{k,k}=F_{i,k}F_{j,k}=F_{k,i}F_{k,j}
=F_{i,i}F_{j,i}F_{i,j}F_{j,j}=\omega_M^{\alpha_{i,i}+\alpha_{j,j}+2\alpha_{i,i}},
\]
and $\alpha_{k,k}\equiv \alpha_{i,i}+\alpha_{j,j}+2\alpha_{i,j}\hspace{-0.06cm}\pmod{M}$
  implies that $0\equiv 0+0+2\alpha_{i,j}\hspace{-0.06cm}\pmod{p}$.
Since $p\ne 2$ and $p\nmid \alpha_{i,j}$, we get a contradiction.
%and $p\nmid \alpha_{k,k}$ since $p\hspace{0.06cm}|\hspace{0.06cm}
%  \alpha_{1,1}$, $p\hspace{0.06cm}|\hspace{0.06cm} \alpha_{2,2}$,
%   and $p\ne 2$.
%This contradicts our assumption.
%{\color{red} (JYC: of course we redo the proof with a little fix, no need to say why we do this little cover-up)}
\end{proof}

The next lemma is the symmetric version of Lemma \ref{decomp2-ours} showing
  that when there exists a diagonal entry $F_{k,k}$ such that
  $p\nmid \alpha_{k,k}$, $\FF$ is the tensor product
  of a Fourier matrix and a discrete unitary matrix satisfying the
group condition \emph{(}\GC\emph{)}.
Note that this lemma also applies to the case when $p=2$.
So the only case left is when $p=2$ but
  $2\hspace{0.06cm}|\hspace{0.06cm} \alpha_{i,i}$ for all $i\in [0:m-1]$.

\begin{lemma}[Lemma 14.4 from~\cite{Cai-Chen-Lu-2013}]\label{t-decomp2-ours}
Suppose $\FF\in \mathbb{C}^{m\times m}$ is  symmetric and $M$-discrete unitary
  and satisfies \emph{(}\GC\emph{)}. Moreover, $M=p^{\beta}$ is
  a prime power.
If there exists a $k\in [0:m-1]$ such that $F_{k,k}=\oo_M^{\alpha}$
  and $p\nmid \alpha$, then there exists a permutation $\Pi$
   such that
$
\FF_{\Pi,\Pi}=\boldsymbol{\calF}_{M,\alpha}\otimes \FF'
$, 
where $\FF'$ is a symmetric and $M'$-discrete unitary matrix that satisfies
  condition \emph{(}\GC\emph{)} with $M'\mid M$.  %$M'=p^{\beta'}$
%  for some $\beta'\le \beta$.
\end{lemma}
\begin{proof}
The proof is the same as that of Lemma 14.4 from~\cite{Cai-Chen-Lu-2013}.
%{\color{red} (JYC: the proof in CCL uses formula (9.12) coming from Lemma 9.7, for us it's Lemma~\ref{decomp2-ours} whose proof is the same as that of Lemma 9.7 from CCL, as we said)}
\end{proof}

Finally, we deal with the case when $p=2$ and $2\hspace{0.06cm}
  |\hspace{0.06cm} \alpha_{i,i}$ for all $i\in [0:m-1]$.

\begin{lemma}[Lemma 14.5 from~\cite{Cai-Chen-Lu-2013}]\label{t-decomp3-ours}
Suppose $\FF\in \mathbb{C}^{m\times m}$ is symmetric and $M$-discrete unitary and
  satisfies \emph{(}\GC\emph{)}, with
  $M=2^{\beta}$ and
  $2\hspace{0.06cm}|\hspace{0.06cm} \alpha_{i,i}$ for all $i\in [0:m-1]$.
Then there exist a  permutation $\Pi$ and a $2\times 2$ symmetric nondegenerate matrix
  $\WW$ over $\mathbb{Z}_M$
  %\emph{(see Definition~\ref{GeneralizedFourierMatrix})} such that
  \emph{(see Section~\ref{reference-for-sec-6.3.2-ours} and
        Definition~\ref{GeneralizedFourierMatrix-ours})}, such that
$ 
\FF_{\Pi,\Pi}=\boldsymbol{\calF}_{M,\WW}\otimes \FF',
$ 
where $\FF'$ is a symmetric, $M'$-discrete unitary 
  matrix that satisfies \emph{(}\GC\emph{)} with $M'\mid M$. %=2^{\beta'}$
  %for some $\beta'\le \beta$, .
\end{lemma}
\begin{proof}
The proof is the same as that of Lemma 14.5 from~\cite{Cai-Chen-Lu-2013}.
\end{proof}

%Theorem \ref{t-step-3-ours} then follows from Lemma \ref{t-decomp1-ours},
%  Lemma \ref{t-decomp2-ours}, and Lemma \ref{t-decomp3-ours}.
Theorem \ref{t-step-3-ours} then follows from Lemmas \ref{t-decomp1-ours},
  \ref{t-decomp2-ours}, and \ref{t-decomp3-ours}.

%{\color{red} (new text)}
\section{Proofs of Theorem \ref{t-step-4-ours} and Theorem \ref{nonbi-step-5-ours}}

Suppose $((M,N),\FF,\fD,(\dd,$ $\calW,\pp,\bft,\calQ,\calK))$ satisfies
  condition $(\calR')$.
We  prove Theorem \ref{t-step-4-ours}: either there exists an edge gadget $\Gamma$ such that $\gm_{\Gamma, \FF, \fD}$ is not \modblrkone/ or
  $\fD$ satisfies conditions $(\calL_1')$ and $(\calL_2')$. 

Suppose for any edge gadget $\Gamma$, $\gm_{\Gamma, \FF, \fD}$ is \modblrkone/.
We use $(\CC,{\frak E})$ to denote the bipartization of $(\FF,\fD)$,
where ${\frak E}=(\EE^{[0]}, \ldots,\EE^{[N-1]})$.
The plan is to show that $(\CC,{\frak E})$ with appropriate
  $\pp',\bft'$, and $\fq'$ satisfies $(\calR)$.

To see this, we permute $\CC$ and ${\frak E}$
  using the following permutation $\Sigma$.
We index the rows and columns of $\CC$ and $\EE^{[r]}$ using
  $\{0,1\}\times \mathbb{Z}^2_{\dd}\times \mathbb{Z}_{\calQ}.$
We set $\Sigma(1,\yy)=(1,\yy)$ for all $\yy\in \mathbb{Z}_{\dd}^2\times \mathbb{Z}_{\calQ}$,
  that is, $\Sigma$ fixes pointwise the second half of the rows and columns,
  and $\Sigma(0,\xx)=(0,\xx')$, where $\xx'$ satisfies 
\[
x_{0,i,1}=W^{[i]}_{1,1}\hspace{0.03cm}x'_{0,i,1}+W^{[i]}_{2,1}\hspace{0.03cm}x'_{0,i,2},\ \ 
x_{0,i,2}=W^{[i]}_{1,2}\hspace{0.03cm}x'_{0,i,1}+W^{[i]}_{2,2}\hspace{0.03cm}x'_{0,i,
  2}\ \ \ \text{for all $i\in [g]$,} 
\]
and 
$x_{1,i,j}=k_{i,j}\cdot x'_{1,i,j}$ {for all $i\in [s],j\in [t_i]$.} 
See ($\calR'$)  for the definitions of these symbols. 

Before proving properties of $ \CC_{\Sigma,\Sigma}$ and ${\frak E}_{\Sigma}$,
  we need to verify that $\Sigma$ is indeed a permutation.
This follows from the fact that $\WW^{[i]}$, for every $i\in [g]$,
  is nondegenerate over $\mathbb{Z}_{d_i}$, and $k_{i,j}$ for all $i\in [s]$ and $j\in [t_i]$
  satisfies $\gcd(k_{i,j},q_{i,j})=1$ (so $\xx'$ above is unique).
We use $\Sigma_0$ to denote the $(0,*)$-part of $\Sigma$ 
  and $I$ to denote the identity map: \[\Sigma(0,\xx)=(0,\Sigma_0(\xx))
  =(0,\xx')\ \ \ \text{for all $\xx\in \mathbb{Z}_{\dd}^2\times \mathbb{Z}_{\calQ}$.}\]

Now we can write $\CC_{\Sigma,\Sigma}$ and ${\frak E}_{\Sigma}=(\EE^{[0]}_{\Sigma}
  ,\ldots,\EE^{[N-1]}_\Sigma)$ as
\begin{equation}\label{everything-ours}
\CC_{\Sigma,\Sigma}=\left(\begin{matrix}\00\hspace{0.15cm} & \FF_{\Sigma_0,I}\\
\FF_{I,\Sigma_0} & \00\end{matrix}\right)\ \ \ \text{and}\ \ \
\EE^{[r]}_{\Sigma}=\left(\begin{matrix}\DD^{[r]}_{\Sigma_0} & \00\\
\00& \DD^{[r]}\end{matrix}\right) 
\end{equation}
{for all $r\in [0:N-1]$.} We make the following two observations:
%\begin{enumerate}
%\item[] \hspace{-0.6cm}

%\begin{description}
\begin{itemize}
\item\mbox{Observation 1:\,}
%Observation $1$: 
For any edge gadget $\Gamma$,
$\gm_{\Gamma, \CC_{\Sigma,\Sigma},\fE_\Sigma}$ is \modblrkone/ iff $\gm_{\Gamma, \CC, \fE}$ is \modblrkone/,
and for any edge gadget $\Gamma$ such that $\gm_{\Gamma, \CC, \fE}$ is not \modblrkone/
there is an edge gadget $\Gamma'$ such that $\gm_{\Gamma', \FF, \fD}$ is not \modblrkone/;
and, thus for any edge gadget $\Gamma$, $M_{\CC_{\Sigma,\Sigma},\fE_\Sigma}$ is \modblrkone/.
%\item[] \hspace{-0.6cm} 
\item\mbox{Observation 2:\,} 
%$\FF'$ is just $\FF_{\Sigma_0,I}$, where $I$ is
%  the identity map, and $\EE^{[r,1]}=\DD^{[r]}$, for all $r$;
% Observation $2$: 
$\FF_{\Sigma_0,I}$ satisfies 
%(letting $\xx'=\Sigma_0(\xx)$)
\begin{align*}
\big(\FF_{\Sigma_0,I}\big)_{\xx,\yy} %=F_{\Sigma_0(\xx),\yy}
  =F_{\xx',\yy} &= \prod_{i\in [g]}\hspace{0.15cm}
\oo_{d_i}^{(x'_{0,i,1}\hspace{0.07cm} x'_{0,i,2})\cdot \WW^{[i]}
  \cdot (y_{0,i,1}\hspace{0.07cm} y_{0,i,2})^{\text{T}}}
\hspace{-0.2cm}\prod_{i\in [s],j\in [t_i]}
  \omega_{q_{i,j}}^{k_{i,j}\cdot x'_{1,i,j} y_{1,i,j}}\\[0.1ex]
  &= \prod_{i\in [g]}\hspace{0.15cm}
\oo_{d_i}^{x_{0,i,1}y_{0,i,1}+ x_{0,i,2} y_{0,i,2} }
\hspace{-0.2cm}\prod_{i\in [s],j\in [t_i]}
  \omega_{q_{i,j}}^{x_{1,i,j} y_{1,i,j}}.
\end{align*}
%\end{enumerate}
\end{itemize}
%\end{description}

By Observation 2, it is easy to show that
  $\CC_{\Sigma,\Sigma}$ and ${\frak E}_{\Sigma}$
  (together with appropriate $\qq',\bft',\fq'$) satisfy condition $(\calR)$.
Since by Observation 1,
for any edge gadget $\Gamma$, $M_{\CC_{\Sigma,\Sigma},\fE_\Sigma}$ is \modblrkone/,
it follows from Theorem \ref{bi-step-4-ours} and (\ref{everything-ours})
  that (the lower half) $\DD^{[r]}$ satisfies $(\calL_2)$, and then $(\calL_3)$.
This proves Theorem \ref{t-step-4-ours} since $(\calL_1')$
  and $(\calL_2')$ follow from $(\calL_2)$ and $(\calL_3)$, respectively.

We continue to prove Theorem \ref{nonbi-step-5-ours}. 
Suppose for any edge gadget $\Gamma$, $\gm_{\Gamma, \FF, \fD}$ is \modblrkone/. Then the argument above shows that
  $(\CC_{\Sigma,\Sigma},{\frak E}_{\Sigma})$ (with appropriate $\pp',\bft'
  ,\fq'$) satisfies both $(\calR)$ and $(\calL)$.
Since by Observation 1,
for any edge gadget $\Gamma$, $\gm_{\Gamma, \CC_{\Sigma,\Sigma}, \fE_\Sigma}$ is \modblrkone/,
 by Theorem \ref{bi-step-5-ours} and (\ref{everything-ours}),
  $\DD^{[r]}$  satisfies  $(\calD_2)$ and $(\calD_4)$ for every $r\in \calZ$. 
$(\calD_1')$ follows from $(\calD_2)$.

To prove $(\calD_2')$,\vspace{-0.06cm}
   let $\FF'=\FF_{\Sigma_0,I}$. 
By $(\calD_4)$, for any $r\in \calZ$, $k\in [s]$ and \mbox{$\aa\in \Gamma_{r,k}^{\text{lin}}$,}
  there exist \mbox{$\bb\in \tilde{\mathbb{Z}}_{\qq_k}$} and $\alpha\in \mathbb{Z}_N$ such that
\[
\oo_N^{\alpha}\cdot F_{\widetilde{\bb},
  \xx}'=D^{[r]}_{\xx+\widetilde{\aa}}\cdot \overline{
  D^{[r]}_{\xx}}\ \ \ \text{for all $\xx\in \Gamma_r$,\ where\ 
$\FF_{\widetilde{\bb},*}'=
  \FF_{\Sigma_0(\widetilde{\bb}),*}$.}\]
Since $\Sigma_0$ works within each prime factor, there exists
  a $\bb'\in \tilde{\mathbb{Z}}_{\qq_k}$ such that
  $\Sigma_0(\widetilde{\bb})=\widetilde{\bb}'$ and ($\calD_2'$) follows.
%and $\Sigma_0(\widetilde{\bb})\in \hat{\mathbb{Z}}_{\qq_k}$ by the
%  construction of $\Sigma$.

%% file: tractability-unipartite.tex
%%%%%%%%%%%%%%%%%%%%%%%%%%%%%%%%%%%%%%%%%%%%%%%%%%%%%%%%%%%%%%%%%%%%
\begin{comment}

{\color{blue} (old text)
\section*{Tractability: Proof of Theorem \ref{tractable-2}}

The proof of Theorem \ref{tractable-2} 
  is~si\-milar to that of Theorem \ref{tractable-1}
  for the bipartite case presented in Section \ref{tractabilitysec}.

Let $((M,N),\FF,\fD,(\dd,\cal{W},\pp,\bft,\calQ,\calK))$ be a tuple that satisfies
    $(\calR'), (\calL')$ and ($\calD'$).
The proof has two steps.
First we use $(\calR'), (\calL'), (\calD')$
  to decompose $\eval(\FF,\fD)$
  into $s$ subproblems (recall $s$ is the length of the sequence $\pp$),
  denoted by
$\eval(\FF^{[i]},\fD^{[i]})$, $i\in [s]$,
such that, if every $\eval(\FF^{[i]},\fD^{[i]})$ is tractable then 
  so is $\eval(\FF,\fD)$.
Second, we reduce each $\eval(\FF^{[i]},\fD^{[i]})$ 
  to $\eval(\pi)$ for some prime power $\pi$.
%%%Recall $\eval(\pi)$ is the following problem: Given a quadratic polynomial
%%%  $f(x_1,\ldots,x_n)$ over $\mathbb{Z}_\pi$,
%%%  compute
%%%$$
%%%Z_\pi(f)=\sum_{x_1,\ldots,x_n\in \mathbb{Z}_\pi} \oo_{\pi}^{f(x_1,\ldots,x_n)}.
%%%$$

By Theorem \ref{pinyan}, $\eval(\pi)$ can be solved in polynomial 
  time for any fixed prime power $\pi$.
Thus, $\eval(\FF^{[i]},\fD^{[i]})$ is tractable for all $i\in [s]$, and so is
  $\eval(\FF,\fD)$.
}

{\color{red} (new text)}
\end{comment}
%%%%%%%%%%%%%%%%%%%%%%%%%%%%%%%%%%%%%%%%%%%%%%%%%%%%%%%%%%%%%%%%%%%%

\section{Tractability: Proof of Theorem \ref{tractable-2-ours}}\label{sec:Tractability-uni}

The statement of Theorem~\ref{tractable-2-ours} is exactly
the same as that of Theorem 6.10 in~\cite{Cai-Chen-Lu-2013}
whose proof is given in Section 16 in~\cite{Cai-Chen-Lu-2013}.
We just repeat a short outline given at the beginning of Section~16 in~\cite{Cai-Chen-Lu-2013}.

%{\color{red} (original ccl text)}

The proof of Theorem \ref{tractable-2-ours} 
  is~si\-milar to that of Theorem \ref{tractable-1-ours}
  for the bipartite case presented in Section \ref{tractabilitysec-ours}.

Let $((M,N),\FF,\fD,(\dd,\cal{W},\pp,\bft,\calQ,\calK))$ be a tuple that satisfies
    $(\calR'), (\calL')$, and ($\calD'$).
The proof has two steps.
First we use $(\calR'), (\calL'), (\calD')$
  to decompose $\eval(\FF,\fD)$
  into $s$ subproblems (recall $s$ is the length of the sequence $\pp$),
  denoted by
$\eval(\FF^{[i]},\fD^{[i]})$, $i\in [s]$,
such that if every $\eval(\FF^{[i]},\fD^{[i]})$ is tractable, then 
  so is $\eval(\FF,\fD)$.
Second, we reduce each $\eval(\FF^{[i]},\fD^{[i]})$ 
  to $\eval(\pi)$ for some prime power $\pi$.
%%%Recall $\eval(\pi)$ is the following problem: Given a quadratic polynomial
%%%  $f(x_1,\ldots,x_n)$ over $\mathbb{Z}_\pi$,
%%%  compute
%%%$$
%%%Z_\pi(f)=\sum_{x_1,\ldots,x_n\in \mathbb{Z}_\pi} \oo_{\pi}^{f(x_1,\ldots,x_n)}.
%%%$$

By Theorem \ref{pinyan-ours}, $\eval(\pi)$ can be solved in polynomial 
  time for any fixed prime power $\pi$.
Thus, $\eval(\FF^{[i]},\fD^{[i]})$ is tractable for all $i\in [s]$, and so is
  $\eval(\FF,\fD)$.

%{\color{red} (added here, see also in Section~\ref{sec:constructivity} (constructivity)}

\begin{remark}
The difference of the proof of tractability for the nonbipartite case
versus the bipartite case resides in what we will state
in Section~\ref{sec:constructivity} for the constructivity of our results.
The proof for the bipartite case
(Theorem~5.10) in~\cite{Cai-Chen-Lu-2013} involves
the second pinning lemma (Lemma 4.3 from~\cite{Cai-Chen-Lu-2013}, of which
we do not have a constructive proof, and we avoid in this paper),
while in the nonbipartite case
(Theorem 6.10)   the proof in~\cite{Cai-Chen-Lu-2013} does not.
%
%The proof for the non-bipartite case 
%(Theorem 6.10) given in~\cite{Cai-Chen-Lu-2013} does not involve
%the second pinning lemma (Lemma 4.3 from~\cite{Cai-Chen-Lu-2013}, of which
%we do not have a constructive proof, and we avoid in this paper),
%while in the bipartite case 
%(Theorem~5.10) the proof in~\cite{Cai-Chen-Lu-2013} does.
%%% lm 12.2. p 1002  (basically CCL claims full EVAL (C, D) tractable
%%% althjugh the algorithmic part relaly gives the arrowed verison also
%%% of EVAL (C, D) is tractable.  but formally, in CCL, it does
%%% from tractability of  EVAL (C, D) to 
% tractability of the arrowed version of  EVAL (C, D). this requires 2nd pinning lm.
Consequently, here for the nonbipartite case
we can simply use the proof in~\cite{Cai-Chen-Lu-2013}.
\end{remark}

%% file: simple-gadget-meta-replacement.tex
\section{Dichotomy over simple graphs of bounded degree}\label{sec:GH-dich-simp}

%%%%%%%%%%%%%%%%%%%%%%%%%%%%%%%%%%%%%%%%%%%%%%%%%%%%%%%%%%%%%%%%%%%%
\begin{comment}

{\color{red} (some more preliminaries copied from GH-dim paper, is it possible to omit these defintions somehow?
Definitely we need to recap some notation such as $\hom$, $\hom_\phi$, labeled graphs, and explain the twin compression procedure)}

{\color{red} We recall some definitions from~\cite{Cai-Govorov-GH-hom-type}. (added)
\end{comment}
%%%%%%%%%%%%%%%%%%%%%%%%%%%%%%%%%%%%%%%%%%%%%%%%%%%%%%%%%%%%%%%%%%%%

%{\color{red} 
%%%% gone
%We need the notion of $k$-labeled graphs.
%%
%%We first recap the notion of weighted graph 
%%homomorphisms~\cite{Freedman-Lovasz-Schrijver-2007}, 
%%but state it for an arbitrary field $\bbF$.
%%
%Let $[k] = \{ 1, \ldots, k \}$ for integer $k \ge 0$, 
%where we denote $[0] = \emptyset$.
%%%%%%%% i think no need to say this here
%%By covention $\mathbb{C}^0  = \{\emptyset\}$,
%%and $0^0 = 1$.
%% in $\bbZ$, $\bbF$, etc.
%%Often we discuss both directed and undirected graphs together.
%(JYC: The whole line can go or should I say something else, as for the second sentence, the notation was already introduced in the preliminaries. As for the first sentence, we do need it but we first introduce weighted graphs.)}
We need to introduce additional definitions. Recall that $[k] = \{ 1, \ldots, k \}$ for integer $k \ge 0$, 
where we denote $[0] = \emptyset$.

A \textit{weighted graph} $H = H_{\AA, \DD}$ \label{H_{A,D}}
 is determined by
a symmetric matrix $\AA \in \mathbb{C}^{m \times m}$ for edge weights
and a vector (or equivalently a diagonal  matrix) $\DD
= (D_1, \ldots, D_m)$ for vertex weights. %{\color{red} (Here $m \ge 0$.)}
In this section, we assume
%We assume
all $D_i \not = 0$, otherwise we delete the
corresponding row and column of $\AA$. %{\color{red} (JYC: the real reason is that we need to apply results from itcs, just a comment)}
%%%%%%%%%%%%%%%%%%%%%%%%%%%%%%%%%%%%%%%%%%%%%%%%%%%%%%%%%%%%%%%%%%%%
\begin{comment}

%n ($\bbF$-)
\textit{weighted graph} $H$ is a finite 
%(di)graph with a 
graph with a
weight $\alpha_H(i) \in \bbF \setminus \{0\}$ associated with 
each vertex $i$ ($0$-weighted vertices can be deleted) 
and a weight $\beta_H(i, j) \in \bbF$ associated with 
each edge $ij$ (or loop if $i=j$). For undirected graphs,
$\beta_H(i, j)= \beta_H(j, i)$.
It is convenient to assume that
$H$ is a complete graph with a loop at all nodes by adding
all missing edges and loops with weight $0$.
Then $H$ is described by an integer $q = |V(H)| \ge 0$ ($H$ can be the empty graph), a nowhere zero vector $\alpha = (\alpha_H(1), \ldots, \alpha_H(q)) \in \bbF^q$
and a matrix $B = (\beta_H(i, j)) \in \bbF^{q \times q}$.
An isomorphism from $H$ to $H'$ is a bijective map
from $V(H)$ to $V(H')$ that preserves vertex and edge weights.

\end{comment}
%%%%%%%%%%%%%%%%%%%%%%%%%%%%%%%%%%%%%%%%%%%%%%%%%%%%%%%%%%%%%%%%%%%%
The following notation from~\cite{Freedman-Lovasz-Schrijver-2007}
is often convenient.
%According to~\cite{Freedman-Lovasz-Schrijver-2007},
Let $G$ be an unweighted graph (with possible 
multiple edges, but no loops) and $H$ a weighted graph given by $(\AA, \DD)$,
we define (see Definition~\ref{def:EVAL(A,D)})
%\begin{equation}\label{eqn:hom-def}
%\hom(G, H) = \sum_{\small {\phi \colon V(G) \to V(H)}}
%\alpha_\phi \hom_\phi(G, H) =
%\sum_{\small {\phi \colon V(G) \to V(H)}}
%\prod_{u \in V(G)} \alpha_H(\phi(u))
% \prod_{u v \in E(G)} \beta_H(\phi(u), \phi(v)).
%\end{equation}
%%% alternative formula
\begin{equation}\label{eqn:hom-def}
%\begin{aligned}
\hom(G, H) = Z_{\AA, \DD}(G)
= \sum_{\small {\phi \colon V(G) \to V(H)}}
D_\phi  \hom_\phi(G, H),
\end{equation}
where for $\phi \colon V(G) \to V(H)$, %{\color{red} (removed the comma in the equation below, don't like it)}
\[
D_\phi = \prod_{w \in V(G)} D_{\phi(w)}~~~~\mbox{and}~~~~
\hom_\phi(G, H) = \prod_{u v \in E(G)} A_{\phi(u), \phi(v)}.
\]
%\alpha_\phi \hom_\phi(G, H) \\
%&= \sum_{\small {\phi \colon V(G) \to V(H)}} \prod_{u \in V(G)} \alpha_H(\phi(u))
 %\prod_{u v \in E(G)} \beta_H(\phi(u), \phi(v)).
%\end{aligned}
%The unweighted case is when all vertex weights are $1$ and all edge weights
%are $0$-$1$ in $H$,
%and $\hom(G, H)$ is the number of homomorphisms from $G$ into $H$.
%
When all $D_i =1$, we say this is the edge-weighted case,
and we denote
$H$ by $H_\AA$. \label{H_A}

A $k$-labeled graph ($k \ge 0$) is a finite graph in which $k$ nodes are labeled by $1, 2, \ldots, k$ (the graph can have any number of unlabeled nodes). 
Two $k$-labeled graphs are isomorphic if there is a label-preserving isomorphism between them. 
$U_k$ denotes the $k$-labeled graph on $k$ nodes with no edges. In particular, 
$U_0$ is the empty graph with no nodes and no edges.
The  \textit{product} $G_1 G_2$ of  two $k$-labeled graphs $G_1$ and $G_2$
 is defined as follows: take their disjoint union, and then identify nodes with the same label. Hence for two $0$-labeled graphs, $G_1 G_2$
% = G_1 \sqcup G_2$ (disjoint union).
is just the disjoint union of $G_1$ and $G_2$.
 Clearly, the graph product is associative and commutative
 with the identity $U_k$, so the set of all (isomorphism classes) of $k$-labeled graphs togegher with the product operation forms a commutative monoid which we denote by $\calPLG[k]$. \label{PLG[k]}
We denote by $\calPLG_{\simp}[k]$ the submonoid of \label{PLGsimp[k]}
\emph{simple} graphs in $\calPLG[k]$; these are graphs with no loops, at most one
edge between any two vertices $i$ and $j$, and \emph{no edge} between labeled vertices.
%A directed labeled graph is simple if its underlying undirected one
%is simple; 
%in particluar, for any $i$ and $j$, we require that
%if $i \rightarrow j$ is an edge then $j \rightarrow i$ is not an edge.
Clearly,  $\calPLG_{\simp}[k]$  is closed under the product operation.
%(for both directed and undirected types).

Fix a weighted graph $H$ given by $(\AA, \DD)$,
and let $G$ be a $k$-labeled graph.
Let $\psi \colon [k] \to V(H)$. 
We say $\phi \colon V(G) \to V(H)$
extends $\psi$,
if $\phi(u_i) = \psi(i)$ for the $i$th labeled vertex $u_i \in V(G)$, $i \in [k]$. %{\color{red} (for the vertex $u_i \in V(G)$ labeled $i$, it the meaning clear otherwise? if replacing we have $i$, $i \in [k]$, not sure it's good)}
If $\phi$ extends $\psi$, we write
$D_{\phi/\psi} = \prod_{w \in V(G) \setminus \{u_i: i \in [k]\}}
D_{\phi(w)}$ to denote the product of vertex weights  
 \emph{other than} $D_{\psi(i)}$, $i \in [k]$, and
\begin{equation}\label{eqn:partial-hom-dec}
\hom_\psi(G, H) = \sum_{\small \substack{\phi \colon V(G) \to V(H) \\ \phi \text{ extends } \psi}} D_{\phi/\psi} \hom_\phi(G, H).
%\prod_{w \in V(G) \setminus \{\psi(i): i \in [k]\}} D_{\phi(w
%\frac{\alpha_\phi}{\alpha_\psi} \hom_\phi(G, H), %\tiny
\end{equation}

Note that 
a 2-labeled graph is synonymous  with an edge gadget.
If $\firstdv, \seconddv$ are the (ordered) two  distinguished vertices
of an edge gadget $\Gamma$, then
%{\color{red} (by extending Definition~\ref{def:M(Gamma,A)} according to Definition~\ref{def:EVAL(A,D)}) (JYC: this addition is out of place better to introduce it under Definition~\ref{def:M(Gamma,C,D)}, see the red text there)}
% $\gm_{\Gamma, \AA}$ or
$\gm_{\Gamma, \AA, \DD}$ is the matrix with
$\hom_\psi(\Gamma, H)$ as its $(i,j)${th} entry, where
$\psi(1) = i$ and $\psi(2) = j$.
We also denote this as $\hom_{(i,j)} (\Gamma, H)$. \label{hom_{(i,j)}(G,H)}

%%%%%%%%%%%%%%%%%%%%%%%%%%%%%%%%%%%%%%%%%%%%%%%%%%%%%%%%%%%%%%%%%%%%
\begin{comment}
where $\phi  \text{ extends } \psi$ means that
 if $u_i \in V(G)$ is labeled by $i \in [k]$
then $\phi(u_i) = \psi(i)$, and
 $\alpha_\psi= \prod_{i=1}^k \alpha_H(\psi(i))$, $\alpha_\phi
= \prod_{v \in V(G)} \alpha_H(\phi(v))$, so $\frac{\alpha_\phi}{\alpha_\psi}$
is the product of vertex weights of $\alpha_\phi$ \emph{not} in $\alpha_\psi$. 
Then
\begin{equation}\label{eqn:full-hom-dec}
\hom(G, H) = \sum_{\psi \colon [k] \to V(H)} \alpha_\psi \hom_\psi(G, H).
\end{equation}
When $k = 0$, we only have the empty map $\emptyset$ with the domain $\emptyset$.
Then $\hom(G, H) = \hom_\emptyset(G, H)$ for every $G \in \calPLG[k]$.
The functions $\hom_\psi(\cdot, H)$ where $\psi \colon [k] \to V(H)$ and $k \ge 0$ satisfy
\begin{equation}\label{eq:part-hom-multiplicativity}
\begin{lcases}
&\hom_\psi(G_1 G_2, H) = \hom_\psi(G_1, H) \hom_\psi(G_2, H), \quad G_1, G_2 \in \calPLG[k], \\
&\hom_\psi(U_k, H) = 1.
\end{lcases}
\end{equation}

\end{comment}
%%%%%%%%%%%%%%%%%%%%%%%%%%%%%%%%%%%%%%%%%%%%%%%%%%%%%%%%%%%%%%%%%%%%

Given a 
%directed or undirected $\bbF$-
weighted graph $H$ specified by $(\AA, \DD)$,
we call two vertices $i, j \in V(H)$ \textit{twins} if 
the $i$th row and $j$th row of $\AA$ are identical
(by symmetry, the $i$th column and $j$th column of $\AA$ are identical as well).
%for every vertex $\ell \in V(H)$,
%$\beta_H(i, \ell) = \beta_H(j, \ell)$ and $\beta_H(\ell, i) = \beta_H(\ell, j)$.
Note that the vertex weights
% $\alpha_H(w)$ 
$\DD$ do not participate in this definition.
If $H$ has no twins, we call it twin-free.

The twin relation 
partitions $V(H)$ into nonempty equivalence classes,
$I_1, \ldots, I_s$ where $s \ge 0$.
We can define a \textit{twin contraction} graph  $\widetilde H$,
having $I_1, \ldots, I_s$  as vertices,
with vertex weight $\sum_{t \in I_r} D_t$ for $I_r$,
and edge  weight between $I_r$ and $I_q$ to be $\AA_{u, v}$ %$\AA(u, v)$ {\color{red} (JYC: better than $A_{u, v}$ because $u, v$ are vertices?)}
for some arbitrary $u \in I_r$ and $v \in I_q$.
After that, we remove all vertices in  $\widetilde H$ 
with zero vertex weights together with all incident edges (still called 
$\widetilde H$).
This defines a twin-free $\widetilde H$.
Clearly, $\hom(G, H) = \hom(G, \widetilde H)$ for all $G$.

An isomorphism from a weighted graph $H$ to another $H'$ is a bijection
$\sigma \colon V(H) \to V(H')$
that preserves vertex and edge  weights.
It is obvious that for any weighted graphs $H$ and $H'$,
and  maps $\varphi \colon [k] \to V(H)$ and $\psi \colon [k] \to V(H')$ such that $\psi = \sigma \circ \varphi$
for some isomorphism $\sigma \colon V(H) \to V(H')$ from $H$ to $H'$,
we have $\hom_\varphi(G, H) = \hom_\psi(G, H')$ for every $G \in \calPLG[k]$.
%}
When $H' = H$, we denote by 
%$\Isom(H, H')$ the set of weighted graph isomorphisms from $H$ to $H'$,
%and by 
$\Aut(H)$ the group of weighted graph automorphisms of $H$,
i.e.,  isomorphisms  from $H$ to itself.

%{\color{red} (state the main theorem from that paper, otherwise it's not clear what we will use)

We will need the following result from~\cite{Cai-Govorov-GH-hom-type},
which is proved for directed and undirected graphs with weights from
any field of characteristic $0$. We will only  use it 
for $\bbC$-weighted undirected graphs.
%but will use it in the undirected graph setting.

\begin{theorem}\label{thm:gh-main}
%Let $\bbF$ be a field of characteristic $0$.
Let $H, H'$ be twin-free $\bbC$-weighted graphs.
%such that $H$ is twin-free and $m = |V(H)| \ge m' = |V(H')|$.
Suppose $\varphi \colon [k] \to V(H)$ and $\psi \colon [k] \to V(H')$ where $k \ge 0$.
If $\hom_\varphi(G, H) = \hom_\psi(G, H')$ for every $G \in \calPLG_{\simp}[k]$, %(for every directed simple loopless $k$-labeled graph $G$),
then $|V(H)|= |V(H')|$, and there exists an isomorphism
% of $\bbF$-weighted graphs
$\sigma \colon V(H) \to V(H')$ from $H$ to $H'$
such that $\psi = \sigma \circ \varphi$.
%(a fortiori, $H'$ is twin-free and $m = m'$).
\end{theorem}

Note that in Theorem~\ref{thm:gh-main} all vertex weights are nonzero;
this is by our definition of weighted graphs in this section.
So
when we apply this theorem we need to make sure that this is satisfied.
%that all vertex weights are nonzero.

If $H$ is a weighted graph, $\varphi \colon [k] \to V(H)$ where $k \ge 0$
and none of the vertices from $\varphi([k])$ are removed
during the twin reduction of $H$ to $\widetilde H$,
then by defining $\widetilde \varphi(i) = \widetilde{\varphi(i)}$ 
for each $i \in [k]$, where $\widetilde{\varphi(i)}$ denotes the
equivalence class that $\varphi(i)$ belongs to in $V(\widetilde H)$,
we get a well-defined map $\widetilde \varphi \colon [k] \to V(\widetilde H)$.
Then it is easy to check that $\hom_\varphi(G, H) = \hom_{\widetilde \varphi}(G, \widetilde H)$
for any $G \in \calPLG_{\simp}[k]$.
% (directed or undirected depending on the context).

%%%%%%%%%%%%%%%%%%%%%%%%%%%%%%%%%%%%%%%%%%%%%%%%%%%%%%%%%%%%%%%%%%%%
\begin{comment}

In this paper $k$-labeled graphs (i.e., from $\calPLG[k]$) correspond to graphs
(or gadgets, but we define only edge-gadgets, e.g., $k = 2$)
with $k$ labeled vertices corresponding to distinguished vertices.
Also simple $k$-labebeled graphs (i.e., from $\calPLG_{\simp}[k]$) correspond to simple graphs
with $k$ distinguished vertices with no edge between any two.

\end{comment}
%%%%%%%%%%%%%%%%%%%%%%%%%%%%%%%%%%%%%%%%%%%%%%%%%%%%%%%%%%%%%%%%%%%%

%We also need to generalize the notion of
%the tensor product of graphs to $\bbF$-weighted graphs.
Given 
%$\bbF$-
weighted graphs $H = H_{\AA, \DD}$ and $H' = H_{\AA', \DD'}$,
the \emph{tensor product} of $H$ and $H'$ is a graph denoted by $H \times H'$
with vertex set $V(H) \times V(H')$,  each vertex $(u, u')$
has weight $D_u D'_{u'}$, and the edge between
$(u, u')$ and $(v, v')$ has weight $A_{u, v} A'_{u', v'}$.
Clearly, if $H$ and $H'$ have no zero vertex weights, then neither does $H \times H'$.
Also note that if $H$ and $H'$ are edge-weighted graphs (i.e.,
all vertex weights are 1), then so is $H \times H'$.
%It is easy to see that the tensor product of $\bbF$-weighted graphs
%is commutative and associative in the sense that for $\bbF$-weighted 
%graphs $H$ and $H'$
%the $\bbF$-weighted graphs $H \times H'$ and $H' \times H$ are isomorphic
%via $(u, u') \times (u', u)$) and for $\bbF$-weighted graphs $H, H'$ and $H'$
%the $\bbF$-weighted graphs $(H \times H') \times H''$ and $H \times (H' \times H'')$ are isomorphic.
%
Clearly, $H \times H'$ and $H' \times H$ are isomorphic,
and also $(H \times H') \times H''$ and $H \times (H' \times H'')$ are isomorphic.
%{\color{red}
%
%If $\AA, \AA' \in \bbF^{m \times m}$,
%then the edge weight matrix corresponding
The edge weight matrix corresponding to $H_\AA \times H_{\AA'}$ is $\AA \otimes \AA'$,
and
%More generally, 
the pair corresponding to $H_{\AA, \DD} \times H_{\AA', \DD'}$
is $(\AA \otimes \AA', \DD \otimes \DD')$.

%to $H_\AA \times H_{\AA'}$ is $\AA \otimes \AA'$.
%Additionally, if $\DD, \DD' \in \bbF^{m \times m}$ are diagonal,
%then the pair corresponding to $H_{\AA, \DD} \times H_{\AA', \DD'}$
%is $(\AA \otimes \AA', \DD \otimes \DD')$.
%
%Next, if $\AA, \AA' \in \bbF^{m \times m}$,
%then $H_\AA \times H_{\AA'}$ is an $\bbF$-edge-weighted graph
%with the edge weight matrix $\AA \otimes \AA'$.
%Additionally, if $\DD, \DD' \in \bbF^{m \times m}$ are diagonal,
%$H_{\AA, \DD} \times H_{\AA', \DD'}$ is an $\bbF$-weighted graph
%with the edge weight matrix $\AA \otimes \AA'$
%and the vertex weight matrix $\DD \otimes \DD'$.
%{\color{red} (feels better)}

%{\color{red} 
%%%%%%%%%%%%%%%%% the following paragraph should be enough
%Given $f \colon X \to Y$ and $g \colon X' \to Y'$,
%define $f \times g \colon X \times X' \to Y \times Y'$
%by $(f \times g)(x, x') = (f(x), f'(x'))$.
%where $(x, x') \in X \times X'$.

Now if $H$ and $H'$ are weighted graphs,
$\varphi \colon [k] \to V(H)$ and $\psi \colon [k] \to V(H')$, then
\[
\hom_{\varphi \times \psi}(G, H \times H') = \hom_\varphi(G, H) \cdot \hom_\psi(G, H')
\]
for any $G \in \calPLG_{\simp}[k]$.
%(directed or undirected depending on the context)
Here $\varphi \times \psi \colon [k] \to V(H \times H')$
is defined by $(\varphi \times \psi)(i) = (\varphi(i), \psi(i))$. 

Similarly, we can define the tensor product of 
any finite number of graphs
%an arbitrary number of graphs
and show similar properties.

Now we are ready to prove that the \#P-hardness part of
Theorem~\ref{thm:main-dich-bd} holds even when restricted to
\emph{simple graphs} of bounded degree.
 %the formulate and prove a statement that
%is instrumental in transitioning from Theorem~\ref{thm:main-dich-bd-edge-gadget} 
%to Theorem~\ref{thm:main-dich-bd-simp-edge-gadget} (resulting in Theorem~\ref{thm:main-dich-bd-simp}),
%and which together with Corollary~\ref{cor:non-mult-bl-rk-1-gadget-hardness-bd-simp}
%which will yield a generalization of our \#P-hardness part to bounded degree simple graphs.
 
\begin{theorem}\label{thm:gadget-simple-non-mult-bl-rk-1}
Let $\AA \in \bbC^{m \times m}$ be a symmetric matrix
and $\Gamma$ be an edge gadget.
If $\gm_{\Gamma, \AA}$ is not \multblrkone/,
then there exists  a \emph{simple} edge gadget $\Gamma'$
%(i.e., no multiple edges and no edges between the distinguished vertices)
such that the matrix $\gm_{\Gamma', \AA}$ is not \multblrkone/.
\end{theorem}
\begin{proof}
Let $\firstdv, \seconddv$ be the distinguished vertices of $\Gamma$ (in this order).
Let $\CC = \gm_{\Gamma, \AA}$.
Then
\[
C_{i, j} = \sum_{\substack{\xi \colon V(\Gamma) \to [m] \\ \xi(\firstdv) = i,~ \xi(\seconddv) = j}} \prod_{(u, v) \in E(\Gamma)} A_{\xi(u), \xi(v)} = \hom_{(i, j)}(\Gamma, H_\AA)
\]
for $1 \le i, j \le m$.
By Lemma~\ref{lem:fin-many-roots-of-unity-full},
in the field $\bbQ(\{A_{i, j}\}_{i, j = 1}^m)$
the roots of unity form a finite cyclic group.
Suppose $R$ is (or any positive multiple of) this order. %(a positive integer multiple of)
We can fix a root of unity  $\varepsilon_R \in \bbQ(\{A_{i, j}\}_{i, j = 1}^m)$
of order $R$.
% that generates this cyclic group.

%{\color{red}
%We are given that $C$ is not \multblrkone/.
%This means that there exist $1 \le i_1 < i_2 \le m$ and $1 \le j_1 < j_2 \le m$ such that
%either the matrix (the following submatrix of $\gm_{\Gamma, B)}$)
%\[
%C_{i_1, i_2; j_1, j_2} =
%\begin{bmatrix}
%C_{i_1, j_1} & C_{i_1, j_2} \\
%C_{i_2, j_1} & C_{i_2, j_2}
%\end{bmatrix}
%\]
%has exactly one zero element or has no zero elements
%but for any $n \ge 1$, its $n$th Hadamard power,
%i.e., the matrix
%\[
%\begin{bmatrix}
%C_{i_1, j_1}^n & C_{i_1, j_2}^n \\
%C_{i_2, j_1}^n & C_{i_2, j_2}^n
%\end{bmatrix}
%\]
%is nondegenerate.
%It is easy to see that this is equivalent
%to saying that $C_{i_1, i_2; j_1, j_2}$
%contains at least three nonzero elements
%and for every $n \ge 1$,
%its $n$th Hadamard power is nondegenerate
%so
%\[
%C_{i_1 j_1}^n C_{i_1 j_2}^n - C_{i_2 j_1}^n C_{i_2 j_2}^n \ne 0.
%\]
%}

Since $\CC$ is not \multblrkone/,
for some $1 \le i_1 < i_2 \le m$ and $1 \le j_1 < j_2 \le m$,
%we have that $(i_1, i_2, j_1, j_2)$ witness
%the fact that $C$ is not \multblrkone/
%(the non-\multblrkone/-ness of $C$).
%In other words,
the $2 \times 2$ submatrix
\[
%\CC_{i_1, i_2; j_1, j_2} =
\begin{pmatrix}
C_{i_1, j_1} & C_{i_1, j_2} \\
C_{i_2, j_1} & C_{i_2, j_2}
\end{pmatrix}
\]
contains at least three nonzero entries
and its $R$th Hadamard power 
%$\CC_{i_1, i_2; j_1, j_2}^{\odot R}$ 
is nondegerate, i.e.,
\[
 C_{i_1, j_1}^R C_{i_2, j_2}^R - C_{i_1, j_2}^R C_{i_2, j_1}^R \ne 0.
\]
Let $(x, y) \in \{i_1, i_2\} \times \{j_1, j_2\}$
be so that for
%any entry of $\CC_{i_1, i_2; j_1, j_2}$
%other than $C_{x, y}$ is nonzero.
%($C_{x, y}$ may or may not be zero.)
%In other words, letting
\[
I = (\{i_1, i_2\} \times \{j_1, j_2\}) \setminus \{x, y\} = \{ (a_1, b_1), (a_2, b_2), (a_3, b_3) \},
\]
we have $C_{i, j} \ne 0$ for each $(i, j) \in I$.
Then
\[
(\prod_{(i, j) \in I} C_{i, j}) C_{i_1, j_1}^R C_{i_2, j_2}^R \ne (\prod_{(i, j) \in I} C_{i, j}) C_{i_1, j_2}^R C_{i_2, j_1}^R
\]
which is the same as
\begin{multline}\label{eq:part-hom-noneq}
(\prod_{(i, j) \in I} \hom_{(i, j)}(\Gamma, H_\AA)) (\hom_{(i_1, j_1)}(\Gamma, H_\AA))^R (\hom_{(i_2, j_2)}(\Gamma, H_\AA))^R \\
\ne (\prod_{(i, j) \in I} \hom_{(i, j)}(\Gamma, H_\AA)) (\hom_{(i_1, j_2)}(\Gamma, H_\AA))^R (\hom_{(i_2, j_1)}(\Gamma, H_\AA))^R.
\end{multline}
Let
\begin{align*}
x_1 &= (a_1, a_2, a_3, \underbrace{i_1, \ldots, i_1}_{R \text{ times}}, \underbrace{i_2, \ldots, i_2}_{R \text{ times}}), \quad x_2 = (a_1, a_2, a_3, \underbrace{i_1, \ldots, i_1}_{R \text{ times}}, \underbrace{i_2, \ldots, i_2}_{R \text{ times}})\\
y_1 &= (b_1, b_2, b_3, \underbrace{j_1, \ldots, j_1}_{R \text{ times}}, \underbrace{j_2, \ldots, j_2}_{R \text{ times}}), \quad 
y_2 = (b_1, b_2, b_3, \underbrace{j_2, \ldots, j_2}_{R \text{ times}}, \underbrace{j_1, \ldots, j_1}_{R \text{ times}})
\end{align*}
so $x_1, x_2, y_1, y_2 \in [m]^{2 R + 3}$ (here $x_1 = x_2$). 
%{\color{red} (JYC: yes $x_1 = x_2$, if we do $C_{i_1, j_2}^R C_{i_2, j_1}^R$ instead of $C_{i_2, j_1}^R C_{i_1, j_2}^R$, but it wasn't a goal)}
%{\color{red} (JYC: of course it's clear just by looking, but shoudn't we say it in () to show that yes we are aware, definitely I like the above definitions and don't want to smash $x_1, x_2$ together)}
Next, put $H' = H_{\AA^{\otimes(2 R + 3)}}$ so $V(H) = [m]^{2 R + 3}$.
Then (\ref{eq:part-hom-noneq}) becomes
\begin{equation}\label{eq:part-hom-noneq-tensor}
\hom_{(x_1, y_1)}(\Gamma, H') \ne \hom_{(x_2, y_2)}(\Gamma, H').
\end{equation}
Consider the graph $\widetilde{H'}$
obtained from $H'$ after the twin reduction step.
Since $H' = H_{\AA^{\otimes(2 R + 3)}}$ is
only edge-weighted (all vertex weights are $1$),
during the twin
 reduction step the vertices corresponding to the twin equivalence classes
have positive integer weights, so none of them are removed.
Thus every $u \in V(H')$ maps to some well-defined $\widetilde u \in V(\widetilde{H'})$
under this reduction step.
%(note that $\alpha_{\widetilde{H'}}(u)$ is a positive integer).
Then $\widetilde{x_1}, \widetilde{x_2}, \widetilde{y_1}, \widetilde{y_2} 
\in V(\widetilde{H'})$
%{\color{red}, so
%%^{2 R + 3}$ so
%%{\color{red} (JYC: 
%if we view $x_1, x_2, y_1, y_2 \colon [2 R + 3] \to V(H')$
%% (even by writing $\hom_{x_1}(\Gamma, H')$)
%then $\widetilde{x_1}, \widetilde{x_2}, \widetilde{y_1}, \widetilde{y_2} 
%\colon [2 R + 3] \to V(\widetilde{H'})$,(incorrect, just remove)} 
and
\[
\hom_{(x_1, y_1)}(\Gamma, H') = \hom_{(\widetilde{x_1}, \widetilde{y_1})}(\Gamma, \widetilde{H'}), \quad \hom_{(x_2, y_2)}(\Gamma, H') = \hom_{(\widetilde{x_2}, \widetilde{y_2})}(\Gamma, \widetilde{H'})
\]
and therefore
\begin{equation}\label{eq:part-hom-noneq-tensor-twin-compressed}
\hom_{(\widetilde{x_1}, \widetilde{y_1})}(\Gamma, \widetilde{H'}) \ne \hom_{(\widetilde{x_2}, \widetilde{y_2})}(\Gamma, \widetilde{H'}).
\end{equation}
It follows that there is no automorphism $\varphi \in \Aut(\widetilde{H'})$
such that $\varphi(\widetilde{x_1}) = \widetilde{x_2}$ and $\varphi(\widetilde{y_1}) = \widetilde{y_2}$,
otherwise~(\ref{eq:part-hom-noneq-tensor-twin-compressed}) is not true.
Now applying Theorem~\ref{thm:gh-main} we get that for some $\Gamma' \in \calPLG_{\simp}[2]$,
\[
\hom_{(\widetilde{x_1}, \widetilde{y_1})}(\Gamma', \widetilde{H'}) \ne \hom_{(\widetilde{x_2}, \widetilde{y_2})}(\Gamma', \widetilde{H'}).
\]
The 2-labeled simple graph $\Gamma'$ is just a simple edge gadget with
two (ordered)
 distinguished vertices (and no edge between these two vertices).
Because
\[
\hom_{(x_1, y_1)}(\Gamma', H') = \hom_{(\widetilde{x_1}, \widetilde{y_1})}(\Gamma', \widetilde{H'}), \quad \hom_{(x_2, y_2)}(\Gamma', H') = \hom_{(\widetilde{x_2}, \widetilde{y_2})}(\Gamma', \widetilde{H'})
\]
we get
\[
\hom_{(x_1, y_1)}(\Gamma', H') \ne \hom_{(x_2, y_2)}(\Gamma', H')
\]
and therefore
\begin{multline}\label{eq:part-hom-noneq-simp-gr}
(\prod_{(i, j) \in I} \hom_{(i, j)}(\Gamma', H_\AA)) (\hom_{(i_1, j_1)}(\Gamma', H_\AA))^R (\hom_{(i_2, j_2)}(\Gamma', H_\AA))^R \\
\ne (\prod_{(i, j) \in I} \hom_{(i, j)}(\Gamma', H_\AA)) (\hom_{(i_1, j_2)}(\Gamma', H_\AA))^R (\hom_{(i_2, j_1)}(\Gamma', H_\AA))^R.
\end{multline}
Putting $\BB = \gm_{\Gamma', \AA}$,
we obtain
\[
B_{i, j} = \sum_{\substack{\xi \colon V(\Gamma') \to [m] \\ \xi(\firstdv) = i,~ \xi(\seconddv) = j}} \prod_{(u, v) \in E(\Gamma')} A_{\xi(u), \xi(v)} = \hom_{(i, j)}(\Gamma', H_\AA)
\]
for $1 \le i, j \le m$.
Then (\ref{eq:part-hom-noneq-simp-gr}) rewrites as
\begin{equation}\label{eq:noneq-B-entries}
(\prod_{(i, j) \in I} B_{i, j}) B_{i_1, j_1}^R B_{i_2, j_2}^R \ne (\prod_{(i, j) \in I} B_{i, j}) B_{i_1, j_2}^R B_{i_2, j_1}^R.
\end{equation}
This means that (1) $B_{i, j} \ne 0$ for each $(i, j) \in I$,
so the $2 \times 2$ submatrix
\[
%\BB_{i_1, i_2; j_1, j_2} =
\begin{pmatrix}
B_{i_1, j_1} & B_{i_1, j_2} \\
B_{i_2, j_1} & B_{i_2, j_2}
\end{pmatrix}
\]
has at least three nonzero entries,
and (2) its $R$th Hadamard power
% $\BB_{i_1, i_2; j_1, j_2}^{\odot R}$ 
is nondegenerate.
This certifies that $\BB^{\odot R}$ is not block-rank-$1$.
Since $B_{i, j} = \hom_{(i, j)}(\Gamma', H_\AA) \in \bbQ(\{A_{i, j}\}_{i, j = 1}^m)$ for $i, j \in [m]$,
from Lemma~\ref{lem:mubr1-br1-equivalence} it follows that
$\BB$ is not \multblrkone/ and we are done.
%
%If $\BB_{i_1, i_2; j_1, j_2}$ has exactly three nonzero entries,
%then $\BB$ is not \multblrkone/ and we are done.
%Assume $\BB_{i_1, i_2; j_1, j_2}$ has no zero entries.
%If $\BB$ is multiplicatibely block-rank-$1$,
%then for some $n \ge 1$, $\BB^{\odot n}$ is block-rank-$1$
%Since $\BB_{i_1, i_2; j_1, j_2}$ has no zero elements,
%neither does $\BB_{i_1, i_2; j_1, j_2}^{\odot n}$
%and therefore $\BB_{i_1, i_2; j_1, j_2}^{\odot n}$ is degenerate so
%\[
%B_{i_1, j_1}^n B_{i_2, j_2}^n - B_{i_1, j_2}^n B_{i_2, j_1}^n = 0.
%\]
%Then $((B_{i_1, j_1} B_{i_2, j_2}) / (B_{i_1, j_2} B_{i_2, j_1}))^n = 1$
%so $(B_{i_1, j_1} B_{i_2, j_2}) / (B_{i_1, j_2} B_{i_2, j_1})$ is a root of unity.
%Since $F_s = \gm_{\Gamma, A^{\odot s})}$,
%we get $(B_{i_1, j_1} B_{i_2, j_2}) / (B_{i_1, j_2} B_{i_2, j_1}) \in Q(\{A_{i j}\}_{i, j = 1}^m)$
%but as shown above all roots of unity in $\bbQ(\{A_{i j}\}_{i, j = 1}^m)$ belong to
%a cyclic multiplicative group of order $R$
%and therefore
%\[
%(B_{i_1, j_1} B_{i_2, j_2}) / (B_{i_1, j_2} B_{i_2, j_1})^R = 1.
%\]
%Thus
%\[
%B_{i_1, j_1}^R B_{i_2, j_2}^R = B_{i_1, j_2}^R B_{i_2, j_1}^R
%\]
%contradicting (\ref{eq:noneq-B-entries}).
%Hence $\BB$ is not \multblrkone/ in this case
%which completes the proof.
\end{proof}
%\begin{remark}
%%In the defintion of the polynomials $p_k(x_1, x_2, x_3, x_4) \in$,
%In the proof above,
%$R$ can be replaced by any $R r$ where $r \ge 1$
%and the proof still goes through. {\color{red} (might need later for constructivity)}
%\end{remark}

\begin{remark}
Using the proof of Theorem~\ref{thm:gh-main} in~\cite{Cai-Govorov-GH-hom-type},
we can actually construct such $\Gamma' \in \calPLG_{\simp}[2]$ explicitly 
%(given $H'$ ($\AA$ and $R$)).
given $\AA$.
%%%  $H'$ and $R$ are from  $\AA$.
%{\color{red} (In fact it can be done in finite time double exponential in the size of $\widetilde{H'}$, (JYC: I mean double exponential in poly of size of $\widetilde{H'})$)}
%{\color{red} (JYC: the size of $\widetilde{H'}$ is the form $O(m^R R (\operatorname{size}(\AA) + C))$ where $C > 0$ which is $O(2^{\operatorname{poly}(R) \operatorname{poly}(\log (\operatorname{size}(\AA))))})$ (I hope I'm right)
%%is exponential in $R$ (with the base being the size of $\AA$ which we assume fixed)
%but $R$ is exponential in the size of $\AA$,
%so do we get a quadruple exponential bound in the size of $\AA$?, JYC: remove the red text, it's only for you, for constructivity we don't the exact time bound, we can
%say that this time bound can be given by an elementary formula)}
%%By a remark following the proof of Theorem~\ref{thm:gh-main} in~\cite{Cai-Govorov-GH-hom-type},
%%such $\Gamma' \in \calPLG_{\simp}[2]$ can be constructed from $\widetilde{H'}$, $(x_1, y_1)$ and $(x_2, y_2)$ in finite time
%%(double exponential).
\end{remark}

%%%
Now we prove the following statement which follows from Corollary~\ref{cor:non-mult-bl-rk-1-hardness}.
We use a  subscript $\simp$ and a superscript $(\Delta)$
to  denote the restriction of an 
$\EVAL$ problem to simple graphs of maximum degree at most $\Delta$.
%{\color{red} (JYC: do we need to reiterate what simple means here, even though we already said it in intro, and for $\calPLG_{\simp}[k]$)}
\begin{corollary}\label{cor:non-mult-bl-rk-1-gadget-hardness-bd-simp}
Let $\AA \in \bbC^{m \times m}$ a symmetric matrix.
Let $\Gamma$ be a simple 
%directed (undirected) 
edge gadget.
%(i.e., no multiple edges and no edges between the distinguished vertices),
%i.e., $\Gamma$ is simple as a graph and has no edges between
%the distinguished endpoints.
If $\gm_{\Gamma, \AA}$ is not \multblrkone/,
then for some $\Delta > 0$, the problem $\EVAL_{\simp}^{(\Delta)}(\AA)$ is \#P-hard.
\end{corollary}
\begin{proof}
We repeat the proof of Corollary~\ref{cor:non-mult-bl-rk-1-gadget-hardness-bd}.
Because the edge gadget $\Gamma$ is simple,
the graph $G'$ obtained by substituting the simple edge gadget
for every edge is also simple in addition to having bounded degree.
The Corollary follows.
% $\Delta(G') \le \Delta$.
%Again, if $\Delta$ becomes $0$, just replace $\Delta$ by $1$,
%the conclusion does not change.
%Thus it gives us $\EVAL^{(\Delta')}(\gm_{\Gamma, \AA}) \le \EVAL_{\simp}^{(\Delta)}(\AA)$
%(but $\EVAL^{(\Delta')}(\gm_{\Gamma, A})$ is \#P-hard by Corollary~\ref{cor:non-mult-bl-rk-1-hardness})
%and therefore $\EVAL_{\simp}^{(\Delta)}(\AA)$ is \#P-hard.
\end{proof}

The \#P-hardness proof of
%Theorem~\ref{thm:main-dich-bd-edge-gadget},
Theorem~\ref{thm:main-dich-bd} has been shown by finding
an edge gadget $\Gamma$ such that 
$\gm_{\Gamma, \AA}$ is not \multblrkone/.
% multiplicative-block-rank-$1$,
Applying Theorem~\ref{thm:gadget-simple-non-mult-bl-rk-1}
and Corollary~\ref{cor:non-mult-bl-rk-1-gadget-hardness-bd-simp}, we
have proved the following theorem.
% implies Theorem~\ref{thm:main-dich-bd-simp-edge-gadget}
%which in turn implies Theorem~\ref{thm:main-dich-bd-simp}.

\begin{theorem}\label{thm:main-dich-bd-simp}
Let $\AA$ be a symmetric and algebraic complex matrix.
Then either $Z_\AA(\cdot)$ can be computed in polynomial time on arbitrary graphs
$G$,
or for some $\Delta > 0$ depending on $\AA$,
it is \#P-hard on \emph{simple} graphs of  degree  at most $\Delta$.
\end{theorem}

%% file: decidability.tex
\section{Decidability in polynomial time of Theorems~\ref{thm:main-dich-bd}
%{thm:main-dich-bd-decidability} 
and~\ref{thm:main-dich-bd-simp}}\label{sec:dec-ours}
%{thm:main-dich-bd-simp-decidability}}\label{sec:dec-ours}

%{\color{red} (did the best I could, the rest is yours)}
%
%{\color{red} (basically we want to say that they are essentially the same)}
%
%{\color{red} (maybe say something about symbolic substitutions $\mu_i \to \pur \mu_i$ and $\nu_i \to \pur \nu_i$)}
%
In this section, we prove that
% Theorems~\ref{thm:main-dich-bd-decidability} and~\ref{thm:main-dich-bd-simp-decidability}, i.e.,
the following decision problem is computable
 in polynomial time in the size of $\AA$:
Given a symmetric $\AA \in {\mathbb C}^{m \times m}$
with algebraic entries,
decide whether $\AA$ satisfies the tractability criterion
in Theorem~\ref{thm:main-dich-bd} (which is the same as
in Theorem~\ref{thm:main-dich-bd-simp}). Thus  in polynomial time 
we can decide in which
category does $\EVAL(\AA)$ fall, either  $\EVAL(\AA)$ is tractable for 
arbitrary graphs, or  for some $\Delta > 0$
 the problem $\EVAL_{\simp}^{(\Delta)}(\AA)$ is \#P-hard.

There are two ways to establish  this.
The first way is to show that the tractability
criterion of the dichotomy in~\cite{Cai-Chen-Lu-2013}
is really equivalent to the tractability 
criterion in this paper (for both
Theorems~\ref{thm:main-dich-bd}
and~\ref{thm:main-dich-bd-simp}).
One can go through the proof of Theorems~\ref{thm:main-dich-bd} 
and~\ref{thm:main-dich-bd-simp} and verify the above claim,
and then appeal to the polynomial time decidability of
the main dichotomy theorem in~\cite{Cai-Chen-Lu-2013}. 
Of course, this is more of a claim that such a proof exists.
To make this a valid proof, 
 one must actually go through the proof in this paper,
and verify the above claim. 

In Appendix~\ref{appendixsubsec1:dec-ours}, %In the Appendix,
we will present the second way to establish this,
which actually goes through the proof of Theorems~\ref{thm:main-dich-bd} 
and~\ref{thm:main-dich-bd-simp} and directly verifies that
the tractability 
criterion is polynomial time decidable.

%% file: constructivity.tex
\section{Constructivity}\label{sec:constructivity}

%{\color{red} (did the best I could, the rest is yours, you can restructure it any way you want)}

By now, we have obtained the dichotomy for $\EVAL(\AA)$ (where $\AA$ is a complex symmetric matrix with algebraic entries) for bounded degree graphs, Theorem~\ref{thm:main-dich-bd}, and its extension to bounded degree 
simple graphs, Theorem~\ref{thm:main-dich-bd-simp}.
We will argue that this dichotomy and its extension are both construcitve.
This notion of constructivity is understood in the following sense.

\begin{theorem}\label{thm:constructivity-eval(A)}
There is an algorithm such that on input a symmetric and algebraic complex matrix $\AA$,
% either
\begin{enumerate}
\item \emph{if} $\EVAL(\AA)$ is tractable by the tractability criterion in this paper,
\emph{then} outputs an algorithm that computes $G \mapsto Z_\AA(G)$ in polynomial time in the size of the input graph $G$;
%if $\EVAL(\AA)$ is tractable by the tractability criterion in this paper; \\ or
\item \emph{else}, outputs a polynomial-time reduction from a canonical \#P-hard problem e.g., \textsc{\#SAT},
%otherwise, 
to $\EVAL_{\simp}^{(\Delta)}(\AA)$ for some $\Delta > 0$ depending on $\AA$. 
%(This $\Delta$ is computable from $\AA$.)
\end{enumerate}
\end{theorem}
%{\color{red} (Theorem~\ref{thm:constructivity-eval(A)} will follow from the arguments in Section~\ref{sec:constr-main-dich-bd}, \ref{sec:constr-main-dich-bd-simp} and~\ref{sec:constr-ccl}.) (probably no need to say it)}
%
%{\color{red} (again the constructivity is actually not just for the original theorems (Theorems~\ref{thm:main-dich-bd} and \ref{thm:main-dich-bd-simp}) but the gadget versions (Theorems~\ref{thm:main-dich-bd-edge-gadget} and \ref{thm:main-dich-bd-simp-edge-gadget}) that we proved)}

By contrast, the existing proof of the dichotomy in~\cite{Cai-Chen-Lu-2013} is not constructive in this sense.
We first argue why Theorem~\ref{thm:main-dich-bd} is constructive
and then we do the same for Theorem~\ref{thm:main-dich-bd-simp}.
%The former is independent of the results from~\cite{Cai-Govorov-GH-hom-type}.
%{\color{red} (By the former, I mean the consrtuctivity part of Theorem~\ref{thm:main-dich-bd}.)}
After that, we will show that we can also make the main dichotomy, Theorem 1.1 from~\cite{Cai-Chen-Lu-2013},
constructive by making some adjustments to its proof.
%%%%%%%%%%%%%%%%%%%%%%%%%%%%%%%%%%%%%%%%%%%%%%%%%%%%%%%%%%%%%%%%%%%%
\begin{comment}

{\color{red} (this will justify our constructivity contribution from the itcs paper)}

{\color{red} (I added as much material as possible so it's rife with details. We don't elaborate for example why each and every step is constructive, so if you don't like something you can remove anything you wish)}

{\color{red} (the dichotomy for / from / no preposition Theorem, actually there's no tangible dichotomy in the statement of our main theorems, but the theorems evidence the dichotomies)}

{\color{red} (JYC: general comment, when citing the itcs paper, I don't cite specific theorems, actually it may not be a bad idea.
In any case for you reference, in the itcs paper, the main theorem is Theorem 2 and the first pinning lemma from CCL and the promised constructivity are discussed in Effective GH Dichotomies)}

\end{comment}
%%%%%%%%%%%%%%%%%%%%%%%%%%%%%%%%%%%%%%%%%%%%%%%%%%%%%%%%%%%%%%%%%%%%

\subsection{Constructivity of the dichotomy for Theorem~\ref{thm:main-dich-bd}}\label{sec:constr-main-dich-bd}

%{\color{red} (basically, here I have coherent paragraphs more or less structured)}

%Theorem~\ref{thm:main-dich-bd} is a simple corollary of Theorem~\ref{thm:main-dich-bd-edge-gadget},
%and we will show how to make the proof of the latter constructive.

In order to have a constructive proof of Theorem~\ref{thm:main-dich-bd}, we entirely
avoided using the three pinning lemmas from~\cite{Cai-Chen-Lu-2013}
(Lemmas 4.1, 4.3 and Corollary 8.4 of~\cite{Cai-Chen-Lu-2013}, respectively).
All our proofs have been intentionally designed to avoid these pinning steps wherever~\cite{Cai-Chen-Lu-2013} made use of them.
%{\color{red} (JYC: we have another invisible pinning step when we invoke the Bulatov-Grohe dichotomy, in any of its proofs)}
%We will explain why our proof is constructive (elaborating on the most nontrivial steps) {\color{red} (but who is to say that they the most nontrivial ones)}.
%
%For example, whenever we say that there exists an edge gadget $\Gamma$ from~\cite{Cai-Chen-Lu-2013}
%whose signature matrix (e.g., $\gm_{\Gamma, \AA}$, $\gm_{\Gamma, \CC, \fD}$, etc.) is not \multblrkone/
%or not modular-block-rank-$1$, such an edge gadget can always be found in finite time, possibly via an open ended search.
%We will elaborate on this later in this section.
%This way we can assume that all base edge gadgets $\Gamma$ needed in the proofs of this paper can be computed.
%{\color{red} (back to the other arguments, first the limiting arguments)}
%
%{\color{red} (JYC: not sure if we should partition it into subsubsections)}

First, we show that all edge gadgets we used in our proofs through the meta-arguments Arguments~\ref{meta1}, \ref{meta2}, and~\ref{meta3},
i.e., ({\sl Meta}$_1$), ({\sl Meta}$_2$) and ({\sl Meta}$_3$) can be obtained constructively.

This starts with all the reused edge gadgets from~\cite{Cai-Chen-Lu-2013}.
%
%{\color{red} (First: computing ``base''gadgets $\Gamma$, which are then thickened)}
%
Whenever we said that there exists an edge gadget $\Gamma$ obtained by adapting a proof from~\cite{Cai-Chen-Lu-2013}
whose signature matrix (e.g., $\gm_{\Gamma, \AA}$, $\gm_{\Gamma, \CC, \fD}$, etc.) 
is not \multblrkone/ or \modblrkone/,
% modular-block-rank-$1$,
%(not \multblrkone/ or) not modular-block-rank-$1$ {\color{red} (only the last one as they come from~\cite{Cai-Chen-Lu-2013})},
such an edge gadget can always be found in finite time. %possibly via an open ended search.
A seemingly questionable point related to this is the use of limiting arguments
in some of the proofs from~\cite{Cai-Chen-Lu-2013}.
%{\color{red} (Another seemingly questionable point may be the use of limiting arguments.)}
(More precisely, it is done in Section 8.4.2 of~\cite{Cai-Chen-Lu-2013}
 on pp. 966 -- 977,
as part of the proof of Lemma 8.12,
and in Section 9.2 on pp. 983 -- 988, as part of the proof of Theorem 5.4.)
All such arguments can be captured by the following statement:
%Assuming the initial matrix $\AA$ (resp. a pair $(\CC, \fD)$) does not satisfy
Assuming some of the tractability conditions imposed at the corresponding step
are not satisfied,
for the constructed sequence of gadgets say $(\Gamma_n)_{n = 1}^\infty$,
there is an $n \ge 1$ such that its signature matrix $M_n$
(e.g., $M_n = \gm_{\Gamma_n, \pur \AA}$) is not
\multblrkone/ or \modblrkone/.
% modular-block-rank-$1$.
%%%%%%%%%%%%%%%%%%%%%%%%%%%%%%%%%%%%%%%%%%%%%%%%%%%%%%%%%%%%%%%%%%%%
\begin{comment}

{\color{red} (This involves verifying that $0 \le |a|$ (equivalent to $a \ne 0$ which is trivial to verify),
$|a| \ne |b|$ (equivalent to $a \overline a \ne b \ne \overline b$, verifying it is trivial by what follows)
or $|a| < |b|$ for constructed algebraic numbers $a, b$.
The latter is equivalent to $c = b \overline b - a \overline a > 0$.
 Clearly $\overline a, \overline b$ can be constructed 
the conjugation of a complex algebraic number is specified by the same minimal irreducible polynomial over $\bbQ$,
plus we can approximate numbers with a constructed number with an arbitrary precision
(in polynomial time in the length of the description of a number for an exponentially small error).
Then the number $c = b \overline b - a \overline a$ can also be constructed.
Then to check if $c > 0$ we first check if $c = 0$ which is done
by making sure that its minimal irreducible polynomial over $\bbQ$ is not $x \in \bbQ[x]$
and then we find a complex rational number $f_1 + f_2 i$ (from $\bbQ[i]$) approximating $c$ well.
After that, we verify that $f_1 > 0$. (stupid, just say it's computable, I'm justifying bullshit here))}
\end{comment}
%%%%%%%%%%%%%%%%%%%%%%%%%%%%%%%%%%%%%%%%%%%%%%%%%%%%%%%%%%%%%%%%%%%%
But for  a fixed $n$ this 
non-\multblrkone/ or non-\modblrkone/ condition
 is computable in our strict
Turing model of computation,
and so by an open-ended search we are guaranteed to find an $n$
for which $M_n$ is not 
\multblrkone/ or \modblrkone/.
 %modular-block-rank-$1$. {\color{red} (better alternative to the previous stuff in red)}
%In this case, the time is still finite.
%
The other edge gadget constructions from~\cite{Cai-Chen-Lu-2013} are easily seen to be constructive,
even if they are parametrized.
% (have parameters in their definitions) {\color{red} (I mean by $p \ge 1$ and we use different values, e.g., for Vandermonde determinants, I don't consider $r \in [N - 1]$ a parameter here)}, 
As these parameters can be explicitly bounded,
 they are computable in finite time as well.

This way we can assume that all reused edge gadgets  
from~\cite{Cai-Chen-Lu-2013} can be computed
as well as those obtained
through the meta-arguments Arguments~\ref{meta1}, \ref{meta2},
 and~\ref{meta3}, i.e.,  ({\sl Meta}$_1$), ({\sl Meta}$_2$) and ({\sl Meta}$_3$).
Thus all gadgets are
constructively obtained before the thickening stage
from Theorem~\ref{thm:gadget-thickening-non-mult-bl-rk-1} is applied.
Now we argue the constructivity of Theorem~\ref{thm:gadget-thickening-non-mult-bl-rk-1}.

%{\color{red} (Second: getting that $p$ in Theorem~\ref{thm:gadget-thickening-non-mult-bl-rk-1})}

%It is important to point out that for the \#P-hardness part
We point out that all invocations of Theorem~\ref{thm:gadget-thickening-non-mult-bl-rk-1} in this paper
eventually arise from the applications of the meta-arguments Arguments~\ref{meta1}, \ref{meta2}, and~\ref{meta3}, i.e.,  ({\sl Meta}$_1$), ({\sl Meta}$_2$) and ({\sl Meta}$_3$),
and the nonbipartite case invokes the bipartite case.
In order to apply Theorem~\ref{thm:gadget-thickening-non-mult-bl-rk-1},
we have to work with the order $R$ of the finite cyclic group
of roots of unity in $\bbQ(\{A_{i, j}\}_{i, j = 1}^m)$
%{\color{red} ($\bbQ(\alpha_1, \ldots, \alpha_n)$ (old))}
(see Corollary~\ref{cor:fin-gen-field-roots-of-unity-full-fin-cyc-gr}).
%It is computable
%However, we may not be able to compute $R$. {\color{red} (JYC: it may be trivial but we don't care)}
%{\color{red} (it may be possible but is actually of no concern to us even if its possible it's good to show how to avoid working with $R$ directly)}
%We can still do the following.
%{\color{red}
%(We assume the entries of $\AA \in \bbC^{m \times m}$ come from the field $\bbQ(\alpha_1, \ldots, \alpha_n)$
%where $\alpha_1, \ldots, \alpha_n$ are algebraic over $\bbQ$.
%Let $k = [\bbQ(\alpha_1, \ldots, \alpha_n) \colon \bbQ]$
%so $k$ can be computed. (old)) %{(or use $\bbQ(\{A_{i, j}\}_{i, j = 1}^m)$)}
%}
Since 
%(we assume that) 
the entries of $\AA \in \bbC^{m \times m}$
are algebraic over $\bbQ$,
we have $k = [\bbQ(\{A_{i, j}\}_{i, j = 1}^m) \colon \bbQ] < \infty$
and it can be computed.
%{\color{red} (can also use $\mathscr A$ to denote the set of entries of $\AA$ as in CCL not sure it's a good idea)}

We know that $\sqrt{R / 2} \le \varphi(R) \le [\bbQ(\{A_{i, j}\}_{i, j = 1}^m) \colon \bbQ] = k$ % \alpha_1, \ldots, \alpha_n
and therefore $R \le 2 k^2$.
In the proof of Theorem~\ref{thm:gadget-thickening-non-mult-bl-rk-1},
%we note that its proof gives us 
an explicitly bounded $p \ge 1$ is chosen based on 
Corollary~\ref{cor:vand-pur-ind-simp} applied to the polynomial
$q_{a, b}(\XX)$.
This $p$ depends 
 only on $m$, $R$ and $\Gamma$.
We fix $\Gamma$ in our reasoning because we can compute it as justified above.
%(Namely, $p$ does not exceed the number of terms in the expansion of $q_{a, b}((X_{i, j})_{i, j = 1}^m)$.)
%We now have two alternatives.
%\begin{enumerate}
%\item It is easy to see that this explicit bound for $p$ can be assumed to be monotone in both $m$ and $R$.
%{\color{red} (The number of terms in $p$ depends of $R$,
%and the number of terms in each polynomials $f_{i, j}((X_{i', j'})_{i', j' = 1}^m)$ depends on $m$ and $V(\Gamma)$ but the latter is fixed,
%and substituting $x_{i, j} = f_{i, j}((X_{i', j'})_{i', j' = 1}^m)$ into $p_{a, b}(x_{i_1, j_2}, x_{i_1, j_2}, x_{i_2, j_1}, x_{i_2, j_2})$
%gives us an additional dependency on $R$ for upper-bounding the number of terms in $q_{a, b}((X_{i, j})_{i, j = 1}^m)$.)
%(best not say it, too repetitive)}
%However, $R \le 2 k^2$, so we can explicitly bound $p$
%and then checking if $\gm_{T_p(\Gamma), \AA} = \gm_{\Gamma, \AA^{\odot p}}$
%is not \multblrkone/ can be easily done
%once we have found a generating set
%of the set of nonzero entries of $\gm_{\Gamma, \AA^{\odot p}}$ (instead of $\AA$ this time).
%(This can be done as discussed in Section~\ref{sec:dec-ours}.)
%{\color{red} (I find this approach worse)}
%
%\item 
To determine $R$, one can use a P-time factoring algorithm for polynomials over
an algebraic number field and factor all cyclotomic polynomials of degree
up to $2k^2$~\cite{LENSTRA-LENSTRA-LOVASZ,a-k-Lenstra,susan-landau}.
%From $R \le 2 k^2$ it follows
Alternatively to just get a computable upper bound,
one can also avoid factoring polynomials
and note that $R \mid (2 k^2)!$. Then we use 
a positive multiple $(2 k^2)! = R r$ of $R$,
where $r \ge 1$ is integer.
As we noted in the proof of Theorem~\ref{thm:gadget-thickening-non-mult-bl-rk-1}
the same proof works 
with $R r$ in place of $R$.
% (see the remark after the proof of this theorem).
Then $p \ge 1$ can be explicitly bounded.
% as $R r = (2 k^2)!$ is explicit.
%(This is because $q_{a, b}((X_{i, j})_{i, j = 1}^m)$ (arising from $R r$ instead of $R$ now) is explicitly constructed.)
At this point, it only suffices to check that
$\gm_{\Gamma, \AA^{\odot p}}^{\odot (R r)}$ is not block-rank-$1$.
%{\color{red} (I prefer this much more intuitive approach)}
%\end{enumerate}
%In any case, 
%This allows us to find such $p \ge 1$ without an open-ended search.
%Without the above argument,  $p$ can still be found by an open-ended search.
%The termination (i.e., the existence of such $p$) is guaranteed by Theorem~\ref{thm:gadget-thickening-non-mult-bl-rk-1}.
%{\color{red} (but this is too stupid not to say how to bound $p$.)}
%
All other steps concerning edge gadgets can be easily seen to be constructive.
%{\color{red} (I mean taking connected components, going from the inner to the outer part of the tensor product, and so on)}

This way we have justified that all edge gadgets $\Gamma$ (including the ones obtained 
after applying Theorem~\ref{thm:gadget-thickening-non-mult-bl-rk-1}) can be computed.

When we regularize a matrix for both the
bipartite case and the nonbipartite cases, we invoke Theorem~\ref{thm:non-mult-bl-rk-1-hardness-sym} from Theorem~\ref{thm:gadget-thickening-non-mult-bl-rk-1} directly
if $\AA$ is not \multblrkone/. In this case an edge is our gadget.
% so this theorem is invoked in not only {\sl Meta}$_1$, {\sl Meta}$_2$, {\sl Meta}$_3$)}
Now consider the constructivity of Theorem~\ref{thm:non-mult-bl-rk-1-hardness-sym}, 
Corollaries~\ref{cor:non-mult-bl-rk-1-hardness} and
\ref{cor:non-mult-bl-rk-1-gadget-hardness-bd} (excluding the Bulatov-Grohe
dichotomy Theorem~\ref{BulatovGrohe}).
%%%%%%%%%%%%%%%%%%%%%%%%%%%%%%%%%%%%%%%%%%%%%%%%%%%%%%%%%%%%%%%%%%%%
\begin{comment}

%{\color{red} (additional stuff, for making sure that the $p$ in Theorem~\ref{thm:gadget-thickening-non-mult-bl-rk-1})}

{\color{red} (Third: now for the \#P-hardness for Theorem~\ref{thm:non-mult-bl-rk-1-hardness-sym} and Corollary~\ref{cor:non-mult-bl-rk-1-hardness} except the dichotomy by Bulatov and Grohe)}

% Corollary~\ref{cor:non-mult-bl-rk-1-gadget-hardness-bd-simp}
%we need to invoke Corollary~\ref{cor:non-mult-bl-rk-1-hardness}
%whose proof in turn invokes Theorem~\ref{thm:non-mult-bl-rk-1-hardness-sym}

{\color{red} (JYC: when we regularize a matrix for both bipartite case and nonbipartite cases, we invoke Theorem~\ref{thm:non-mult-bl-rk-1-hardness-sym} from Corollary~\ref{thm:gadget-thickening-non-mult-bl-rk-1} directly
because if $A$ is not \multblrkone/, then an edge is our gadget, so this theorem is invoked in not only {\sl Meta}$_1$, {\sl Meta}$_2$, {\sl Meta}$_3$)}

\end{comment}
%%%%%%%%%%%%%%%%%%%%%%%%%%%%%%%%%%%%%%%%%%%%%%%%%%%%%%%%%%%%%%%%%%%%

The \#P-hardness conclusions for Theroem~\ref{thm:main-dich-bd} are granted by Corollary~\ref{cor:non-mult-bl-rk-1-gadget-hardness-bd}. %Corollary~\ref{cor:non-mult-bl-rk-1-hardness}.
Namely, this corollary is subsequently used in the regularization steps Theorems \ref{bi-step-1-ours} and~\ref{t-step-1-ours} (see the respective
 paragraphs after their statements),
and
in the meta-arguments Arguments~\ref{meta1}, \ref{meta2}, and~\ref{meta3}, i.e.,  ({\sl Meta}$_1$), ({\sl Meta}$_2$) and ({\sl Meta}$_3$).
The proof of Corollary~\ref{cor:non-mult-bl-rk-1-gadget-hardness-bd}
is a simple invocation of Corollary~\ref{cor:non-mult-bl-rk-1-hardness}
whose proof in turn is a simple invocation of Theorem~\ref{thm:non-mult-bl-rk-1-hardness-sym},
so provided the proof of Theorem~\ref{thm:non-mult-bl-rk-1-hardness-sym} is constucitive,
the proof Corollary~\ref{cor:non-mult-bl-rk-1-gadget-hardness-bd} is constructive as well.
We now argue that the proof of Theorem~\ref{thm:non-mult-bl-rk-1-hardness-sym} is indeed constructive.
%except one step which will be handled later.
%Corollary~\ref{thm:non-mult-bl-rk-1-hardness-sym} is in particular used in the meta-arguments Arguments~\ref{meta1}, \ref{meta2}, and~\ref{meta3} ({\sl Meta}$_1$, {\sl Meta}$_2$, {\sl Meta}$_3$).

The construction in the proof of Theorem~\ref{thm:non-mult-bl-rk-1-hardness-sym}
depends on two parameters, first $p \ge 1$ and then $\ell \ge 1$.
The parameter
$p \ge 1$ is picked using the proof of Lemma~\ref{lem:ADconjA-nondeg-thick}.
In fact, it is easy to get an explicit polynomial bound for $p$
in the size of $\AA$ and $\DD$.
% ($m$ and) the length the description of entries of $\AA$ and $\DD$,
%5i.e., in the size of $\AA$ and $\DD$ {\color{red} (actually, 
%this description is of length at least $m$ because the matrices are $m \times m$)}.
We can take the first $p \ge \lfloor \ln(2 m) / \ln(1 / \gamma) \rfloor + 1$ 
in the proof of Lemma~\ref{lem:ADconjA-nondeg-thick},
and $\gamma$ can be constructed from the entries of $\AA$ and $\DD$ 
in polynomial time.
% {\color{red} (I suppose I'm correct here, $p$ is polynomial still)}.
This shows that this lower bound for $p$ is polynomial in the size of $\AA$ and $\DD$,
and so $p$ can be computed in polynomial time in the size of $\AA$ and $\DD$.
%{\color{red} (I hope I'm correct, basically we need to find $p$ such that $a^p \ge n!$ where $a \ge 1$ and is constructed from the entries of $\AA$ in polynotmial time: roughly smth like $min(A_{i i} A_{j j} / (A_{i j} A_{j i}))$ if you recall the proof but we make divisions)}
%Alternatively, by using this $p \ge 1$ and then adapting the Vandermonde argument Lemma~\ref{lem:simpleVandermonde}
%{\color{red} (basically, we can define $f(x) = x^p$ for this $p$ that was obtained from the norm inequiality)},
%it is also possible get another such $p \ge 1$ which is bounded by the number of terms
%in the expression of $\det((\AA \DD \overline \AA)^{\odot p})$ which is (at most) $m!$.
%{\color{red} (if counting in possibly zero terms, this precisely $m!$)}
%This way we get a higher non-polynomial (exponential) bound in $m$ but it is now independent of the entries of $\AA$ and $\DD$.
Having chosen $p \ge 1$, we can further bound $\ell \ge 1$ using
the proof of Corollary~\ref{lem:vand-conj} whose proof in turn invokes Lemma~\ref{lem:vand-conj-simp}.
%More precisely, such $\ell \ge 1$ that does not exceed
%the number of terms in the expansion of $f_1(\XX, \yy) \prod_{i = 1}^k (f_{2, i}(\XX, \yy) f_{3, i}(\XX, \yy))$.

Next at the end of the proof of Theorem~\ref{thm:non-mult-bl-rk-1-hardness-sym},
we invoke Corollary~\ref{cor:non-mubr1-hard-unbounded}
whose proof invokes the purification step. This step  is constructive:
we use Lemma~\ref{lem:eval-A-B-equivalence}, which is 
Lemma 7.4 from~\cite{Cai-Chen-Lu-2013} and its proof in~\cite{Cai-Chen-Lu-2013}
is construcitive.
%%% I am hiding the COUNT version
%%which in turn invokes Lemma~\ref{count-ours}, both of which are construcitive
%{\color{red} (which can be seen from their proofs in~\cite{Cai-Chen-Lu-2013} (if we don't repeat their proofs in our paper (e.g., put them in an appendix))}.
%We also note that it is not hard to show that computing a purification of a matrix is constructive task.
%{\color{red} (
In fact, this can done in polynomial time in the size of the matrix to be purified,
which is $\AA \in \bbC^{m \times m}$ in the context of 
Lemmas~\ref{lem:eval-A-B-equivalence}.
% and Lemma~\ref{count-ours}), as both operations of finding the first $d$ prime numbers and a generating set of the set
%of nonzero entries of a matrix can both be done in polynomial time (in $d \le m^2$ and the size of $\AA$, respectively) as discussed in Section~\ref{sec:dec-ours}.)}
%{\color{red} (JYC: you may want to remove this text in red, afraid of scratching that polynomial time decidability thing, my only worry is that it may not be obvious how to find a generating set without this lattice result constructively, in which case we must keep it here)}
The proof of Corollary~\ref{cor:non-mubr1-hard-unbounded}
also invokes dichotomy by Bulatov and Grohe (Theorem~\ref{BulatovGrohe}).
%Assuming the proof of Theorem~\ref{basicsharp-ours}
%can be made constructive, the proof of Theorem~\ref{thm:non-mult-bl-rk-1-hardness-sym}
%becomes constructive as well.
Now we argue how to make Theorem~\ref{BulatovGrohe} constructive.
This is the most nontrivial step for the constructivity of 
Theorem~\ref{thm:main-dich-bd}.

In order to discuss the constructivity of Theorem~\ref{BulatovGrohe}, we need to introduce a very important pinning step.
%%%
%{\color{red} (copied from CCL, this lemma is formulated for complex $\AA$)}

Let $\AA$ be an $m\times m$ symmetric complex matrix.
We define a new problem $\evalp(\AA)$, which is the pinned version of $\eval(\AA)$:
The input is a triple $(G,w,i)$,
where $G \in \calPLG[1]$ is a 1-labeled graph
%$(G,w)$ is a 1-labeled graph  in $\calPLG[1]$ 
with a distinguished vertex $w \in V(G)$,
and a domain element $i\in [m]$;
the output is
\[\hom_{(i)}(G, H_{\AA}) = 
\sum_{\substack{\xi:V(G) \rightarrow [m],~\xi(w)=i}}
~\prod_{u v \in E(G)} A_{\xi(u), \xi(v)}.
\]
(In~\cite{Cai-Chen-Lu-2013} this is denoted as $Z_{\AA}(G,w,i)$.) 
%$$
%Z_{\AA}(G,w,i)\hspace{0.036cm}=\sum_{\substack{\xi:V\rightarrow [m]\\\xi(w)=i}}
  %\twt_{\AA}(\xi).
%$$
It is easy to see that $\eval(\AA)\le \evalp(\AA)$.
%$$
%Z_{\AA}(G)=\sum_{i\in [m]} Z_{\AA}(G,w,i),\ \ \ \text{for any $w\in V$}.
%$$
The other direction also holds.

\begin{lemma}[First pinning lemma, Lemma 4.1 from~\cite{Cai-Chen-Lu-2013}]\label{pinning1-ours}
$\evalp(\AA)\equiv \eval(\AA)$.
\end{lemma}
%%%
%{\color{red} (end of the copied fragment)}

The proof of the first pinning lemma in~\cite{Cai-Chen-Lu-2013} is nonconstructive.
%because for the reason we are about to describe. 
%{\color{red} (this lemma is formulated for complex symmetric $\AA$)}
%
To describe this, we define an equivalence relation $\sim$ over $[m]$:
$i\sim j$ if for any  $G \in \calPLG[1]$,
%undirected graph $G=(V,E)$ and $w\in V$,
$\hom_{(i)}(G, H_{\AA}) = \hom_{(j)}(G, H_{\AA})$.
  %$Z_{\AA}(G,w,i)=Z_{\AA}(G,w,j)$.}

%Note that we do not know, given $\AA$, how to compute $\sim$ 
%  efficiently, 
%although this is possible using the new results of Schrijver~\cite{Schrijver222}. 
Given a symmetric $\AA \in \bbC^{m \times m}$, it was not known
 how to compute $\sim$.
(In~\cite{Cai-Chen-Lu-2013}  it was noted that it may be possible 
to compute this using~\cite{Schrijver-2009},
but in fact the techniques given in that paper appear to be insufficient.
This can in fact be computed using results in~\cite{Cai-Govorov-GH-hom-type},
in particular, the proof of Theorem~\ref{thm:gh-main} which is
constructive.)
%
%In~\cite{Schrijver-2009} it was noted that it may be possible using~\cite{Schrijver-2009},
%but in fact the techniques given in that paper are insufficient for our goal).
% 
%{\color{red} (in CCL, you said ``although this is possible using ...'', trying to alleviate it)}
%
It can be checked that other than $\sim$, 
the first pinning lemma (Lemma 4.1 in~\cite{Cai-Chen-Lu-2013}) 
is proved constructively.
%It is not hard to check that provided $\sim$ can be computed efficiently,
%the first pinning lemma (Lemma 4.1 in~\cite{Cai-Chen-Lu-2013}) can be easily made constructive.
%{\color{red} (We would like to add that the rest of the proof can be further simplified using the Vandermonde argument fully described in~\cite{Cai-Govorov-GH-hom-type}.
%(here I mean the multidimestional one, you can freely remove that)} 
%{\color{red} (JYC: I don't cite particular theorems here because the itcs version numbers statements with integers, in any case you can remove this red text))}
%%%

Concerning the nonnegative case, it can be checked that,
with the exception of the so-called pinning steps,
 the proofs of the Bulatov-Grohe
dichotomy Theorem~\ref{BulatovGrohe}
in~\cite{Bulatov-Grohe-2005, Thurley-2009, Grohe-Thurley-2011}
are constructive in the above sense.
% with the exception of the so-called pinning steps.
%In what follows $\AA$ is not the original matrix $\AA$.
In the terminology of~\cite{Cai-Chen-Lu-2013}, this corresponds to the first pinning lemma
restricted to the case when $\AA$ is a nonnegative symmetric matrix 
%($\AA \in \bbR_{\ge 0}^{m \times m}$) 
with algebraic entries.
%{\color{red} (Here $\AA$ is a local notation and not directly related to the original matrix $\AA$ in this paper.)}
%
%%%%%%%%%%%%%%%%%%%%%%%%%%%%%%%%%%%%%%%%%%%%%%%%%%%%%%%%%%%%%%%%%%%%
\begin{comment}

We note that some of these proofs (of Theorem~\ref{basicsharp-ours}) also invoke the Dyer-Greenhill dichotomy~\cite{Dyer-Greenhill-2000, Dyer-Greenhill-corrig-2004}
which corresponds to the case when $\AA$ is a $0$-$1$ symmetric matrix,
but this is less general than the previous case.
{\color{red} Some invocations may seemingly involve the case when $\AA$ is $0$-$1$ is not necessarily symmetric
(in which case we talk about directed GH if $\AA$ is not symmetric) but the actual usages within such a proof of Theorem~\ref{thm:Bulatov-Grohe}
are only for the symmetric $\AA$. (JYC: I mean rectangular relations and so on, I hope you understand, best to omit it)}

\end{comment}
%%%%%%%%%%%%%%%%%%%%%%%%%%%%%%%%%%%%%%%%%%%%%%%%%%%%%%%%%%%%%%%%%%%%
%
Therefore it suffices to show how to compute $\sim$ in the nonnegative
case in order to make the proof of Theorem~\ref{BulatovGrohe} constructive.

%%%%%%%%%%%%%%%%%%%%%%%%%%%%%%%%%%%%%%%%%%%%%%%%%%%%%%%%%%%%%%%%%%%%
\begin{comment}

the proof of the first pinning lemma for a nonnegative symmetric $\AA$ constructive
which will then make the proof of Theorem~\ref{basicsharp-ours} constructive.

{\color{red} (now discussing how to make it constructive)}

\end{comment}
%%%%%%%%%%%%%%%%%%%%%%%%%%%%%%%%%%%%%%%%%%%%%%%%%%%%%%%%%%%%%%%%%%%%

%The first pinning lemma for a symmetric nonnegative matrix $\AA$ can be
This can be done by applying the results in~\cite{Lovasz-2006} 
(see also~\cite{Lovasz-loops-correction} 
for a small correction suggested by Martin Dyer which needs to be applied 
to this paper). However, this method 
can involve an open-ended search, with no a priori time bounds.

%made constructive either by~\cite{Lovasz-2006} (see also~\cite{Lovasz-loops-correction}
%for a small correction which needs to be applied to this paper (as suggested by Martin Dyer)),
%although it can involve an open-ended search (with no precise time bounds).

To be more concrete, 
given $i, j \in [m]$ and a symmetric nonnegative $\AA \in \bbR^{m \times m}$,
we determine whether $i \sim j$ as follows.
%For convenience, we assume that $H_\AA$ is on the vertex set $[m]$.
%
%More precisely, we can do one of the following:
%\begin{quote}
Recall that $\widetilde{H_\AA}$ denotes the twin-free graph obtained after applying the twin compression step to $\AA$.
As all vertices in $H_\AA$ have weight $1$, during the twin compression step
no vertex is removed, so each $i \in [m]$ maps to $\widetilde i \in \widetilde{H_\AA}$
having a positive integer vertex weight.
%It is easy to see that $Z_\AA(G, u, \ell) = 
Then $\hom_{(i)}(G, H_{\AA}) = \hom_{(\widetilde i)}(G, \widetilde{H_\AA})$,
for any $G \in \calPLG[1]$.
%where on the RHS $G_u \in \calPLG[1]$ is viewed as the $1$-labeled graph with the underlying graph $G$ in which $u$ is has label $1$
%and in the evaluation of $\hom_{(\widetilde \ell)}$, $u$ is maps to $\widetilde \ell \in V(\widetilde{H_\AA})$.
Then for any $i, j \in [m]$, $i \sim j$ iff $\hom_{(\widetilde i)}(G, \widetilde{H_\AA}) = \hom_{(\widetilde j)}(G, \widetilde{H_\AA})$
for all $G \in \calPLG[1]$.
Because $\widetilde{H_\AA}$ is twin-free, by Lemma 2.4 from~\cite{Lovasz-2006}
this is equivalent to the existence of a $\widetilde \varphi \in \Aut(\widetilde{H_\AA})$
such that $\widetilde \varphi(\widetilde i) = \widetilde j$.

%Alternatively, it can be easily checked that any $\widetilde \varphi \in \Aut(\widetilde{H_\AA})$
%such that $\widetilde \varphi(\widetilde i) = \widetilde j$
%``lifts'' to $\varphi \in \Aut(H_\AA)$ such that $\varphi(i) = j$.
%Here the fact that all vertices in $H_\AA$ have weight $1$ is essential.
%\end{quote}
%
%Each of these options can be easily carried out in exponential time in the size of $\AA$.
%{\color{red} (or in $2^{\operatorname{poly}(m)} \cdot \operatorname{poly}(\AA)$ (we can't say in $m$ only))}.

However, in case $i \not \sim j$, finding a separating ``witness'' $G \in \calPLG[1]$ for which $\hom_{(\widetilde i)}(G, \widetilde{H_\AA}) \ne \hom_{(\widetilde j)}(G, \widetilde{H_\AA})$
may require an open-ended search because the proof of Lemma 2.4 from~\cite{Lovasz-2006}
 is nonconstructive.
%does not provide an explicit set of witnesses to check.
In this context, this means that we can enumerate all graphs from $\calPLG[1]$
and search for such a separating ``witness ''; 
we know that eventually we will find one.
% such $G$ but we do not know how soon,
%i.e., we do not have any concrete time bound.

We point out that the result and proof in~\cite{Cai-Govorov-GH-hom-type}
do provide an explicit finite set of witnesses to check.

This completes the discussion on the constructivity for the \#P-hardness case.

Finally we justify that the tractability proofs for both
the bipartite case (Theorem~\ref{tractable-1-ours}) and
the nonbipartite case (Theorem~\ref{tractable-2-ours}) are constructive.
For the bipartite case, we carefully
designed the proof so that we avoided using the second pinning lemma.
(This is why we considered the 
 problem $\EVAL^{\leftrightarrow}(\CC, \fD)$ instead of just $\EVAL(\CC, \fD)$.
%{\color{red} (by considering a problem $\EVAL^{\leftrightarrow}(\CC, \fD)$ instead of just $\EVAL(\CC, \fD)$ (actually remove it, no need to excuse ourselves)}.
%{\color{red} (JYC: they are all equivalent with or without $\leftrightarrow$ for a pair $(\CC, \fD)$ that satisfies the conditions needed for the second pinning lemma,
These problems are equivalent, but the equivalence proof is not constructive.)
%but it's not known in the constructive scene)}
In the nonbipartite case, the tractability proof is the same as in~\cite{Cai-Chen-Lu-2013} (Section 16)
and does not invoke the second pinning lemma (Lemma 4.3 from~\cite{Cai-Chen-Lu-2013}). 
We note that both proofs are reliant of Theorem~\ref{pinyan-ours} (Theorem 12.3 from~\cite{Cai-Chen-Lu-2013})
and its proof is partitioned into Lemma 12.7 (for the case $q = p^k$ where $p$ is an odd prime and $k$ is a positive integer)
and Lemma 12.9 (for the case $q = 2^k$ where $k$ is a positive integer) from Section 12.3 in~\cite{Cai-Chen-Lu-2013}).
We note that the proof of Lemma 12.9 for the case $k = 1$ (so $q = 2$) uses the tractability result from~\cite[Theorem~6.30]{Carlitz} or~\cite{field}
but they are both constructive.
The rest can be easily seen to be constructive.
We can conclude that our tractability proof is indeed constructive.

We have proved Theorem~\ref{thm:constructivity-eval(A)},
the constructivity of Theorem~\ref{thm:main-dich-bd}.

% obtain the following theorem (statement).
%%%%%%%%%%%%%%%%%%%%%%%%%%%%%%%%%%%%%%%%%%%%%%%%%%%%%%%%%%%%%%%%%%%%
\begin{comment}

\begin{theorem}
There is an algorithm that on input a symmetric and algebraic complex matrix $\AA$ either
\begin{enumerate}
\item outputs a polynomial-time algorithm that computes $Z_\AA(\cdot)$ in polynomial time \emph(in the size of the input graph $G$\emph)
if $\EVAL(\AA)$ is tractable by the tractability criterion obtained in this paper; or
\item outputs a polynomial-time reduction from a canonical \#P-hard problem \emph(e.g., \textsc{\#SAT}, \textsc{\#Permanent}, etc.\emph), it is of no importance which one as they are all polynomial-time equivalent)
to $\EVAL^{(\Delta)}(\AA)$ for some $\Delta > 0$ depending on $\AA$. (This $\Delta$ is computable from $\AA$.)
\end{enumerate}
\end{theorem}

\end{comment}
%%%%%%%%%%%%%%%%%%%%%%%%%%%%%%%%%%%%%%%%%%%%%%%%%%%%%%%%%%%%%%%%%%%%

\subsection{Constructivity of the dichotomy for Theorem~\ref{thm:main-dich-bd-simp}}\label{sec:constr-main-dich-bd-simp}

The dichotomy in Theorem~\ref{thm:main-dich-bd-simp} is a continuation of the dichotomy from Theorem~\ref{thm:main-dich-bd}.
We  show how to make constructive
 transfers to simple graphs for non-\multblrkone/
or non-\modblrkone/ gadgets.
%for  transition from Theorem~\ref{thm:main-dich-bd-edge-gadget} to Theorem~\ref{thm:main-dich-bd-simp-edge-gadget},
%and we will show how to make this transition constructive.
We only need to take care of Theorem~\ref{thm:gadget-simple-non-mult-bl-rk-1} and Corollary~\ref{cor:non-mult-bl-rk-1-gadget-hardness-bd-simp}.
%and we only need to make some adjustments.

%We need to show how the following step can be made constructive:
%Given an edge gadget $\Gamma$ such that $\gm_{\Gamma, \AA}$ is not \multblrkone/,
%we wish to produce an simple edge gadget $\Gamma'$ such that $\gm_{\Gamma', \AA}$ is not \multblrkone/.

Theorem~\ref{thm:gadget-simple-non-mult-bl-rk-1} is used
to construct simple edge gadgets.
The constructivity of this theorem entirely depends on
that of Theorem~\ref{thm:gh-main} from~\cite{Cai-Govorov-GH-hom-type}
and the ability to compute $R$, the order of the finite cyclic group
of roots of unity in $\bbQ(\{A_{i, j}\}_{i, j = 1}^m)$.

The proof of Theorem~\ref{thm:gh-main} in~\cite{Cai-Govorov-GH-hom-type}
actually produces an explicit finite list
% (containing double exponentially many elements in $m$ and $k$)
of graphs $G \in \calPLG_{\simp}[k]$, such that if there is no isomorphism 
from $H$ to $H'$, 
then $\hom_\varphi(G, H) \ne \hom_\psi(G, H')$ for some $G$ on this list.
%{\color{red} (JYC: despite the fact that the theorem is not formulated as such)}
When applied inside the proof of 
Theorem~\ref{thm:gadget-simple-non-mult-bl-rk-1} with $k = 2$,
this results in an explicit finite list of edge gadgets
% to check through,
 that contains  an edge gadget $\Gamma'$ promised by
Theorem~\ref{thm:gadget-simple-non-mult-bl-rk-1}.
% will be found.
%The number of elements (edge gadgets) in this list is double exponential in $|V(\widetilde{H_\AA^{\otimes R}})|$.
%%The proof Theorem~\ref{thm:gh-main} given in~\cite{Cai-Govorov-GH-hom-type}
%%is constructive and in fact gives us an explicit finite list of edge gadgets to check, in case
%%(double (triple) exponential) time bound to find such an edge gadget $\Gamma'$.
%%This involves checking double (triple) exponentially (or just finitely) many witnesses (edge gadgets) resulting from the proof in~\cite{Cai-Govorov-GH-hom-type}.
%{\color{red} (We note that perform the twin-compression step
%in the proof of Theorem~\ref{thm:gadget-simple-non-mult-bl-rk-1} before invoking Theorem~\ref{thm:gh-main},
%so we do not need to repeat it. (silly comment))}
%Next, if $R$ cannot be computed directly,
(One can also avoid running the polynomial 
factoring algorithm~\cite{LENSTRA-LENSTRA-LOVASZ,a-k-Lenstra,susan-landau}
by using a positive multiple  $R r = (2 k^2)!$ instead of $R$.)
%Here $r \ge 1$ is integer such that $R r = (2 k^2)!$ from Section~\ref{sec:constr-main-dich-bd}.
%In this case, the number of witnesses to check may become even larger (not it is double exponential in $|V(\widetilde{H_\AA^{\otimes ((2 k)!)}})|$)
%but the list of witnesses to check still finite explicit.

The \#P-hardness for Theorem~\ref{thm:main-dich-bd-simp} is derived from
 Corollary~\ref{cor:non-mult-bl-rk-1-gadget-hardness-bd-simp}
(instead of Corollary~\ref{cor:non-mult-bl-rk-1-gadget-hardness-bd} for Theorem~\ref{thm:main-dich-bd}).
%Next, when we move on to the proof of Corollary~\ref{cor:non-mult-bl-rk-1-gadget-hardness-bd-simp},
%inside of it, we need to invoke Corollary~\ref{cor:non-mult-bl-rk-1-hardness}
Corollary~\ref{cor:non-mult-bl-rk-1-gadget-hardness-bd-simp}
uses Corollary~\ref{cor:non-mult-bl-rk-1-hardness}
whose proof invokes Theorem~\ref{thm:non-mult-bl-rk-1-hardness-sym}.
The proof of 
Theorem~\ref{thm:non-mult-bl-rk-1-hardness-sym}
 invokes the Bulatov-Grohe dichotomy (Theorem~\ref{BulatovGrohe}),
but it has been shown already in Section~\ref{sec:constr-main-dich-bd}
how to make the proof of Theorem~\ref{BulatovGrohe} constructive.
Thus the proof of Corollary~\ref{cor:non-mult-bl-rk-1-gadget-hardness-bd-simp}
can be made constructive.

We remark that the tractability criterion for Theorem~\ref{thm:main-dich-bd-simp}
is the same as that for Theorem~\ref{thm:main-dich-bd}
whose constructivity has already been argued in Section~\ref{sec:constr-main-dich-bd}.
Therefore we do not need to deal with the constructivity issue for tractability
in Theorem~\ref{thm:main-dich-bd-simp}. 
%{\color{red} (additional remark, can be removed)}

Summing up, we conclude that the dichotomy for Theorem~\ref{thm:main-dich-bd-simp}
is constructive. This concluides the proof of
Theorem~\ref{thm:constructivity-eval(A)}.

In the appendix, we will
describe the modifications one can make
to make the dichotomy in~\cite{Cai-Chen-Lu-2013} constructive.

%% file: appendix1.tex
\appendix
\section{Appendix} %\section*

\subsection{P-time decidability of Theorems~\ref{thm:main-dich-bd} %\subsection*
%{thm:main-dich-bd-decidability} 
and~\ref{thm:main-dich-bd-simp}}\label{appendixsubsec1:dec-ours}
%{thm:main-dich-bd-simp-decidability}}\label{sec:dec-ours}

%{\color{red} (did the best I could, the rest is yours)}
%
%{\color{red} (basically we want to say that they are essentially the same)}
%
%{\color{red} (maybe say something about symbolic substitutions $\mu_i \to \pur \mu_i$ and $\nu_i \to \pur \nu_i$)}
%
We go through the proof of Theorems~\ref{thm:main-dich-bd},
%{thm:main-dich-bd-decidability}
and verify that the
tractability criterion is polynomial time decidable.

In our proof of Theorems~\ref{thm:main-dich-bd} and~\ref{thm:main-dich-bd-simp},
the case  $\EVAL_{\simp}^{(\Delta)}(\AA)$ is \#P-hard
for some $\Delta > 0$
is when we can construct a gadget $\Gamma$ for which $\gm_{\Gamma, \AA}$ is not \multblrkone/.
On the other hand,
the tractability criterion described in this paper, when applied to $\AA$,
is essentially the same as that in~\cite{Cai-Chen-Lu-2013}. 
Therefore, the polynomial time decidability of
Theorems~\ref{thm:main-dich-bd} and~\ref{thm:main-dich-bd-simp}
will basically follow that of the dichotomy in~\cite{Cai-Chen-Lu-2013}. 
We only point out a few slight differences.

The first step is the reduction to connected components done in Section 4.3 in~\cite{Cai-Chen-Lu-2013}.
We perform a corresponding step in Section~\ref{sec:pinning-free}.
Then the tractability criteria in both papers
consist of working with each connected component of $\AA$ separately
so we may assume $\AA$ is connected.
Then we are either in the bipartite or the nonbipartite case.
From Theorem 17.1 (Section 17) in~\cite{Cai-Chen-Lu-2013} which uses the results from~\cite{Ge1,Ge2} (formulated as Theorem 17.3 in~\cite{Cai-Chen-Lu-2013}),
it follows that for a complex matrix $\AA$ a generating set $\mathscr G = \{g_1, \ldots, g_d\} \subset \bbQ(\mathscr A)$
of the set of nonzero entries $\mathscr A$ of $\AA$ can be found in polynomial time. 
%{\color{red} (we may omit it)}
Then a purification $\pur \AA$ of $\AA$ (obtained by going from $(g_1, \ldots, g_d)$ to $(p_1, \ldots, p_d)$)
can also be computed in polynomial time.
%First, 
Here we use the fact that we can get $d \le m^2$ efficiently.
We note that this inequality is true because the group generated by the nonzero elements of $\AA$
is a finitely generated abelian group generated by at most $m^2$ generators
and it has rank at most $m^2$,
and the approach described in the proof of Theorem 17.1 (Section 17) in~\cite{Cai-Chen-Lu-2013} uses this principle.
%{\color{red} (can we say that if we have a large generating set then we can get a small one as its subset?)}
Also, we use the fact the $d$ smallest primes $p_1 < \ldots < p_d$
can be computed in polynomial time in $d$.~\footnote{This statement follows from the prime number theorem, i.e., $\pi(n) \sim n / \ln(n)$ for large $n$.
However, we can use an explicit bound
$\log n + \log \log n -1 < \frac{p_n}{n} < \log n + \log \log n$ for
$n \ge 6$~\cite{rosser,dusart}. }
%it suffices to have a weaker property that $\pi(n) > C n / \ln(n)$ for sufficiently large $n$
%where $0 < C < 1$ is some constant. For an elementary proof, see~\cite{foo} {\color{red} (I can add a reference to that elementary proof, basically it's based on the Erd\H{o}s argument)}}.

%To prove Theorem \ref{lattice}, we will also need the following theorem by Ge \cite{Ge1,Ge2}:
%
%\begin{theorem}[\hspace{0.03cm}\cite{Ge1,Ge2}\hspace{0.03cm}]\label{ge}
%Given a finite set of non-zero algebraic numbers $\mathscr{A}=\{a_1,\ldots,a_n\}$
%  \emph{(under the model of computation described in Section \ref{complexmodel})},
%  %an algebraic number field ${\bf F}$ and non-zero elements
%  %$a_1,\ldots,a_n\in {\bf F}$,\\
%  one can in polynomial time find a lattice basis for the lattice $L$ given by
%\[L = \Big\{ \xx=\big(x_1,\ldots,x_n\big) \in
%{\mathbb Z}^n \hspace{0.1cm} \Big| \hspace{0.1cm} a_1^{x_1}\cdots a_n^{x_n}=1\Big\}.\]
%\end{theorem} 

The difference between~\cite{Cai-Chen-Lu-2013} and our paper
is that in~\cite{Cai-Chen-Lu-2013} a purification was applied immediately
to the connected $\AA$ (in both bipartite and nonbipartite cases)
so from that point $\AA$ was assumed to be already purified,
whereas in our paper we compute its purification $\pur \AA$
and then both $\AA$, $\pur \AA$ are passed on to the next step.
This subtle difference is indicated in conditions $(\calS_1)$--$(\calS_3)$ (resp. $(\calS_1')$--$\calS_3'$) between the two papers.
(Of course, for this to happen we need $\AA$ to be 
%either
a regularized matrix,
% (resp., purified in~\cite{Cai-Chen-Lu-2013}) bipartite or
%regularized (resp., purified in~\cite{Cai-Chen-Lu-2013}) nonbipartite matrix
otherwise we terminate and the corresponding problem is \#P-hard.)
Next, when arriving at the ({\sl Shape}) (resp. ({\sl Shape}$'$)) conditions,
the difference is that in~\cite{Cai-Chen-Lu-2013}
the pair $(\CC, \fD)$ is obtained from the already purified $\AA$,
whereas in our paper $(\CC, \fD)$ results from the nonpurified $\AA$
and $(\pur \CC, \fD)$ from its purification $\pur \AA$.
Finally, the tractability criteria in both papers converge
when we get to conditions ($\calU_1$--$\calU_4$) (resp. ($\calU_1'$--$\calU_4'$))
for the inner pairs.
At this point, $(\CC, \fD)$ denotes the normalized inner
pair $(\XX, \fY)$ in both papers.
%Now we have two alternatives:
%%% I have no idea what "two alternatives" you are talking bout.
%\begin{enumerate}
%\item It is easy to check that the segments of the tractability criterion in our paper
%before the convergence point at the conditions ($\calU_1$--$\calU_4$) (resp. ($\calU_1'$--$\calU_4'$))
%can be done in polynomial time, and then we can use the fact that the tractability
%criterion as shown in Section 17 from~\cite{Cai-Chen-Lu-2013} and
%summarized in Theorem 1.2 from~\cite{Cai-Chen-Lu-2013} is decidable in polynomial time.
%{\color{red} (I mean the rest of the tractability criterion from that point)}
%
%\item 

We note that
%It can be noticed that 
the tractability criterion from~\cite{Cai-Chen-Lu-2013},
when applied to our matrix $\AA$,
is actually consistent with the tractability criterion from this paper.
% and yields the correct result for our dichotomies.
In~\cite{Cai-Chen-Lu-2013},
%Here, 
it will first convert $\AA$ to $\pur \AA$ and then produce $(\pur \CC, \fD)$ (if possible),
and after that it goes to the inner  pair, renamed  $(\CC, \fD)$,
% (formerly, $(\XX, \fY)$, 
if possible, and the rest is trivial.

%% file: appendix2.tex
\subsection{Making the dichotomy in~\cite{Cai-Chen-Lu-2013} constructive}\label{sec:constr-ccl} %\subsection*

%{\color{red} (a few steps to show how to 2nd pinning lemma can be dodged in CCL)}
%
%{\color{red} (can be put in the appendix, actually we can put the entire contsructivity part into the appendix)}

Now we remark how to make the dichotomy in~\cite{Cai-Chen-Lu-2013} (Theorem 1.1) constructive.
%(Page numbers below refer to~\cite{Cai-Chen-Lu-2013}.)
The nonconstructive steps for the proof are the following
(page numbers below refer to~\cite{Cai-Chen-Lu-2013}):
\begin{enumerate}
\item The first pinning lemma (Lemma 4.1) from Section 4.1 on p. 937. We need to apply it for complex symmetric $\AA$ (with algebraic entries).
Its usages are
\begin{enumerate}
\item in Section 4.1 on p. 940 to reduce to connected matrices;
\item in the proof of the third pinning lemma (Corollary 8.4).
\end{enumerate}

\item The second pinning lemma (Lemma 4.3) from Section 4.1 on p. 938 (and its direct Corollary~4.4 on p. 938).
Its usages are
\begin{enumerate}
\item in the proof of Lemma 8.24 from Section 8.5 (Step 2.4) on p. 978 (lines -17 -- -15):
it is used to show that the problems of computing $Z^\rightarrow_{\CC'', \fL}$ and $Z^\leftarrow_{\CC'', \fL}$ are reducible to $\EVAL(\CC'', \fL)$;
\item in Section 8.6 (Step 2.5 which is the normalization step for the bipartite case) on p.~980 (lines -17 -- -13):
it is used (implicitly)
%the actual usage is (not quite stated)) 
to show that the problems of computing $Z^\rightarrow_{\CC, \fD}$ and $Z^\leftarrow_{\CC, \fD}$
(resp. $Z^\rightarrow_{\XX, \fY}$ and $Z^\leftarrow_{\XX, \fY}$) are reducible to $\EVAL(\CC, \fD)$ (resp. $\EVAL(\XX, \fY)$);
\item in Section 12.1 (inside the tractability part for the bipartite case) in the proof of Lemma~12.2 on p. 1002 (lines -14 -- -13):
it is used to show that $Z^\rightarrow_{\CC^{[i]}, \fD^{[i]}}$ and $Z^\leftarrow_{\CC^{[i]}, \fD^{[i]}}$
are reducible to $\EVAL(\CC^{[i]}, \fD^{[i]})$ for $i \in \{1, 2\}$.
\end{enumerate}

We see that all applications of the second pinning lemma are limited to its weaker form, namely Corollary 4.4.
We also note that we do not need to use the second pinning lemma
\begin{enumerate}
\item in Section 13.3 (Step 2.4, inside the reduction part for the nonbipartite case)
to prove Lemma 13.6 ($\EVAL(\FF, \fD) \equiv \EVAL(\HH, \fL)$) from Section 13.3 (Step 2.4) on p. 1016;
\item in Section 13.4 (Step 2.5 which is the normalization step for the nonbipartite case) on p. 1016 (lines 7 -- 9).
\end{enumerate}

\item The third pinning lemma (Corollary 8.3) from Section 8.1 on p. 954.
Its usages are
\begin{enumerate}
\item in the proof of Lemma 8.24 from Section 8.5 (Step 2.4) on p. 978 (lines -14 -- -11):
it is used to show that the problems of computing $Z^\rightarrow_{\CC, \fD}$ and $Z^\leftarrow_{\CC, \fD}$ are reducible to $\EVAL(\CC, \fD)$;
\item we need it to prove Lemma 13.6 ($\EVAL(\FF, \fD) \equiv \EVAL(\HH, \fL)$) from Section 13.3 (Step 2.4) on p. 1016.
\end{enumerate}

\item The pinning steps resulting from the applications of the Bulatov-Grohe dichotomy (Theorem 2.4 in~\cite{Cai-Chen-Lu-2013})
which, as discussed earlier in Section~\ref{sec:constr-main-dich-bd} all boil down to the first pinning lemma but for nonnegative symmetric $\AA$.
\end{enumerate}

The proof of the first pinning lemma can be made constructive
by the results from~\cite{Cai-Govorov-GH-hom-type} for complex symmetric $\AA$.
%(actually for complex (not necessarily symmetric) $\AA$).
As a consequence, this makes the proof of the third pinning lemma constructive
since it only invokes a single application of the first pinning lemma
and this invocation is its only nonconstructive step.
In addition, this makes the applications of the Bulatov-Grohe dichotomy constructive.

We do not know how to make the proof of the second pinning lemma constructive
but we will show how to avoid it entirely
while preserving the validity or the dichotomy from~\cite{Cai-Chen-Lu-2013} (Theorem 1.1)
by making slight modifications to its proof.

\begin{enumerate}
\item
%%%%%%%%%%%%%%%%%%%%%%%%%%%%%%%%%%%%%%
\begin{comment}
 In Section 8.5 on p. 978, we reformulate Lemma 8.24 to state
$\EVAL(\CC, \fD) \equiv \EVAL^\leftrightarrow(\CC'', \fL)$.
By the arguments in the proof of Lemma 8.24,
$\EVAL(\CC, \fD) \le \EVAL^\leftrightarrow(\CC'', \fL)$ is obvious.
%(Here we do not need to use Claim 8.25.)
As for the other direction, $\EVAL^\leftrightarrow(\CC, \fD) \le \EVALP(\CC, \fD)$ is clear,
and then by the third pinning lemma whose proof has been made constructive,
the pinned version $\EVALP(\CC, \fD) \equiv \EVAL(\CC, \fD)$,
resulting in $\EVAL^\leftrightarrow(\CC, \fD) \le \EVAL(\CC, \fD)$.
%The reasoning preceeding Claim 8.25 allows us to conclude
%$\EVAL^\leftrightarrow(\CC, \fD) \le \EVAL^\leftrightarrow(\CC'', \fL)$;
Combining it with Claim 8.25 (and similar reasoning for the other direction for $Z^\leftarrow$)
yields $\EVAL^\leftrightarrow(\CC'', \fL) \le \EVAL^\leftrightarrow(\CC, \fD)$.
Therefore $\EVAL^\leftrightarrow(\CC'', \fL) \le \EVAL^\leftrightarrow(\CC, \fD)$.
We conclude that $\EVAL(\CC, \fD) \equiv \EVAL^\leftrightarrow(\CC'', \fL)$,
which implies our reformulation of Lemma 8.24.
This way we do not use the second pinning lemma here.
\end{comment}
%%%%%%%%%%%%%%%%%%%%%%%%%%%%%%%%%%%%%
In Section 8.5 on p. 978, we reformulate Lemma 8.24 to state
$\EVAL(\CC, \fD) \equiv \EVAL^\leftrightarrow(\CC'', \fL)$.
To prove this, we reason as follows:
For one direction, by the argument in the  proof of Lemma~8.24,
$\EVAL(\CC, \fD) \le \EVAL^\leftrightarrow(\CC'', \fL)$.
% is obvious.
%(Here we do not need to use Claim 8.25.)
For the other direction,
the main consequence of Claim 8.25 is that 
$\EVAL^\leftrightarrow(\CC'', \fL) \le \EVAL^\leftrightarrow(\CC, \fD)$.
Also it is quite clear that 
$\EVAL^\leftrightarrow(\CC, \fD) \le \EVALP(\CC, \fD)$,
the pinned version of $\EVAL(\CC, \fD)$.
Then, by the third pinning lemma whose proof has been made constructive,
we have  $\EVALP(\CC, \fD) \equiv \EVAL(\CC, \fD)$.
We conclude that $\EVAL(\CC, \fD) \equiv \EVAL^\leftrightarrow(\CC'', \fL)$,
which implies our reformulation of Lemma 8.24.
This way we do not use the second pinning lemma here.

\item Continuing the previous step, in Section 8.6 we now have an equivalence to $\EVAL^\leftrightarrow(\CC'', \fL)$.
Then the normalization step described on p. 980 results in a pair $(\XX, \fY)$
for which we now have $\EVAL^\leftrightarrow(\CC'', \fL) \equiv \EVAL^\leftrightarrow(\XX, \fY)$.
This way we do not use the second pinning lemma here either.
Note that we still have access to $\EVAL(\XX, \fY)$
as $\EVAL(\XX, \fY) \le \EVAL^\leftrightarrow(\XX, \fY)$.
Thus we can still use $\EVAL(\XX, \fY)$ for the \#P-hardness arguments as was done in~\cite{Cai-Chen-Lu-2013}
but for the tractability part we will have to work with $\EVAL^\leftrightarrow(\XX, \fY)$ instead of $\EVAL(\XX, \fY)$.
%for the tractability ones.
The rest of the \#P-hardness arguments for the bipartite case are constructive and go through,
so we only need to deal with the tractability part for this case.

\item This reasoning allows us to have $\EVAL(\AA) \le \EVAL^\leftrightarrow(\CC, \fD)$
constructively. (Note that here $(\CC, \fD)$ is renamed for the pair $(\XX, \fY)$
at the beginning of Section 9 on p.~980 and is different from the previous one.)
However, regarding tractability for the bipartite case we need to show that $\EVAL^\leftrightarrow(\CC, \fD)$ is tractable, instead
of just $\EVAL(\CC, \fD)$ specified in Theorem 5.3 on p. 941, without involving nonconstructive steps.
We show how to do it by making adjustment to the proof of this theorem given in Section 12.
%In the tractability part for the bipartite case
In Section 12.1, we reformulate Lemma 12.2,
replacing $\EVAL(\CC^{[i]}, \fD^{[i]})$ by $\EVAL^\leftrightarrow(\CC^{[i]}, \fD^{[i]})$ for $i \in \{0, 1, 2\}$, respectively.
With these adjustments the proof 
%becomes trivial
goes through  without involving the second pinning lemma.
%and does not involve the second pinning lemma.
Finally, we note that in Section 12.2,
it was actually proved that $\EVAL^\rightarrow(\CC, \fD) \le \EVAL(\widehat \pi)$
(i.e., $Z^\rightarrow_{\CC, \fD}(G, u^*)$ is reducible to
the problem $\EVAL(\widehat \pi)$ (see p.~980 line~-4).
As mentioned in~\cite{Cai-Chen-Lu-2013}, 
 $\EVAL^\leftarrow(\CC, \fD) \le \EVAL(\widehat \pi)$ can be shown similarly.
%The latter was shown to be tractable in Section 12.3 (pp. 1008 -- 1013).
Combining these we get $\EVAL^\leftrightarrow(\CC, \fD) \le \EVAL(\widehat \pi)$ constructively.
The tractability of $\EVAL(\widehat \pi)$ is evidenced by Theorem 12.1 on p. 1001
which is shown in Section 12.3 on p. 1008 and is constructive
as noted in Section~\ref{sec:constr-main-dich-bd}.

\item As noted earlier for the nonbipartite case (both \#P-hardness and tractability parts), we do not need to use the second pinning lemma.
But we still use the third pinning lemma which admits a constructive proof now.
The \#P-hardness for the nonbipartite case with this adjustment becomes constructive.
The tractability part also becomes constructive,
in particular, the proof in Section 16 needs no change.
\end{enumerate}
This shows how to make the dichotomy from~\cite{Cai-Chen-Lu-2013} (Theorem 1.1) constructive.
\begin{theorem}[Constructive version of Theorem 1.1~\cite{Cai-Chen-Lu-2013}]\label{thm:constructivity-eval(B)}
There is an algorithm such that on input a symmetric and algebraic complex matrix $\AA$,
% either
\begin{enumerate}
\item \emph{if} $\EVAL(\AA)$ is tractable by the tractability criterion in~\cite{Cai-Chen-Lu-2013},
\emph{then} outputs an algorithm that computes $G \mapsto Z_\AA(G)$ in polynomial time in the size of the input graph $G$;
%if $\EVAL(\AA)$ is tractable by the tractability criterion in this paper; \\ or
\item \emph{else}, outputs a polynomial-time reduction from a canonical \#P-hard problem e.g., \textsc{\#SAT},
%otherwise, 
to $\EVAL(\AA)$.
\end{enumerate}
\end{theorem}

%%%%%%%%%%%%%%%%%%%%%%%%%%%%%%%%%%%%%%%%%%%%%%%%%%%%%%%%%%%%%%%%%%%%
\begin{comment}

\begin{theorem}\label{thm:construcitity-ccl}
There is an algorithm that on input a symmetric and algebraic complex matrix $\AA$ either
\begin{enumerate}
\item outputs a polynomial-time algorithm that computes $Z_\AA(\cdot)$ in polynomial time \emph(in the size of the input graph $G$\emph)
if $\EVAL(\AA)$ is tractable by the tractability criterion obtained in~\cite{Cai-Chen-Lu-2013} {\color{red} (actually same as ours)}; or
\item outputs a polynomial-time reduction from a canonical \#P-hard problem \emph(e.g., \textsc{\#SAT}, \textsc{\#Permanent}, etc.\emph)
to $\EVAL(\AA)$.
\end{enumerate}
\end{theorem}

\end{comment}
%%%%%%%%%%%%%%%%%%%%%%%%%%%%%%%%%%%%%%%%%%%%%%%%%%%%%%%%%%%%%%%%%%%%

%{\color{red} (the remark shows that the CCL proof doesn't extend to simple graphs,
%because it breaks the gadget structure, also the time bounds don't work anymore, can remove it)}
\begin{remark}
However, the proof in~\cite{Cai-Chen-Lu-2013} does not extend to simple graphs.
We cannot apply the results from~\cite{Cai-Govorov-GH-hom-type}
%in the same way as we did 
to get simple edge gadgets in Section~\ref{sec:constr-main-dich-bd-simp}
to avoid an open-ended search in the limiting arguments
%(such as those in Section 8.4.2 on pp. 966 -- 977 or in Section 9.2 on pp. 983 -- 988).
(such as those in Section 8.4.2 on pp. 966 -- 977 (as part of the proof of Lemma 8.12)
and in Section 9.2 on pp. 983 -- 988 (as part of the proof of Theorem 5.4)).
The way we constructed the proof in this paper,
we always end up with a gadget $\Gamma$ in the original framework $\EVAL(\AA)$
for which $\gm_{\Gamma, \AA}$ is not \multblrkone/,
%mutliplicatively-block-rank-$1$, 
otherwise $\EVAL(\AA)$ is tractable.
Then we use Theorem~\ref{thm:gadget-simple-non-mult-bl-rk-1} to get a simple egde gadget $\Gamma'$
such that $\gm_{\Gamma', \AA}$ is not \multblrkone/,
%mutliplicatively-block-rank-$1$
and as mentioned before $\Gamma'$ can be constructed in bounded time,
and we do not need to construct $\Gamma$.
But the reductions in~\cite{Cai-Chen-Lu-2013} do not translate
to easily preserve this edge gadget property.
For this reason, the modifications in this section only make the dichotomy (Theorem 1.1) in~\cite{Cai-Chen-Lu-2013} construcitve
but the actual algorithm from~\cite{Cai-Govorov-GH-hom-type} 
%Theorem~\ref{thm:construcitity-ccl}
 taking $\AA$ as input still involves open-ended searches for the \#P-hardness part. 
%without providing any actual time bounds.
\end{remark}

%% file: main.bbl
\begin{thebibliography}{10}

\bibitem{Barvinok-book}
A.~I. Barvinok.
\newblock {\em Combinatorics and Complexity of Partition Functions}, volume~30
  of {\em Algorithms and combinatorics}.
\newblock Springer, 2017.

\bibitem{Barvinok-Soberon-2017}
A.~I. Barvinok and P.~Sober\'on.
\newblock Computing the partition function for graph homomorphisms.
\newblock {\em Combinatorica}, 37(4):633--650, 2017.

\bibitem{Baxter-book}
R.~J. Baxter.
\newblock {\em Exactly Solved Models in Statistical Mechanics}.
\newblock Academic Press, London, 1982.

\bibitem{Bulatov-Grohe-2005}
A.~Bulatov and M.~Grohe.
\newblock The complexity of partition functions.
\newblock {\em Theor. Comput. Sci.}, 348(2-3):148--186, 2005.
\newblock A preliminary version appeared in ICALP 2004: 294--306.

\bibitem{cai-chen-book}
J.-Y. Cai and X.~Chen.
\newblock {\em Complexity Dichotomies for Counting Problems}, volume 1: Boolean
  Domain.
\newblock Cambridge University Press, 2017.
\newblock \href {https://doi.org/10.1017/9781107477063}
  {\path{doi:10.1017/9781107477063}}.

\bibitem{Cai-Chen-Lu-2013}
J.-Y. Cai, X.~Chen, and P.~Lu.
\newblock Graph homomorphisms with complex values: A dichotomy theorem.
\newblock {\em SIAM J. Comput.}, 42(3):924--1029, 2013.

\bibitem{Cai-Govorov-GH-hom-type}
J.-Y. Cai and A.~Govorov.
\newblock On a theorem of {L}ov\'asz that $\hom(\cdot, {H})$ determines the
  isomorphism type of ${H}$.
\newblock In {\em Proceedings of the 11th Innovations in Theoretical Computer
  Science (ITCS)}, pages 17:1--17:15, 2020.
\newblock Full version available at \url{https://arxiv.org/abs/1909.03693}.

\bibitem{Carlitz}
L.~Carlitz.
\newblock Kloosterman sums and finite field extensions.
\newblock {\em Acta Arithmetica}, 16(2):179--193, 1969.

\bibitem{dusart}
P.~Dusart.
\newblock The $k^{\text{th}}$ prime is greater than $k(\log k + \log \log k -
  1)$ for $k \ge 2$.
\newblock {\em Mathematics of Computation}, 68(225):411--415, 1999.

\bibitem{DyerFJ02}
M.~E. Dyer, A.~M. Frieze, and M.~Jerrum.
\newblock On counting independent sets in sparse graphs.
\newblock {\em {SIAM} J. Comput.}, 31(5):1527--1541, 2002.

\bibitem{Dyer-Greenhill-2000}
M.~E. Dyer and C.~S. Greenhill.
\newblock The complexity of counting graph homomorphisms.
\newblock {\em Random Struct. Algorithms}, 17(3-4):260--289, 2000.
\newblock A preliminary version appeared in SODA 2000: 246--255.

\bibitem{Dyer-Greenhill-corrig-2004}
M.~E. Dyer and C.~S. Greenhill.
\newblock Corrigendum: The complexity of counting graph homomorphisms.
\newblock {\em Random Struct. Algorithms}, 25(3):346--352, 2004.

\bibitem{freedman-l-s}
M.~Freedman, L.~Lov\'{a}sz, and A.~Schrijver.
\newblock Reflection positivity, rank connectivity, and homomor\-phism of
  graphs.
\newblock {\em Journal of the American Mathematical Society}, 20:37--51, 2007.

\bibitem{Freedman-Lovasz-Schrijver-2007}
M.~Freedman, L.~Lov\'asz, and A.~Schrijver.
\newblock Reflection positivity, rank connectivity, and homomorphism of graphs.
\newblock {\em J. Amer. Math. Soc.}, 20(1):37--51, 2007.

\bibitem{Ge1}
G.~Ge.
\newblock {\em Algorithms related to multiplicative representations of
  algebraic numbers}.
\newblock PhD thesis, University of California, Berkeley, 1993.

\bibitem{Ge2}
G.~Ge.
\newblock Testing equalities of multiplicative representations in polynomial
  time.
\newblock In {\em Proceedings of the 34th Annual IEEE Symposium on Foundations
  of Computer Science (FOCS)}, pages 422--426, 1993.

\bibitem{Goldberg-et-al-2010}
L.~A. Goldberg, M.~Grohe, M.~Jerrum, and M.~Thurley.
\newblock A complexity dichotomy for partition functions with mixed signs.
\newblock {\em SIAM J. Comput.}, 39(7):3336--3402, 2010.

\bibitem{Goldberg-Guo-2017}
L.A. Goldberg and H.~Guo.
\newblock The complexity of approximating complex-valued {I}sing and {T}utte
  partition functions.
\newblock {\em Computational Complexity}, 26(4):765--833, 2017.

\bibitem{Goldberg-Jerrum-2012}
L.A. Goldberg and M.~Jerrum.
\newblock Approximating the partition function of the ferromagnetic {P}otts
  model.
\newblock {\em J. ACM}, 59(5):25:1--25:31, 2012.

\bibitem{Govorov-Cai-Dyer}
A.~Govorov, J.-Y. Cai, and M.~Dyer.
\newblock A dichotomy for bounded degree graph homomorphisms with nonnegative
  weights.
\newblock {\em arXiv:2002.02021}, 2020.
\newblock URL: \url{https://arxiv.org/abs/2002.02021}.

\bibitem{Grohe-Thurley-2011}
M.~Grohe and M.~Thurley.
\newblock Counting homomorphisms and partition functions.
\newblock In M.~Grohe and J.~Makowsky, editors, {\em Model Theoretic Methods in
  Finite Combinatorics}, volume 558 of {\em Contemporary Mathematics}, pages
  243--292. American Mathematical Society, 2011.

\bibitem{HellBook}
P.~Hell and J.~Ne\v{s}et\v{r}il.
\newblock {\em Graphs and Homomorphisms}.
\newblock Oxford University Press, 2004.

\bibitem{HelmuthPR19}
T.~Helmuth, W.~Perkins, and G.~Regts.
\newblock Algorithmic {P}irogov-{S}inai theory.
\newblock In {\em Proceedings of the 51st Annual ACM SIGACT Symposium on Theory
  of Computing (STOC)}, pages 1009--1020, 2019.

\bibitem{Ising-1925}
E.~Ising.
\newblock Beitrag zur {T}heorie des {F}rromagnetismus.
\newblock {\em Z. Phys.}, 31(1):253--258, 1925.

\bibitem{Jerrum-Sinclair-1993}
M.~Jerrum and A.~Sinclair.
\newblock Polynomial-time approximation algorithms for the {I}sing model.
\newblock {\em SIAM J. Comput.}, 22(5):1087--1116, 1993.

\bibitem{Lenstra-1992}
H.~W.~Lenstra Jr.
\newblock Algorithms in algebraic number theory.
\newblock {\em Bull. Amer. Math. Soc.}, 26(2):211--244, 1992.

\bibitem{susan-landau}
S.~Landau.
\newblock Factoring polynomials over algebraic number fields.
\newblock {\em SIAM Journal on Computing}, 14(1):184--195, 1985.

\bibitem{a-k-Lenstra}
A.~K. Lenstra.
\newblock Factoring polynomials over algebraic number fields.
\newblock In J.~A. van Hulzen, editor, {\em Computer Algebra}, volume 162 of
  {\em Lecture Notes in Computer Science}, pages 245--254. Springer Berlin
  Heidelberg, 1983.

\bibitem{LENSTRA-LENSTRA-LOVASZ}
A.~K. Lenstra, H.~W.~Lenstra Jr., and L\'aszl\'o Lov\'asz.
\newblock Factoring polynomials with rational coefficients.
\newblock {\em Mathematische Annalen}, 261(4):515--534, 1982.

\bibitem{LiLY13}
L.~Li, P.~Lu, and Y.~Yin.
\newblock Correlation decay up to uniqueness in spin systems.
\newblock In {\em Proceedings of the 24th Annual ACM-SIAM Symposium on Discrete
  Algorithms (SODA)}, pages 67--84, 2013.

\bibitem{field}
R.~Lidl and H.~Niederreiter.
\newblock {\em Finite fields}.
\newblock volume 20 of Encyclopedia of Mathematics and its Applications.
  Cambridge University Press, Cambridge, 1997.

\bibitem{lovasz67}
L.~Lov\'{a}sz.
\newblock Operations with structures.
\newblock {\em Acta Mathematica Hungarica}, 18:321--328, 1967.

\bibitem{Lovasz-2006}
L.~Lov\'asz.
\newblock The rank of connection matrices and the dimension of graph algebras.
\newblock {\em Eur. J. Comb.}, 27(6):962--970, 2006.

\bibitem{Lovasz-loops-correction}
L.~Lov\'asz.
\newblock On the dimension of graph algebras for homomorphism functions, 2013.
\newblock \url{https://web.cs.elte.hu/~lovasz/book/homnotes-6-4-1.pdf}.

\bibitem{Peters-Regts-2018}
H.~Peters and G.~Regts.
\newblock Location of zeros for the partition function of the {I}sing model on
  bounded degree graphs.
\newblock {\em arXiv:1810.01699}, 2018.
\newblock URL: \url{https://arxiv.org/abs/1810.01699}.

\bibitem{Potts-1952}
R.~B. Potts.
\newblock Some generalized order-disorder transformations.
\newblock {\em Mathematical Proceedings of the Cambridge Philosophical
  Society}, 48(1):106--109, 1952.

\bibitem{Potts-Ward-1955}
R.~B. Potts and J.~C. Ward.
\newblock The combinatrial method and the two-dimensional {I}sing model.
\newblock {\em Progress of Theoretical Physics}, 13(1):38--46, 1955.

\bibitem{rosser}
B.~Rosser.
\newblock Explicit bounds for some functions of prime numbers.
\newblock {\em American Journal of Mathematics}, 63(1):211--232, 1941.

\bibitem{Schrijver-2009}
A.~Schrijver.
\newblock Graph invariants in the spin model.
\newblock {\em J. Comb. Theory Ser. B}, 99(2):502--511, 2009.

\bibitem{SinclairST12}
A.~Sinclair, P.~Srivastava, and M.~Thurley.
\newblock Approximation algorithms for two-state anti-ferromagnetic spin
  systems on bounded degree graphs.
\newblock In {\em Proceedings of the 23rd Annual ACM-SIAM Symposium on Discrete
  Algorithms (SODA)}, pages 941--953, 2012.

\bibitem{Sly10}
A.~Sly.
\newblock Computational transition at the uniqueness threshold.
\newblock In {\em Proceedings of the 51st Annual IEEE Symposium on Foundations
  of Computer Science (FOCS)}, pages 287--296, 2010.

\bibitem{Thurley-2009}
M.~Thurley.
\newblock {\em The Complexity of Partition Functions}.
\newblock PhD thesis, Humboldt Universit\"at zu Berlin, 2009.

\bibitem{Weitz06}
D.~Weitz.
\newblock Counting independent sets up to the tree threshold.
\newblock In {\em Proceedings of the 38th Annual ACM SIGACT Symposium on Theory
  of Computing (STOC)}, pages 140--149, 2006.

\end{thebibliography}
